\documentclass[final,3p,times,nopreprintline]{elsarticle}
\usepackage{amsmath,amssymb,amsfonts,dsfont}
\usepackage{graphicx}
\usepackage{slashed}
\usepackage{booktabs,tabulary}
\usepackage{placeins}

\newcommand{\Fpi}{F_\pi}
\newcommand{\mpi}{M_{\pi}}
\newcommand{\mpii}{M_{\pi^0}}
\newcommand{\ga}{g_A}
\newcommand{\Order}{\mathcal{O}}
\newcommand{\muu}{m_u}
\newcommand{\md}{m_d}
\newcommand{\mpp}{m_p}
\newcommand{\mn}{m_n}
\newcommand{\mN}{m_N}
\newcommand{\mK}{M_K}
\newcommand{\xip}{\xi_p}
\newcommand{\xid}{\xi_d}
\newcommand{\tpi}{t_\pi}
\newcommand{\tK}{t_K}
\newcommand{\tN}{t_N}
\newcommand{\diff}{\text{d}}
\newcommand{\eps}{\epsilon}
\newcommand{\sm}{s_\text{m}}
\newcommand{\tm}{t_\text{m}}
\newcommand{\Wm}{W_\text{m}}
\newcommand{\Wa}{W_\text{a}}
\newcommand{\MeV}{\,\text{MeV}}
\newcommand{\GeV}{\,\text{GeV}}
\newcommand{\beq}{\begin{equation}}
\newcommand{\eeq}{\end{equation}}
\newcommand{\Amp}{\mathcal{A}}
\newcommand{\qq}{\mathbf{q}}
\newcommand{\pp}{\mathbf{p}}

\newcommand{\unity}{\mathds{1}}
\newcommand{\ff}{\mathbf{f}}
\newcommand{\athr}{a_{l\pm}^I}
\newcommand{\bthr}{b_{l\pm}^I}
\newcommand{\cthr}{c_{l\pm}^I}
\newcommand{\dthr}{d_{l\pm}^I}
\providecommand{\tdelta}{\tilde{\delta}}
\providecommand{\tgamma}{\tilde{\gamma}}
\renewcommand{\Re}{\text{Re}\,}
\renewcommand{\Im}{\text{Im}\,}

\DeclareMathOperator{\Res}{\text{Res}}

\newcommand{\toright}[1]{\hspace*{\fill}{\footnotesize{#1}}}

\def\Xint#1{\mathchoice
   {\XXint\displaystyle\textstyle{#1}}%
   {\XXint\textstyle\scriptstyle{#1}}%
   {\XXint\scriptstyle\scriptscriptstyle{#1}}%
   {\XXint\scriptscriptstyle\scriptscriptstyle{#1}}%
   \!\int}
\def\XXint#1#2#3{{\setbox0=\hbox{$#1{#2#3}{\int}$}
     \vcenter{\hbox{$#2#3$}}\kern-0.5\wd0}}
\providecommand{\dashint}[1][0pt]{\Xint{\hspace{#1}-}}
 
\providecommand{\ste}[1]{\left[#1\right]_{(0,0)}} 
\providecommand{\zet}[1]{\left[#1\right]_{t=0}}
\providecommand{\zeq}[1]{\left[#1\right]_{\qq^2=0}}
\providecommand{\zetq}[1]{\left[#1\right]_{t=0,\qq^2=0}}
\providecommand{\te}[1]{\left[#1\right]_{(\mpi,0)}}
\providecommand{\swt}[1]{\left[#1\right]_{s=s(\omega),t=0}}

\allowdisplaybreaks[1]

\begin{document}

\renewcommand{\theequation}{\arabic{equation}}

\numberwithin{equation}{section}
 
\setcounter{tocdepth}{2}

\begin{frontmatter}

\title{\toright{\textnormal{INT-PUB-15-050}}\\Roy--Steiner-equation analysis of pion--nucleon scattering}

\author[Seattle]{Martin Hoferichter}
\author[Bonn]{Jacobo Ruiz de Elvira}
\author[Bonn]{Bastian Kubis}
\author[Bonn,Juelich]{Ulf-G.\ Mei{\ss}ner}

\address[Seattle]{Institute for Nuclear Theory, University of Washington, Seattle, WA 98195-1550, USA}
\address[Bonn]{Helmholtz-Institut f\"ur Strahlen- und Kernphysik (Theorie) and
   Bethe Center for Theoretical Physics, Universit\"at Bonn, D--53115 Bonn, Germany}
\address[Juelich]{Institut f\"ur Kernphysik, Institute for Advanced Simulation, 
   J\"ulich Center for Hadron Physics, JARA-HPC, and JARA-FAME,\\  Forschungszentrum J\"ulich, D--52425  J\"ulich, Germany}

\begin{abstract}
 We review the structure of Roy--Steiner equations for pion--nucleon scattering, the solution for the partial waves of the $t$-channel process $\pi\pi\to \bar N N$, as well as the high-accuracy extraction of the pion--nucleon $S$-wave scattering lengths from data on pionic hydrogen and deuterium. We then proceed to construct solutions for the lowest partial waves of the $s$-channel process $\pi N\to \pi N$ and demonstrate that accurate solutions can be found if the scattering lengths are imposed as constraints. 
 Detailed error estimates of all input quantities in the solution procedure are performed and
 explicit parameterizations for the resulting low-energy phase shifts as well as results for subthreshold parameters and higher threshold parameters are presented.
Furthermore, we discuss the extraction of the pion--nucleon $\sigma$-term via the Cheng--Dashen low-energy theorem, including the role of isospin-breaking corrections, to obtain a precision determination consistent with all constraints from analyticity, unitarity, crossing symmetry, and pionic-atom data.
We perform the matching to chiral perturbation theory in the subthreshold region and 
detail the consequences for the chiral convergence of the threshold parameters and the nucleon mass.
\end{abstract}


\end{frontmatter}

\tableofcontents

\section{Introduction}

Pion--nucleon ($\pi N$) scattering is one of the fundamental processes of low-energy QCD. Its low-energy parameters, most notably the scattering lengths, encode crucial information about the spontaneous and explicit breaking of chiral symmetry as realized in the nucleon sector~\cite{Weinberg:1966kf,Tomozawa:1966jm}. Indeed, the pattern in different isospin channels could hardly be more distinct.
While in the isovector channel there is a low-energy theorem (LET) that determines the scattering length, $a^-$, solely in terms of masses and the pion decay constant, its isoscalar counterpart, $a^+$, is poorly constrained by chiral symmetry and vanishes at leading order. This expansion around the chiral limit of QCD in terms of momenta and quark masses can be performed systematically in the framework of Chiral Perturbation Theory (ChPT)~\cite{Weinberg:1978kz,Gasser:1983yg,Gasser:1984gg}, once suitably extended towards the single-baryon sector~\cite{Gasser:1987rb,Jenkins:1990jv,Bernard:1992qa,Ellis:1997kc,Becher:1999he,Goity:2001ny,Lehmann:2001xm,Schindler:2003xv,Schindler:2003je,Bernard:2007zu}, wherein $\pi N$ scattering constitutes one of the most important applications~\cite{Mojzis:1997tu,Fettes:1998ud,Buettiker:1999ap,Fettes:2000gb,Fettes:2000xg,Becher:2001hv,Gasparyan:2010xz,Alarcon:2012kn,Chen:2012nx,Siemens:2016hdi}.
In the case of the scattering lengths the chiral corrections~\cite{Bernard:1993fp,Bernard:1995pa,Fettes:2000xg} reveal further striking differences. While for $a^-$ next-to-leading-order (NLO) and next-to-next-to-next-to-leading-order (N$^3$LO) contributions vanish, the chiral expansion of $a^+$ starts already with a combination of low-energy constants (LECs) not predicted by chiral symmetry and is afflicted by substantial cancellations at subleading orders.  

As argued in~\cite{Weinberg:1977hb}, the vanishing of the leading-order isoscalar amplitude makes $\pi N$ scattering an ideal testing ground for isospin violation~\cite{Meissner:1997ii,Fettes:1998wf,Muller:1999ww,Fettes:2000vm,Fettes:2001cr,Gasser:2002am,Hoferichter:2009ez,Hoferichter:2009gn}, as any small correction by either source of isospin breaking, the mass difference between up- and down-quark and electromagnetic interactions, will be strongly magnified. In fact, since the physical value of $a^+$ stays close to zero, the difference of the scattering lengths in the $\pi^0p$ and $\pi^0n$ channels is of similar size as their central values~\cite{Baru:2011bw}.
In the last decade precision spectroscopy~\cite{Gotta:2008zza,Strauch:2010vu,Hennebach:2014lsa} in pionic hydrogen ($\pi H$) and deuterium ($\pi D$), electromagnetic bound states of a $\pi^-$ and a proton or deuteron core, have pushed the determination of the $\pi N$ scattering lengths to a level where isospin-breaking corrections, in addition to few-body corrections in $\pi D$, become absolutely critical in the extraction. A combined analysis of $\pi H$ and $\pi D$ now constrains them at percent-level accuracy~\cite{Baru:2010xn,Baru:2011bw}.  

Most of the recent interest in $\pi N$ scattering has been triggered by its connection to the $\pi N$ $\sigma$-term $\sigma_{\pi N}$, which measures the portion of the nucleon mass generated by the up- and down-quarks. More generally, knowledge of the $\sigma$-term completely determines the corresponding scalar matrix elements $\langle N|m_q\bar q q|N\rangle$ for $q=u,d$~\cite{Crivellin:2013ipa}, which makes it a crucial input quantity for the interpretation of dark-matter searches in direct-detection experiments~\cite{Bottino:1999ei,Bottino:2001dj,Ellis:2008hf,Belanger:2008sj,Hill:2011be,Cirigliano:2012pq,Belanger:2013oya,Beane:2013kca,Hill:2014yxa,Crivellin:2015bva,Hoferichter:2015ipa,Ellis:2015rya}, searches for lepton-flavor violation in $\mu\to e$ conversion~\cite{Cirigliano:2009bz,Crivellin:2014cta}, the determination of $CP$-violating $\pi N$ couplings in the context of electric dipole moments~\cite{Crewther:1979pi,Mereghetti:2010tp,Bsaisou:2012rg,Engel:2013lsa,deVries:2015una,deVries:2015gea}, as well as for nuclear-matter applications~\cite{Cohen:1991nk,Li:1994mq,Drukarev:2001wd,Meissner:2001gz,Kaiser:2007nv,Kaiser:2008qu,Lacour:2010ci,Fiorilla:2012bc,Kruger:2013iza,Kruger:2014caa,Gubler:2015yna}.
The relation between $\sigma_{\pi N}$ and the $\pi N$ scattering amplitude proceeds by means of the Cheng--Dashen LET~\cite{Cheng:1970mx,Brown:1971pn}, which requires an analytic continuation of the Born-term-subtracted isoscalar amplitude into the unphysical region.
This task is difficult to perform in ChPT alone.
Moreover, to determine the LECs that appear in the chiral representation experimental input is required, typically in the form of partial-wave analyses (PWAs) for the phase shifts. However, there is a dearth of low-energy data, which, in combination with inconsistencies in the $\pi N$ data base, has led to contradictory PWAs, 
the Karlsruhe--Helsinki~\cite{Koch:1980ay,Hoehler:1983,Koch:1985bp} and the GWU/SAID solutions~\cite{Arndt:2006bf,Arndt:2008zz,Workman:2012hx}.

One of the most powerful traits of effective field theories (EFTs) is that LECs once determined in one process can subsequently be used to predict others. 
For $\pi N$ scattering this implies applications that reach far into the domain of nuclear physics, where the same LECs that appear in the $\pi N$ amplitude govern the long-range part of the nucleon--nucleon ($NN$) potential and the three-nucleon force. This connection can be made precise in Nuclear Chiral Effective Field Theory (ChEFT)~\cite{Weinberg:1990rz,Weinberg:1991um,Weinberg:1992yk,vanKolck:1994yi}, the extension of ChPT to the few-nucleon sector (see~\cite{Epelbaum:2008ga,Machleidt:2011zz} for recent reviews). Although the LECs can be determined either by fits to $\pi N$ scattering itself~\cite{Krebs:2012yv,Wendt:2014lja} or by including few-nucleon observables~\cite{Rentmeester:2003mf,Perez:2013jpa,Carlsson:2015vda}, in the general spirit of EFTs one would expect the extraction from the simplest process to provide the best sensitivity and reliability, which makes a precision determination of the $\pi N$ amplitude in the kinematic range best suited for the matching to ChPT paramount.

Finally, also the partial waves for the crossed channel $\pi\pi\to\bar N N$ enter applications that extend beyond the $\pi N$ system. 
The response of the nucleon to external currents can be analyzed via a $t$-channel dispersion relation, and depending on the quantum numbers $\pi\pi$ intermediate states frequently provide the dominant contribution to the integral. In this way, the $S$-wave for $\pi\pi\to\bar N N$ is related to the scalar form factor of the nucleon~\cite{Gasser:1990ap,Hoferichter:2012wf,Hoferichter:2012tu}, the $P$-waves to electromagnetic form factors~\cite{Frazer:1960zzb,Hohler:1974eq,Hohler:1976ax,Mergell:1995bf,Belushkin:2005ds,Belushkin:2006qa,Hill:2010yb,Lorenz:2012tm,Granados:2013moa}, and the $D$-waves to generalized parton distribution functions~\cite{Pasquini:2014vua}.

Roy--Steiner (RS) equations are a set of coupled partial-wave dispersion relations (PWDRs), constructed in such a way that all constraints from analyticity, unitarity, and crossing symmetry are fulfilled. In contrast to $\pi\pi$ Roy equations~\cite{Roy:1971tc}, they are derived from hyperbolic dispersion relations (HDRs)~\cite{Hite:1973pm}, which automatically relate the different channels in the $\pi N$ system. 
In close analogy to similar analyses of the $\pi\pi$~\cite{Ananthanarayan:2000ht} and the $\pi K$ system~\cite{Buettiker:2003pp}, solving the RS equations for $\pi N$, in particular once combined with the pionic-atom constraints on the scattering lengths, can provide a remarkably precise representation of the $\pi N$ amplitude at low energies, with all the implications listed above. In this paper we review the structure of the RS equations as well as the solution of the $t$-channel equations~\cite{Ditsche:2012fv,Hoferichter:2012wf,Hoferichter:2012tu,Ditsche:2012ja} and provide the details of the $s$-channel solution (partial results were already discussed in~\cite{Elvira:2014wma,Elvira:2014lta}) as well as the determination of $\sigma_{\pi N}$~\cite{Hoferichter:2015dsa} and the matching to ChPT~\cite{Hoferichter:2015tha}.

The structure of the paper is as follows: we first provide a general introduction to dispersion theory and Roy equations in Sect.~\ref{sec:disp_rel}, before turning to the $\pi N$ case in Sect.~\ref{sec:RSsystem}. The solution of the $t$-channel partial waves is reviewed in Sect.~\ref{sec:tchannel_sol}. We then proceed to lay out the solution strategy for the $s$-channel equations in Sect.~\ref{sec:schannel_sol}, review the determination of the scattering lengths from pionic atoms that feature prominently therein in Sect.~\ref{sec:pionic_atom}, and present the results for the low-energy phase shifts as well as subtraction constants in Sect.~\ref{sec:results}. Consequences for threshold parameters, the $\sigma$-term, and the matching to ChPT are discussed in Sects.~\ref{sec:threshold}, \ref{sec:sigma_term}, and \ref{sec:ChPT}, respectively, before we close with a summary in Sect.~\ref{sec:summary}. Various details of the calculation are provided in the appendices.

\section{Dispersion relations, Roy equations, and chiral perturbation theory}
\label{sec:disp_rel}

Since ChPT is an effective field theory, the chiral expansion of a given quantity can, in principle, be systematically improved by including higher and higher orders in the calculation. 
However, as alluded to in the introduction, the chiral expansion may not converge equally well in all parts of the low-energy region, for instance due to low-lying resonances, and the analytic continuation towards the unphysical region or even into the complex plane may not be sufficiently stable when relying solely on the chiral representation.
Moreover, the number of LECs, which parameterize the effect of degrees of freedom not included in the theory, increases rapidly at subleading orders. 
In recent years, it has become apparent that the predictive power of chiral symmetry can be vastly increased by combining ChPT with dispersive techniques, which exploit analyticity to arrive at a representation that relates the amplitude at an arbitrary point in the complex plane to an integral over its imaginary part. While the latter can be constrained by the respective unitarity relation, convergence of the dispersive integral often requires a certain number of a priori undetermined subtraction constants that, in turn, can frequently be pinned down by matching to ChPT. 
Once the subtraction constants are fixed, a dispersive representation provides the ideal framework to reliably perform an analytic continuation into the complex plane, which becomes of fundamental importance for broad resonances situated far away from the real axis. 
In this section, we will illustrate these ideas for the case of $\pi\pi$ scattering.

\subsection{Fixed-$t$ dispersion relations} 
\label{sec:fixed_t}

\begin{figure}[t!]
\centering
\includegraphics[width=0.5\linewidth]{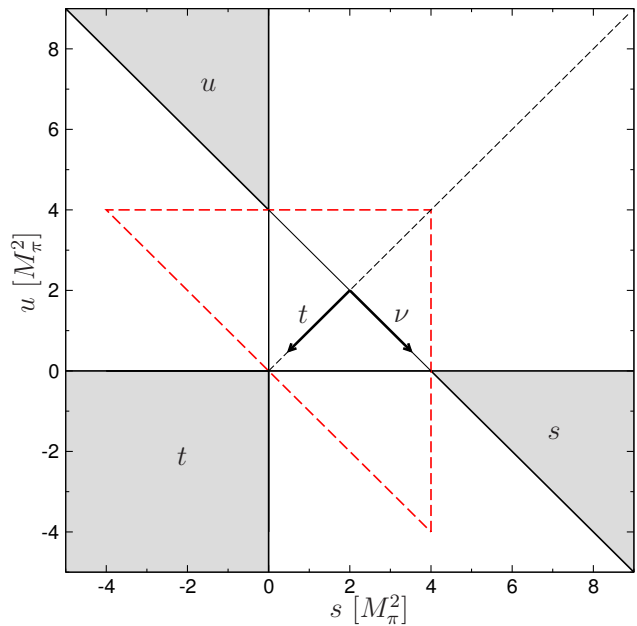}
\caption{Mandelstam plane for $\pi\pi$ scattering. The filled areas mark the $s$-, $t$-, and $u$-channel physical regions, the red dashed line the subthreshold triangle, and the arrows the orientation of the plane in $t$ and $\nu=s-u$.}
\label{fig:mandelstam_pipi}
\end{figure}

For simplicity, we consider the process
$\pi(p_1)+\pi(p_2)\to\pi(p_3)+\pi(p_4)$ (ignoring isospin labels for the time being) with Mandelstam variables
\beq
s=(p_1+p_2)^2,\qquad t=(p_1-p_3)^2,\qquad u=(p_1-p_4)^2.
\eeq
On the mass shell, they fulfill the relation
\beq
\label{onshell}
s+t+u=4\mpi^2,
\eeq
with the result that the scattering amplitude $T(s,t)$ reduces to a function of only two independent variables. The basic assumption in the construction of dispersion relations can be summarized as the principle of maximal analyticity: the amplitude $T(s,t)$ is represented by a complex function that exhibits no further singularities except for those required by general principles such as unitarity and crossing symmetry. The amplitude in the physical regions of the Mandelstam plane (cf.\ Fig.~\ref{fig:mandelstam_pipi}) is given as a particular limit of $T(s,t)$, e.g.\ for fixed $t=t_0$ the physical $s$-channel amplitude on the right-hand cut is defined as the limit from the upper half of the complex $s$-plane
\beq
\label{phys_lim}
T(s,t_0)=\lim_{\eps\to 0}T(s+i\eps,t_0).
\eeq  
These assumptions can be justified in the framework of perturbation theory, since the definition of the physical limit~\eqref{phys_lim} corresponds to the $i\eps$-prescription in Feynman propagators (see, e.g.,~\cite{Eden:1966}). We will comment below on the issue to what extent analyticity can even be vindicated from axiomatic field theory. 

\begin{figure}
\centering
\includegraphics[width=0.5\linewidth]{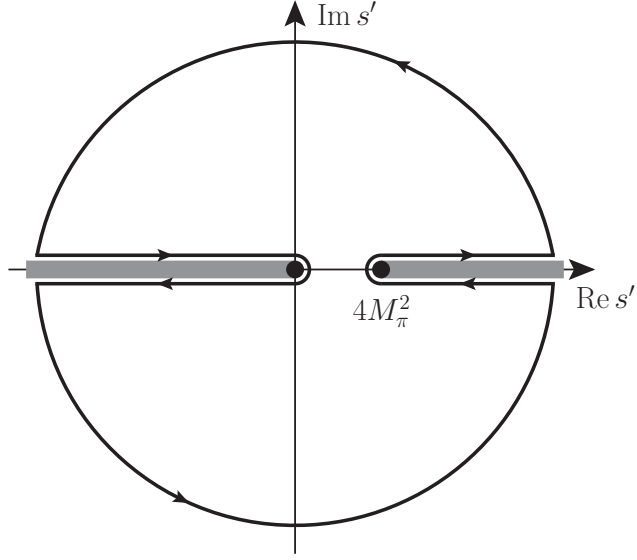}
\caption{Integration contour $\mathcal{C}$ in the complex $s'$-plane for fixed $t=0$. The gray bands denote left- and right-hand cuts, respectively, and the black dots the corresponding branch points.}
\label{fig:disp_rel}
\end{figure}

Once analyticity is established, the powerful machinery of complex analysis may be invoked, primarily by means of Cauchy's integral formula. The corresponding integral equation for the scattering amplitude, itself a function of the \emph{external kinematics} $(s,t,u)$, will involve integrals over the \emph{internal kinematics} $(s',t',u')$, which, a priori, can take arbitrary values in the Mandelstam plane. However, in order to write down a single-variable integral equation the allowed range of these internal variables needs to be restricted appropriately. The standard choice that the on-shell condition~\eqref{onshell} be valid for the internal kinematics as well is universal to all dispersion relations, while the
second condition, relating external and internal kinematics, distinguishes different kinds thereof, e.g.\ the fixed-$t$ version is characterized by 
$t'=t$ (but, in principle, any path through the Mandelstam plane would be adequate). In this case, Cauchy's theorem yields 
\beq
T(s,t)=\frac{1}{2\pi i}\oint_{\mathcal{C}}\diff s'\frac{T(s',t)}{s'-s},
\eeq
where the integration proceeds along the contour $\mathcal{C}$ as indicated in Fig.~\ref{fig:disp_rel}. If $T(s',t)$ vanishes for $|s'|\to\infty$, the contribution from the circle will vanish as well, as soon as its radius is taken to infinity. The remaining integration around the cuts can be expressed in terms of the discontinuity
\beq
\text{disc}\, T(s',t)=\lim_{\eps\to 0}\big[T(s'+i\eps,t)-T(s'-i\eps,t)\big],
\eeq
which, by virtue of hermitian analyticity~\cite{Olive:1962}, directly follows from unitarity. More precisely, hermitian analyticity---itself a fundamental consequence of the CPT theorem of quantum field theory---states that if the amplitude $T_{ab}$ for a process $a\to b$ is the boundary value of an analytic function from above, cf.~\eqref{phys_lim}, the amplitude $T_{ba}^*$ will be given by the limit of the same function from below. For time-invariant interactions this property permits the identification\footnote{We exclude the possibility that a particle in the initial or final state of the reaction is kinematically allowed to decay into the other particles involved. In such a case, the discontinuity is not purely imaginary and cannot be simply related to the imaginary part~\cite{Khuri:1960zz,Aitchison:1977ej,Kambor:1995yc,Anisovich:1996tx,Niecknig:2012sj,Schneider:2012ez,Danilkin:2014cra,Guo:2014vya,Guo:2015zqa}.} 
\beq
\text{disc}\, T(s',t)=2i\,\Im T(s',t),
\eeq
and thus leads to
\beq
\label{fixedt_unsub}
T(s,t)=\frac{1}{\pi}\int\limits_{4\mpi^2}^\infty\diff s'\bigg\{\frac{1}{s'-s}+\frac{1}{s'-u}\bigg\}\Im T(s',t).
\eeq

In practice, the asymptotic behavior of $T(s',t)$ for large $|s'|$ does not allow for an unsubtracted dispersion relation, since the contribution from the contour at infinity cannot be discarded. Provided that $T(s',t)$ does not grow faster than a polynomial, this obstacle may be overcome by introducing so-called subtractions, i.e.\ by considering dispersion relations not for $T(s',t)$, but for $T(s',t)/P_n(s')$ instead, where
\beq
P_n(s')=\prod\limits_{i=1}^{n}(s'-s_i)
\eeq  
 involves the subtraction points $\{s_i\}$, and $n$ is chosen sufficiently large to ensure convergence of the dispersive integral. In the application of Cauchy's
theorem, the poles introduced by dividing by $P_n(s')$ can be dealt with using the residue theorem. Eventually, $n$ additional powers of $s'$ appear in the denominator of~\eqref{fixedt_unsub}, but at the same time one also incurs a subtraction polynomial of degree $n-1$ with a priori unknown coefficients, which, in the case of fixed-$t$ dispersion relations, will actually depend on the value of $t$ chosen. Therefore, we will refer to these coefficients as subtraction functions in the following.

The maximal number of subtractions necessary for the dispersive integrals to converge and the contour at infinity to be irrelevant is restricted by
 the Froissart--Martin bound~\cite{Froissart:1961ux,Martin:1962rt}, which requires the total cross section not to increase faster than $\log^2s$ for $s\to\infty$,
 and, by means of the optical theorem,
 implies that at most two subtractions are needed.\footnote{The appearance of the logarithm may be understood intuitively from a classical example already given in~\cite{Froissart:1961ux}: suppose, the scattering of two particles were described by a Yukawa-type interaction with probability density function $P(r)=P_0e^{-b r}$ and typical range $b$. Suppose further, that the energy dependence of the interaction probability were limited by a polynomial in $s$, i.e.\ $P(s,r)<P_0(s/s_0)^Ne^{-b r}$. Then, the interaction would be exponentially suppressed for $r>r_0=N/b\log s/s_0$ and the cross section bounded by $\sigma<\pi r_0^2=\pi N^2/b^2\log^2 s/s_0$.}
However, one may perform further subtractions to reduce the sensitivity of the integrals to the high-energy regime, where the imaginary part is often poorly known, of course at the expense of introducing additional undetermined parameters.
Subtracting twice at $s=0$, \eqref{fixedt_unsub} becomes
\beq
\label{fixedt_2sub}
T(s,t)=c(t)+\frac{1}{\pi}\int\limits_{4\mpi^2}^\infty\diff s'\bigg[\frac{s^2}{s'^2(s'-s)}+\frac{u^2}{s'^2(s'-u)}\bigg]\Im T(s',t).
\eeq
Here, we have taken advantage of crossing symmetry to discard terms proportional to $s-u$ in the subtraction polynomial, while terms proportional to $s+u=4\mpi^2-t$ can be absorbed into $c(t)$. Indeed, the validity of a twice-subtracted dispersion relation for $|t|<4\mpi^2$ has been established from axiomatic field theory~\cite{Jin:1964zza}, which together with~\cite{Martin:1962rt} for $t<0$ rigorously vindicates~\eqref{fixedt_2sub} for all $t<4\mpi^2$.

Finally, one can try to go another step forward and drop the restriction of single-variable dispersion relations (keeping the internal kinematics on-shell). The corresponding assumption that $T(s,t)$ can be expressed in terms of double-spectral density functions $\rho_{su}$, $\rho_{tu}$, and $\rho_{st}$ by double dispersive integrals of the form
\beq
\label{mandelstam}
T(s,t)=\frac{1}{\pi^2}\iint\diff s'\diff u'\frac{\rho_{su}(s',u')}{(s'-s)(u'-u)}+ 
\frac{1}{\pi^2}\iint\diff t'\diff u'\frac{\rho_{tu}(t',u')}{(t'-t)(u'-u)}
+\frac{1}{\pi^2}\iint\diff s'\diff t'\frac{\rho_{st}(s',t')}{(s'-s)(t'-t)},
\eeq
where the integration ranges extend over those regions in the Mandelstam plane where the corresponding double-spectral functions have support, is referred to as Mandelstam analyticity~\cite{Mandelstam:1958xc}. In either case, this concept can be justified in perturbation theory~\cite{Mandelstam:1958xc,Mandelstam:1959bc,Mandelstam:1959}, but while for $\pi\pi$ scattering the validity of the Mandelstam representation can at least be derived rigorously in a finite domain~\cite{Martin:1965jj,Martin:1966}, for $\pi N$ scattering only the uniqueness of amplitudes satisfying Mandelstam analyticity has been proven~\cite{Macdowell:1973je,Cheung:1972tt}.

\subsection{Roy equations}

\begin{figure}
\centering
\includegraphics[width=0.3\linewidth]{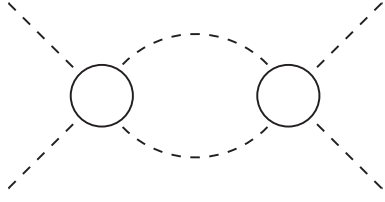}
\caption{Elastic unitarity for $\pi\pi$ scattering. Dashed lines denote pions and the spheres the $\pi\pi$ scattering amplitude.}
\label{fig:pipi_unitarity}
\end{figure}

Roy equations are a coupled system of PWDRs that respect analyticity, unitarity, and crossing symmetry of the scattering amplitude~\cite{Roy:1971tc}. The starting point in the construction of these equations is the twice-subtracted fixed-$t$ dispersion relation~\eqref{fixedt_2sub}. First, Roy realized that the subtraction function $c(t)$ may be determined by means of $s\leftrightarrow t$ crossing symmetry 
\begin{align}
\label{Roy_crossing}
 T(0,t)&=c(t)+\frac{1}{\pi}\int\limits_{4\mpi^2}^\infty\diff s'\frac{(4\mpi^2-t)^2}{s'^2(s'-4\mpi^2+t)}\Im T(s',t)\notag\\
=T(t,0)&=c(0)+\frac{1}{\pi}\int\limits_{4\mpi^2}^\infty\diff s'\bigg[\frac{t^2}{s'^2(s'-t)}+\frac{(4\mpi^2-t)^2}{s'^2(s'-4\mpi^2+t)}\bigg]\Im T(s',0).
\end{align}
Second, the remaining subtraction constant $c(0)$ is intimately related to the amplitude at threshold, and thus to the scattering length, via
\beq
T(4\mpi^2,0)=c(0)+\frac{1}{\pi}\int\limits_{4\mpi^2}^\infty\diff s'\frac{16\mpi^4}{s'^2(s'-4\mpi^2)}\Im T(s',0).
\eeq
Third, the imaginary part of the amplitude that appears inside the dispersive integrals is expanded in partial waves, and finally the partial-wave projection of the resulting equation is performed. Retrieving isospin indices again, one thus arrives at a system of integral equations for the $\pi\pi$ amplitudes $t_J^I(s)$ 
\beq
\label{Royeq}
t_J^I(s)=k_J^I(s)+\frac{1}{\pi}\int\limits_{4\mpi^2}^\infty \diff s' \sum\limits_{I'=0}^{2}\sum\limits_{J'=0}^\infty K_{JJ'}^{II'}(s,s')\Im t_{J'}^{I'}(s'),
\eeq
that relates a partial wave of given angular momentum $J$ and isospin $I$ to all other partial waves via analytically calculable kinematic kernel functions $K_{JJ'}^{II'}(s,s')$. These kernels are composed of a singular Cauchy kernel and a regular remainder according to
\beq
K_{JJ'}^{II'}(s,s')=\frac{\delta_{JJ'}\delta_{II'}}{s'-s-i\epsilon}+\bar K_{JJ'}^{II'}(s,s').
\eeq
In particular, the construction that led to~\eqref{Royeq} ensures that the $t_J^I(s)$ automatically fulfill the analytic properties expected for the partial waves: while the Cauchy kernel implements the right-hand cut, $\bar K_{JJ'}^{II'}(s,s')$ will incorporate all analytic structure required from the left-hand cut. The only free parameters of the approach are hidden in the subtraction term $k_J^I(s)$ that depends on the $S$-wave scattering lengths $a_0^0$ and $a_0^2$.
As long as elastic unitarity holds, i.e.\ only $\pi\pi$ intermediate states enter the unitarity relation (see Fig.~\ref{fig:pipi_unitarity}), the $\pi\pi$ partial waves may be parameterized as
\beq
\label{pipi_PW}
t_{J}^{I}(s)=\frac{e^{2i\delta^I_J(s)}-1}{2i\sigma_s^\pi},\qquad\sigma_s^\pi=\sqrt{1-\frac{4\mpi^2}{s}},
\eeq 
with the result that the Roy equations~\eqref{Royeq} reduce to coupled integral equations for the phase shifts $\delta^I_J(s)$.

An important issue is the range of validity of the Roy equations. While the convergence of the fixed-$t$ dispersion relations is guaranteed for all $t<4\mpi^2$~\cite{Froissart:1961ux,Martin:1962rt,Jin:1964zza}, the reduction to partial waves imposes further constraints on the domain of validity of the system. As a matter of fact, the partial-wave expansion of the imaginary part in the dispersive integral converges only for scattering angles $z'$ that lie within the large Lehmann ellipse~\cite{Lehmann:1958}. It has been derived from axiomatic field theory that this condition is met for all $s'\in[4\mpi^2,\infty)$ if $-28\mpi^2\leq t\leq 4\mpi^2$~\cite{Martin:1965jj,Martin:1966}. By virtue of Bose symmetry, the partial-wave projection of the equations can be restricted onto $0\leq z\leq 1$, which translates into a range in $t$ of
\beq
-\frac{s-4\mpi^2}{2}\leq t \leq 0.
\eeq 
Consequently, the Roy equations can be established from axiomatic field theory up to~\cite{Roy:1971tc}
\beq
\label{range_of_conv_pipi_ax}
s_\text{max}=60\mpi^2=(1.08\GeV)^2.
\eeq
It is crucial to observe that the derivation of this result heavily relies on the fact that internal and external kinematics are related by $t'=t$, which allows for the translation of constraints originating from the Lehmann ellipse into a range of convergence in $s$. This is the essential part of the derivation that needs to be generalized in the analysis of different kinds of dispersion relations.

\begin{figure}
\centering
\includegraphics[width=0.5\linewidth]{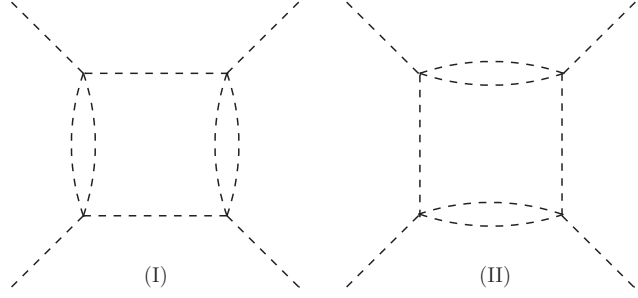}
\caption{Box graphs constraining the boundaries of the double-spectral functions for $\pi\pi$ scattering.}
\label{fig:boundary_pipi}
\end{figure}

Since rigorous results from axiomatic field theory are rarely available for processes other than $\pi\pi$ scattering, we will assume that the analytic properties of the $\pi N$ amplitude are correctly reproduced by Mandelstam analyticity. Therefore, it is instructive to compare the consequences of this relaxed assumption also for the $\pi\pi$ case to the axiomatic-field-theory result~\eqref{range_of_conv_pipi_ax}. The central objects of the analysis are the boundaries of the support of the double-spectral functions that determine the integration range in~\eqref{mandelstam}. These boundaries can be inferred from the box diagrams depicted in Fig.~\ref{fig:boundary_pipi}, which are to be understood as generalizations of four-propagator box diagrams (see, e.g.,~\cite{Itzykson:1980rh}), with one or more lines replaced by a particle whose mass is equal to the input mass of the lowest-lying intermediate state accessible to the interacting particles. 

\begin{figure}
\centering
\includegraphics[width=0.49\linewidth]{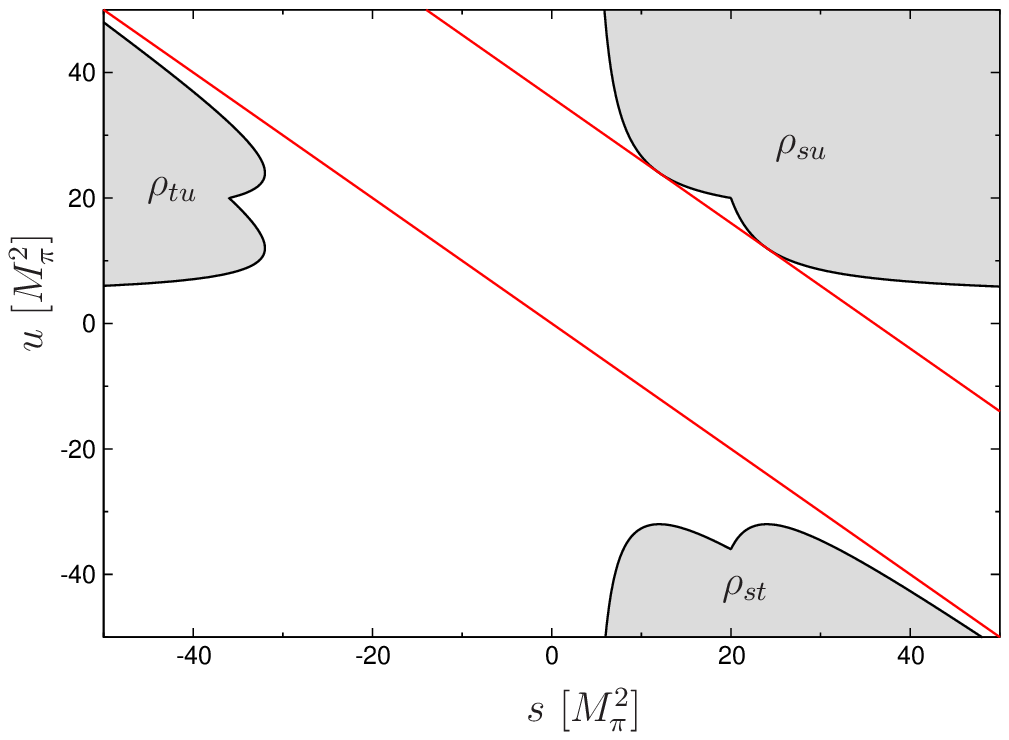}
\includegraphics[width=0.49\linewidth]{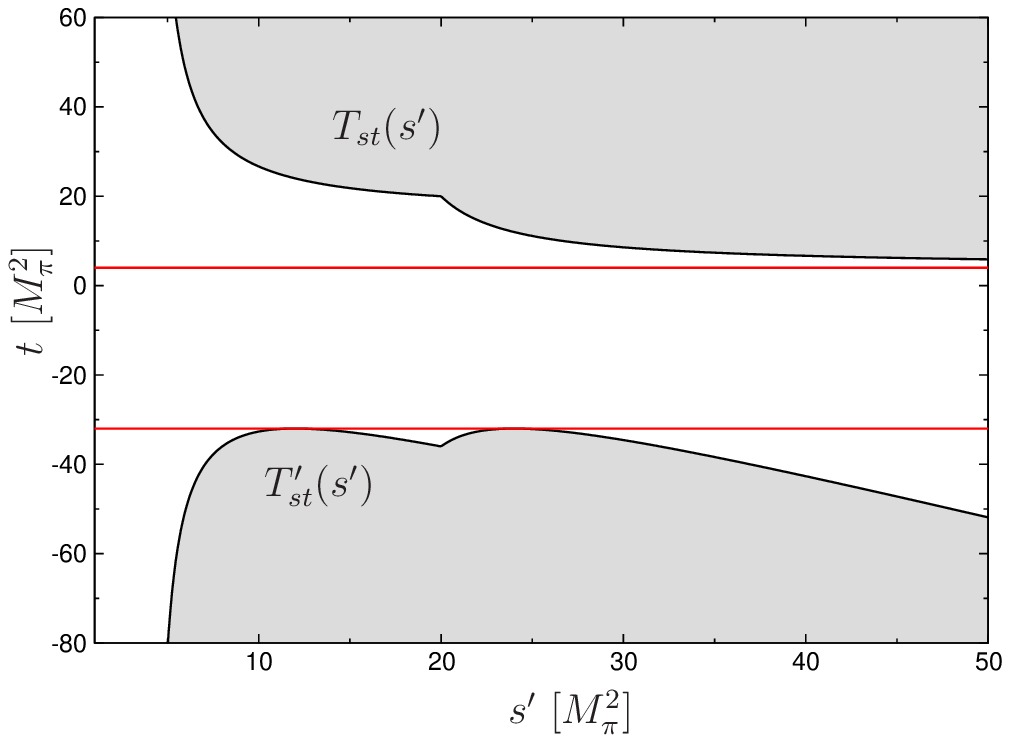}
\caption{Double-spectral regions (left) and allowed range of $t$ (right) for  $\pi\pi$ scattering. The red lines refer to $t=4\mpi^2$ and $t=-32\mpi^2$, respectively.}
\label{fig:t_mandelstam_pipi}
\end{figure}

Due to crossing symmetry, it suffices to investigate the boundary of $\rho_{st}$. From diagrams $(\text{I})$ and $(\text{II})$ in Fig.~\ref{fig:boundary_pipi} we find that this boundary is defined by
\beq
\label{b_pipi}
 b_\text{I}(s,t)=t(s-4\mpi^2)-16\mpi^2 s=0,\quad
 b_\text{II}(s,t)=t(s-16\mpi^2)-4\mpi^2 s=0,
\eeq
and thus obeys
\beq
t=T_{st}(s)=\min\big\{T_\text{I}(s),T_\text{II}(s)\big\},
\eeq  
where $T_\text{I}$ and $T_\text{II}$ follow from solving~\eqref{b_pipi} for $t$. The corresponding double-spectral regions, defined as the portions of the Mandelstam plane that obey $s + t + u=4 \mpi^2$ and where any one of the functions $\rho_{st}$, $\rho_{su}$, $\rho_{tu}$ has support, are shown in the left panel of Fig.~\ref{fig:t_mandelstam_pipi}. By definition, the line in the Mandelstam plane corresponding to a fixed value of $t$ must not enter the double-spectral regions if a single-variable dispersion relation with this value of $t$ is supposed to hold. Moreover, the maximally allowed value of $z'$ becomes
\beq
\label{zs_pipi}
z'_\text{max}=1+\frac{2T_{st}(s')}{s'-4\mpi^2},
\eeq 
and hence the Lehmann-ellipse constraint in the form
$-z'_\text{max}\leq z'\leq z'_\text{max}$ restricts the allowed values of $t$ to
\beq
T'_{st}(s')\leq t \leq T_{st}(s'), \quad
T'_{st}(s')=4\mpi^2-s'-T_{st}(s'),\qquad \forall\, s'\in\big[4\mpi^2,\infty\big).
\eeq
As illustrated in Fig.~\ref{fig:t_mandelstam_pipi}, both constraints actually yield the same range $-32\mpi^2\leq t\leq 4\mpi^2$, and thus
\beq
\label{range_of_conv_pipi_Mandelstam}
s_\text{max}=68\mpi^2=(1.15\GeV)^2,
\eeq
slightly larger than~\eqref{range_of_conv_pipi_ax}. Irrespective of the analyticity assumptions, the range of validity of the Roy equations can be extended significantly, at least up to $s_\text{max}=165\mpi^2$, if dispersion relations in the manifestly crossing-symmetric variables
\beq
x=st+tu+us,\qquad y=stu,
\eeq  
instead of fixed-$t$ dispersion relations are employed~\cite{Mahoux:1974ej,Auberson:1974in}, however, at the expense of a substantial increase in complexity of the equations. 

Due to the finite domain of validity, the Roy equations cannot be used up to infinity. Above a certain energy, referred to as the matching point $\sm$, input from experiment for the imaginary parts of the partial waves is required, so that in practice the equations are solved between threshold and $\sm$. Furthermore, the partial-wave expansion will be truncated at a certain angular momentum $J$ and higher partial waves treated on the same footing as the lower partial waves above $\sm$. In fact, the existence and uniqueness of a solution depends on the value of the phases $\delta_i$ of the partial waves dynamically included in the calculation at the matching point~\cite{Epele:1977un,Epele:1977um,Gasser:1999hz,Wanders:2000mn}. More precisely, the situation is characterized by the multiplicity index $m$, which is given by
\beq
\label{m_index}
m=\sum_i m_i,\qquad m_i=\begin{cases}\left\lfloor\frac{2\delta_i(\sm)}{\pi}\right\rfloor \qquad&\text{if }\delta_i(\sm)>0,\\
-1 \qquad&\text{if }
\delta_i(\sm)<0,\end{cases}
\eeq
where $\lfloor x\rfloor$ denotes the largest integer $\leq x$ and we have assumed that $\delta_i(\sm)>-\pi/2$ for all partial waves. If $m=0$, a unique solution exists, while for $m>0$ the neighborhood of each solution contains an $m$-parameter family of solutions, and for $m<0$ only for a specific choice (constrained by $|m|$ conditions) of the input, i.e.\ subtraction constants, imaginary parts above $\sm$, and higher partial waves, a solution can be found.  

\subsection{Matching to chiral perturbation theory}
\label{sec:roy+ChPT}

Shortly after Roy's article~\cite{Roy:1971tc}, a comprehensive phenomenological analysis of the $\pi\pi$ data available at that time was performed using the Roy-equation formalism~\cite{Basdevant:1972gk,Basdevant:1972uv,Basdevant:1972uu,Basdevant:1973ru,Pennington:1973xv,Pennington:1973hs,Froggatt:1975me,Froggatt:1977hu}. Later, there was renewed interest in the Roy equations~\cite{Ananthanarayan:1996gj,Ananthanarayan:1996iq,Ananthanarayan:1997yi,Ananthanarayan:1998hj,Ananthanarayan:2000ht,DescotesGenon:2001tn,Pelaez:2004vs,Kaminski:2006yv,Kaminski:2006qe,GarciaMartin:2011cn,Moussallam:2011zg,Caprini:2011ky}, mainly triggered by recognizing the full potential of the approach in combination with effective field theory, which has led to a determination of the low-energy $\pi\pi$ scattering amplitude with unprecedented accuracy. In the following, we will briefly summarize the strategy for the solution of the equations, especially focusing on the interplay with ChPT.\footnote{Formulated in terms of a suitable set of isospin amplitudes, one can derive a variant of the Roy equations that allows one to start with a once-subtracted fixed-$t$ dispersion relation instead of~\eqref{fixedt_2sub}. These GKPY equations, in addition to the original Roy equations, are used in~\cite{Pelaez:2004vs,Kaminski:2006yv,Kaminski:2006qe,GarciaMartin:2011cn} to perform a constrained fit to data without relying on ChPT at the same time.}  

A typical truncation scheme for the numerical solution of the Roy equations proceeds as follows: in~\cite{Ananthanarayan:2000ht},
the system was truncated at $J=1$ and the matching point chosen as $\sm=(0.8\GeV)^2$, implying that $m=0$. The effects of higher partial waves as well as higher energies, i.e.\ the part of~\eqref{Royeq} with $J'\geq 2$ and $s'\geq \sm$, are accounted for in so-called driving terms, which are determined from experimental input for the intermediate-energy regime and from Regge theory for the high-energy behavior. A crucial result of~\cite{Ananthanarayan:2000ht} was that the amplitudes in the low-energy regime  are remarkably insensitive to the details of the contributions from higher partial waves and higher energies, so that a very precise representation of the $S$- and $P$-wave amplitudes at low energies in terms of the scattering lengths $a_0^0$ and $a_0^2$ could be provided. 

The simplest matching procedure between Roy equations and ChPT, matching at threshold, would amount to inserting the two-loop ChPT result for the scattering lengths into the Roy equations. However, this approach is unfavorable, since the chiral expansion at threshold converges rather slowly, caused by the onset of the unitarity cut, and further constraints in the whole low-energy region, where both the Roy-equation and the chiral parameterization are valid, would be ignored. 
Instead, the strategy put forward in~\cite{Colangelo:2001df,Colangelo:2000jc} relies on the fact that ultimately the subtraction constants in the dispersive calculation and the LECs in the chiral expansion can be identified. To this end, both parameterizations are brought into a form that proves that agreement at low energies is ensured if the polynomial parts match. In this way, by requiring consistency of both representations in the full low-energy regime, the slow convergence  at threshold is avoided and the sensitivity to terms in the chiral expansion beyond two-loop order diminished. Thus, since the Roy equations have elastic unitarity fully built in---in contrast to ChPT that restores unitarity only perturbatively---they can be regarded as a means to unitarize the chiral expansion. 
Retaining only the LECs $\bar l_3$ and $\bar l_4$ (the latter eliminated in favor of the scalar pion radius $\langle r^2\rangle^S_\pi$), which 
measure the quark-mass dependence and thus
cannot be determined dynamically in the matching of the polynomials parts, the scattering lengths can be expressed as~\cite{Colangelo:2001df,Colangelo:2000jc} 
\begin{align}
 a_0^0 &= 0.198 \pm 0.001 + 0.0443\,\text{fm}^{-2} \langle r^2\rangle^S_\pi - 0.0017\, \bar l_3,\notag\\
a_0^2 &= -0.0392 \pm 0.0003 - 0.0066\,\text{fm}^{-2} \langle r^2\rangle^S_\pi - 0.0004\, \bar l_3,
\end{align}
which together with $\bar l_3=2.9\pm 2.4$, $\langle r^2\rangle^S_\pi=(0.61\pm 0.04)\,\text{fm}^2$
finally led to a very precise prediction of the $\pi\pi$ scattering lengths~\cite{Colangelo:2001df,Colangelo:2000jc}
\beq
\label{pipi_scatt_lengths}
a_0^0=0.220\pm 0.005, \qquad a_0^2=-0.0444\pm 0.0010.
\eeq
Recently, these predictions have been tested in various high-precision experiments. First, the decay width of pionium is sensitive to $|a_0^0-a_0^2|$, and has been measured by the DIRAC collaboration~\cite{Adeva:2005pg,Adeva:2011tc}. Next, a measurement of $K_{l4}$ decays~\cite{Pislak:2003sv,Batley:2007zz,Batley:2010zza} yields access to the phase-shift difference $\delta^0_0-\delta^1_1$ of $\pi\pi$ $S$- and $P$-waves, which, combined with a numerical solution of the Roy equations, determines $a_0^0$ and $a_0^2$. Lastly, the scattering lengths may be extracted from a high-statistics analysis of $K\to 3\pi$ decays~\cite{Batley:2005ax,Batley:2000zz}, where the rescattering of pions in the final state generates a cusp whose strength relates to the pertinent $\pi\pi$ scattering amplitude at threshold~\cite{Cabibbo:2004gq,Cabibbo:2005ez,Colangelo:2006va,Gasser:2011ju}. Presently, the most stringent constraints on $a_0^0$ and $a_0^2$ originate from $K_{l4}$ and $K\to 3\pi$ decays.
The NA48/2 collaboration quotes for the combination of both measurements~\cite{Batley:2010zza}
\beq
\label{NA48}
a_0^0=0.2210\pm 0.0047_\text{stat}\pm 0.0040_\text{syst}, \qquad a_0^2=-0.0429\pm 0.0044_\text{stat}\pm 0.0028_\text{syst},
\eeq
in beautiful agreement with~\eqref{pipi_scatt_lengths}. In fact, at this level of accuracy it is critical that isospin-violating corrections specific to each experiment be properly taken into account~\cite{Knecht:1997jw,Gasser:2001un,Bissegger:2008ff,Colangelo:2008sm,Colangelo:2009zza}.
In recent years, lattice calculations have reached a level of accuracy that 
warrants their inclusion in a comprehensive global analysis of all constraints in the $a_0^0$--$a_0^2$ plane. For $a_0^2$ lattice results agree well with data and Roy equations, while for $a_0^0$ direct calculations including disconnected diagrams are still called for. Alternatively, the required LECs can be calculated on the lattice and the scattering lengths reconstructed indirectly in this way. 
For a graphical representation of the results~\eqref{pipi_scatt_lengths} and~\eqref{NA48} also in comparison with the cornucopia of lattice
determinations we refer to~\cite{Leutwyler:2015jga}. 

The importance of an accurate knowledge of the $\pi\pi$ scattering lengths, and thus of the low-energy phase shifts, is hard to overestimate. First of all, the scattering lengths are central parameters of low-energy QCD themselves, intimately related to the pattern of chiral symmetry breaking. Indeed, their precise determination was essential to confirm the role of the quark condensate as the leading order parameter of the spontaneous breaking of chiral symmetry~\cite{Colangelo:2001df,Colangelo:2000jc}. 
Moreover, once the subtraction constants are fixed, the Roy equations automatically provide the analytic continuation of the $\pi\pi$ amplitude beyond the physical region. The domain of validity of the equations reaches sufficiently far into the complex plane to encompass the $\sigma$ pole $m_\sigma=M_\sigma-i\Gamma_\sigma/2$, with the result that the combination of Roy equations and ChPT allowed for the first model-independent determination of the pole parameters of the $\sigma$-meson~\cite{Caprini:2005zr}
\beq
\label{sigma}
M_\sigma=441^{+16}_{-8}\MeV,\qquad \Gamma_\sigma=544^{+18}_{-25}\MeV.
\eeq
These numbers were recently confirmed in~\cite{GarciaMartin:2011jx,Moussallam:2011zg} (and extended to the $f_0(980)$), with the result that the range for the pole parameters quoted in~\cite{Agashe:2014kda} could be reduced substantially, including a change of name from $f_0(600)$ to $f_0(500)$~\cite{Pelaez:2015qba}.

\subsection{Beyond $\pi\pi$ scattering}
\label{sec:beyond_pipi}

Evidently, it would be of high interest to extend the successful program of a combined framework of Roy or Roy-like equations and ChPT to processes other than $\pi\pi$ scattering. An important step forward in this direction was taken in~\cite{Ananthanarayan:2000cp,Ananthanarayan:2001uy,Buettiker:2003pp}, where RS equations for $\pi K$ scattering were constructed. Unfortunately, the generalization beyond $\pi\pi$ scattering comes with plenty of complications, mainly rooted in unequal masses and more involved crossing properties, as we will demonstrate in the following using the example of $\pi K$ scattering~\cite{Ananthanarayan:2000cp,Ananthanarayan:2001uy,Buettiker:2003pp}.

First of all, a full system of PWDRs will include dispersion relations for two distinct physical processes, $\pi K$ scattering ($s$-channel) and $\pi\pi\to\bar K K$ ($t$-channel). An immediate consequence concerns the applicability of fixed-$t$ dispersion relations and the use of crossing symmetry to determine the subtraction function $c(t)$ in~\eqref{Roy_crossing}: $s\leftrightarrow t$ crossing symmetry will intertwine $s$- and $t$-channel equations, so that the equations for the $s$-channel partial waves will also involve $t$-channel dispersive integrals that extend over $t'\geq 4\mpi^2$. Accordingly, the determination of the $t$-channel partial waves will require the partial-wave projection for $t>4\mpi^2$, which lies beyond the range of validity of fixed-$t$ dispersion relations. 

A convenient choice of dispersion relations that evade these limitations are HDRs~\cite{Hite:1973pm},\footnote{Even more,
HDRs are the unique choice if one demands that the curves pass through all kinematic channels, avoid double-spectral regions, do not introduce ostensible kinematic cuts into the partial-wave amplitudes, and still yield manageable kernel functions~\cite{Hite:1973pm}. In view of the efforts~\cite{Hite:1973pm,Baacke:1970mi,Steiner:1970mh,Steiner:1971ms} that led to $s$-channel PWDRs for $\pi N$ scattering and thus provided the first step towards the construction of a Roy-equation analog for processes with $\pi N$ crossing properties, the resulting full system of partial-wave hyperbolic dispersion relations is referred to as Roy--Steiner equations.} defined by
\beq
\label{hyperbola}
(s-a)(u-a)=(s'-a)(u'-a)\equiv b.
\eeq 
While $b=b(s,t,a)$ is fixed by the external kinematics, the hyperbola parameter $a$ can be freely chosen. In particular, it can be used to optimize the range of validity of the resulting system of RS equations. 
HDRs are particularly suitable for processes such as $\pi K$ scattering, since $s\leftrightarrow u$ crossing is manifest, so that all constraints by crossing symmetry are automatically fulfilled. 
The unsubtracted version of HDRs for a crossing-symmetric amplitude $T^+(s,t)$ reads
\beq
\label{HDR_cross_even}
T^+(s,t)=\frac{1}{\pi}\int\limits_{s_+}^\infty\diff s'\bigg[\frac{1}{s'-s}+\frac{1}{s'-u}-\frac{1}{s'-a}\bigg]\Im T^+(s',t')
+\frac{1}{\pi}\int\limits_{4\mpi^2}^\infty\diff t'\frac{\Im T^+(s',t')}{t'-t},
\eeq
where
\beq
s_+=(\mpi+\mK)^2,
\eeq
and $\mK$ denotes the mass of the kaon. The above integrals are understood in such a way that the integrands shall be expressed in terms of the integration variable and the external kinematics by virtue of~\eqref{hyperbola} and
\beq
s'+t'+u'=2\mpi^2+2\mK^2.
\eeq
The first integral in~\eqref{HDR_cross_even} is reminiscent of fixed-$t$ dispersion relations, but in that case $\Im T^+(s',t') \to  \Im T^+(s',t)$ and the last term is removed. Thus, the key difference here is that $t'$ depends not only on $t$, but on $s$ and $s'$ as well. The dispersion relation for an amplitude $T^-(s,t)$ that is odd under crossing may be constructed by considering $T^-(s,t)/\nu$, $\nu=s-u$, yielding
\beq
\label{HDR_cross_odd}
T^-(s,t)=\frac{1}{\pi}\int\limits_{s_+}^\infty\diff s'\bigg[\frac{1}{s'-s}-\frac{1}{s'-u}\bigg]\Im T^-(s',t')
+\frac{1}{\pi}\int\limits_{4\mpi^2}^\infty\diff t'\frac{\nu}{\nu'}\frac{\Im T^-(s',t')}{t'-t}.
\eeq
In~\cite{Buettiker:2003pp}, a combination of fixed-$t$ and hyperbolic dispersion relations (with $a=0$) was used, arguing that the range of validity of fixed-$t$ dispersion relations in the $s$-channel is slightly larger than for HDRs. We suspect that this advantage would dissolve if HDRs with general $a$ were allowed (as is true, e.g., for $\pi N$ scattering). Hence, we will bypass the fixed-$t$ step and solely consider HDRs here.

\begin{figure}
\centering
\includegraphics[width=0.3\linewidth]{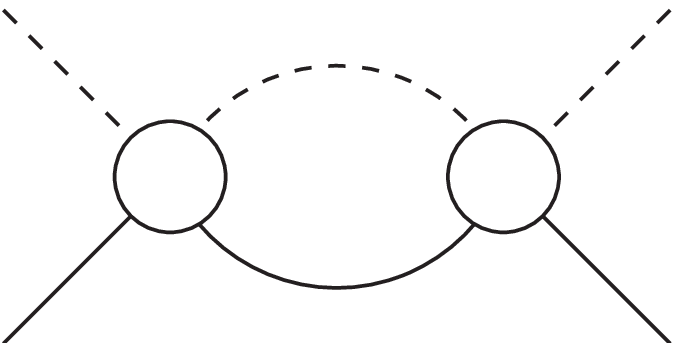}
\qquad\qquad
\includegraphics[width=0.3\linewidth]{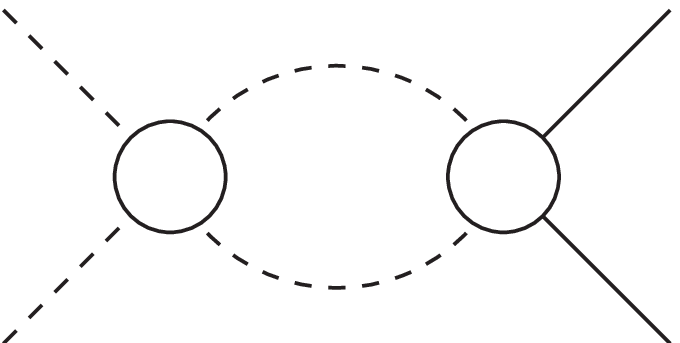}
\caption{Elastic unitarity for $\pi K$ scattering (left) and $\pi\pi\to \bar K K$ (right). Dashed/solid lines denote pions/kaons, and the spheres refer to the pertinent scattering amplitudes.}
\label{fig:piK_unitarity}
\end{figure}

The second major impediment concerns unitarity in the $t$-channel. While $s$-channel unitarity corresponds exactly to unitarity in $\pi\pi$ scattering, the $t$-channel unitarity relation does not. In particular, it is linear in the $\pi\pi\to\bar K K$ partial waves, and thus far less restrictive than $s$-channel unitarity, cf.\ Fig.~\ref{fig:piK_unitarity}. The resulting equations for the $t$-channel partial waves take the form of a Muskhelishvili--Omn\`es (MO) problem~\cite{Muskhelishvili:1953,Omnes:1958hv}, whose solution will require input for the $\pi\pi$ partial waves. However, once the $t$-channel equations are solved, the remaining $s$-channel problem will be amenable to the same methods that can be used to solve $\pi\pi$ Roy equations.

Last, $\pi\pi$ scattering is also exceptional in the matching to ChPT, since the number of LECs is small and neither potentially large $SU(3)$ corrections nor the presence of baryons threaten the rapid convergence of the chiral expansion. In contrast, the comparison of RS equations and ChPT in $\pi K$ scattering is indeed hampered by large uncertainties in the chiral series~\cite{Ananthanarayan:2000cp,Ananthanarayan:2001uy,Buettiker:2003pp,Bernard:1990kx,Bernard:1990kw,Bijnens:2004bu}.
Despite all drawbacks in the non-identical-particle case one should note that the important feature of the Roy-equation approach that the kernel functions will correctly incorporate the analytic properties of the partial waves prevails in the general case. As long as the dispersion relations on the amplitude level hold, the correct analytic structure of the partial waves will emerge automatically. Although the complications besetting the generalization of $\pi\pi$ Roy equations to other processes lead to a considerable increase in complexity, a full solution of the corresponding system of RS equations is highly rewarding nonetheless. After all, the result will maintain analyticity, unitarity, crossing symmetry, and, by matching to ChPT, chiral symmetry, and thus all symmetries of the underlying quantum field theory. In fact, RS equations for $\pi K$ scattering have provided invaluable information on the $\pi K$ scattering lengths, low-energy phase shifts, and the pole position of the $\kappa$-meson~\cite{Buettiker:2003pp,DescotesGenon:2006uk}.
Similarly, an analysis of $\gamma\gamma\to\pi\pi$ based on RS equations~\cite{Hoferichter:2011wk}, in combination with input from $2$-loop ChPT~\cite{Gasser:2005ud,Gasser:2006qa},
provides access to both the low-energy partial waves as well as the two-photon coupling of the $\sigma$~\cite{GarciaMartin:2010cw,Moussallam:2011zg,Hoferichter:2011wk}.
Roy-like equations have also been derived for $\gamma\pi\to\pi\pi$ in~\cite{Hannah:2001ee}, based on fixed-$t$ dispersion relations, but in this case the equations simply amount to a MO problem in all channels~\cite{Truong:2001en,Hoferichter:2012pm}. The generalization to the virtual processes $\gamma^*\gamma^{(*)}\to\pi\pi$~\cite{Moussallam:2013una,Hoferichter:2013ama}
and $\gamma^*\pi\to\pi\pi$~\cite{Niecknig:2012sj,Schneider:2012ez,Danilkin:2014cra,Hoferichter:2014vra} has become of vital interest recently
in the context of a dispersive approach towards hadronic light-by-light scattering in the anomalous magnetic moment of the muon~\cite{Colangelo:2014dfa,Colangelo:2014pva,Pauk:2014rfa,Colangelo:2015ama}.

In the case of $\pi N$ scattering, many of the challenges present for $\pi K$ scattering persist, in fact, some of them are even aggravated: the pseudophysical region in the $t$-channel is much larger and the number of amplitudes relevant at low energies increases further. 
As concerns the matching to ChPT, the chiral amplitude is known at $1$-loop level only, so in a direct matching one would have to try to quantify the uncertainties induced by higher-order corrections.
On the other hand, the $\pi N$ scattering lengths are known very precisely from pionic atoms~\cite{Gotta:2008zza,Strauch:2010vu,Hennebach:2014lsa,Baru:2010xn,Baru:2011bw}, tightly constraining the low-energy $\pi N$ amplitude. For these reasons, we will pursue the following strategy for the solution of $\pi N$ RS equations: first, the equations are solved by imposing the scattering lengths as additional constraint, without any ChPT input. The matching to ChPT proceeds as a second step by matching the polynomial that describes the subthreshold expansion of the amplitude, thus in the kinematic region where ChPT is expected to converge best. Anticipating this step, we subtract the RS equations at the subthreshold point, see Sect.~\ref{sec:RSsystem}, another difference to the original $\pi\pi$ system~\cite{Roy:1971tc} (or the $\pi K$ system~\cite{Buettiker:2003pp}), where the subtraction constants are expressed in terms of scattering lengths.

\section{Roy--Steiner equations for $\boldsymbol{\pi N}$ scattering}
\label{sec:RSsystem}

The construction of a complete system of RS equations for $\pi N$ scattering has been presented in detail in~\cite{Ditsche:2012fv}, here we review its salient features. 
As anticipated in Sect.~\ref{sec:beyond_pipi}, the starting point in the derivation is provided by HDRs for the invariant $\pi N$ amplitudes, which, in combination with the pertinent partial-wave expansions as well as unitarity relations, are used to derive a closed system of PWDRs that fully respects analyticity, unitarity, and crossing symmetry. 
Subtractions will be performed at the so-called subthreshold point, which proves convenient for the matching to ChPT and for the extrapolation to the Cheng--Dashen point~\cite{Cheng:1970mx}, and thus for establishing the relation to $\sigma_{\pi N}$ by means of a low-energy theorem~\cite{Cheng:1970mx,Brown:1971pn,Bernard:1996nu,Becher:2001hv}. In fact, it has been pointed out previously that a reliable extrapolation to the subthreshold region requires additional input from the $t$-channel ($\pi\pi\to\bar N N$) partial waves~\cite{Stahov:1999,Stahov:2002,Hite:2005tg}, a requirement that is straightforward to comply with in the RS formalism, as HDRs by construction intertwine all physical regions. The PWDRs for the $s$-channel ($\pi N$) partial waves in their unsubtracted form were already written down in~\cite{Hite:1973pm}, while the $t$-channel equations that are necessary to obtain a closed system of equations were omitted, as was the issue of subtractions. In the end, both the $s$- and $t$-channel equations will involve the subtraction constants and the $\pi N$ coupling constant as free parameters, see~\cite{Ditsche:2012fv}. In this paper, we introduce additional subtractions compared to the system put forward in~\cite{Ditsche:2012fv} in order to match exactly the number of degrees of freedom in the RS equations.

\begin{figure}[!t]
\centering
\includegraphics[width=\textwidth]{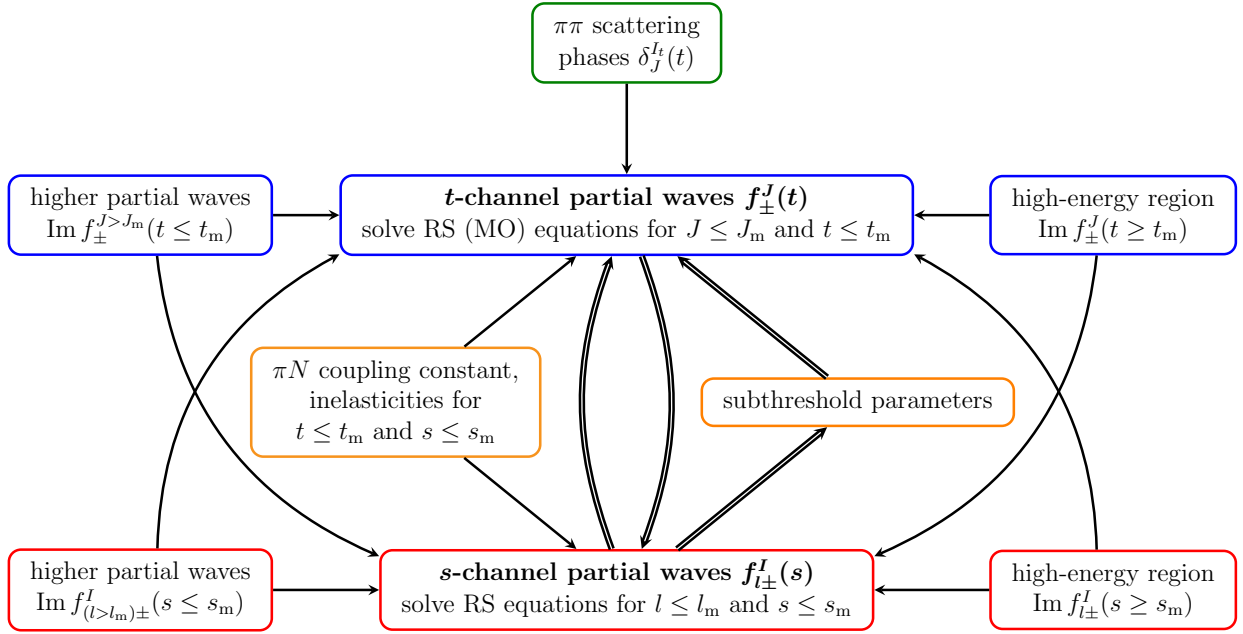}
\caption{Solution strategy for RS equations in $\pi N$ scattering. The $s$- and $t$-channel partial waves will be solved for up to angular momenta $l_\text{m}=1$ and $J_\text{m}=2$, respectively.}
\label{fig:flowchart}
\end{figure} 

The strategy for the solution of the RS equations is outlined in Fig.~\ref{fig:flowchart}: in the $s$-channel, the six $S$- and $P$-waves $f_{l\pm}^I$, with $I=\pm$ for the isospin index, orbital angular momentum $l$, and total angular momentum
$j = |l\pm 1/2|$, are considered dynamically below the matching point $\sm$, whereas the imaginary parts of higher partial waves for all $s$, the imaginary parts of the $S$- and $P$-waves above $\sm$, and, potentially, inelasticities below $\sm$ are required as input. In practice, we will choose the matching point at its optimal value $\sm = (1.38\GeV)^2$ as argued below~\cite{Ditsche:2012fv}.
In contrast to the six $s$-channel amplitudes, there are only three $S$- and $P$-waves in the $t$-channel, $f^J_\pm$, with total angular momentum $J$ 
and the subscript referring to parallel/antiparallel antinucleon--nucleon helicities. The equations for the $t$-channel partial waves take the form of a MO problem~\cite{Muskhelishvili:1953,Omnes:1958hv}, whose solution requires---in addition to higher partial waves and the imaginary parts above the matching point $\tm$---input for the $\pi\pi$ phase shifts. 
This step has been carried out in~\cite{Ditsche:2012fv,Hoferichter:2012wf}.
Once the $t$-channel problem is solved, the resulting $t$-channel partial waves are used as input for the $s$-channel problem, which then reduces to the form of conventional $\pi\pi$ Roy equations, rendering known results on the existence and uniqueness of solutions~\cite{Epele:1977um,Epele:1977un,Gasser:1999hz,Wanders:2000mn} as well as known solution techniques~\cite{Ananthanarayan:2000ht} applicable. Eventually, a full solution of the system can be obtained by iterating this procedure until all partial waves and parameters are determined self-consistently. In practice, virtually all interdependence proceeds via the subtractions constants, so that the need for an iterative procedure can be avoided if the corresponding terms are included explicitly in the $s$-channel fit, as will be explained in detail in Sect.~\ref{sec:schannel_sol}.

Next, we comment on the structure of the equations. The $s$-channel unitarity relation in $\pi N$ scattering is dominated by elastic unitarity at low energies, so that the RS self-consistency condition will translate to constraints on the low-energy $\pi N$ phase shifts. However, inelasticities due to $\pi\pi N$ intermediate states set in rather early, especially in the partial wave corresponding to the Roper resonance $P_{11}(1440)$\footnote{In the following, we will make use of the spectroscopic notation $L_{2I_s2J}$ to label the partial waves.} the inelasticity cannot be neglected and has to be taken as input. While the $s$-channel partial waves are all mutually coupled, the $t$-channel problem actually decouples to a certain degree, as the equation for an amplitude with even/odd $J$ only depends on partial waves with even/odd $J'$ larger than $J$. Nonetheless, the solution of the $t$-channel equations is subject to an additional complication as compared to $\pi\pi\to\bar K K$ that is related to the large pseudophysical region in $\pi\pi\to\bar N N$ and the fact that it is advantageous to choose $\tm = 4\mN^2$ (the maximal allowed value is $\tm = (2.00\GeV)^2$, see Sect.~\ref{sec:HDR} and~\cite{Ditsche:2012fv}).  In either case, intermediate states besides $\pi\pi$ become relevant in the unitarity relation around $1\GeV$, most notably in the $S$-wave, where $\bar K K$ intermediate states account for the occurrence of the $f_0(980)$ resonance. While in $\pi\pi\to\bar K K$ these effects can simply be included by choosing $\tm$ around $1\GeV$ and using phase-shift solutions above, physical input for $\pi\pi\to\bar N N$ becomes only available at the two-nucleon threshold, which leaves a large fraction of the pseudophysical region unconstrained by the single-channel approximation. Since for similar reasons no reliable input information for higher partial waves is available, we also solve for $D$-waves in the $t$-channel problem. In the end, we consider a full two-channel MO problem for the $S$-wave to reproduce the $f_0(980)$ dynamics, a single-channel solution for $P$- and $D$-waves, and put higher partial waves as well as the imaginary parts above $\tm$ to zero~\cite{Ditsche:2012fv,Hoferichter:2012wf}. The stability of this approximation will be checked by studying the role of $F$-waves, the impact from PWAs above the two-nucleon threshold~\cite{Anisovich:2011bw}, as well as the effect of varying the input in the calculation of the partial waves with $J\leq 2$.    

\subsection{Kinematics and conventions}
\label{sec:conventions}

\begin{figure}[t!]
\centering
\includegraphics[width=0.5\linewidth]{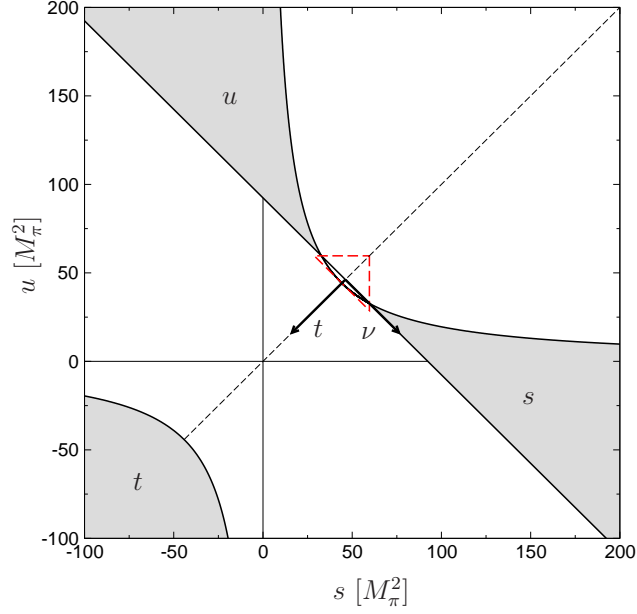}
\caption{Mandelstam plane for $\pi N$ scattering. The filled areas mark the $s$-, $t$-, and $u$-channel physical regions, the red dashed line the subthreshold triangle, and the arrows the orientation of the plane in $t$ and $\nu$.}
\label{fig:mandelstam_piN}
\end{figure}

We consider the reaction\footnote{For more details on $\pi N$ kinematics and conventions we refer to~\cite{Hoehler:1983,Ditsche:2012fv}. Most of the variables defined here will appear in a primed version as well to denote the corresponding quantity in the internal kinematics, e.g.\ $z_s'$, $q_t'$, $\nu'$, etc.} 
\beq
\pi^a(q)+N(p)\rightarrow \pi^b(q')+N(p'),
\eeq
with pion isospin labels $a$, $b$, and Mandelstam variables
\beq
s=(p+q)^2,\qquad t=(q-q')^2,\qquad u=(p-q')^2,
\eeq
which fulfill
\beq
s+t+u=2\mN^2+2\mpi^2\equiv\Sigma.
\eeq
The masses of the nucleon ($\mN$), the pion ($\mpi$), and, later, the kaon ($\mK$) are identified with the charged-particle masses~\cite{Agashe:2014kda}. Furthermore, we will need the definitions
\beq
s=W^2,\qquad \nu=\frac{s-u}{4\mN},\qquad s_\pm=W_\pm^2=(\mN\pm\mpi)^2,\qquad s_0=\frac{\Sigma}{2}.
\eeq
The kinematics for the $s$-channel reaction may be described by the center-of-mass (CMS) momentum $\qq$, nucleon energy $E$, and scattering angle $z_s=\cos\theta_s$
\beq
\qq^2=\frac{(s-s_+)(s-s_-)}{4s}, \qquad E=\sqrt{\mN^2+\qq^2}=\frac{s+\mN^2-\mpi^2}{2W},\qquad
z_s=1+\frac{t}{2\qq^2}.
\eeq
Similarly, the $t$-channel reaction is determined by CMS momenta $q_t=|\qq_t|$, $p_t=|\pp_t|$, and scattering angle $z_t=\cos\theta_t$
\begin{align}
q_t&=\sqrt{\frac{t}{4}-\mpi^2}=\sqrt{\frac{t-\tpi}{4}}=\frac{\sqrt{t}}{2}\sigma^\pi_t=iq_-, \qquad
z_t=\frac{s-u}{4p_tq_t}=\frac{\mN\nu}{p_tq_t},\notag\\
p_t&=\sqrt{\frac{t}{4}-\mN^2}=\sqrt{\frac{t-\tN}{4}}=\frac{\sqrt{t}}{2}\sigma^N_t=ip_-,
\end{align}
where the phases between $q_t$, $p_t$, and
\beq
\label{pmqm}
q_-=\sqrt{\mpi^2-\frac{t}{4}}, \qquad
p_-=\sqrt{\mN^2-\frac{t}{4}},
\eeq
have been fixed by convention. 
The physical regions for $\pi N$ scattering, determined by the requirement that the Kibble function $\Phi$~\cite{Kibble:1960zz}
\beq
\Phi=t\Big[s u-(\mN^2-\mpi^2)^2\Big]
\eeq
be non-negative, are shown in Fig.~\ref{fig:mandelstam_piN}. Points of special interest in the Mandelstam plane are the Cheng--Dashen point at $(s=u=\mN^2,t=2\mpi^2)$ for the relation to the $\sigma$-term, the subthreshold point at $(s=u=s_0,t=0)$ as the starting point for the subthreshold expansion, and the $s$-channel threshold point $(s=s_+,t=0,u=s_-)$ for the definition of the threshold expansion.

The scattering amplitude may be expressed in terms of Lorentz-invariant amplitudes $A$, $B$, and $D$ according to
\begin{align}
\label{genlorentzinvamps}
T^{ba}(s,t)&=\delta^{ba}T^+(s,t)+\frac{1}{2}[\tau^b,\tau^a]T^-(s,t),\notag\\
T^I(s,t)&=\bar{u}(p')\bigg\{A^I(s,t)+\frac{\slashed q'+\slashed q}{2}B^I(s,t)\bigg\}u(p)
=\bar{u}(p')\bigg\{D^I(s,t)-\frac{[\slashed q',\slashed q]}{4\mN}B^I(s,t)\bigg\}u(p),\notag\\
D^I(s,t)&=A^I(s,t)+\nu B^I(s,t),
\end{align}
where $I=\pm$, $\tau^a$ denotes the Pauli matrices, and the spinors are normalized as $\bar u u=2\mN$. The pertinent crossing properties become most transparent when the amplitudes are written as functions of $\nu$ and $t$
\beq
\label{crossingamps}
A^\pm(\nu,t)=\pm A^\pm(-\nu,t), \qquad B^\pm(\nu,t)=\mp B^\pm(-\nu,t),\qquad D^\pm(\nu,t)=\pm D^\pm(-\nu,t).
\eeq
Moreover, isospin symmetry leaves only two independent amplitudes that are needed to describe
 all eight $\pi N$ scattering reactions, characterized by total $s$-channel isospin $I_s\in\{1/2,3/2\}$,
\begin{align}
\label{sampisorels}
\Amp(\pi^+ p\to\pi^+ p)&=\Amp(\pi^-n\to\pi^-n)=\Amp^+-\Amp^-=\Amp^{3/2},\notag\\
\Amp(\pi^- p\to\pi^- p)&=\Amp(\pi^+n\to\pi^+n)=\Amp^++\Amp^-=\frac{1}{3}(2\Amp^{1/2}+\Amp^{3/2}),\notag\\
\Amp(\pi^-p\to\pi^0n)&=\Amp(\pi^+n\to\pi^0p)=-\sqrt{2}\Amp^-=-\frac{\sqrt{2}}{3}(\Amp^{1/2}-\Amp^{3/2}),\notag\\
\Amp(\pi^0p\to\pi^0p)&=\Amp(\pi^0n\to\pi^0n)=\Amp^+=\frac{1}{3}(\Amp^{1/2}+2\Amp^{3/2}),
\end{align}
where $\Amp\in\{A,B,D\}$.
These relations can be summarized in matrix notation as
\beq
\label{schannelcrossing}
\begin{pmatrix}\Amp^+\\\Amp^-\end{pmatrix}=C_{\nu s}\begin{pmatrix}\Amp^{1/2}\\\Amp^{3/2}\end{pmatrix}, \qquad \begin{pmatrix}\Amp^{1/2}\\\Amp^{3/2}\end{pmatrix}=C_{s\nu}\begin{pmatrix}\Amp^+\\\Amp^-\end{pmatrix}, \qquad C_{\nu s}=\frac{1}{3}C_{s\nu}=\frac{1}{3}\begin{pmatrix}1&2\\1&-1\end{pmatrix}.
\eeq
Similarly, the $t$-channel amplitudes read
\begin{align}
\label{tampisorels}
\Amp(\bar{p}p\to\pi^+\pi^-)& =-\Amp^++\Amp^- =-\Amp^{3/2} =-\frac{1}{\sqrt{6}}\Amp^0+\frac{1}{2}\Amp^1,\notag\\
\Amp(\bar{p}p\to\pi^-\pi^+)& =-\Amp^+-\Amp^- =-\frac{1}{3}(2\Amp^{1/2}+\Amp^{3/2})=-\frac{1}{\sqrt{6}}\Amp^0-\frac{1}{2}\Amp^1,\notag\\
\Amp(\bar{n}p\to\pi^+\pi^0)& =\sqrt{2}\Amp^- =\frac{\sqrt{2}}{3}(\Amp^{1/2}-\Amp^{3/2}) =\frac{1}{\sqrt{2}}\Amp^1,\notag\\
\Amp(\bar{p}p\to\pi^0\pi^0)& =\Amp^+ =\frac{1}{3}(\Amp^{1/2}+2\Amp^{3/2})=\frac{1}{\sqrt{6}}\Amp^0,
\end{align}
which can be summarized in the $s\leftrightarrow t$ crossing relations
\beq
\label{tchannelcrossing}
\begin{pmatrix}\Amp^{1/2}\\\Amp^{3/2}\end{pmatrix}=C_{st}\begin{pmatrix}\Amp^0\\\Amp^1\end{pmatrix}, \qquad
\begin{pmatrix}\Amp^0\\\Amp^1\end{pmatrix}=C_{ts}\begin{pmatrix}\Amp^{1/2}\\\Amp^{3/2}\end{pmatrix}, \qquad
\begin{pmatrix}\Amp^+\\\Amp^-\end{pmatrix}=C_{\nu t}\begin{pmatrix}\Amp^0\\\Amp^1\end{pmatrix},
\eeq
with crossing matrices
\beq
\label{ampspmtoampsIt01}
C_{st}=\begin{pmatrix}\frac{1}{\sqrt{6}}&1\\\frac{1}{\sqrt{6}}&-\frac{1}{2}\end{pmatrix},\qquad
C_{ts}=\frac{2}{3}\begin{pmatrix}\sqrt{\frac{3}{2}}&\sqrt{6}\\1&-1\end{pmatrix},\qquad
C_{\nu t}=C_{\nu s}C_{st}=\begin{pmatrix}\frac{1}{\sqrt{6}}&0\\0&\frac{1}{2}\end{pmatrix}.
\eeq
This result shows that the amplitudes $\Amp^+$ and $\Amp^-$ have well-defined $t$-channel isospin $I_t=0$ and $I_t=1$, respectively.
Since the $G$-parity of the pion is negative,
the antinucleon--nucleon initial state in $\bar{N}N\to\pi\pi$ has to have positive $G$-parity, so that this state can only couple to an even number of pions. Moreover, since
\beq
G|\bar{N}N\rangle=(-1)^{J+I_t}|\bar{N}N\rangle,
\eeq
 only the combinations of even angular momentum $J$ with $I_t=0$ and odd $J$ with $I_t=1$ are permitted. 
The same selection rules follow from the $\pi\pi$ system, where Bose symmetry gives rise to the same factor $(-1)^{J+I_t}$ due to the parity $(-1)^L$ for an orbital state with total angular momentum $J=L$ and the symmetry/antisymmetry of the pion isosinglet/isotriplet state.
These observations are crucial for the $t$-channel partial-wave expansion of $\Amp^{I=\pm}$ or $\Amp^{I_t=0,1}$, which contains only partial waves with even/odd $J$. According to~\eqref{ampspmtoampsIt01}, the transition between 
the $I=\pm$ or $I_t=0,1$ bases involves the isospin crossing coefficients $c_J$
\beq
\label{crossingcoefficients}
c_J=\begin{cases}\frac{1}{\sqrt{6}}\qquad\text{for even }J,\\\frac{1}{2}\,\,\,\,\qquad\text{for odd }J.\end{cases}
\eeq

\subsection{Hyperbolic dispersion relations}
\label{sec:HDR}

\begin{figure}
\centering
\includegraphics[height=\linewidth,angle=-90]{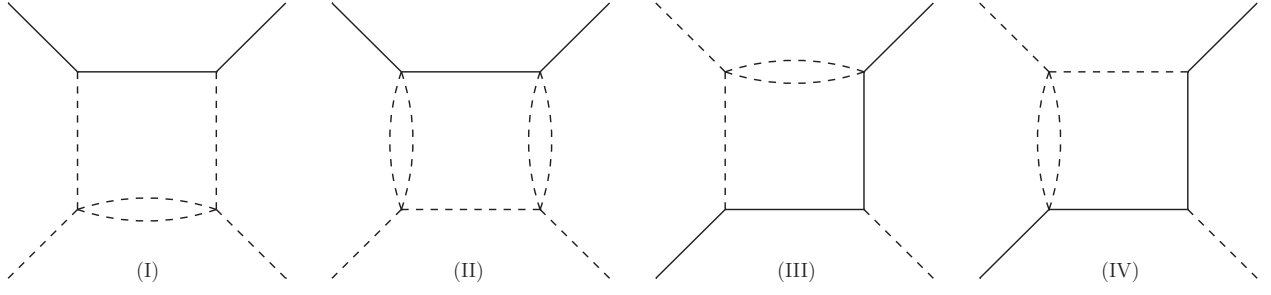}
\caption{Box graphs constraining the boundaries of the double-spectral regions for $\pi N$ scattering. Solid/dashed lines denote nucleons/pions.}
\label{fig:boundary_piN}
\end{figure}

The unsubtracted set of HDRs for the $\pi N$ amplitudes as derived in~\cite{Hite:1973pm} reads
\begin{align}
\label{hdr}
A^+(s,t)&=\frac{1}{\pi}\int\limits_{s_+}^\infty\diff s'\left[\frac{1}{s'-s}+\frac{1}{s'-u}-\frac{1}{s'-a}\right]\Im A^+(s',t')
 +\frac{1}{\pi}\int\limits_{\tpi}^\infty\diff t'\;\frac{\Im A^+(s',t')}{t'-t},\notag\\
A^-(s,t)&=\frac{1}{\pi}\int\limits_{s_+}^\infty\diff s'\left[\frac{1}{s'-s}-\frac{1}{s'-u}\right]\Im A^-(s',t')
 +\frac{1}{\pi}\int\limits_{\tpi}^\infty\diff t'\;\frac{\nu}{\nu'}\frac{\Im A^-(s',t')}{t'-t},\notag\\
B^+(s,t)&=N^+(s,t)+\frac{1}{\pi}\int\limits_{s_+}^\infty\diff s'\left[\frac{1}{s'-s}-\frac{1}{s'-u}\right]\Im B^+(s',t')
 +\frac{1}{\pi}\int\limits_{\tpi}^\infty\diff t'\;\frac{\nu}{\nu'}\frac{\Im B^+(s',t')}{t'-t},\notag\\
B^-(s,t)&=N^-(s,t)+\frac{1}{\pi}\int\limits_{s_+}^\infty\diff s'\left[\frac{1}{s'-s}+\frac{1}{s'-u}-\frac{1}{s'-a}\right]\Im B^-(s',t')
+\frac{1}{\pi}\int\limits_{\tpi}^\infty\diff t'\;\frac{\Im B^-(s',t')}{t'-t},
\end{align}
cf.~\eqref{HDR_cross_even} and~\eqref{HDR_cross_odd}, with Born-term contributions\footnote{Note that we dropped the $1/(\mN^2-a)$ term in $N^-(s,t)$ from~\cite{Hite:1973pm}. Such a term indeed appears when taking the residue in the derivation of the HDRs, but is canceled by the dispersive integral over the circle at infinity, since the Born terms do not vanish sufficiently fast that this contribution can be ignored. A similar issue was already observed before  in the context of the Bernabeu--Tarrach sum rule, see~\cite{Bernabeu:1974zu,Bernabeu:1977hp}.}
\beq
\label{hdrnucleonpoleterms}
N^+(s,t)=g^2\left[\frac{1}{\mN^2-s}-\frac{1}{\mN^2-u}\right],\qquad
N^-(s,t)=g^2\left[\frac{1}{\mN^2-s}+\frac{1}{\mN^2-u}\right].
\eeq
We will take $g^2/4\pi=13.7\pm 0.2$ as derived from the Goldberger--Miyazawa--Oehme (GMO) sum rule~\cite{Goldberger:1955zza} with input for the scattering lengths from pionic atoms~\cite{Baru:2010xn,Baru:2011bw}.
Before turning to the explicit structure of the RS equations, we briefly comment on the range of convergence of the final system. Assuming Mandelstam analyticity, the investigation of the box diagrams depicted in Fig.~\ref{fig:boundary_piN} shows that for the $s$-channel
\begin{align}
\label{a_schannel}
a=-23.2\mpi^2 \qquad&\Rightarrow\qquad
s\in\big[s_+=(\mN+\mpi)^2,97.3\mpi^2\big]=\big[59.6\mpi^2,97.3\mpi^2\big],\notag\\
&\Leftrightarrow\qquad
W\in[W_+=1.08\GeV,1.38\GeV],
\end{align}
 and for the $t$-channel
\begin{align}
\label{a_tchannel}
a=-2.7\mpi^2 \qquad&\Rightarrow\qquad
t\in[\tpi=4\mpi^2,205.5\mpi^2],\notag\\
&\Leftrightarrow\qquad
\sqrt{t}\in[\sqrt{\tpi}=0.28\GeV,2.00\GeV],
\end{align}
yield the largest domain of validity (for details of the derivation we refer to~\cite{Ditsche:2012fv}). Accordingly, the hyperbola parameter $a$ will be constrained to the values given in~\eqref{a_schannel} and \eqref{a_tchannel} for the $s$- and $t$-channel part of the system, respectively.

\subsection{Partial-wave projection}

The $s$-channel partial-wave amplitudes $f^I_{l\pm}(W)$ with isospin index $I=\pm$, orbital angular momentum $l$, and total angular momentum $j=|l\pm1/2|$ are defined as~\cite{Frazer:1960zz}
\begin{align}
\label{sprojform}
f^I_{l\pm}(W)&=\frac{1}{16\pi W}\Big\{(E+\mN)\big[A^I_l(s)+(W-\mN)B^I_l(s)\big]
+(E-\mN)\big[-A^I_{l\pm1}(s)+(W+\mN)B^I_{l\pm1}(s)\big]\Big\},
\end{align}
where
\beq
\label{sprojinvampl}
X^I_l(s)=\int\limits_{-1}^1\diff z_s\;P_l(z_s)X^I(s,t)\Big|_{t=-2q^2(1-z_s)}, \qquad X\in\{A,B\}.
\eeq
They fulfill the MacDowell symmetry relation~\cite{MacDowell:1959zza} in the complex $W$-plane
\beq
\label{macdowell}
f^I_{l+}(W)=-f^I_{(l+1)-}(-W),
\eeq
which allows us to restrict the analysis to $f^I_{l+}(W)$.
Performing the partial-wave expansion of the integrands of~\eqref{hdr} and subsequent projection onto the $s$-channel partial waves yields~\cite{Hite:1973pm}
\begin{align}
\label{spwhdr}
f^I_{l+}(W)&=N^I_{l+}(W)
+\frac{1}{\pi}\int\limits^\infty_{\tpi}\diff t'\;\sum\limits_J
 \Big\{G_{lJ}(W,t')\,\Im f^J_+(t')+H_{lJ}(W,t')\,\Im f^J_-(t')\Big\}\notag\\
&+\frac{1}{\pi}\int\limits^\infty_{W_+}\diff W'\;\sum\limits_{l'=0}^\infty
 \Big\{K^I_{ll'}(W,W')\,\Im f^I_{l'+}(W')+K^I_{ll'}(W,-W')\,\Im f^I_{(l'+1)-}(W')\Big\},
\end{align}
where $N^I_{l\pm}(W)$ corresponds to the projection of the nucleon pole terms. Each partial wave $f^I_{l\pm}(W)$ is coupled to all other $s$-channel partial waves by means of the kernel functions $K^I_{ll'}(W,W')$. Moreover, the equations involve the $t$-channel partial waves $f^J_\pm(t)$, with subscript $\pm$ for parallel/antiparallel antinucleon--nucleon helicities and total momentum $J$, via the kernels $G_{lJ}(W,t')$ and $H_{lJ}(W,t')$. The summation is restricted to even/odd values of $J$ for $I=\pm$.
Explicit expressions for $N^I_{l\pm}(W)$ and the kernel functions are provided in~\ref{app:kernels_schannel}.

The $t$-channel partial-wave projection reads~\cite{Frazer:1960zza}
\begin{align}
\label{tprojform}
f^J_+(t)&=-\frac{1}{4\pi}\int\limits^1_0\diff z_t\;P_J(z_t)\bigg\{\frac{p_t^2}{(p_tq_t)^J}A^I(s,t)\Big|_{s=s(t,z_t)}-\frac{\mN}{(p_tq_t)^{J-1}}z_tB^I(s,t)\Big|_{s=s(t,z_t)}\bigg\},\notag\\
f^J_-(t)&=\frac{1}{4\pi}\frac{\sqrt{J(J+1)}}{2J+1}\frac{1}{(p_tq_t)^{J-1}}\int\limits^1_0\diff z_t\Big[P_{J-1}(z_t)-P_{J+1}(z_t)\Big]B^I(s,t)\Big|_{s=s(t,z_t)},
\end{align}
where the integration has been restricted to half the angular interval by virtue of Bose symmetry, and again $I=\pm$ corresponds to even/odd $J$. The result for the $t$-channel equations takes the form
\begin{align}
\label{tpwhdr}
f^J_+(t)&=\tilde N^J_+(t)+\frac{1}{\pi}\int\limits^{\infty}_{W_+}\diff W'\sum\limits^\infty_{l=0}\Big\{
\tilde G_{J l}(t,W')\,\Im f^I_{l+}(W')+\tilde G_{J l}(t,-W')\,\Im f^I_{(l+1)-}(W')\Big\}\notag\\
&\qquad+\frac{1}{\pi}\int\limits^{\infty}_{\tpi}\diff t'\sum\limits_{J'}\Big\{
\tilde K^1_{J J'}(t,t')\,\Im f^{J'}_+(t')+\tilde K^2_{J J'}(t,t')\,\Im f^{J'}_-(t')\Big\},\notag\\
f^J_-(t)&=\tilde N^J_-(t)+\frac{1}{\pi}\int\limits^{\infty}_{W_+}\diff W'\sum\limits^\infty_{l=0}\Big\{
\tilde H_{J l}(t,W')\,\Im f^I_{l+}(W')+\tilde H_{J l}(t,-W')\,\Im f^I_{(l+1)-}(W')\Big\}\notag\\
&\qquad+\frac{1}{\pi}\int\limits^{\infty}_{\tpi}\diff t'\sum\limits_{J'}\tilde K^3_{J J'}(t,t')\,\Im f^{J'}_-(t'),
\end{align}
where the sum over $J'$ runs over even/odd values of $J'$ if $J$ is even/odd.
In contrast to the $s$-channel equations, we have $\tilde K^{1,2,3}_{JJ'}(t,t')=0$ for all $J'<J$, which implies that a given $t$-channel partial wave only receives contributions from the absorptive parts of higher partial waves. Explicit expressions for Born terms and kernel functions are relegated to~\ref{app:kernels_tchannel}.

In order to combine analyticity with partial-wave unitarity, the equations have to be cast into a form that permits diagonal $s$-channel
partial-wave unitarity relations (see Sect.~\ref{sec:RS:unirel}). Therefore, we work in the  $s$-channel isospin basis $I_s\in\{1/2,3/2\}$ instead of the $I=\pm$ basis. Using the definitions
\beq
\begin{pmatrix}X^{1/2}\\X^{3/2}\end{pmatrix}=C_{s\nu}\begin{pmatrix}X^+\\X^-\end{pmatrix}, \qquad
\begin{pmatrix}X^+\\X^-\end{pmatrix}=C_{\nu s}\begin{pmatrix}X^{1/2}\\X^{3/2}\end{pmatrix}, \qquad
X\in\{f_{l\pm},N_{l\pm},K_{ll'}\},
\eeq
and
\beq
K_{ll'}^{1/2+3/2}(W,W')=K_{ll'}^{1/2}(W,W')+K_{ll'}^{3/2}(W,W')=2K_{ll'}^+(W,W')+K_{ll'}^-(W,W'),
\eeq
the complete system of RS equations becomes
\begin{align}
\label{sRSpwhdr}
f^{1/2}_{l+}(W)&=N^{1/2}_{l+}(W)
+\frac{1}{\pi}\int\limits^{\infty}_{W_+}\diff W'\sum\limits^\infty_{l'=0}\frac{1}{3}
\Big\{K_{ll'}^{1/2}(W,W')\,\Im f^{1/2}_{l'+}(W')+2K_{ll'}^{3/2}(W,W')\,\Im f^{3/2}_{l'+}(W')\notag\\
&\qquad+K_{ll'}^{1/2}(W,-W')\,\Im f^{1/2}_{(l'+1)-}(W')+2K_{ll'}^{3/2}(W,-W')\,\Im f^{3/2}_{(l'+1)-}(W')\Big\}\notag\\
&+\frac{1}{\pi}\int\limits^{\infty}_{\tpi}\diff t'\sum\limits_{J=0}^{\infty}\frac{\big(3-(-1)^J\big)}{2}
\Big\{G_{lJ}(W,t')\,\Im f^J_+(t')+H_{lJ}(W,t')\,\Im f^J_-(t')\Big\},\notag\\
f^{3/2}_{l+}(W)&=N^{3/2}_{l+}(W)
+\frac{1}{\pi}\int\limits^{\infty}_{W_+}\diff W'\sum\limits^\infty_{l'=0}\frac{1}{3}
\Big\{K_{ll'}^{3/2}(W,W')\,\Im f^{1/2}_{l'+}(W')+K_{ll'}^{1/2+3/2}(W,W')\,\Im f^{3/2}_{l'+}(W')\notag\\
&\qquad+K_{ll'}^{3/2}(W,-W')\,\Im f^{1/2}_{(l'+1)-}(W')+K_{ll'}^{1/2+3/2}(W,-W')\,\Im f^{3/2}_{(l'+1)-}(W')\Big\}\notag\\
&+\frac{1}{\pi}\int\limits^{\infty}_{\tpi}\diff t'\sum\limits_{J=0}^{\infty}(-1)^J
\Big\{G_{lJ}(W,t')\,\Im f^J_+(t')+H_{lJ}(W,t')\,\Im f^J_-(t')\Big\},\notag\\
f^{1/2}_{(l+1)-}(W)&=N^{1/2}_{(l+1)-}(W)
-\frac{1}{\pi}\int\limits^{\infty}_{W_+}\diff W'\sum\limits^\infty_{l'=0}\frac{1}{3}
\Big\{K_{ll'}^{1/2}(-W,W')\,\Im f^{1/2}_{l'+}(W')+2K_{ll'}^{3/2}(-W,W')\,\Im f^{3/2}_{l'+}(W')\notag\\
&\qquad+K_{ll'}^{1/2}(-W,-W')\,\Im f^{1/2}_{(l'+1)-}(W')+2K_{ll'}^{3/2}(-W,-W')\,\Im f^{3/2}_{(l'+1)-}(W')\Big\}\notag\\
&-\frac{1}{\pi}\int\limits^{\infty}_{\tpi}\diff t'\sum\limits_{J=0}^{\infty}\frac{\big(3-(-1)^J\big)}{2}
\Big\{G_{lJ}(-W,t')\,\Im f^J_+(t')+H_{lJ}(-W,t')\,\Im f^J_-(t')\Big\},\notag\\
f^{3/2}_{(l+1)-}(W)&=N^{3/2}_{(l+1)-}(W)
-\frac{1}{\pi}\int\limits^{\infty}_{W_+}\diff W'\sum\limits^\infty_{l'=0}\frac{1}{3}
\Big\{K_{ll'}^{3/2}(-W,W')\,\Im f^{1/2}_{l'+}(W')+K_{ll'}^{1/2+3/2}(-W,W')\,\Im f^{3/2}_{l'+}(W')\notag\\
&\qquad+K_{ll'}^{3/2}(-W,-W')\,\Im f^{1/2}_{(l'+1)-}(W')+K_{ll'}^{1/2+3/2}(-W,-W')\,\Im f^{3/2}_{(l'+1)-}(W')\Big\}\notag\\
&-\frac{1}{\pi}\int\limits^{\infty}_{\tpi}\diff t'\sum\limits_{J=0}^{\infty}(-1)^J
\Big\{G_{lJ}(-W,t')\,\Im f^J_+(t')+H_{lJ}(-W,t')\,\Im f^J_-(t')\Big\},
\end{align}
and
\begin{align}
\label{tRSpwhdr}
f^J_+(t)&=\tilde N^J_+(t)+\frac{1}{\pi}\int\limits^{\infty}_{W_+}\diff W'\sum\limits^\infty_{l=0}\frac{1}{3}
\bigg\{\tilde G_{J l}(t,W')\bigg[\Im f^{1/2}_{l+}(W')+\frac{1+3(-1)^J}{2}\Im f^{3/2}_{l+}(W')\bigg]\notag\\
&\qquad+\tilde G_{J l}(t,-W')\bigg[\Im f^{1/2}_{(l+1)-}(W')+\frac{1+3(-1)^J}{2}\Im f^{3/2}_{(l+1)-}(W')\bigg]\bigg\}\notag\\
&+\frac{1}{\pi}\int\limits^{\infty}_{\tpi}\diff t'\sum\limits^\infty_{J'=J}\frac{1+(-1)^{J+J'}}{2}
\Big\{\tilde K^1_{J J'}(t,t')\,\Im f^{J'}_+(t')+\tilde K^2_{J J'}(t,t')\,\Im f^{J'}_-(t')\Big\},\notag\\
f^J_-(t)&=\tilde N^J_-(t)+\frac{1}{\pi}\int\limits^{\infty}_{W_+}\diff W'\sum\limits^\infty_{l=0}\frac{1}{3}
\bigg\{\tilde H_{J l}(t,W')\bigg[\Im f^{1/2}_{l+}(W')+\frac{1+3(-1)^J}{2}\Im f^{3/2}_{l+}(W')\bigg]\notag\\
&\qquad+\tilde H_{J l}(t,-W')\bigg[\Im f^{1/2}_{(l+1)-}(W')+\frac{1+3(-1)^J}{2}\Im f^{3/2}_{(l+1)-}(W')\bigg]\bigg\}\notag\\
&+\frac{1}{\pi}\int\limits^{\infty}_{\tpi}\diff t'\sum\limits^\infty_{J'=J}\frac{1+(-1)^{J+J'}}{2}
\tilde K^3_{J J'}(t,t')\,\Im f^{J'}_-(t').
\end{align}
In these equations, all sums run over both even and odd values, but those over $J'$ in the $t$-channel part~\eqref{tRSpwhdr} are restricted to $J'\geq J$.
The equations for $f^{I}_{(l+1)-}(W)$ are given here merely for convenience, as they could be obtained by means of MacDowell symmetry~\eqref{macdowell}.

\subsection{Subtractions}
\label{sec:piN_subtractions}

The introduction of subtractions is essential to suppress the high-energy regime and derive constraints that solely involve low-energy physics. The most convenient choice for the subtraction point is provided by the subthreshold point at $(\nu=0,t=0)$, since the relation to the Cheng--Dashen point and the $\pi N$ $\sigma$-term can be established in a straightforward manner, while the subtraction constants are given directly in terms of subthreshold parameters. In addition, this choice of the subtraction point proves favorable for matching to ChPT, which is expected to be most reliable in the subthreshold region.   

The subthreshold expansion in $\pi N$ scattering is conventionally applied to the pseudovector-Born-term-subtracted amplitudes
\begin{align}
\label{def_PV_pole}
\bar A^+(s,t)&=A^+(s,t)-\frac{g^2}{\mN},\qquad & 
\bar B^+(s,t)&=B^+(s,t)-g^2\left[\frac{1}{\mN^2-s}-\frac{1}{\mN^2-u}\right],\notag\\
\bar A^-(s,t)&=A^-(s,t), \qquad& 
\bar B^-(s,t)&=B^-(s,t)-g^2\left[\frac{1}{\mN^2-s}+\frac{1}{\mN^2-u}\right]+\frac{g^2}{2\mN^2}.
\end{align}
Separating factors of $\nu$ that are required by crossing symmetry, these amplitudes permit the expansions
\begin{align}
\bar A^+(\nu,t)&=\sum\limits_{m,n=0}^\infty a_{mn}^+\nu^{2m}t^{n}, & 
\bar B^+(\nu,t)&=\sum\limits_{m,n=0}^\infty b_{mn}^+\nu^{2m+1}t^{n},\notag\\
\bar A^-(\nu,t)&=\sum\limits_{m,n=0}^\infty a_{mn}^-\nu^{2m+1}t^{n}, & 
\bar B^-(\nu,t)&=\sum\limits_{m,n=0}^\infty b_{mn}^-\nu^{2m}t^{n},
\end{align}
and similarly for $\bar D^\pm=\bar A^\pm+\nu\bar B^\pm$. The corresponding subthreshold parameters fulfill the relations
\beq
d_{mn}^+=a_{mn}^++b_{m-1,n}^+, \qquad d_{mn}^-=a_{mn}^-+b_{mn}^-,\qquad d_{0n}^+=a_{0n}^+.
\eeq

The implementation of subtractions into the RS system proceeds as follows: expanding~\eqref{hdr} around $(\nu=0,t=0)$ and equating the coefficients to the subthreshold expansion yields sum rules for the subthreshold parameters, see \ref{app:subthr_sum_rules} for explicit expressions. 
Once subtracted from~\eqref{hdr}, these sum rules lead to a subtracted version of the HDRs and, after the expansion into partial waves, of the RS equations. For instance, if only $d_{00}^+$ were introduced as subtraction constant, the HDR for $A^+$ would become
\begin{align}
A^+(s,t)&=d_{00}^++\frac{g^2}{\mN}
+\frac{1}{\pi}\int\limits_{\tpi}^\infty\diff t'
\Bigg\{\frac{\Im A^+(s',t')}{t'-t}-\frac{1}{t'}\ste{\Im A^+(s',t')}\Bigg\}\notag\\
&+\frac{1}{\pi}\int\limits_{s_+}^\infty\diff s'
\Bigg\{\bigg[\frac{1}{s'-s}+\frac{1}{s'-u}-\frac{1}{s'-a}\bigg]\Im A^+(s',t')
-\bigg[\frac{2}{s'-s_0}-\frac{1}{s'-a}\bigg]\ste{\Im A^+(s',t')}\Bigg\},
\end{align}
where the subscript $(0,0)$ indicates evaluation at $(\nu=0,t=0)$. 
The corresponding modifications of the kernel functions required by 
the new terms in the dispersive integrals, for the number of subtractions used in practice, are listed in \ref{app:kernel_subtractions}.

For the $s$-channel equations we consider a system with $10$ subtraction constants, which can be identified with the parameters listed in Table~\ref{table:subtractions_piN}.
The motivation for this choice is provided by the mathematical properties of the RS equations, i.e.\ general statements about existence and uniqueness of solutions, see Sect.~\ref{sec:existence_uniqueness}. 
Table~\ref{table:subtractions_piN} also delineates the dimension of the various subthreshold parameters, typically given in inverse powers of $\mpi$. 
For the $t$-channel problem we use a variant of the system with two subtractions less ($a_{10}^-$ and $b_{10}^-$ are not included). The reason for this choice is that implementing those two parameters does not further improve the convergence of the PWDRs for the $t$-channel partial waves, so that their inclusion does not provide a numerical advantage.

\begin{table}
\centering
\renewcommand{\arraystretch}{1.3}
\begin{tabular}{cl}
\toprule
dimension & subthreshold parameters\\\midrule
$\mpi^{-1}$ & $a_{00}^+=d_{00}^+$ \\
$\mpi^{-2}$ & $a_{00}^-=d_{00}^--b_{00}^-$, $b_{00}^-$\\
$\mpi^{-3}$ & $a_{01}^+=d_{01}^+$, $a_{10}^+=d_{10}^+-b_{00}^+$, $b_{00}^+$\\
$\mpi^{-4}$ & $a_{01}^-=d_{01}^--b_{01}^-$, $a_{10}^-=d_{10}^--b_{10}^-$, $b_{01}^-$, $b_{10}^-$\\
\bottomrule
\end{tabular}
\caption{Subthreshold parameters at different dimensions.}
\label{table:subtractions_piN}
\renewcommand{\arraystretch}{1.0}
\end{table}

\subsection{Partial-wave unitarity relations}
\label{sec:RS:unirel}

Unitarity constraints are most conveniently expressed in terms of partial waves, as separate relations for a given angular momentum $J$.  
In the helicity formalism, the general partial-wave expansion for a process $a+b\to c+d$ with particle helicities $\lambda_P$ reads~\cite{Jacob:1959at}
\beq
\label{diffmodJWpwe}
T^{\lambda_c,\lambda_d;\lambda_a,\lambda_b}_{fi}(s,t)=\sqrt{S_fS_i}16\pi\sum\limits_J(2J+1)T^J_{\lambda_c,\lambda_d;\lambda_a,\lambda_b}(s)d^J_{\lambda_a-\lambda_b,\lambda_c-\lambda_d}(\theta),
\eeq
with CMS scattering angle $\theta$, azimuthal angle $\varphi$ set to zero, Wigner $d$-functions $d^J_{m,m'}(\theta)$, and final-/initial-state symmetry factors $S_f$ and $S_i$.
The ensuing partial-wave unitarity relations become particularly simple if the matrix $T^J(s)$ in helicity space is diagonal, which e.g.\ for $\pi N$ scattering holds true for the $s$-channel isospin basis $I_s\in\{1/2,3/2\}$, but not for the $I=\pm$ basis. 
Elastic unitarity then yields for each diagonal element $T^J_{\lambda}(s)$ of $T^J(s)$
\beq
\label{elpwunitrel}
\Im T^J_{\lambda}(s)=\frac{2|\pp|}{\sqrt{s}}\big|T^J_{\lambda}(s)\big|^2
\eeq
(with CMS momentum $\pp$). In this normalization, $T^J_{\lambda}(s)$ may be expressed in terms of phase shifts $\delta^J_{\lambda}(s)$ according to
\beq
\label{elpwunitnorm}
T^J_{\lambda}(s)=\frac{\sqrt{s}}{2|\pp|}\sin\delta^J_{\lambda}(s)e^{i\delta^J_{\lambda}(s)},
\eeq 
and related to the $S$-matrix elements by
\beq
S^J_{\lambda}(s)=e^{2i\delta^J_{\lambda}(s)}=1+i\frac{4|\pp|}{\sqrt{s}}T^J_{\lambda}(s).
\eeq
Inelastic contributions are accounted for by introducing inelasticity parameters ${0\leq\eta^J_{\lambda}(s)\leq1}$ into these parameterizations. 

Accordingly, the partial waves for $s$-channel $\pi N$ scattering are conventionally parameterized in terms of phase shifts $\delta^{I_s}_{l\pm}$ and inelasticities $\eta^{I_s}_{l\pm}$ as
\beq
\label{eq:s-uni}
f^{I_s}_{l\pm}(W)=\frac{\big[S^{I_s}_{l\pm}(W)\big]_{\pi N\to\pi N}-1}{2i|\qq|}=\frac{\eta^{I_s}_{l\pm}(W)e^{2i\delta^{I_s}_{l\pm}(W)}-1}{2i|\qq|}
\overset{W<W_\text{inel}}{=}\frac{e^{i\delta^{I_s}_{l\pm}(W)}\sin\delta^{I_s}_{l\pm}(W)}{|\qq|},
\eeq
where the first inelastic threshold due to $\pi\pi N$ intermediate states occurs at $W_\text{inel}=W_++\mpi$. In this normalization, the unitarity relation becomes
\beq
\label{selunitrel}
\Im f^{I_s}_{l\pm}(W)=|\qq|\big|f^{I_s}_{l\pm}(W)\big|^2\,\theta\big(W-W_+\big)+\frac{1-\big(\eta^{I_s}_{l\pm}(W)\big)^2}{4|\qq|}\,\theta\big(W-W_\text{inel}\big).
\eeq

The derivation of the $t$-channel unitarity relations is complicated by the fact that the corresponding reaction is necessarily inelastic. For instance, the contribution from $\pi\pi$ intermediate states alone yields
\beq
\label{tunitrel}
\Im f^J_{\pm}(t)=\sigma^\pi_t\big(t^{I_t}_J(t)\big)^*f^J_{\pm}(t)\,\theta\big(t-\tpi\big)
\eeq
(with $\pi\pi$ partial waves $t^{I_t}_J(t)$), which proves Watson's theorem~\cite{Watson:1954uc} for $f^J_{\pm}(t)$ (the phase of $f^J_{\pm}(t)$ equals the $\pi\pi$ phase up to integer multiples of $\pi$). However, \eqref{tunitrel} is invariant under rescaling of $f^J_{\pm}(t)$ with arbitrary real factors, so that the correct normalization and relation to the $S$-matrix element can only be inferred relative to the pertinent elastic reactions. In~\ref{app:tchannel_unitarity} this procedure is illustrated for a coupled system of $\pi\pi$, $\bar K K$, and $\bar N N$ states.
The relations~\eqref{tunitrel} provide the basis for the single-channel treatment of the $t$-channel $P$- and $D$-waves, while for the $S$-wave we use the coupled-channel extension discussed in~\ref{app:tchannel_unitarity}.

\subsection{Existence and uniqueness of solutions}
\label{sec:existence_uniqueness}

The mathematical properties of $\pi\pi$ Roy equations have been investigated in great detail in the literature~\cite{Epele:1977un,Epele:1977um,Gasser:1999hz,Wanders:2000mn}. The central result is that the existence and uniqueness of solutions can be characterized by a single number, the multiplicity index $m$ as defined in~\eqref{m_index}, which is solely determined by the value of the phase shift at the matching point $\Wm=\sqrt{\sm}$. In the multi-channel case, the multiplicity of the full system can be found by adding the multiplicity indices for each partial wave included in the system.
A unique solution corresponds to $m=0$, while $m>0$ implies that there is an $m$-dimensional space of solutions and additional input is required to identify the physical one. If $m<0$, there is in general no solution, only if $|m|$ constraints on the input are fulfilled can a solution be found. From a practical point of view, the ideal situation therefore appears for $m=0$. Since these results only depend on the mathematical properties of the Roy equations, not the specific form of the driving terms (higher partial waves, imaginary parts above $\Wm$, and, in the RS case, $t$-channel integrals), they apply to the case of $\pi N$ RS equations after the $t$-channel problem has been solved.

Besides these mathematical properties, a physical solution is characterized by the requirement of a smooth matching, i.e.\ the phase shift at the matching point must not display a cusp. This no-cusp condition provides an additional constraint for each partial wave included in the solution. It is instructive to see how these considerations affect the final number of degrees of freedom in previous Roy-equation studies. In the $\pi\pi$ analysis of~\cite{Ananthanarayan:2000ht} the matching point, $\sqrt{\sm}=0.8\GeV$, was chosen in such a way that $m=0$. With three partial waves considered, isospin-$0$ and -$2$ $S$-waves and the $P$-wave, one would thus actually expect three additional conditions on the two subtraction constants, the isospin-$0$ and -$2$ $S$-wave scattering lengths, from the no-cusp conditions. Indeed, tuning away the $P$-wave cusp reproduces the ``universal band,'' a relation between the two scattering lengths discovered much earlier~\cite{Morgan:1969ca,Morgan:1970zza}, but the remaining $S$-wave cusps are too small to provide a meaningful constraint. The reason for this behavior could be identified as the large separation between threshold and matching point, made possible by the comparatively large domain of validity of $\pi\pi$ Roy equations. For a typical choice of the matching point the no-cusp conditions are too weak for a precision determination of the subtraction constants, so that 
matching to ChPT was required as an additional step to pin down the scattering lengths to the accuracy quoted in~\eqref{pipi_scatt_lengths}.

In the $\pi K$ RS analysis of~\cite{Buettiker:2003pp} the matching point was chosen at $\sqrt{\sm}=0.97\GeV$, right at the border of the domain of validity of the equations, and thus necessarily much closer to threshold. The multiplicity index for a system of isospin-$1/2$ and -$3/2$ $S$-waves and isospin-$1/2$ $P$-wave is again $m=0$, but the constraints from the no-cusp conditions were observed to be much stronger than in $\pi\pi$: to remove the cusps in the isospin-$1/2$ partial waves both subtraction constants (identified with the isospin-$1/2$ and -$3/2$ $S$-wave scattering lengths) needed to be tuned, and even then a small remaining cusp survived in the isospin-$3/2$ $S$-wave. To exactly fulfill all no-cusp conditions, a third subtraction constant would have had to be introduced. In practice, however, the corresponding phase shift above $\sm$ is not known sufficiently precisely to actually render the remaining cusp relevant. Given the fact that the constraints from ChPT are much less rigorous than in $\pi\pi$ scattering~\cite{Ananthanarayan:2000cp,Ananthanarayan:2001uy,Buettiker:2003pp,Bernard:1990kx,Bernard:1990kw,Bijnens:2004bu}, the final results for the $\pi K$ scattering lengths were solely derived from the no-cusp conditions.

In our $\pi N$ RS analysis, we aim for a solution of the $S$- and $P$-waves, in total six amplitudes. The multiplicity index as a function of the matching point is shown in Table~\ref{tab:m} (using the phase shifts from~\cite{Arndt:2006bf}). For the maximally allowed value $\Wm=1.38\GeV$ we therefore have $m=-2$, and, in addition, there are six more constraints from the no-cusp conditions, since,
given that the matching point is still relatively low and that the data situation in $\pi N$ is much better than in $\pi K$, there is no justification for not implementing all of them.
To make sure that a unique solution exists, we would therefore need to formulate a version of the RS equations with $8$ free parameters.
However, due to the larger number of partial waves and free parameters, we found that the solution is stabilized substantially when the $S$-wave scattering lengths are imposed as constraints on the system instead of trying to predict them from the RS solution. Since the scattering lengths are already known very precisely from pionic atoms, see Sect.~\ref{sec:pionic_atom}, a prediction from the RS solution, if it could be extracted in a reliable manner at all, would be extremely unlikely to be able to compete in accuracy. 
For these reasons, we consider a RS system with $10$ subtraction constants, as anticipated in Sect.~\ref{sec:piN_subtractions}, which matches exactly the number of degrees of freedom for $\Wm=1.38\GeV$ once 
the scattering lengths are implemented as additional constraints.

\begin{table}
\renewcommand{\arraystretch}{1.3}
\centering
\begin{tabular}{crrrrrrr}
\toprule
 & $m_{f_{0+}^{1/2}}$ & $m_{f_{0+}^{3/2}}$ & $m_{f_{1+}^{1/2}}$ & $m_{f_{1+}^{3/2}}$ & $m_{f_{1-}^{1/2}}$ & $m_{f_{1-}^{3/2}}$ & $m$ \\
\midrule
$W_+\leq \Wm\leq 1.20\GeV$ & $0$ & $-1$ & $-1$ & $0$ & $-1$ & $-1$ & $-4$\\
$1.20\GeV\leq \Wm\leq 1.23\GeV$ & $0$ & $-1$ & $-1$ & $0$ & $0$ & $-1$ & $-3$\\
$1.23\GeV\leq \Wm\leq 1.52\GeV$ & $0$ & $-1$ & $-1$ & $1$ & $0$ & $-1$ & $-2$\\
$1.52\GeV\leq \Wm\leq 1.69\GeV$ & $0$ & $-1$ & $-1$ & $1$ & $1$ & $-1$ & $-1$\\
$1.69\GeV\leq \Wm\leq 1.80\GeV$ & $1$ & $-1$ & $-1$ & $1$ & $1$ & $-1$ & $0$\\
\bottomrule
\end{tabular}
\caption{Multiplicity index $m$ as a function of $\Wm$.}
\label{tab:m}
\renewcommand{\arraystretch}{1.0}
\end{table}

\section{Solution of the $\boldsymbol{t}$-channel equations}
\label{sec:tchannel_sol}

Given that data in the $t$-channel reaction $\pi\pi\to\bar N N$ become available only above the two-nucleon threshold, the amplitudes in the pseudophysical region $\tpi\leq t\leq \tN$ required for the $t$-channel integrals need to be reconstructed from unitarity. While for every partial wave $\pi\pi$ intermediate states generate by far the dominant contribution, significant inelasticities do emerge. Most relevant are $\bar K K$ intermediate states related to the $f_0(980)$ in the $S$-wave, but in all partial waves inelasticities from the $4\pi$ channel start setting in before the two-nucleon threshold is reached. An explicit coupled-channel framework is only feasible if the corresponding $S$-matrix is known sufficiently accurately, a requirement that in the present application is only  met for the $\pi\pi/\bar K K$ $S$-wave system. For this reason, we adopt a single-channel framework for $P$- and $D$-waves, estimating the impact of $4\pi$ inelasticities by appropriate variations of the input. Similarly, we include the $\bar K K$ channel explicitly in the $S$-wave, while accounting for effects from higher channels in the uncertainty estimate.  

\subsection{Single-channel Muskhelishvili--Omn\`es solution}

In the single-channel approximation where only $\pi\pi$ intermediate states are considered in the unitarity relation the MO solution for $f^J_\pm(t)$ can be derived from the RS equations~\eqref{tRSpwhdr} in the following way. First, we rearrange the equations in such a way that the behavior of $f^J_\pm(t)$ at the two-nucleon threshold is properly taken into account.
Starting from the partial-wave expansion~\eqref{tprojform} one can show that the $S$-wave vanishes for $t\to\tN$ according to
\beq
\label{threshold_f0p}
f^0_+(t)=\Order\big(p_t^2\big).
\eeq  
Although higher partial waves with $J\geq 1$ individually take a finite value,
the linear combination
\beq
\label{Gamma_def}
\Gamma^J(t)=\mN\sqrt{\frac{J}{J+1}}f^J_-(t)-f^J_+(t)
\eeq
again vanishes at threshold~\cite{Frazer:1960zzb} 
\beq
\label{threshold_Gamma_J}
\Gamma^J(t)=\Order(p_t^2).
\eeq  
Indeed, the leading terms in the $t$-channel partial-wave expansion~\eqref{texpform}
\begin{align}
\frac{A^+(\nu,t)}{4\pi}&=-\frac{f^0_+(t)}{p_t^2}+\frac{15}{2}\mN^2\nu^2\frac{\Gamma^2(t)}{p_t^2}+\frac{5}{2}q_t^2f^2_+(t)+\dots,&
\frac{A^-(\nu,t)}{4\pi}&=3\mN\nu\frac{\Gamma^1(t)}{p_t^2}+\dots,\notag\\
\frac{B^+(\nu,t)}{4\pi}&=\frac{15}{\sqrt{6}}\mN\nu f^2_-(t)+\dots,&
\frac{B^-(\nu,t)}{4\pi}&=\frac{3}{\sqrt{2}}f^1_-(t)+\dots,
\end{align}
already demonstrate that~\eqref{threshold_f0p} and~\eqref{threshold_Gamma_J} need to hold for the invariant amplitudes to remain finite. 

Before introducing subtractions, the $t$-channel part of the RS system~\eqref{tRSpwhdr} can be recast as
\begin{align}
\label{MOGamsub0}
f^0_+(t)&=\Delta^0_+(t)
+\frac{t-\tN}{\pi}\int\limits^\infty_{\tpi}\diff t'\frac{\Im f^0_+(t')}{(t'-\tN)(t'-t)},\qquad
f^J_-(t)=\Delta^J_-(t)
+\frac{1}{\pi}\int\limits^\infty_{\tpi}\diff t'\frac{\Im f^J_-(t')}{t'-t},\notag\\
\Gamma^J(t)&=\Delta_\Gamma^J(t)
+\frac{t-\tN}{\pi}\int\limits^\infty_{\tpi}\diff t'\frac{\Im\Gamma^J(t')}{(t'-\tN)(t'-t)},\qquad
\Delta_\Gamma^J(t)=\mN\sqrt{\frac{J}{J+1}}\Delta^J_-(t)-\Delta^J_+(t),
\end{align}
where
\begin{align}
\label{Deltadef}
\Delta^J_\pm(t)&=\tilde N^J_\pm(t)+\bar\Delta^J_\pm(t),\notag\\
\bar\Delta^J_+(t)&=\frac{1}{\pi}\int\limits^{\infty}_{W_+}\diff W'\sum\limits^\infty_{l=0}
\Big\{\tilde G_{Jl}(t,W')\,\Im f^I_{l+}(W')+\tilde G_{Jl}(t,-W')\,\Im f^I_{(l+1)-}(W')\Big\}\notag\\
&\qquad+\frac{1}{\pi}\int\limits^{\infty}_{\tpi}\diff t'\sum^{\infty}\limits_{J'=J+2}\frac{1+(-1)^{J+J'}}{2}
\Big\{\tilde K^1_{JJ'}(t,t')\,\Im f^{J'}_+(t')+\tilde K^2_{JJ'}(t,t')\,\Im f^{J'}_-(t')\Big\},\notag\\
\bar\Delta^J_-(t)&=\frac{1}{\pi}\int\limits^{\infty}_{W_+}\diff W'\sum\limits^\infty_{l=0}
\Big\{\tilde H_{Jl}(t,W')\,\Im f^I_{l+}(W')+\tilde H_{Jl}(t,-W')\,\Im f^I_{(l+1)-}(W')\Big\}\notag\\
&\qquad+\frac{1}{\pi}\int\limits^{\infty}_{\tpi}\diff t'\sum^{\infty}\limits_{J'=J+2}\frac{1+(-1)^{J+J'}}{2}
\tilde K^3_{JJ'}(t,t')\,\Im f^{J'}_-(t').
\end{align}
Again, the convergence of the integrals in~\eqref{MOGamsub0} is ensured by the threshold behavior~\eqref{threshold_f0p} and~\eqref{threshold_Gamma_J}. In particular, the 
kernel functions obey the general property
\begin{align}
\Res\Big[H_{lJ}(W,t'),t'=\tN\Big]&=-\mN\sqrt{\frac{J}{J+1}}\Res\Big[G_{lJ}(W,t'),t'=\tN\Big],\notag\\
 \Res\Big[\tilde K^2_{JJ'}(t,t'),t'=\tN\Big]&=-\mN\sqrt{\frac{J'}{J'+1}}\Res\Big[\tilde K^1_{JJ'}(t,t'),t'=\tN\Big],
\end{align}
which induces the cancellation of ostensible $p_t'^{-2}$ divergences in the $t$-channel dispersive integrals in~\eqref{sRSpwhdr} and~\eqref{tRSpwhdr}.

The $t$-channel integrals in the definition of the inhomogeneities $\Delta^J_\pm(t)$~\eqref{Deltadef} show that
 the equations decouple in the sense that the integral equation for a given partial-wave amplitude only depends on the $s$-channel partial waves as well as $t$-channel partial waves with higher angular momentum. In this way, they
reduce to a form directly accessible to MO techniques. Once the solutions for $\Gamma^J(t)$ and $f^J_-(t)$ are obtained, the result for $f^J_+(t)$ can be recovered by means of~\eqref{Gamma_def}. With the Omn\`es function defined as~\cite{Omnes:1958hv}
\beq
\label{Omnes_matrix}
\Omega_J(t)=\exp\Bigg\{\frac{t}{\pi}\int\limits_{\tpi}^{\tm}\diff t'\frac{\delta_J(t')}{t'(t'-t)}\Bigg\},
\eeq
in terms of the $\pi\pi$ phase shift $\delta_J(t)$ for angular momentum $J$ (and isospin $0/1$ for even/odd $J$),
the solution for the unsubtracted case becomes
\begin{align}
\label{MO_piN_unsub}
f^0_+(t)&=\Delta^0_+(t)
+\frac{(t-\tN)\Omega_0(t)}{\pi}
\Bigg\{
\int\limits_{\tpi}^{\tm}\diff t'\frac{\Delta^0_+(t')\sin\delta_0(t')}{(t'-\tN)(t'-t)|\Omega_0(t')|}
+\int\limits_{\tm}^{\infty}\diff t'\frac{\Im f^0_+(t')}{(t'-\tN)(t'-t)|\Omega_0(t')|}
\Bigg\},\notag\\
\Gamma^J(t)&=\Delta_\Gamma^J(t)
+\frac{(t-\tN)\Omega_J(t)}{\pi}\Bigg\{\int\limits_{\tpi}^{\tm}\diff t'\frac{\Delta_\Gamma^J(t')\sin\delta_J(t')}{(t'-\tN)(t'-t)|\Omega_J(t')|}
+\int\limits_{\tm}^{\infty}\diff t'\frac{\Im\Gamma^J(t')}{(t'-\tN)(t'-t)|\Omega_J(t')|}\Bigg\},\notag\\
f^J_-(t)&=\Delta^J_-(t)
+\frac{\Omega_J(t)}{\pi}\Bigg\{\int\limits_{\tpi}^{\tm}\diff t'\frac{\Delta^J_-(t')\sin\delta_J(t')}{(t'-t)|\Omega_J(t')|}
+\int\limits_{\tm}^{\infty}\diff t'\frac{\Im f^J_-(t')}{(t'-t)|\Omega_J(t')|}
\Bigg\}.
\end{align}
By separating the integrals into their principal-value and imaginary parts one can show explicitly that these representations fulfill Watson's theorem~\cite{Watson:1954uc}, i.e.\ the phase of $f^J_\pm(t)$ coincides with $\delta_J(t)$. In the practical application the uncertainties below $\tm=\tN$ are much larger than the contribution from the imaginary part above $\tm$, so that we will drop the second integral everywhere. Based on a PWA of available $\pi\pi\to\bar N N$ data~\cite{Anisovich:2011bw} we have checked explicitly that for the number of subtractions we are using these contributions are completely negligible.

Introducing the subtractions as detailed in~\ref{app:subtractions}, specializing the general expressions to $J=1,2$, and dropping the integral above $\tm$, we obtain\footnote{The kernel functions in $\Delta^J_\pm(t)$ also need to be replaced by their respective subtracted versions. In order not to unnecessarily clutter the notation, we continue to use the same symbols as before.}
\begin{align}
\label{MO_piN_sub}
\Gamma^1(t)&=\Delta_\Gamma^1(t)+\frac{p_t^2}{12\pi \mN}\Big\{a_{00}^-\big(1-t\,\dot\Omega_1(0)\big)+a_{01}^-t\Big\}\Omega_1(t)
+\frac{t^2(t-\tN)\Omega_1(t)}{\pi}\int\limits_{\tpi}^{\tm}\diff t'\frac{\Delta_\Gamma^1(t')\sin\delta_1(t')}{t'^2(t'-\tN)(t'-t)|\Omega_1(t')|},\notag\\
f^1_-(t)&=\Delta^1_-(t)+\frac{\sqrt{2}}{12\pi}\bigg\{\bigg(b_{00}^--\frac{g^2}{2\mN^2}\bigg)\big(1-t\,\dot\Omega_1(0)\big)+b_{01}^-t\bigg\}\Omega_1(t)
+\frac{t^2\Omega_1(t)}{\pi}\int\limits_{\tpi}^{\tm}\diff t'\frac{\Delta^1_-(t')\sin\delta_1(t')}{t'^2(t'-t)|\Omega_1(t')|},\notag\\
\Gamma^2(t)&=\Delta_\Gamma^2(t)+\frac{p_t^2}{30\pi \mN^2}a_{10}^+\Omega_2(t)
+\frac{t(t-\tN)\Omega_2(t)}{\pi}\int\limits_{\tpi}^{\tm}\diff t'\frac{\Delta_\Gamma^2(t')\sin\delta_2(t')}{t'(t'-\tN)(t'-t)|\Omega_2(t')|},\notag\\
f^2_-(t)&=\Delta^2_-(t)+\frac{\sqrt{6}}{60\pi \mN}b_{00}^+\Omega_2(t)
+\frac{t\,\Omega_2(t)}{\pi}\int\limits_{\tpi}^{\tm}\diff t'\frac{\Delta^2_-(t')\sin\delta_2(t')}{t'(t'-t)|\Omega_2(t')|},
\end{align}
where $\dot \Omega_J(0)$ denotes the derivative of the Omn\`es function at $t=0$.

\subsection{Two-channel Muskhelishvili--Omn\`es problem}

The $\pi\pi\to\bar N N$ and $\bar K K\to\bar N N$ $S$-waves $f^0_+(t)$ and $h^0_+(t)$ fulfill the unitarity relation
\beq
\label{tchannel_Swave}
\Im \ff(t)=T^*(t)\Sigma(t)\ff(t),\qquad 
\ff(t)=\begin{pmatrix}
               f^0_+(t)\\
\frac{2}{\sqrt{3}}h^0_+(t)
              \end{pmatrix},
\eeq
see~\ref{app:tchannel_unitarity} for its derivation. In addition, the RS equations 
 provide a dispersion relation of the form
\beq
\label{f_disp_rel}
\ff(t)=\boldsymbol{\Delta}(t)+(\mathbf{a}+\mathbf{b}\,t)(t-\tN)+\frac{t^2(t-\tN)}{\pi}\int\limits_{\tpi}^\infty\diff t'\frac{\Im\ff(t')}{t'^2(t'-\tN)(t'-t)},
\eeq
where all bold-faced quantities are two-component vectors in channel space. For instance, the subtraction constants that belong to the $\pi N$ subsystem are related to the subthreshold parameters according to
\beq
a_1=-\frac{1}{16\pi}\Bigg(d_{00}^++\frac{g^2}{\mN}+b_{00}^+\frac{\mpi^2}{3}\Bigg),\qquad 
b_1= -\frac{1}{16\pi}\Bigg(d_{01}^+-\frac{b_{00}^+}{12}\Bigg),
\eeq
and the inhomogeneity $\Delta_1(t)$ equals $\Delta^0_+(t)$ evaluated for the number of subtractions corresponding to~\eqref{f_disp_rel}. Similarly, the second components are related to $KN$ subthreshold parameters and the $S$-wave component of the $KN$ RS equations. In practice, all $KN$ amplitudes are taken as input, with nucleon Born terms replaced by the hyperon-pole contributions, and the required kernel functions can be obtained by means of sending $\mpi\to\mK$~\cite{Hoferichter:2012wf}.

The solution of~\eqref{tchannel_Swave} and~\eqref{f_disp_rel} is still accessible to MO techniques. In the multi-channel case the Omn\`es function is replaced by an Omn\`es matrix~\cite{Donoghue:1990xh,Moussallam:1999aq}, which in general cannot be given in closed form but has to be determined numerically for a given $T$-matrix. In fact,
the major challenge in the RS application concerns the generalization of these methods to the case of a finite matching point $\tm$~\cite{Hoferichter:2012wf}.
The final solution takes the form
\beq
\label{final_omnes_sol}
\ff(t)=\boldsymbol{\Delta}(t)+(t-\tN)\Omega(t)(\unity-t\,\dot\Omega(0))\mathbf{a}+t(t-\tN)\Omega(t)\mathbf{b}
-\frac{t^2(t-\tN)}{\pi}\Omega(t)\int\limits_{\tpi}^{\tm}\diff t'\frac{\Im\Omega^{-1}(t')\boldsymbol{\Delta}(t')}{t'^2(t'-\tN)(t'-t)},
\eeq
where we dropped again the integral above $\tm$ and $\Omega(t)$ is now a $2\times 2$ matrix. While there is no analytic formula for its individual components, the determinant fulfills a constraint very similar to~\eqref{Omnes_matrix}~\cite{Moussallam:1999aq}
\beq
\label{det_fin}
\det \Omega(t)=\exp\Bigg\{\frac{t}{\pi}\int\limits_{\tpi}^{\tm}\diff t'\frac{\psi(t')}{t'(t'-t)}\Bigg\},
\eeq
i.e.\ with the $\pi\pi$ phase shift replaced by the phase $\psi(t)$ of the $\pi\pi\to\bar K K$ partial wave.

\subsection{Numerical results}

According to Fig.~\ref{fig:flowchart} the solution of the full RS system requires an iteration between $s$- and $t$-channel. However, it was shown in~\cite{Ditsche:2012fv} that by far the dominant recoupling between the subsystems proceeds by means of the subtraction constants, while the sensitivity of the $t$-channel solution on the precise input used for the $\pi N$ partial waves was found to be very small.
Therefore, in the end the accuracy that can be reached for the $t$-channel amplitudes is limited by the remaining uncertainty in the subthreshold parameters as determined from the iterated RS solution, and effects significantly below that threshold can be neglected.
For this reason, the iteration can be organized more efficiently in practice by solving the $t$-channel system once with reference values for the subthreshold parameters, and then including the corrections to this starting-point solution directly into the solution procedure for the $s$-channel equations. We use the subthreshold parameters from the KH80 solution~\cite{Koch:1980ay,Hoehler:1983} as our reference point and express the corrections in terms of
\beq
\Delta X_{mn}^\pm=X_{mn}^\pm-X_{mn}^\pm\big|_\text{KH80},\qquad X\in\{a,b,d\}.
\eeq

\begin{figure}[t!]
\centering
\includegraphics[width=0.447\linewidth,clip]{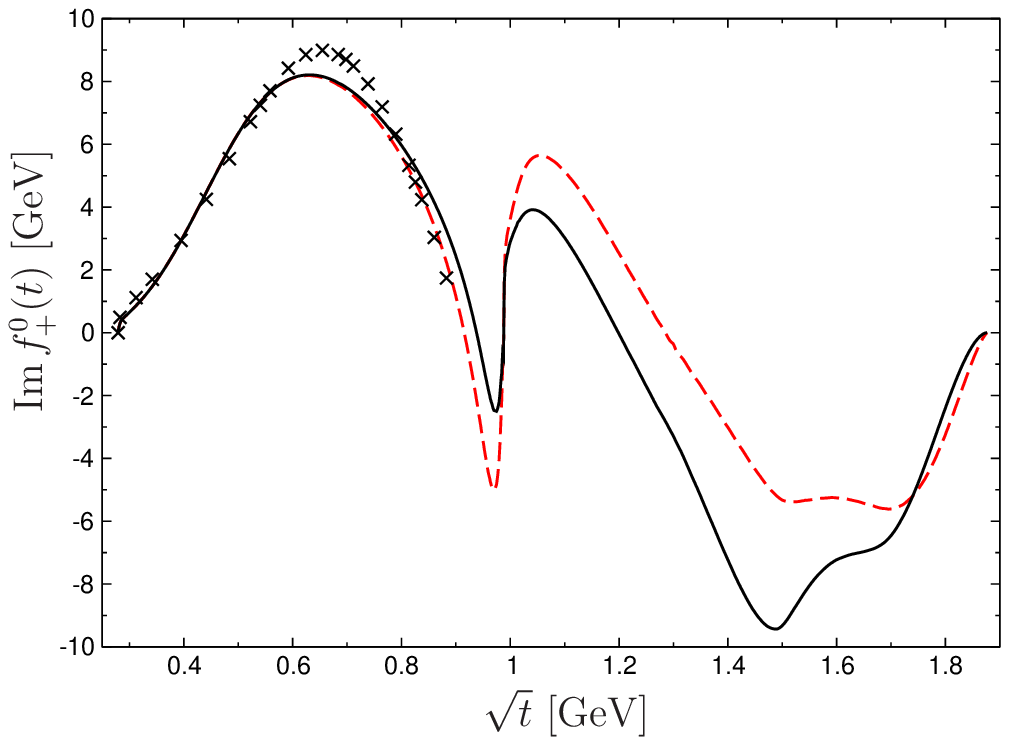}\\[1mm]
\includegraphics[width=0.447\linewidth,clip]{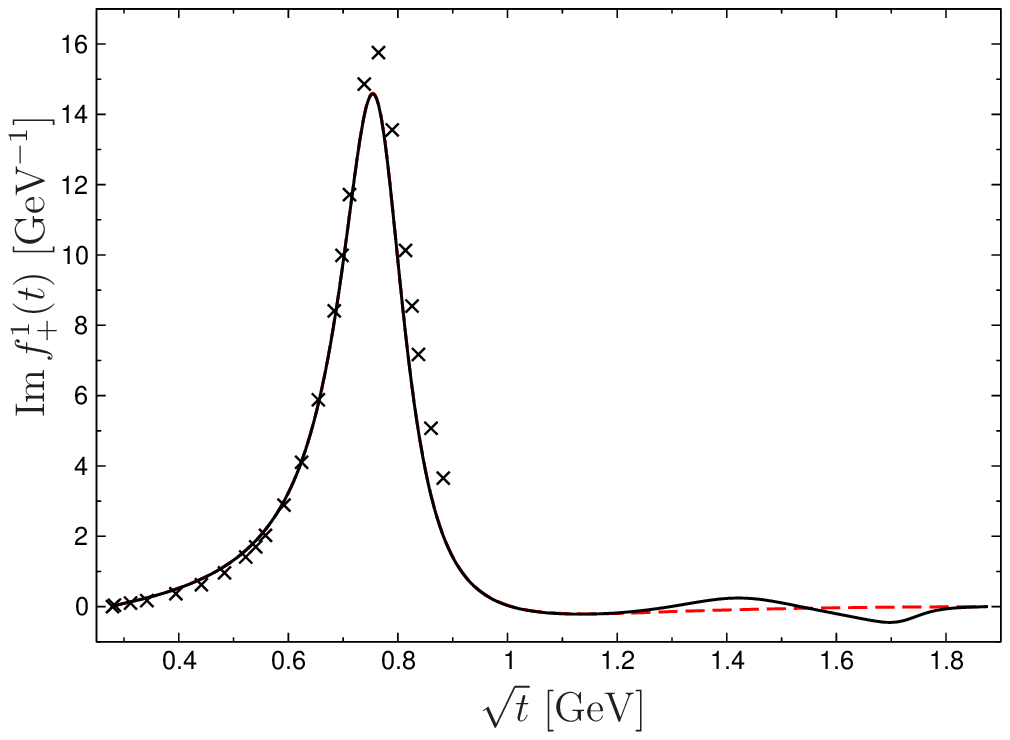}\quad
\includegraphics[width=0.447\linewidth,clip]{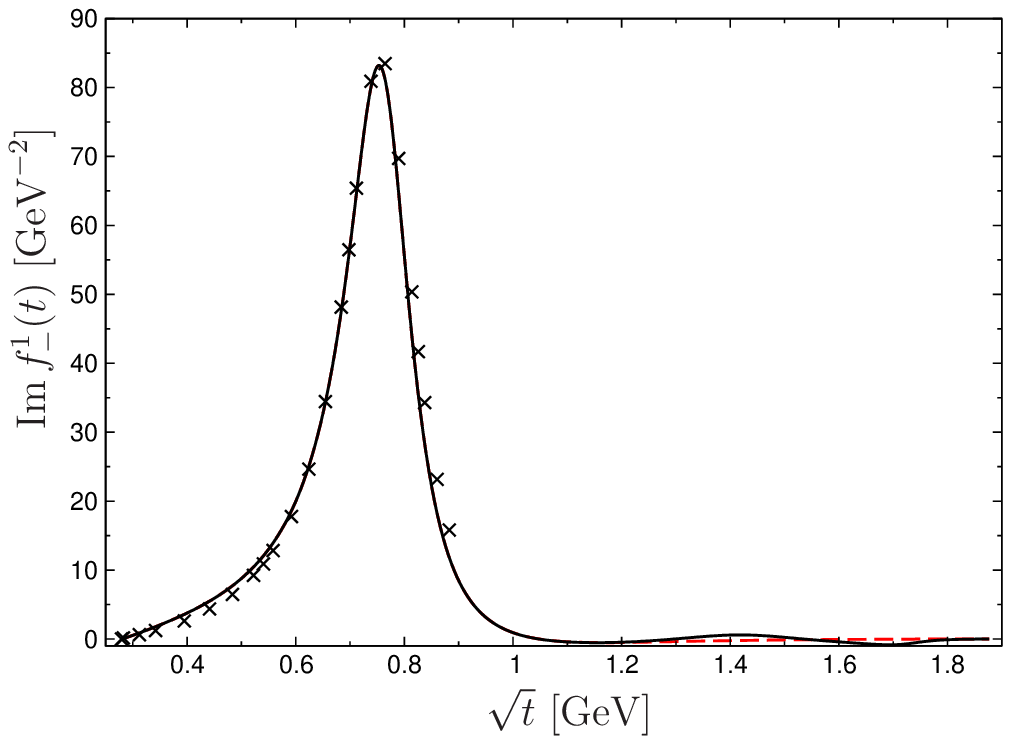}\\[1mm]
\includegraphics[width=0.447\linewidth,clip]{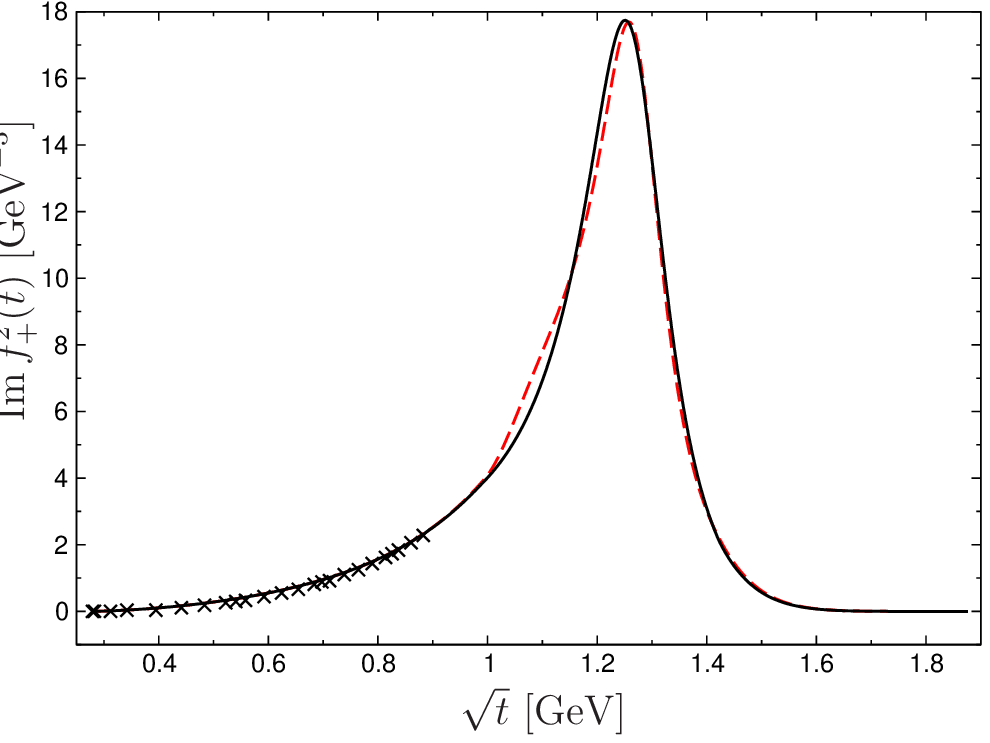}\quad
\includegraphics[width=0.447\linewidth,clip]{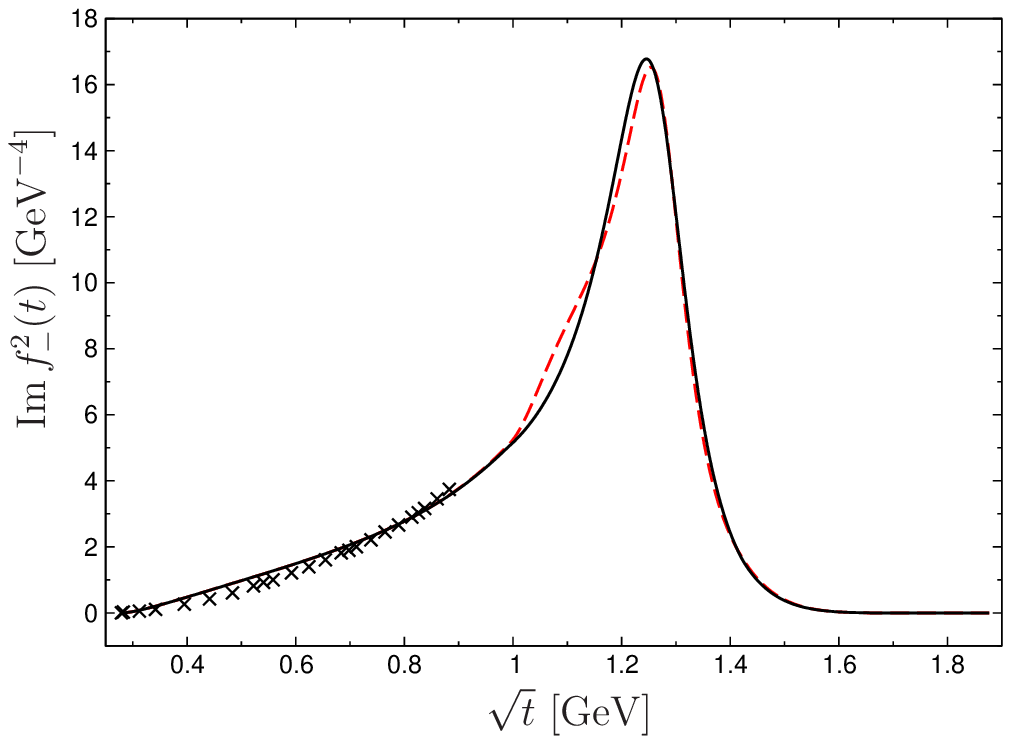}\\[1mm]
\includegraphics[width=0.447\linewidth,clip]{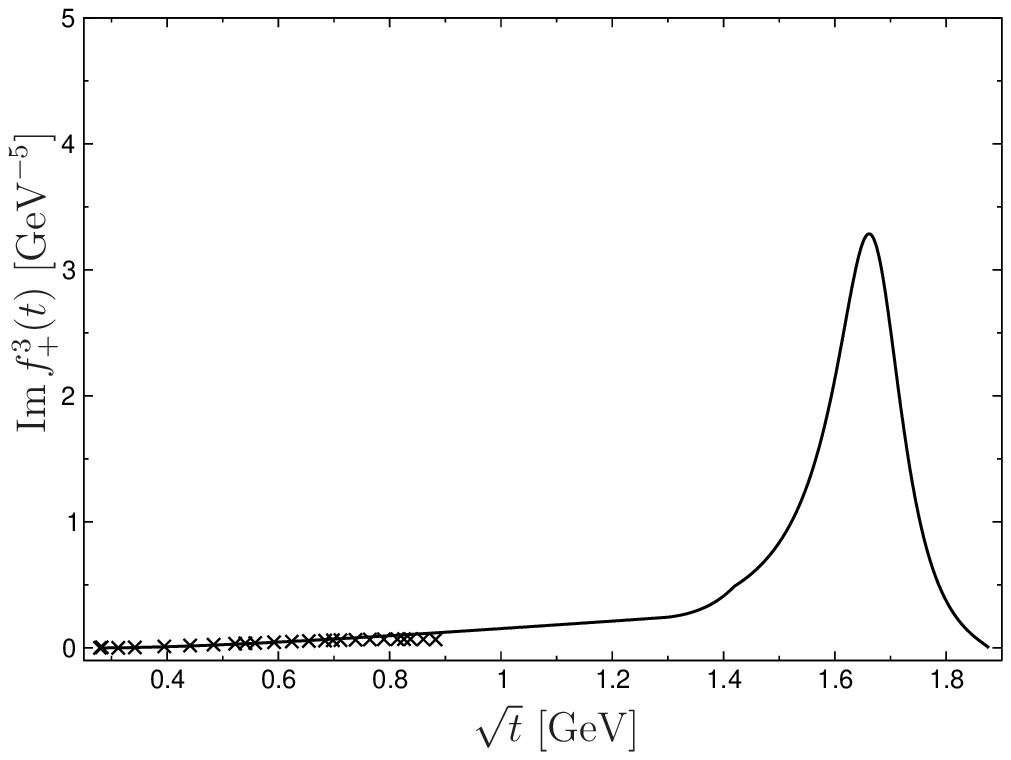}\quad
\includegraphics[width=0.447\linewidth,clip]{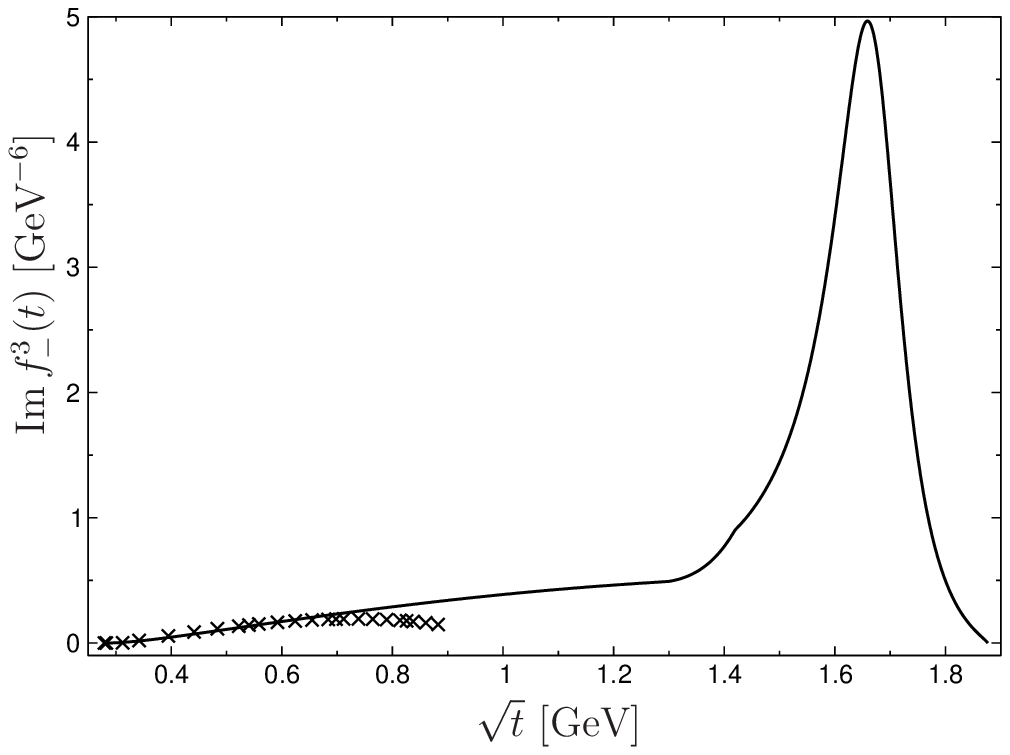}
\caption{Imaginary parts of the $t$-channel partial waves with KH80 subthreshold parameters. The black solid (red dashed) line refers to our central solution (the dominant input variation as described in the main text), while the black crosses indicate the results from~\cite{Hoehler:1983}.}
\label{fig:MO_KH80}
\end{figure}

These corrections translate to shifts in the $t$-channel amplitudes, for $P$- and $D$-waves one finds
\begin{align}
\label{P-wave-subth}
 \Delta\Gamma^1(t)&=\frac{p_t^2}{12\pi \mN}\Big\{\Delta a_{00}^-\big(1-t\,\dot\Omega_1(0)\big)+\Delta a_{01}^-t\Big\}\Omega_1(t), &
 \Delta f^1_-(t)&=\frac{\sqrt{2}}{12\pi}\bigg\{\Delta b_{00}^-\big(1-t\,\dot\Omega_1(0)\big)+\Delta b_{01}^-t\bigg\}\Omega_1(t),\notag\\
 \Delta\Gamma^2(t)&=\frac{p_t^2}{30\pi \mN^2}\Delta a_{10}^+\Omega_2(t), &
 \Delta f^2_-(t)&=\frac{\sqrt{6}}{60\pi \mN}\Delta b_{00}^+\Omega_2(t),
\end{align}
 while the $S$-wave is modified according to
\beq
\label{S-wave-subth}
\Delta f^0_+(t)=-\frac{p_t^2}{4\pi}\bigg\{\bigg[\Delta d_{00}^++\Delta b_{00}^+\frac{\mpi^2}{3}\bigg]\Big(\Omega_{11}(t)\big(1-t\,\dot\Omega_{11}(0)\big)-\Omega_{12}(t)t\,\dot \Omega_{21}(0)\Big)
+\bigg[\Delta d_{01}^+ -\frac{\Delta b_{00}^+}{12}\bigg]t\,\Omega_{11}(t)\bigg\}.
\eeq

In the following, we present the results for the imaginary parts of the $t$-channel partial waves when the subthreshold parameters are fixed at KH80 values, and investigate the sensitivity to the input quantities apart from the subtraction constants. 
First of all, we choose the matching point as $\tm=\tN$, taking advantage of the large domain of validity in the $t$-channel~\eqref{a_tchannel} and of the kinematic zeros in $f^0_+(t)$ and $\Gamma^J(t)$ at $t=\tN$. In general, for the evaluation of the $s$-channel integrals we use the phase shifts from~\cite{Arndt:2006bf} summed up to $l=4$ and $W=2.5\GeV$ as well as the Regge model from~\cite{Huang:2009pv} above (see also~\cite{Mathieu:2015gxa} for a recent Regge parameterization). However, as shown in~\cite{Ditsche:2012fv}, the contribution from the Regge region is completely negligible, and similarly 
truncating the partial-wave expansion at $l=5$ or using input from the KH80 analysis~\cite{Koch:1980ay,Hoehler:1983} generates variations that can be neglected compared to the final uncertainties in the subthreshold parameters. 

For the $S$-wave, we take the $\pi\pi$ phase shift and inelasticity up to $\sqrt{t_0}=1.3\GeV$ from the Roy-equation analysis of~\cite{Caprini:2011ky} and the $\pi\pi\to\bar K K$ partial wave from RS equations~\cite{Buettiker:2003pp}. In the latter analysis, the modulus $|g(t)|$ in the pseudophysical region $\tpi\leq t\leq \tK$ follows from the RS system, while above the two-kaon threshold phase-shift solutions~\cite{Cohen:1980cq,Etkin:1981sg} are used. The $\pi N$ coupling constant is fixed at $g^2/(4\pi)=13.7$~\cite{Baru:2010xn,Baru:2011bw}, the $KN$ partial waves are taken from~\cite{Hyslop:1992cs}, and the hyperon Born terms are evaluated with the couplings from~\cite{Holzenkamp:1989tq}.
The main uncertainty is generated by the fact that around $1.3\GeV$ inelasticities from $4\pi$ intermediate states start to become relevant, so that a two-channel description is not strictly applicable any more and the meaning of the phase shifts $\delta$ and $\psi$ becomes unclear. For the continuation of the phase shifts above $t_0$ we consider two extreme cases: first, we guide the phase shifts smoothly to $2\pi$ above $t_0$ (motivated by the asymptotic behavior), and second, we keep the phase shifts constant, with the results represented by the solid and dashed lines in the first panel of Fig.~\ref{fig:MO_KH80}, respectively.
The variation between these two curves is indeed the largest systematic uncertainty, for instance much larger than the effect of switching off the $KN$ input altogether~\cite{Hoferichter:2012wf}, and will be propagated accordingly in the $s$-channel error analysis later.

For the $P$-waves, the largest uncertainty is again generated by $4\pi$ inelasticities, which manifest themselves in $\rho'$ and $\rho''$ resonances that have a substantial branching fraction to $4\pi$. In order to estimate their effect we follow~\cite{Schneider:2012ez} and use a phase shift constructed in such a way as to reproduce the $\rho'$ and $\rho''$ in the pion vector form factor in an elastic approximation; this input actually corresponds to the phase of the form factor~\cite{Hanhart:2012wi}. The results for $\Im f^1_\pm(t)$ with (solid lines) and without (dashed lines) these resonances built in are shown in the second row of Fig.~\ref{fig:MO_KH80}. The effects are indeed restricted to the energy region above $1\GeV$, so that for the error propagation to the $s$-channel solution the effects are negligible compared to the errors in the subthreshold parameters. The same holds true for the intrinsic uncertainties in the $\pi\pi$ phase shift: using the phase shifts from~\cite{GarciaMartin:2011cn} or \cite{Caprini:2011ky} produces only tiny differences in the $\rho$-peak.

The $D$-waves are dominated by the $f_2(1270)$ resonance, which has a $15\%$ inelasticity to the $4\pi$ and $\bar K K$ channels. One way to estimate the potential impact of these inelasticities is to replace the $\pi\pi$ phase shift in the MO solution by the phase of the $\pi\pi$ partial wave. The former corresponds to the solid, the latter to the dashed lines in the third row of Fig.~\ref{fig:MO_KH80} (we use phase shift and inelasticity from~\cite{GarciaMartin:2011cn}). The effect is again quite moderate and does not need to be propagated to the $s$-channel solution. However, we note that the $f_2(1270)$ resonance itself plays an important role in the interplay between the $s$- and $t$-channel RS subsystems: without including the $f_2(1270)$ in the $t$-channel $D$-waves, we could not find an acceptable solution of the $s$-channel equations.

In view of this surprising role of the $f_2(1270)$, one may obviously wonder about the impact of yet higher partial waves. The first resonance in the $F$-waves is the $\rho_3(1690)$, at significantly higher energies than the $f_2(1270)$. Moreover, it is predominantly inelastic with a $70\%$ branching fraction to the $4\pi$ channel, so that an elastic treatment becomes difficult to justify. However, to get a rough estimate of the expected size of the $F$-waves, we constructed an $F$-wave phase shift by matching the parameterization from~\cite{GarciaMartin:2011cn} to a Breit--Wigner description of the $\rho_3(1690)$, with Breit--Wigner parameters taken from~\cite{Agashe:2014kda} and results shown in the last row of Fig.~\ref{fig:MO_KH80}. Indeed, we find that the $F$-wave contribution is much smaller than the $D$-wave counterpart, even to the extent that the effect can be safely absorbed into the uncertainty estimate altogether. Given that due to the inelastic nature of the $\rho_3(1690)$ our calculation of the $F$-waves is less rigorous than that of the lower partial waves, we will indeed quote central results for $J\leq 2$, and only include the $F$-waves in the error analysis.

Finally, the $t$-channel partial waves are coupled according to~\eqref{Deltadef}, i.e.\ there is a $t$-channel $D$-wave contribution to the $S$-wave MO inhomogeneity, and an $F$-wave contribution to the $P$-waves. However, ultimately the inhomogeneities $\Delta^J_\pm(t)$ are dominated by the Born terms and $s$-channel integrals, while the $t$-channel corrections constitute a very minor effect. In practice, they can be ignored given the remaining uncertainties in the input.  

\FloatBarrier

\section{Strategy for the $\boldsymbol{s}$-channel solution}
\label{sec:schannel_sol}

Once the $t$-channel equations are solved, the structure of the $s$-channel problem resembles the form of $\pi\pi$ Roy equations, 
and it should be amenable to similar solution techniques. The basic idea can be summarized in such a way that the phase shifts at low energies, from the $\pi N$ threshold to the maximum allowed matching point at $\Wm=1.38\GeV$,  are represented in suitable parameterizations whose free parameters, together with the subtraction constants, are determined by minimizing the difference between the left-hand side (LHS) and right-hand side (RHS) of~\eqref{sRSpwhdr}.

As a first step, the $s$-channel RS system is decomposed into equations for the real and imaginary parts. Defining the physical limit by $W\to W+i\epsilon$ and collecting the imaginary pieces that follow from the principal-value prescription for the Cauchy kernels shows that the equations for the imaginary parts are trivially fulfilled, while those for the real parts are identical to~\eqref{sRSpwhdr}  upon replacing $f^{I_s}_{l\pm}(W)$ by $\Re f^{I_s}_{l\pm}(W)$ and the integrals by their principal-value analogs.

\begin{figure}[t!]
\centering
\includegraphics[width=0.45\linewidth,clip]{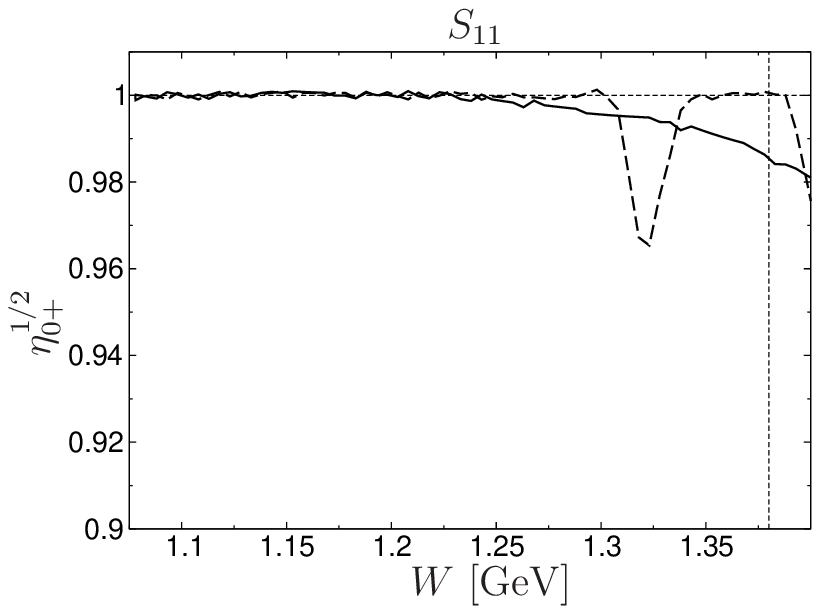}\quad
\includegraphics[width=0.45\linewidth,clip]{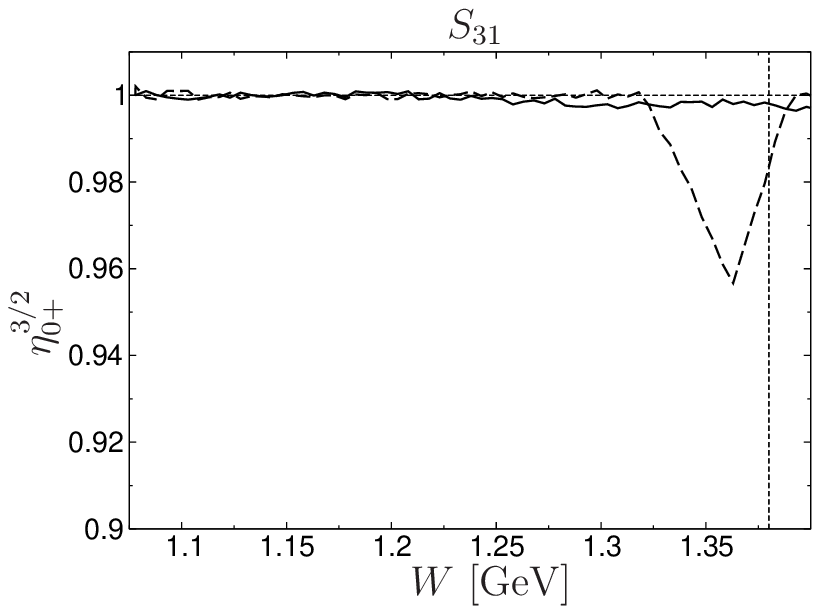}\\[0.1cm]
\includegraphics[width=0.45\linewidth,clip]{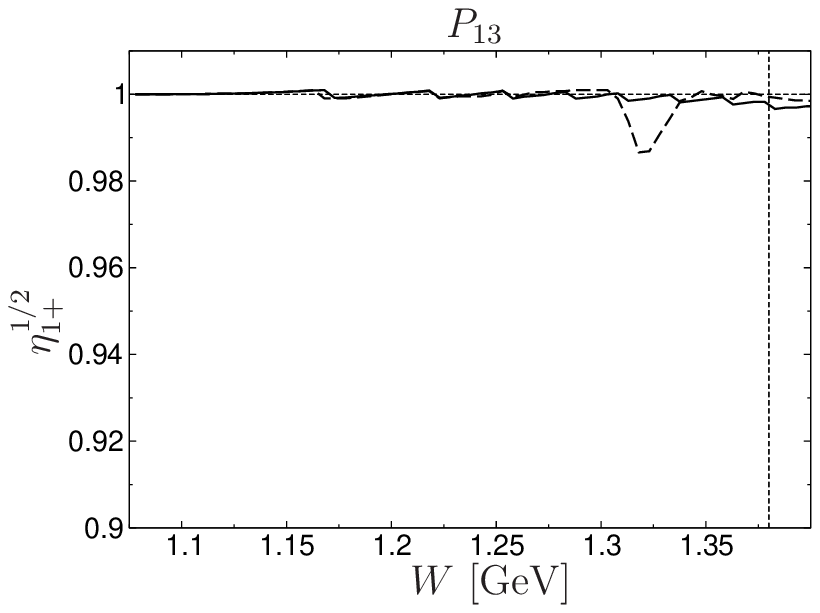}\quad
\includegraphics[width=0.45\linewidth,clip]{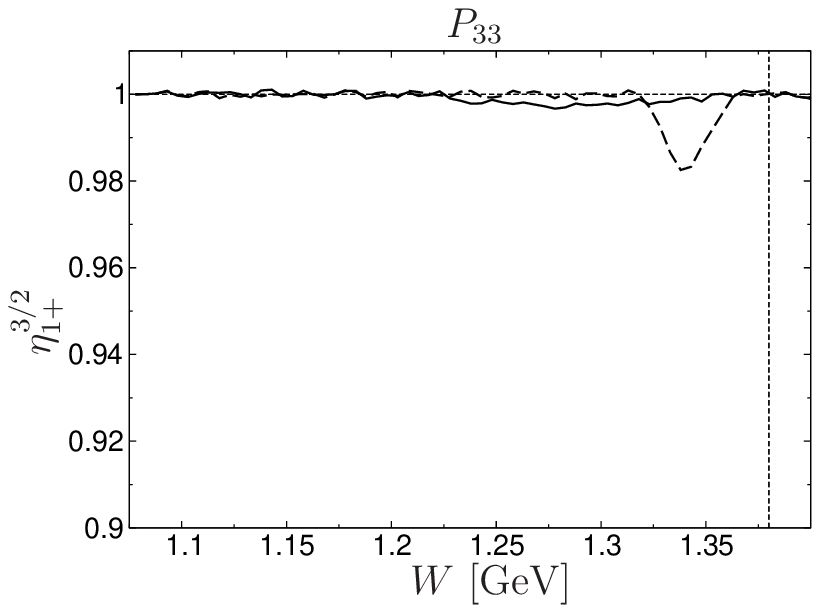}\\[0.1cm]
\includegraphics[width=0.45\linewidth,clip]{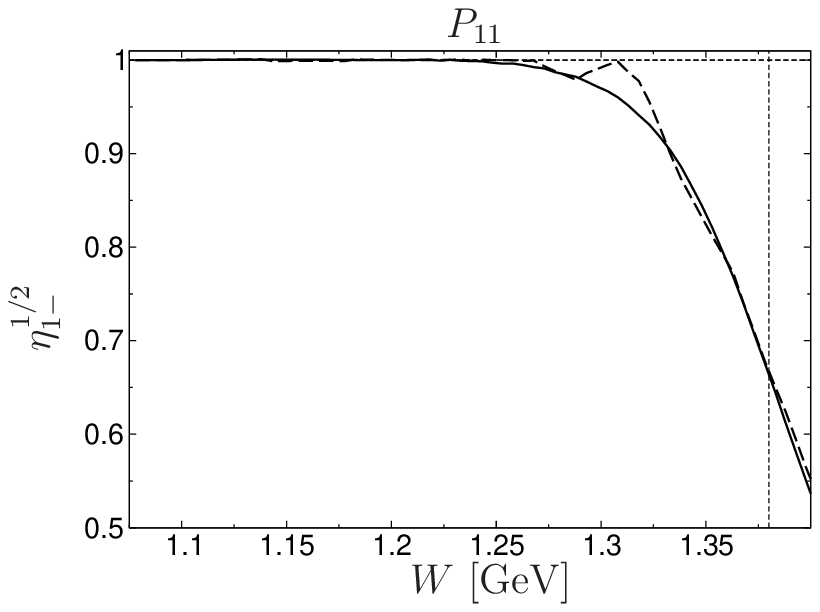}\quad
\includegraphics[width=0.45\linewidth,clip]{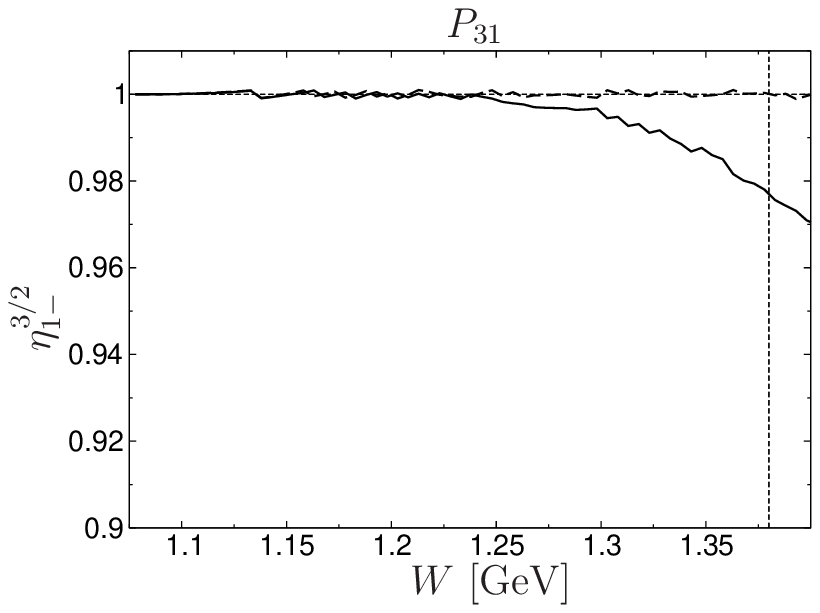}
\caption{Inelasticities of the $s$-channel partial waves from the GWU/SAID~\cite{Arndt:2006bf,Workman:2012hx} (solid line) and KH80~\cite{Koch:1980ay,Hoehler:1983} (dashed line) PWAs in the region below $\Wm$.
The short-dashed lines indicate $\Wm=1.38 \GeV$ and $\eta_{l\pm}^{I_s}=1$.}
\label{fig:etas-PWAs}
\end{figure}

The real and imaginary parts of $f^{I_s}_{l\pm}(W)$ are linked to phase shifts and inelasticities through the unitarity condition given in~\eqref{eq:s-uni}. However, as explained in Sect.~\ref{sec:RSsystem} and indicated in Fig.~\ref{fig:flowchart}, the inelasticities cannot be determined from an iterative solution of the RS system, but have to be taken as input.
The inelasticities $\eta^{I_s}_{l\pm}(W)$ for the $S$- and $P$-waves as given by the GWU/SAID~\cite{Arndt:2006bf,Workman:2012hx} and KH80~\cite{Koch:1980ay,Hoehler:1983} PWAs in the region below $\Wm$ are depicted in Fig.~\ref{fig:etas-PWAs}. In the case of the GWU/SAID solutions, the figure suggests that inelasticities may be ignored for the $S_{31}$-, $P_{13}$-, and $P_{33}$-waves, while the remaining ones can be well described by the parameterization
\beq
\label{eq:eta-par}
\eta_{l\pm}^{I_s}=\frac{1-\alpha_{l\pm}^{I_s}(s-s_{\text{inel}})^{r^{I_s}_{l\pm}} (s-s_{+})^{r^{I_s}_{l\pm}}}{1+\alpha_{l\pm}^{I_s}(s-s_\text{inel})^{r^{I_s}_{l\pm}}(s-s_{+})^{r^{I_s}_{l\pm}}},\quad\quad s_\text{inel}=(\mN+2M_\pi)^2,
\eeq
with $r^{1/2}_{0+}=r^{3/2}_{1-}=3/2$, $r^{1/2}_{1-}=5/2$, and
\beq
\alpha^{1/2}_{0+}=0.0412\GeV^{-6},\qquad\alpha^{3/2}_{1-}=0.066\GeV^{-6},\qquad\alpha^{1/2}_{1-}=3.716\GeV^{-10},
\eeq
which have been determined by fitting to the SAID inelasticities. The powers in~\eqref{eq:eta-par} have been chosen on purely phenomenological grounds to accurately reproduce the experimental inelasticities. 
When analytically continued below $s_\text{inel}$, the inelasticities parameterized as in~\eqref{eq:eta-par} turn into an additional contribution to the phase shifts, which is a convenient way to preserve analyticity in the presence of strong inelasticities, cf.\ the isospin-$0$ $S$-wave in $\pi\pi$ scattering~\cite{Caprini:2011ky}. In our case, the inelasticities are sufficiently smooth that this does not become necessary. 
However, although not considered explicitly here, the threshold behavior of the ensuing phase shifts below $s_\text{inel}$ still suggests to take $r_{l\pm}^{I_s}\geq l+1/2$.

For the KH80 solution, several amplitudes exhibit oscillatory tendencies that should be considered an artifact of the calculation, so ignoring the clearly unphysical peaks, Fig.~\ref{fig:etas-PWAs} seems to indicate 
that the inelasticity might be ignored for all but the $P_{11}$-wave, which can also be described by the functional form given in~\eqref{eq:eta-par}, with $r^{1/2}_{1-}=5/2$ and  $\alpha^{1/2}_{1-}=3.669\GeV^{-10}$. 

\subsection{Parameterizations of the $s$-channel partial waves}
\label{sec:param_s-channel}

In principle, the minimization of the difference between the LHS and RHS of~\eqref{sRSpwhdr} should be carried out over the whole space of admissible functions. 
However, for obvious practical reasons, we restrict ourselves to a set of simple parameterizations.  Following the example of~\cite{Ananthanarayan:2000ht,Buettiker:2003pp}, we use a Schenk-like parameterization~\cite{Schenk:1991xe} of the form
\beq
\label{eq:schenk-par}
\tan{\delta_{l{\pm}}^{I_s}}=|\qq|^{2l+1}\left(A_{l\pm}^{I_s}+B_{l\pm}^{I_s}\qq^2+C_{l\pm}^{I_s}\qq^4+D_{l\pm}^{I_s}\qq^6\right)\frac{s_+-s_{l\pm}^{I_s}}{s-s_{l\pm}^{I_s}},
\eeq
except for the $P_{33}$-wave, where a conformal parameterization~\cite{GarciaMartin:2011cn}
\beq
\label{eq:conf-par}
\cot{\delta_{l\pm}^{I_s}}=\frac{1}{|\qq|^{2l+1}}\frac{s-s_{l\pm}^{I_s}}{s_+-s_{l\pm}^{I_s}}\left(\frac{1}{\tilde A_{l\pm}^{I_s}}+\tilde B_{l\pm}^{I_s}\big[w(s)-w_+\big]+\tilde C_{l\pm}^{I_s}\big[w(s)-w_+\big]^2\right),\qquad
w(s)=\frac{\sqrt{s}-\sqrt{\bar s_{l\pm}^{I_s}-s}}{\sqrt{s}+\sqrt{\bar s_{l\pm}^{I_s}-s}},\qquad
w_+=w(s_+),
\eeq
proves to be more adequate to reproduce the phase shift above the resonance region.
$A_{l\pm}^{I_s}$, $B_{l\pm}^{I_s}$, etc.\ are related to threshold parameters, see Sect.~\ref{sec:threshold}, for instance $A_{l\pm}^{I_s}=a_{l\pm}^{I_s}$ coincides with the scattering length. In case of a resonance, $s_{l\pm}^{I_s}$ can be interpreted as its mass projection onto the real axis. Similarly, $\tilde A_{l\pm}^{I_s}=a_{l\pm}^{I_s}$, while $\bar s_{l\pm}^{I_s}> \sm$ defines the conformal mapping.

Moreover, as previously discussed in Sect.~\ref{sec:existence_uniqueness}, a physical solution is characterized by the requirement of a smooth matching.
In this way, one expects another constraint on the parameterizations given by~\eqref{eq:schenk-par} and \eqref{eq:conf-par} due to the no-cusp condition.
The simplest way to satisfy these conditions is to impose continuity and a continuous derivative for each phase shift at $\sm=\Wm^2$ directly on the phase space of parameters. 
Therefore, defining
\beq
\delta^{I_s}_{\text{m},l\pm}=\lim_{\epsilon\to 0}\delta_{l{\pm}}^{I_s}(\sm+\epsilon),\qquad \delta^{\prime\,I_s}_{\text{m},l\pm}=\lim_{\epsilon\to 0}\frac{\diff \delta_{l{\pm}}^{I_s}(s+\epsilon)}{\diff s}\bigg|_{s=\sm},
\eeq
which are obtained from the $s$-channel input considered above the matching point, 
it is straightforward to re-express the last two parameters of~\eqref{eq:schenk-par} and \eqref{eq:conf-par} 
as a function of the input phase at $\Wm$, $\delta^{I_s}_{\text{m},l\pm}$, and its derivative $\delta^{\prime\,I_s}_{\text{m},l\pm}$, 
so that we restrict ourselves to a set of solutions where the matching conditions are fulfilled automatically. 

In addition, we require that the $S$-wave scattering lengths agree with their pionic-atom values, see Sect.~\ref{sec:pionic_atom}, and accordingly we fix
$A_{0+}^{I_s}=a_{0+}^{I_s}$. For the $P$-waves or the higher Schenk-parameters $B_{l\pm}^{I_s}$, etc.\ we do not enforce agreement with the values predicted from sum rules for the threshold parameters as discussed in Sect.~\ref{sec:threshold}: such sum rules offer a much more reliable access to the threshold parameters, in contrast to the calculation from the phase shifts, which soon becomes very unstable. 
Indeed, all we need for the RS solution is an efficient parameterization of the phase shifts, while the threshold parameters will be determined in a second step.
Given that the first parameter of the $S$-wave phase shifts is fixed by the pionic-atom constraint, 
we will allow for one more free parameter in their Schenk parameterizations~\eqref{eq:schenk-par}, so that the same number of free parameters is considered in all partial waves.
In summary, we will use for the $S$-waves the parameterizations
\begin{align}
\label{eq:schenk-par-Sw}
\tan{\delta_{0+}^{I_s}}&=|\qq|\left(A_{0+}^{I_s}+B_{0+}^{I_s}\qq^2+C_{0+}^{I_s}\qq^4+D_{0+}^{I_s}\qq^6+E_{0+}^{I_s}\qq^8\right)\frac{s_+-s_{0+}^{I_s}}{s-s_{0+}^{I_s}},\notag\\
D_{0+}^{I_s}&=-\frac{1}{q_\text{m}^6}\left\{4 A_{0+}^{I_s}+3 B_{0+}^{I_s} q_\text{m}^2+2 C_{0+}^{I_s} q_\text{m}^4+\frac{\sm-s_{0+}^{I_s}}{2 q^\prime_\text{m} \big(s_+-s_{0+}^{I_s}\big)} \left(\frac{\delta^{\prime\,I_s}_{\text{m},0+}}{ \cos^2\delta^{I_s}_{\text{m},0+}}+\frac{\tan\delta^{I_s}_{\text{m},0+}}{\sm-s_{0+}^{I_s}}-\frac{9 q^\prime_\text{m}}{q_\text{m}}\tan\delta^{I_s}_{\text{m},0+}\right)\right\},\notag\\
E_{0+}^{I_s}&=\frac{1}{q_\text{m}^8}\left\{3 A_{0+}^{I_s}+2 B_{0+}^{I_s} q_\text{m}^2+ C_{0+}^{I_s} q_\text{m}^4+\frac{\sm-s_{0+}^{I_s}}{2 q^\prime_\text{m} \big(s_+-s_{0+}^{I_s}\big)} 
\left(\frac{\delta^{\prime\,I_s}_{\text{m},0+}}{\cos^2\delta^{I_s}_{\text{m},0+}}+\frac{\tan\delta^{I_s}_{\text{m},0+}}{\sm-s_{0+}^{I_s}}-\frac{7 q^\prime_\text{m}}{q_\text{m}}\tan\delta^{I_s}_{\text{m},0+}\right)\right\},
\end{align}
with $q_\text{m}=|\qq|(\sm)$ and $q^\prime_\text{m}=\diff |\qq|(s)/\diff s|_{s=\sm}$.   

In the case of the $P_{13}$-, $P_{11}$-, and $P_{31}$-waves we use the Schenk parameterization of~\eqref{eq:schenk-par}, but with
\begin{align}
\label{eq:schenk-par-Pw}
C_{1\pm}^{I_s}&=-\frac{1}{q_\text{m}^4}\left\{3 A_{1\pm}^{I_s}+2 B_{1\pm}^{I_s} q_\text{m}^2+\frac{\sm-s_{1\pm}^{I_s}}{2q_\text{m}^2 q^\prime_\text{m} \big(s_+-s_{1\pm}^{I_s}\big)} \left(\frac{\delta^{\prime\,I_s}_{\text{m},1\pm}}{ \cos^2\delta^{I_s}_{\text{m},1\pm}}+\frac{\tan\delta^{I_s}_{\text{m},1\pm}}{\sm-s_{1+}^{I_s}}-\frac{9 q^\prime_\text{m}}{q_\text{m}}\tan\delta^{I_s}_{\text{m},1\pm}\right)\right\},\notag\\
D_{1\pm}^{I_s}&=\frac{1}{q_\text{m}^6}\left\{2 A_{1\pm}^{I_s}+ B_{1\pm}^{I_s} q_\text{m}^2+\frac{\sm-s_{1\pm}^{I_s}}{2q_\text{m}^2 q^\prime_\text{m} \big(s_+-s_{1\pm}^{I_s}\big)} \left(\frac{\delta^{\prime\,I_s}_{\text{m},1\pm}}
{\cos^2\delta^{I_s}_{\text{m},1\pm}}+\frac{\tan\delta^{I_s}_{\text{m},1\pm}}{\sm-s_{1+}^{I_s}}-\frac{7 q^\prime_\text{m}}{q_\text{m}}\tan\delta^{I_s}_{\text{m},1\pm}\right)\right\}.
\end{align}

Finally, for the $P_{33}$-wave the last two parameters of the conformal parameterization of~\eqref{eq:conf-par} fixed from the no-cusp condition read
\begin{align}
\label{eq:schenk-par-Pw-conf}
\tilde B^{3/2}_{1+}&=-\frac{2}{\tilde A^{3/2}_{1+} \big[w_\text{m}-w_+\big]}-\frac{q_\text{m}^3 \big(s_+-s_{1+}^{3/2}\big)}{w_\text{m}^\prime \big(\sm-s_{1+}^{3/2}\big)} \left\{\cot\delta_{\text{m},1+}^{3/2} \left(\frac{3 q_\text{m}^\prime}{q_\text{m}}+\frac{1}{s_{1+}^{3/2}-\sm}-\frac{2
   w_\text{m}^\prime}{w_\text{m}-w_+}\right)-\frac{\delta_{\text{m},1+}^{\prime 3/2}}{ \sin^2\delta_{\text{m},1+}^{3/2}}\right\},\notag\\
\tilde C^{3/2}_{1+}&=\frac{1}{\tilde A^{3/2}_{1+} \big[w_\text{m}-w_+\big]^2}+\frac{q_\text{m}^3 \big(s_+-s_{1+}^{3/2}\big)}{w_\text{m}^\prime \big(\sm-s_{1+}^{3/2}\big)
   \big[w_\text{m}-w_+\big]} \left\{\cot\delta_{\text{m},1+}^{3/2} \left(\frac{3
   q_\text{m}^\prime}{q_\text{m}}+\frac{1}{s_{1+}^{3/2}-\sm}-\frac{w_\text{m}^\prime}{w_\text{m}-w_+}\right)-\frac{\delta_{\text{m},1+}^{\prime 3/2}}{ \sin^2\delta_{\text{m},1+}^{3/2}}\right\},
\end{align}
where $w_\text{m}=w(\sm)$ and $w^\prime_\text{m}=\diff w(\sm)/\diff s|_{s=\sm}$.

\begin{figure}[t!]
\centering
\includegraphics[width=0.45\linewidth,clip]{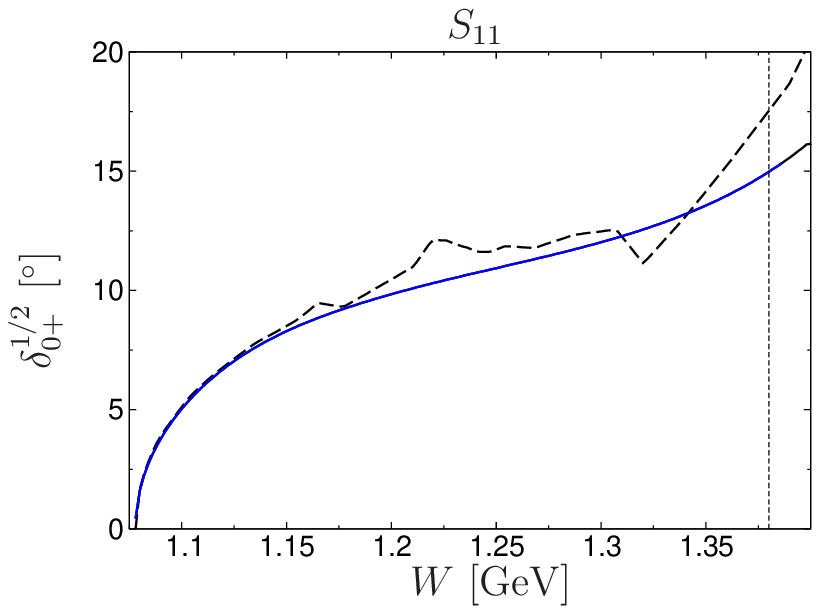}\quad
\includegraphics[width=0.45\linewidth,clip]{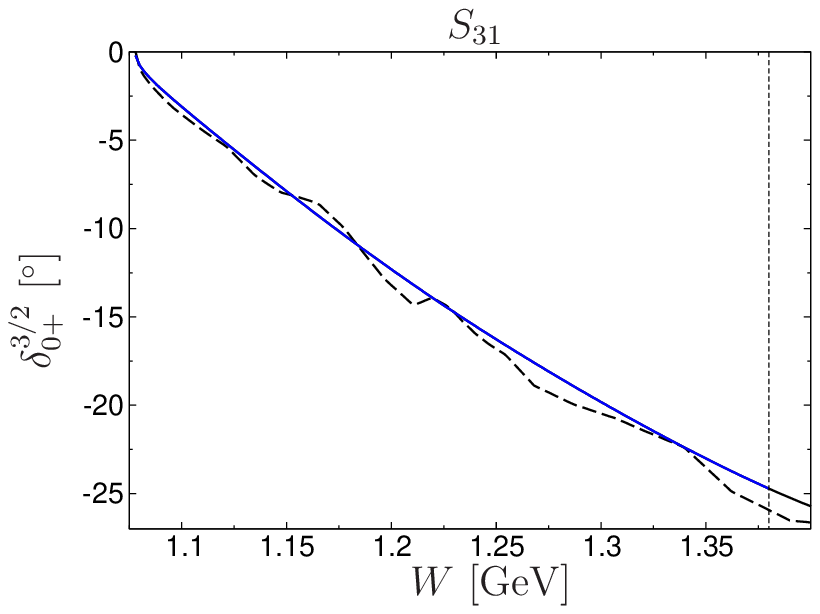}\\[0.1cm]
\includegraphics[width=0.45\linewidth,clip]{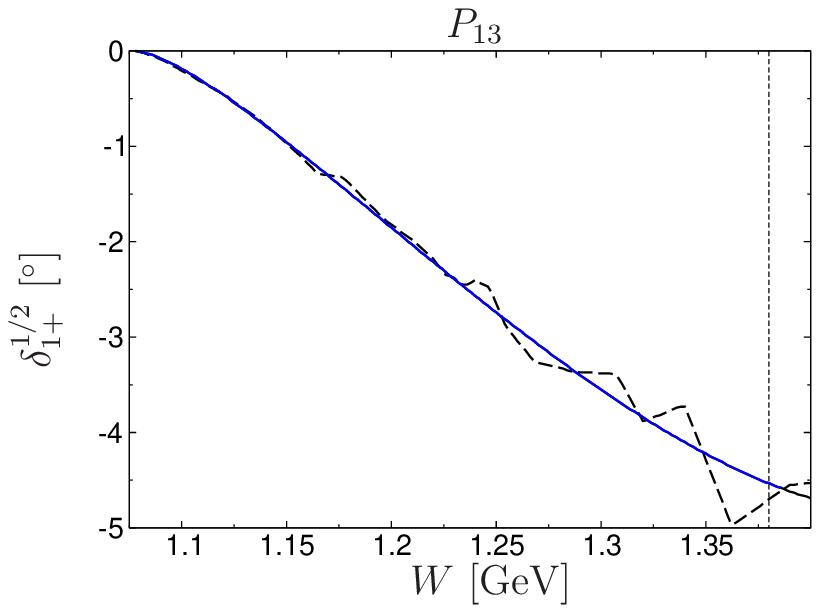}\quad
\includegraphics[width=0.45\linewidth,clip]{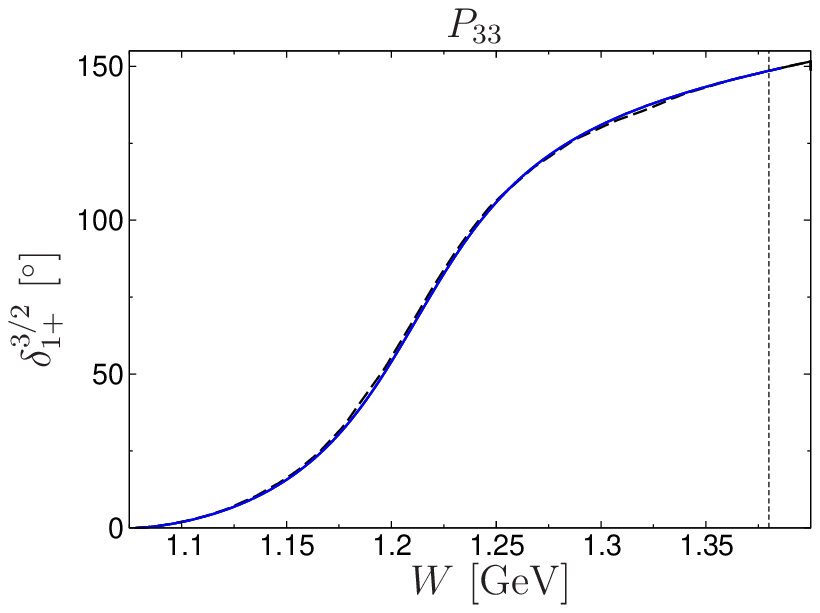}\\[0.1cm]
\includegraphics[width=0.45\linewidth,clip]{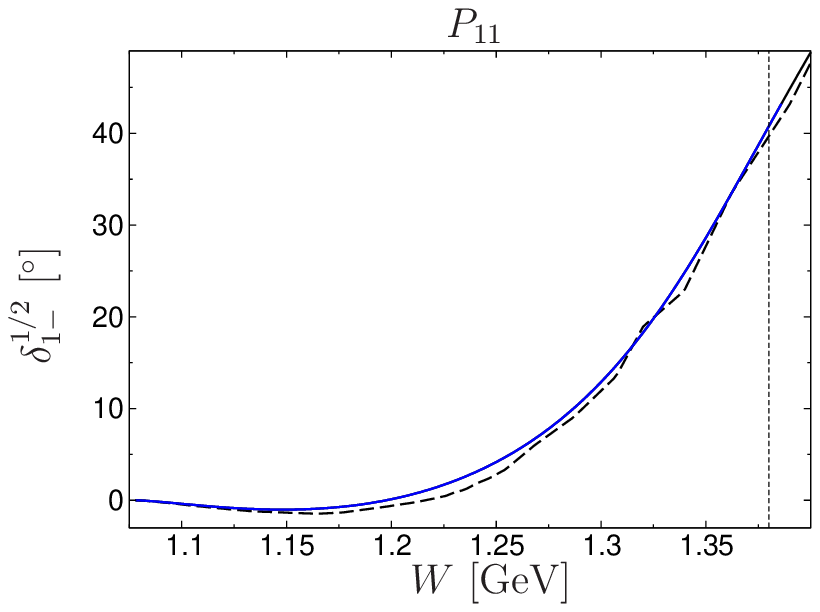}\quad
\includegraphics[width=0.45\linewidth,clip]{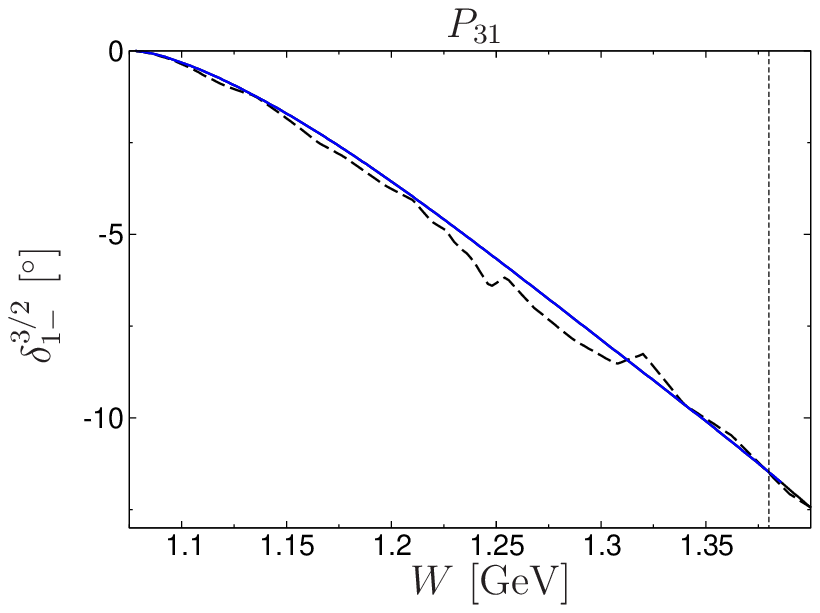}
\caption{Phase shifts of the $s$-channel partial waves from GWU/SAID~\cite{Arndt:2006bf,Workman:2012hx} (solid line) and KH80~\cite{Koch:1980ay,Hoehler:1983} (dashed line) PWAs in the low-energy region. 
On this scale, the original SAID curves and the fits based on the parameterizations described in the main text (blue solid line) are indistinguishable.}
\label{fig:phases-PWAs}
\end{figure}

In Fig.~\ref{fig:phases-PWAs}, SAID and KH80 $S$- and $P$-wave phase shifts are plotted in the low-energy region. In addition, we also include fits to SAID data and SAID matching conditions in this plot, using the parameterizations described above. 
Despite the fact that there are only three free parameters for each partial wave,
the parameterizations considered allow for an accurate description of the data. 
In addition, we can also see in Fig.~\ref{fig:phases-PWAs} that the conditions imposed by~\eqref{eq:schenk-par-Sw}, \eqref{eq:schenk-par-Pw}, and \eqref{eq:schenk-par-Pw-conf} ensure that the SAID matching conditions are fulfilled exactly for each partial wave.
Finally, the $S$-wave phase shifts at threshold are fixed by the scattering lengths, which in this case are taken also from SAID. 
These fits to SAID data will be used as a starting point for the minimization of the difference between the LHS and RHS of~\eqref{sRSpwhdr}, which we will address in the next section.

The information required to solve the RS $s$-channel equations is summarized in the flowchart of Fig.~\ref{fig:flowchart}.
With respect to the $s$-channel input, in addition to the $S$- and $P$-wave inelasticities,  we have to include the imaginary part of $S$- and $P$-waves above $\Wm$ as input, 
as well as the imaginary parts of the $l>1$ partial waves above the $\pi N$ threshold $W_+$.
In the region $W\le \Wa=2.5\GeV$, we will use the solutions of the Karlsruhe--Helsinki~\cite{Koch:1980ay,Hoehler:1983} and the GWU/SAID~\cite{Arndt:2006bf,Workman:2012hx} PWAs as input. 
The effect of taking one or the other will be considered as a source of  uncertainty.
In addition, the  significance of higher partial waves decreases as $l$ increases due to the centrifugal barrier. 
For our central solution we sum up all partial waves up to $l_\text{max}=4$, while the difference to $l_\text{max}=5$ is taken as an indication for the truncation error in the partial-wave expansion and will be included in the final uncertainty estimate.
Above $\Wa=2.5\GeV$ we consider the Regge model from~\cite{Huang:2009pv} based on differential cross section and polarization data of backward  $\pi N$ scattering.

With respect to the $t$-channel contribution, a solution to~\eqref{sRSpwhdr} requires information on the imaginary part of all $t$-channel partial waves.  
We use the solutions of the RS $t$-channel subproblem  discussed in Sect.~\ref{sec:tchannel_sol} as input, 
i.e.\ a two-channel MO problem for the $S$-wave to reproduce the $f_0(980)$ dynamics and a single-channel solution for $P$- and $D$-waves.
The role of the $F$-waves, whose inelastic nature is not fully compatible with the approach followed in Sect.~\ref{sec:tchannel_sol}, will be included only as a source of uncertainty, 
as will be the effect of the different ways to perform the continuation of the $S$-wave phase shift in the region above $\sqrt{t_0}=1.3 \GeV$.

\subsection{Numerical determination of the $s$-channel solution}
\label{sec:num_sol}

In order to obtain a numerical solution to the RS $s$-channel problem
we introduce a set of $N$ mesh points between threshold and $\Wm$ and minimize the square of the difference between the LHS and RHS of~\eqref{sRSpwhdr} for each partial wave
\beq
\label{chi2-pre}
\Delta_\text{RS}^2=\sum_{l,I_s,\pm}\sum_{j=1}^N\left(\Re f_{l\pm}^{I_s}(W_j)-F\left[f_{l\pm}^{I_s}\right](W_j)\right)^2,
\eeq
where $F[f_{l\pm}^{I_s}](W_j)$ denotes the RHS of the RS equations and $N$ should be varied within an interval, thus ensuring a result independent of the particular number of grid points.

In principle, an exact solution fulfills $\Delta_\text{RS}^2=0$, so the specific definition of~\eqref{chi2-pre} would be irrelevant in an ideal world. 
However, the input considered is known only within errors and for discrete values of the energy, 
which in particular affects the boundary conditions and thus introduces an uncertainty to the RS solutions. 
Therefore, the precise definition of the merit function considered for the minimization becomes relevant to achieve an accurate approximate solution.
The absence of statistical error estimates in any $s$-channel input prevents the use of a $\chi^2$, which in principle would be the more appropriate quantity for minimization algorithms.  
Moreover, as follows from Fig.~\ref{fig:phases-PWAs},
the partial waves involve different scales, the $P_{33}$-wave, for instance, is $30$ times bigger than the $P_{13}$-wave,
so that a merit function such as~\eqref{chi2-pre} that only considers absolute differences
cannot be used if a similar level of accuracy for all partial waves is to be achieved.
In practice, a more promising ansatz is given by
\beq
\label{chi2}
\Delta_\text{RS}^2=\sum_{l,I_s,\pm}\sum_{j=1}^N\left(\frac{\Re f_{l\pm}^{I_s}(W_j)-F\left[f_{l\pm}^{I_s}\right](W_j)}{\Re f_{l\pm}^{I_s}(W_j)}\right)^2,
\eeq
which weighs each partial wave in the same way. 
We have checked the stability of the solution with respect to the choice of $\Delta_\text{RS}^2$ as well as the number of grid points, which is varied between $15$ and $30$, but in the end fixed to $N=25$.
We have also checked the impact of the energy regions where the different partial waves approach zero, which in particular occurs for the $P_{11}$-wave in the region close to threshold, so that for a large number of grid points the definition of $\Delta^2_\text{RS}$ in~\eqref{chi2} becomes extremely sensitive to the amplitudes in the vicinity of the zeros. 
In those cases, we considered a modified version of~\eqref{chi2} where the denominator was substituted by $\Re f_{l\pm}^{I_s} +\epsilon$, with $\epsilon$ varied within the interval $[0.1,0.2]\GeV^{-1}$. We found that for the number of grid points we are using, $N<30$, both definitions led to equivalent results.

The requirement that the $S$-wave parameterizations in~\eqref{eq:schenk-par-Sw} fulfill the pionic-atom values for the scattering lengths exactly
already ensures that, after the minimization, the LHS of the $s$-channel RS equations will reproduce these values.    
However, in order to fully exploit the constraints on the $S$-wave scattering lengths from pionic atoms for
a solution of the RS system, we will make use of the RS sum rules for the threshold parameters derived in Sect.~\ref{sec:threshold}, 
and add a further piece to the merit function
\beq
\label{chia0}
\Delta_\text{SL}^2=\left(a_{0+}^{1/2}-F\left[a_{0+}^{1/2}\right]\right)^2+\left(a_{0+}^{3/2}-F\left[a_{0+}^{3/2}\right]\right)^2,
\eeq
where $F\left[a_{0+}^{I_s}\right]$ denotes the value for the RS sum rule of the $S$-wave scattering length with isospin $I_s$ in the $s$-channel basis.
The combined merit function considered for the minimization is then given by
\beq
\label{chitot}
\Delta^2=W_\text{SL}\,\Delta_\text{SL}^2+W_\text{RS}\,\Delta_\text{RS}^2,
\eeq
where the weights $W_\text{SL}=5$ and $W_\text{RS}=100$ have been chosen so that 
the final fit satisfies the pionic-atom constraints to reasonable accuracy, but is still flexible enough to move around easily in the phase space of free parameters.
In this way, $W_\text{RS}\Delta_\text{RS}^2$ could be more intuitively interpreted as a $\chi^2$-like function if we defined the $\chi^2$ divided by the number of data points and if the input error for each partial wave were taken as $1/\sqrt{W_\text{RS} N}=2\%$ of the $\Re f_{l\pm}^{I_s}(W_j)$ value at each energy point.

\begin{figure}[t!]
\includegraphics[width=0.45\linewidth,clip]{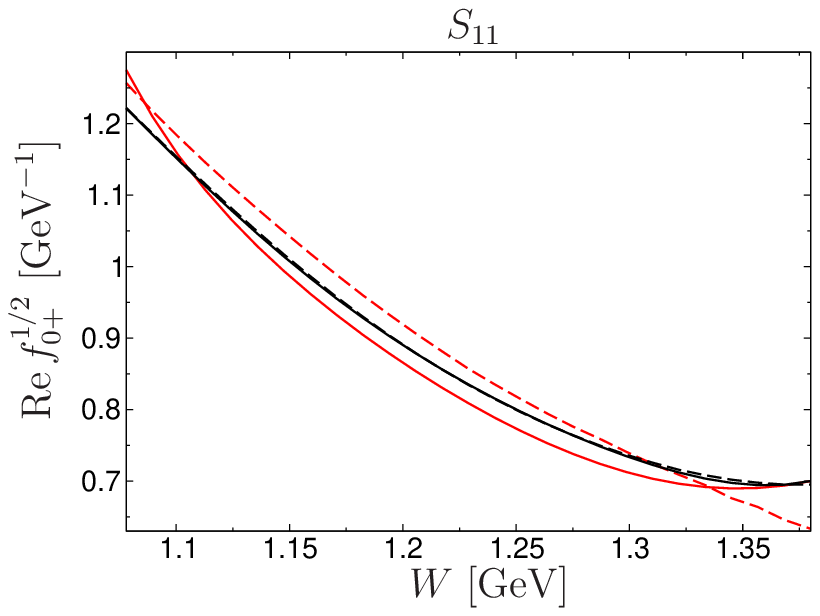}\quad
\includegraphics[width=0.45\linewidth,clip]{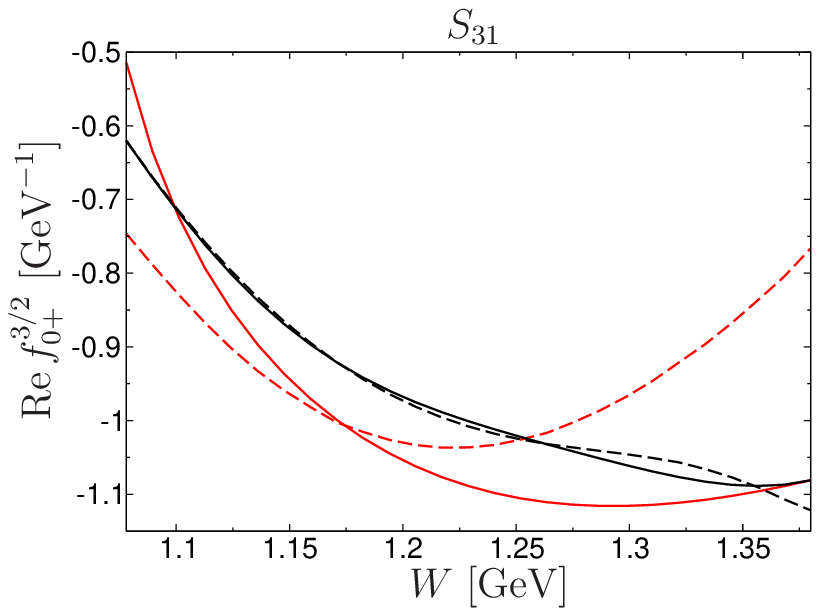}\\[0.1cm]
\includegraphics[width=0.45\linewidth,clip]{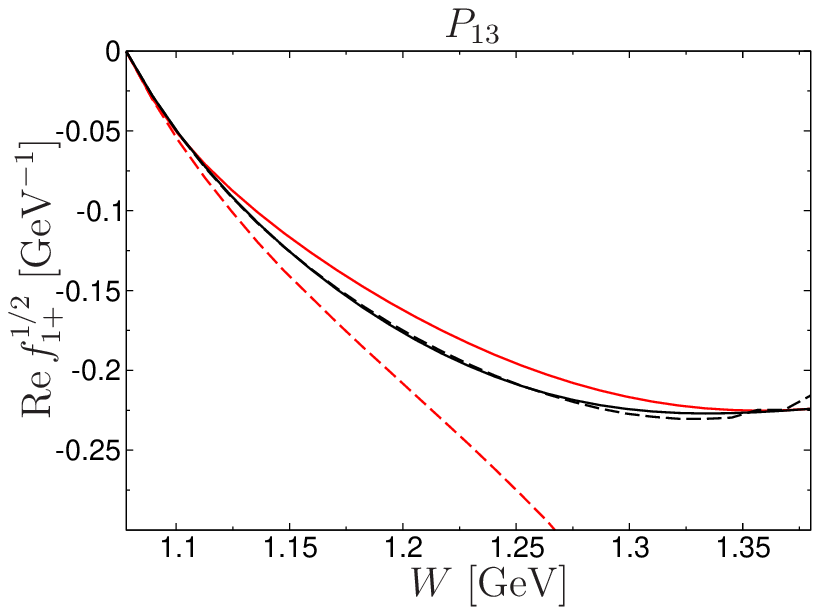}\quad
\includegraphics[width=0.45\linewidth,clip]{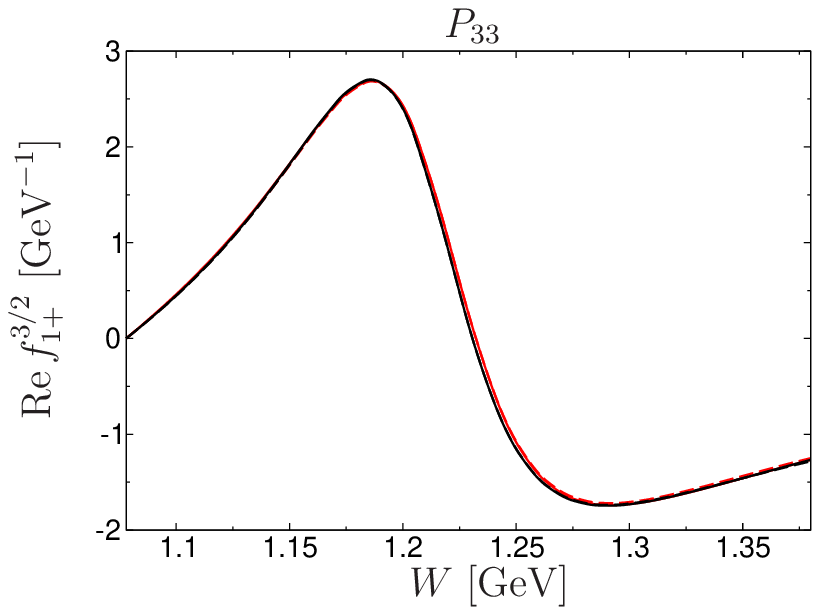}\\[0.1cm]
\includegraphics[width=0.45\linewidth,clip]{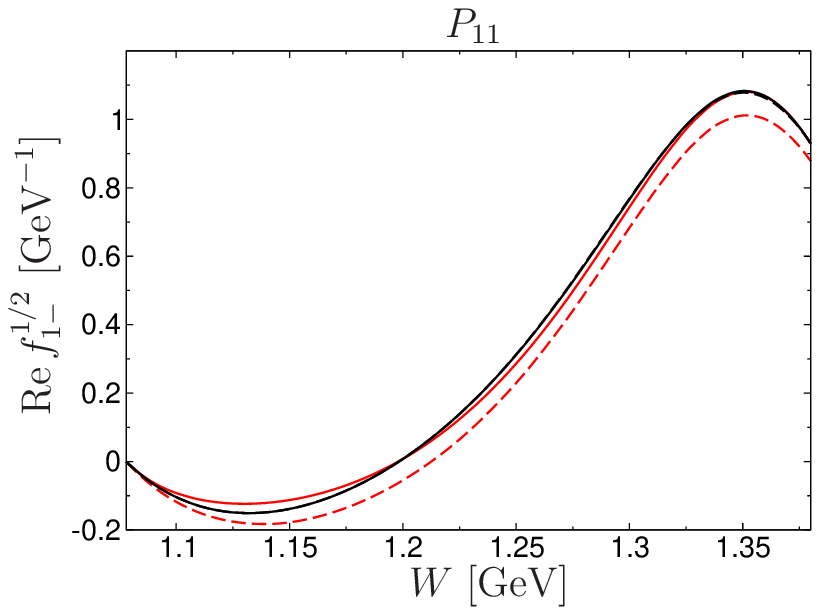}\quad
\includegraphics[width=0.45\linewidth,clip]{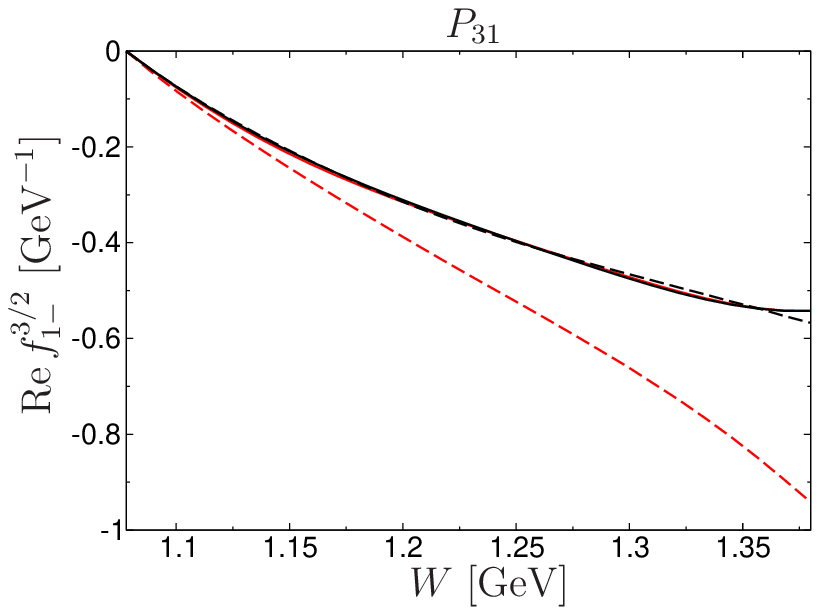}
 \caption{LHS and RHS of the RS equations for $\Re f_{l\pm}^{I_s}$. 
 The red solid curves indicate SAID results~\cite{Arndt:2006bf,Workman:2012hx},
   i.e.\ the LHS of~\eqref{spwhdr} before the fit. 
 Red dashed lines correspond to the RHS of~\eqref{spwhdr} when SAID $s$-channel amplitudes~\cite{Arndt:2006bf,Workman:2012hx} and KH80~\cite{Koch:1980ay,Hoehler:1983} subthreshold parameters are considered.
 Black solid and dashed lines correspond to the LHS and RHS of~\eqref{spwhdr} after the fit.}
 \label{fig:fit-notsub}
 \end{figure}

This concludes the discussion of the input quantities and the numerical procedure for the $s$-channel subproblem. 
As a first step, before facing a full self-consistent solution for both the $s$- and $t$-channel RS problems, 
we will start solving the $s$-channel problem with the $t$-channel results with KH80 subthreshold parameters described in Sect.~\ref{sec:tchannel_sol}.    
As the starting point for the minimization, we use KH80 values for the subthreshold parameters, 
and the fits to the GWU/SAID solutions depicted in Fig.~\ref{fig:phases-PWAs}.
In order to investigate to what extent these equations are fulfilled for the SAID $s$-channel amplitudes,
we compare the LHS and RHS of~\eqref{sRSpwhdr} before starting the minimization in Fig.~\ref{fig:fit-notsub}. 
This figure shows that the equations are fulfilled in the threshold region (except for the $S_{31}$-wave), 
while deviations emerge at higher energies in nearly all partial waves, most notably in the $P_{13}$ and $P_{31}$.
For the $S_{31}$  we also find a significant deviation already in the threshold region, 
but, in fact, this discrepancy is not surprising since the KH80 subthreshold parameters are tailored in such a way as to reproduce the KH80 scattering lengths, 
and the KH80 and SAID values for $a_{0+}^{3/2}$ differ substantially. 

As explained above, the minimization of~\eqref{chitot} provides us with a new set of subthreshold parameters and $S$- and $P$-wave phase shifts.
The results for the LHS and RHS of the $s$-channel RS equations after the fit, also shown in Fig.~\ref{fig:fit-notsub},
demonstrate good agreement, only for the $S_{31}$-wave small deviations are still perceptible close to $\Wm$. 
These remaining discrepancies can be further reduced by allowing for more degrees of freedom in the parameterization for the phase shifts in~\eqref{eq:schenk-par-Sw}. 
In a minimal version, already increasing the number of parameters in the description of the $S_{31}$ partial wave itself reduces the discrepancy appreciably, with
an effect on the phase shift localized to the energy region in question just below the matching point. 
However, compared to the other uncertainties of the input the accuracy reached with the default number of parameters  
proves to be sufficient, so that the shifts when allowing for extra parameters can simply be included in the error estimate.  

In addition, Fig.~\ref{fig:fit-notsub} shows that only for the $S$-waves there is a sizable change between the new solution and the GWU/SAID fit.
For the $P$-waves, the agreement between LHS and RHS of the RS $s$-channel equations is almost entirely due to the change of the subthreshold parameters.
This result justifies the approach already introduced in Sect.~\ref{sec:tchannel_sol}: 
a solution of the full RS system of equations can be achieved by including the dependence of the $t$-channel results on the subthreshold parameters~\eqref{P-wave-subth} and \eqref{S-wave-subth} explicitly in the minimization of~\eqref{chitot},  but neglecting their weak dependence on the $s$-channel input, which, in addition, changes little with respect to the SAID values considered in the first place.   
 
\subsection{Decomposition of the Roy--Steiner $s$-channel equations}
\label{sec:decomp}

Once the $s$-channel RS equations have been solved
and in order to discuss the relative importance of the input included, 
we decompose the $S$- and $P$-wave RS equations into four different parts 
\beq
\label{decomp}
f^{I_s}_{l\pm}(W)=N^{I_s}_{l\pm}(W)+K_sT(W)^{I_s}_{l\pm}+K_tT(W)^{I_s}_{l\pm}+DT(W)^{I_s}_{l\pm},
\eeq
where the last three contributions read
\begin{align}
\label{decomp-terms}
K_sT(W)^{I_s}_{l\pm}&=\frac{1}{\pi}\int\limits^{\Wm}_{W_+}\diff W'\;\sum\limits_{l'=0}^1
 \Big\{K^I_{ll'}(W,W')\,\Im f^I_{l'+}(W')+K^I_{ll'}(W,-W')\,\Im f^I_{(l'+1)-}(W')\Big\},\notag\\
K_tT(W)^{I_s}_{l\pm}&=\frac{1}{\pi}\int\limits^\infty_{\tpi}\diff t'\;\sum\limits_J
 \Big\{G_{lJ}(W,t')\,\Im f^J_+(t')+H_{lJ}(W,t')\,\Im f^J_-(t')\Big\},\notag\\
DT(W)^{I_s}_{l\pm}&=\frac{1}{\pi}\int\limits^{\infty}_{\Wm}\diff W'\;\sum\limits_{l'=0}^1
 \Big\{K^I_{ll'}(W,W')\,\Im f^I_{l'+}(W')+K^I_{ll'}(W,-W')\,\Im f^I_{(l'+1)-}(W')\Big\}\notag\\
 &+\frac{1}{\pi}\int\limits^{\infty}_{W_+}\diff W'\;\sum\limits_{l'=2}^\infty
 \Big\{K^I_{ll'}(W,W')\,\Im f^I_{l'+}(W')+K^I_{ll'}(W,-W')\,\Im f^I_{(l'+1)-}(W')\Big\}.
\end{align}
The first piece $N^{I_s}_{l\pm}$ includes the $s$-channel projection of the nucleon pole,  
as well as the contribution to the partial waves originating from the subtraction constants,
which are written in terms of subthreshold parameters as explained in~\ref{app:kernels_schannel} and \ref{app:kernel_subtractions}.
The low-energy contributions $W_+ \le W \le \Wm$ for $S$- and $P$-waves are collected in the $s$-channel kernel term $K_sT^{I_s}_{l\pm}$, 
whereas $K_tT^{I_s}_{l\pm}$ includes the contribution of the $t$-channel partial waves. 
Finally, the driving terms $DT^{I_s}_{l\pm}$ gather the $S$- and $P$-wave contribution above $\Wm$ and 
the effect of higher partial waves. In particular, the high-energy part above $\Wa=2.5 \GeV$ has been parameterized by means of the Regge representation from~\cite{Huang:2009pv}.

\begin{figure}[t!]
  \centering
 \includegraphics[width=0.45\linewidth,clip]{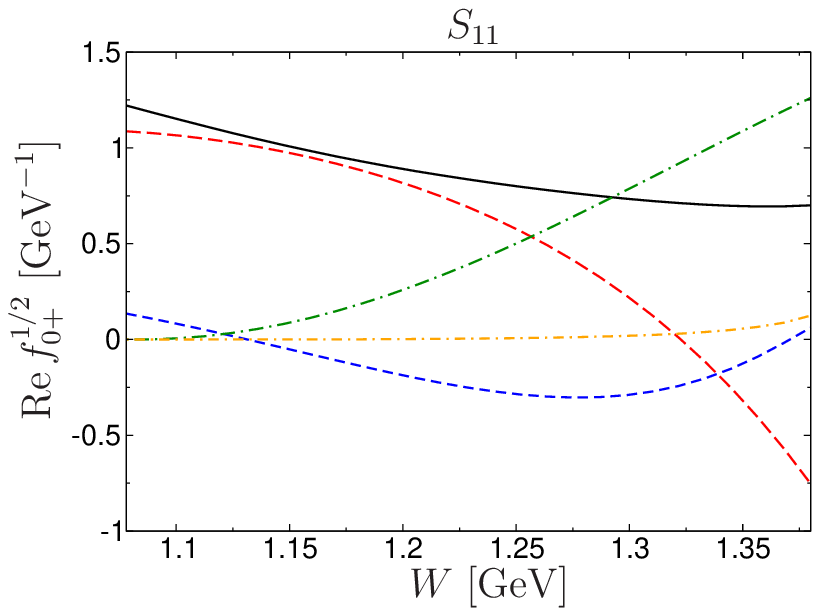}\quad
\includegraphics[width=0.45\linewidth,clip]{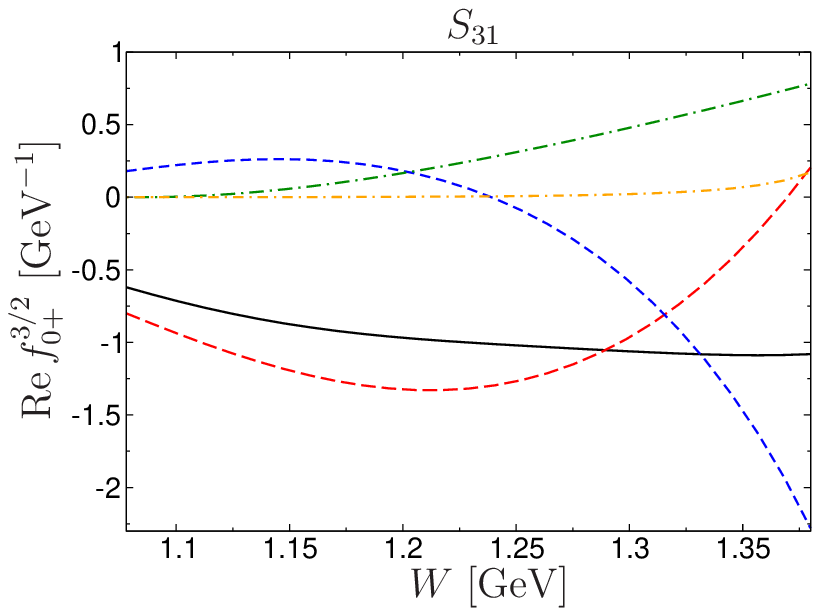}\\[0.1cm]
\includegraphics[width=0.45\linewidth,clip]{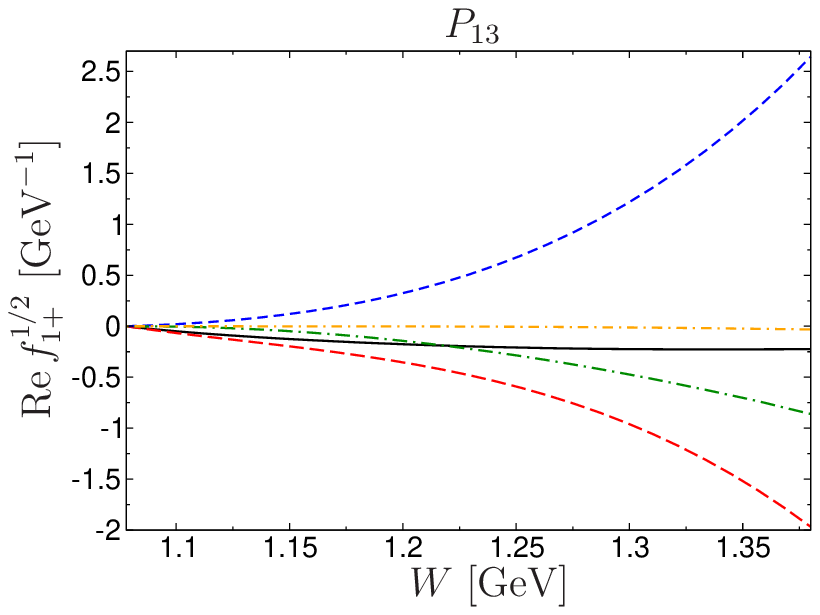}\quad
\includegraphics[width=0.45\linewidth,clip]{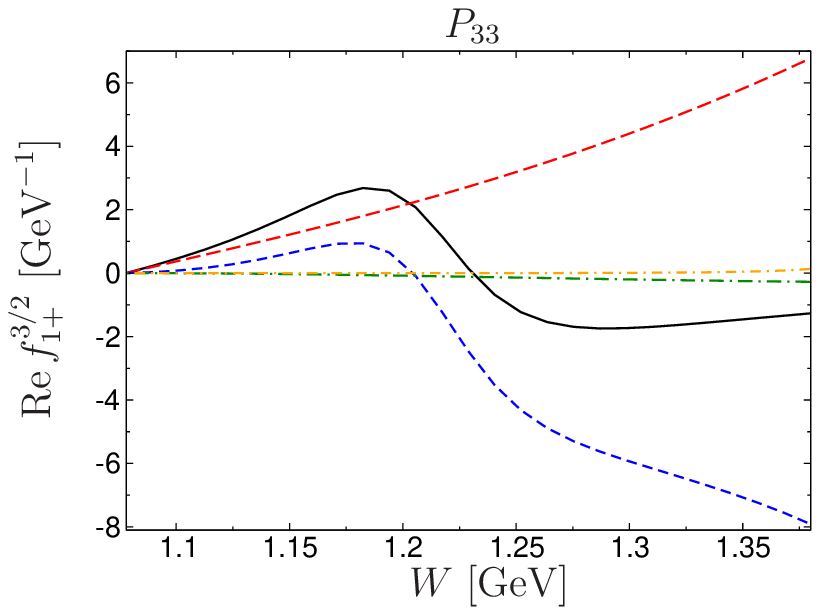}\\[0.1cm]
\includegraphics[width=0.45\linewidth,clip]{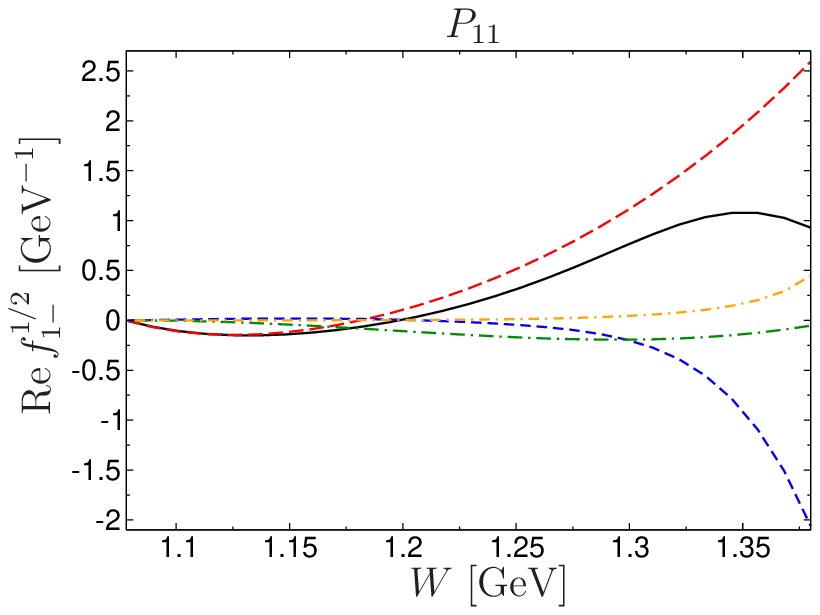}\quad
\includegraphics[width=0.45\linewidth,clip]{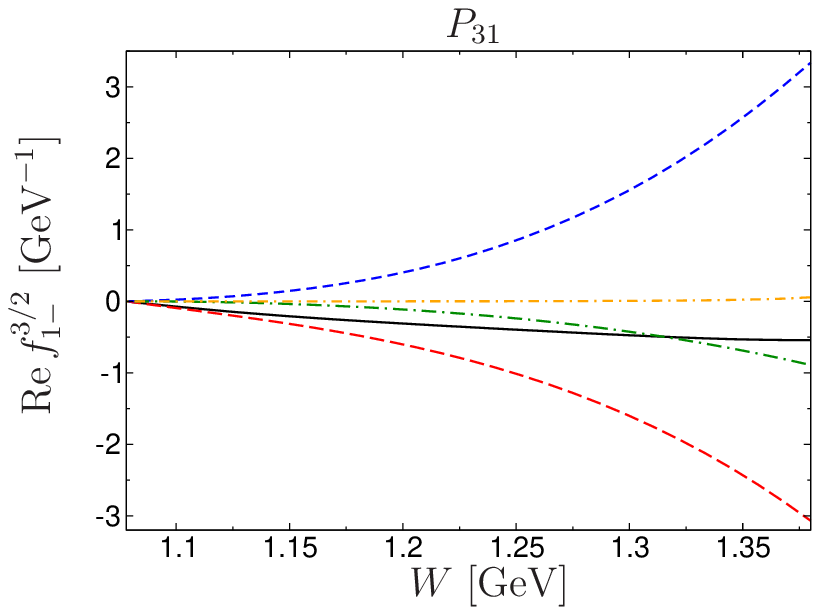}
  \caption{Decomposition of the RS equations into the different contributions given in~\eqref{decomp}. 
    Black solid lines correspond to the LHS of~\eqref{decomp}. 
    Red long-dashed lines denote the nucleon pole plus subtraction constant contribution $N^{I_s}_{l\pm}(W)$,
    whereas the blue short-dashed lines refer to the $s$-channel kernels $K_sT(W)$.
    The $t$-channel kernels $K_tT(W)$ are given by the green dot-long-dashed lines, and finally the driving terms $DT(W)$ 
    are described by the orange dot-short-dashed lines.} 
  \label{fig:decomp}
\end{figure} 

In Fig.~\ref{fig:decomp} we present the decomposition of the solution from Sect.~\ref{sec:num_sol}.
Black solid lines in this figure correspond to the real part of the partial wave, i.e.\ the LHS of~\eqref{spwhdr}.
Note that for the equations to be satisfied exactly this contribution should equal the sum of the other four. 
The pole terms $N^{I_s}_{l\pm}$ correspond to the red long-dashed lines and dominate completely in the region close to the $\pi N$ threshold.
However, as the energy increases they are largely canceled by the $s$-channel kernels, as expected since they also include the subtraction constants.
The only exception occurs for the $S_{11}$-wave, where the main cancellation is due the $t$-channel kernel. 
The $s$-channel kernels $K_sT^{I_s}_{l\pm}$ (blue short-dashed lines in Fig.~\ref{fig:decomp}) are small at low energies, 
but their role becomes dominant at higher energies. These two contributions are fully constrained by the RS solutions, and thus they are independent of the experimental input. 
The $t$-channel kernels $K_tT^{I_s}_{l\pm}$ (green dot-long-dashed lines) are relatively small for the $P$-waves, but meaningful for the $S$-waves. 
Finally, the driving terms $DT^{I_s}_{l\pm}$ (orange dot-short-dashed lines in Fig.~\ref{fig:decomp}) have a tiny effect for every partial wave. They are largely given by $S$- and $P$-wave contributions from $\Wm \leq W \leq \Wa$. 
The contribution from the asymptotic high-energy region is even two order of magnitude smaller, which makes the Regge contribution to~\eqref{spwhdr} totally negligible.
The driving terms are not constrained by RS equations, so their small role ensures a small dependence on the $s$-channel input considered.
These results already anticipate the conclusions of the error analysis that we will carry out in Sect.~\ref{sec:results}: 
the systematic error regarding the sensitivity of our solution to the input quantities is small. 

\subsection{Isospin conventions}

The solution strategy laid out in the present section relies on input for the $S$-wave scattering lengths $a_{0+}^{I_s}$. Indeed, we found that forcing both the RHS (via sum rules) and the LHS (via the Schenk parameter $A_{0+}^{I_s}$) to a given fixed value stabilizes the fit enormously in the space of $6\times 3+10=28$ fit parameters. At the level of accuracy that can be achieved with RS equations it becomes critical that the conventions for isospin breaking be consistently taken into account. For instance, the $\pi\pi$ analysis~\cite{Ananthanarayan:2000ht} was performed in the isospin limit, so that the resulting scattering lengths correspond to the strict isospin limit as well and subsequent comparison to experiment is only meaningful once the pertinent isospin-breaking corrections are applied~\cite{Knecht:1997jw,Gasser:2001un,Bissegger:2008ff,Colangelo:2008sm,Colangelo:2009zza}. These conventions imply that the PWAs used as input above the matching point are assumed to be purified of isospin-breaking effects as well, i.e.\ in principle both electromagnetic corrections and quark-mass effects need to be taken into account. Isospin-breaking corrections are expected to be most important close to threshold, so that once the dominant Coulomb effects are subtracted from the data, further corrections in the input above the matching point did not need to be considered in~\cite{Ananthanarayan:2000ht}.

For the $\pi N$ application we pursue a slightly different approach: we do not work with $f^{I_s}_{l\pm}$ etc.\ as would correspond to the isospin-limit amplitudes $T^\pm$ introduced in~\eqref{genlorentzinvamps}, but, by means of~\eqref{sampisorels}, define the isospin limit in terms of the scattering channels $\pi^\pm p\to\pi^\pm p$ with virtual photons removed (the latter is necessary to avoid photon cuts in the $\pi N$ amplitude, which would invalidate the derivation of the RS equations). This identification differs from the strict isospin-limit amplitudes by terms proportional to $\mpi^2-\mpii^2$, $e^2$ (from hard photons), and, at higher orders, $\mn-\mpp$, a leading-order example of which is given below by the difference between $a^+$ and $\tilde a^+$ in~\eqref{atilde_def}. 

The reason for this strategy is two-fold. First, the $\pi N$ data base is dominated by these channels, so that this isospin convention should be closest to the PWAs used as input. With Coulomb effects removed exactly and subleading electromagnetic effects approximately based on the Tromborg procedure~\cite{Tromborg:1976bi}, the remaining contamination of the PWA input should therefore be minimal. Second, the uncertainties in the scattering lengths extracted from pionic atoms become smallest once expressed in the $\pi^\pm p\to\pi^\pm p$ basis, while the translation to the strict isospin limit involves electromagnetic LECs that are not well determined. 
In these conventions, the RS solution will therefore become least sensitive to missing isospin-breaking corrections while at the same time profiting most from the pionic-atom constraints. In the same way as for the $\pi\pi$ case, the appropriate isospin-breaking corrections might have to be applied when the results are applied elsewhere, as illustrated by the discussion of the Cheng--Dashen LET in
Sect.~\ref{sec:cheng_dashen_theorem}.
Before presenting the full RS solution, we first turn to a review of the scattering-length extraction from pionic atoms in the next section.

\section{Extraction of the $\boldsymbol{\pi N}$ scattering lengths from pionic atoms}
\label{sec:pionic_atom}

In recent years, data on pionic atoms have become the primary source of information on the $\pi N$ scattering lengths~\cite{Gasser:2007zt}. In these systems, a $\pi^-$ and a proton or deuteron bound by electromagnetism, the strong interaction modifies the spectrum compared to pure QED by shifting the energy levels and introducing a finite width to the states, effects that are sensitive to threshold $\pi N$ physics.
In this way, new information on the $\pi N$ scattering lengths can be extracted by performing spectroscopy measurements. The most precise experimental results both for $\pi H$ and $\pi D$ have been obtained at PSI. For the shift of the ground state energies they are~\cite{Strauch:2010vu,Hennebach:2014lsa}
\beq
\label{level_shifts}
\eps_{1s}^{\pi H}=(-7.086\pm 0.009)\,\text{eV},\qquad 
\eps_{1s}^{\pi D}=(2.356\pm 0.031)\,\text{eV},
\eeq
where the sign convention is such that a negative shift corresponds to an attractive interaction. For the width of the $\pi H$ ground state the preliminary value from~\cite{Gotta:2008zza} is
\beq
\label{width}
\Gamma_{1s}^{\pi H}=(0.823\pm 0.019)\,\text{eV},
\eeq
but the error does not yet include all systematic effects. 
In the remainder of this section, we briefly review the theory input needed to extract the $\pi N$ scattering lengths from~\eqref{level_shifts} and \eqref{width} (closely following~\cite{Baru:2010xn,Baru:2011bw}) and present the final numbers to be used as central values in our RS solution. 

\subsection{Formalism}

\subsubsection{Deser formula}

The shift of the ground state level of $\pi H$ is related to the $\pi^-p$ scattering length $a_{\pi^- p}$, the effect corresponds to strong rescattering of the constituents, while the width gives access to the charge-exchange scattering length $a_{\pi^-p}^\text{cex}\equiv a_{\pi^-p\to \pi^0 n}$~\cite{Gasser:2007zt}, reflecting the decay of the $\pi^-p$ system into $\pi^0 n$. More precisely, $\epsilon_{1s}^{\pi H}$ is
related to $a_{\pi^- p}$ through an improved Deser formula~\cite{Lyubovitskij:2000kk} 
\beq
\eps_{1s}^{\pi H}=-2\alpha^3 \mu_H^2 a_{\pi^-p}\Big[1+2\alpha(1-\log \alpha)\mu_Ha_{\pi^-p}+\delta_\eps^\text{vac}\Big],
\label{eq:eps1s}
\eeq
where 
$\alpha=e^2/(4\pi)$ denotes the fine-structure constant, $\mu_H$ the reduced mass of $\pi H$, and 
$\delta_{\eps}^\text{vac}=2\delta \Psi_H(0)/\Psi_H(0)=0.48\%$ is the effect of 
vacuum polarization on the wave function at the origin~\cite{Eiras:2000rh}. 
The width determines $a_{\pi^-p}^\text{cex}$ by means of~\cite{Zemp:2004zz}
\beq
\Gamma_{1s}^{\pi H}=4\alpha^3\mu_H^2p_1\bigg(1+\frac{1}{P}\bigg)\big(a_{\pi^-p}^\text{cex}\big)^2
\Big[1+4\alpha(1-\log\alpha) \mu_Ha_{\pi^-p}+2\mu_H(\mpp+\mpi-\mn-\mpii)(a_{\pi^0n})^2+\delta_\eps^\text{vac}\Big],
\label{eq:Gamma1s}
\eeq
where $\mpp$, $\mn$, $\mpi$, and $\mpii$ refer to the masses of proton, neutron, charged and neutral pions, respectively,
$p_1$ is the momentum of the outgoing $n \pi^0$ pair, and the Panofsky ratio $P$ is given by~\cite{Spuller:1977ve}
\beq
P=\frac{\sigma(\pi^- p\rightarrow \pi^0 n )}{\sigma(\pi^- p\rightarrow n \gamma )}=1.546\pm 0.009.
\eeq 
Similarly, the shift of the $\pi D$ ground state, $\eps_{1s}^{\pi D}$, yields the real part of the $\pi^-d$ scattering length $\Re a_{\pi^-d}$ by means of~\cite{Meissner:2005bz}
\beq
\label{pid_Deser}
\eps_{1s}^{\pi D}=-2\alpha^3 \mu_D^2 \Re a_{\pi^-d}\Big[1+2\alpha(1-\log \alpha)\mu_D\Re a_{\pi^-d}+\delta_{\eps^D}^\text{vac}\Big],
\eeq
with reduced mass $\mu_D$ and vacuum-polarization effect $\delta_{\eps^D}^\text{vac}=2\delta \Psi_D(0)/\Psi_D(0)=0.51\%$~\cite{Eiras:2000rh}. In contrast to the level shift, the dominant decay channels $\pi^-d\to nn$ ($\text{BR}=73.9\,\%$) and $\pi^-d\to nn\gamma$ ($\text{BR}=26.1\,\%$)~\cite{Highland:1980vf} do not provide additional information on threshold $\pi N$ physics.

In view of~\eqref{eq:eps1s}, \eqref{eq:Gamma1s}, and \eqref{pid_Deser}, the experimental results quoted in~\eqref{level_shifts} and \eqref{width} amount to constraints on three different scattering channels, $a_{\pi^-p}$, $a_{\pi^-p}^\text{cex}$, and $\Re a_{\pi^-d}$. The latter is not directly related to $\pi N$ physics, only the two-body ($\pi N$) contribution 
\beq
a_{\pi^- d}^{(2)}=\frac{2\xip}{\xid}\big(a_{\pi^-p}+a_{\pi^-n}\big),
\eeq
with 
\beq
\xip=1+\frac{\mpi}{\mpp},\qquad \xid=1+\frac{\mpi}{\md},
\eeq
and deuteron mass $m_d$, provides the desired independent constraint. Therefore, to isolate $a_{\pi^- d}^{(2)}$, three-body ($\pi NN$) effects need to be carefully taken into account. Following the strategy laid out in~\cite{Baru:2010xn,Baru:2011bw} we decompose the three-body part of $a_{\pi^-d}$ as
\beq
\label{eq:apid3}
a_{\pi^- d}^{(3)}=a^\text{str}+a^{\text{disp}+\Delta}+a^\text{EM},
\eeq
where 
$a^{\text{disp} + \Delta}$ involves two-nucleon or $\Delta$-isobar intermediate states, $a^\text{EM}$ represents virtual-photon
corrections, and $a^\text{str}$ denotes the remaining classes of (strong-interaction) diagrams that enter in ChEFT.
Finally, to evaluate the constraints on the three charge channels $a_{\pi^-p}$, $a_{\pi^-p}^\text{cex}$, $a_{\pi^-n}$ in a global analysis, isospin-breaking corrections need to be applied. In the isospin limit, these channels reduce to $a^++a^-$, $-\sqrt{2}a^-$, and $a^+-a^-$ (where $a^\pm\equiv a_{0+}^\pm$), respectively, so that the two-body part of $\Re a_{\pi^-d}$ provides access to the isoscalar combination $a^+$.
In the following, we will briefly review the various corrections.

\subsubsection{Isospin breaking}
\label{sec:isospin_breaking}

In the isospin limit the eight physical $\pi N$ channels can be expressed by just two amplitudes, see~\eqref{sampisorels}, which at threshold implies
\begin{align}
a_{\pi^- p}&\equiv a_{\pi^- p\rightarrow \pi^- p}=a_{\pi^+ n}\equiv a_{\pi^+ n\rightarrow \pi^+ n}=a^++a^-,\notag\\
a_{\pi^+ p}&\equiv a_{\pi^+ p\rightarrow \pi^+ p}=a_{\pi^- n}\equiv a_{\pi^- n\rightarrow \pi^- n}=a^+-a^-,\notag\\
a_{\pi^- p}^\text{cex}&\equiv a_{\pi^- p\rightarrow \pi^0 n}=a_{\pi^+ n}^\text{cex}\equiv a_{\pi^+ n\rightarrow \pi^0 p}=-\sqrt{2}\,a^-,\notag\\
a_{\pi^0 p}&\equiv a_{\pi^0 p\rightarrow \pi^0 p}=a_{\pi^0 n}\equiv a_{\pi^0 n\rightarrow \pi^0 n}=a^+.
\end{align}
To extract $a^\pm$ from the pionic-atom measurements, we therefore need the corrections
\beq
\Delta a_{\pi^- p}= a_{\pi^- p}- (a^++a^-),\qquad \Delta a_{\pi^- n}= a_{\pi^- n}- (a^+-a^-),\qquad \Delta a_{\pi^- p}^\text{cex}=a_{\pi^- p}^\text{cex}+\sqrt{2}\,a^-,
\eeq 
which arise because of the quark mass difference $\md-\muu$ and electromagnetic interactions, and can be systematically addressed within ChPT. They have been worked out at NLO in the chiral expansion in~\cite{Hoferichter:2009ez,Hoferichter:2009gn,Gasser:2002am}.
Already at LO one encounters isospin-breaking effects that involve the LECs $c_1$, $f_1$, $f_2$~\cite{Meissner:2005ne}
\begin{align}
\label{piN_IV_LO}
\Delta a_{\pi^- p}^\text{LO}&=\frac{1}{4\pi\xip}\bigg\{\frac{4\Delta_\pi}{\Fpi^2}c_1-\frac{e^2}{2}(4f_1+f_2)\bigg\},& 
\Delta a_{\pi^- p}^\text{cex\, LO}&=\frac{\sqrt{2}}{4\pi\xip}\bigg\{\frac{e^2f_2}{2}+\frac{\ga^2\Delta_\pi}{4\Fpi^2\mpp}\bigg\},\notag\\
\Delta a_{\pi^- n}^\text{LO}&=\frac{1}{4\pi\xip}\bigg\{\frac{4\Delta_\pi}{\Fpi^2}c_1-\frac{e^2}{2}(4f_1-f_2)\bigg\},& 
\Delta_\pi&=\mpi^2-\mpii^2,
\end{align}
so that it becomes impossible to extract $a^+$ directly, only the combination
\beq
\label{atilde_def}
\tilde a^+ \equiv a^+ + \frac{1}{4\pi\xip}
\bigg\{\frac{4\Delta_\pi}{\Fpi^2}c_1-2e^2 f_1\bigg\}
\eeq
is accessible. $c_1$ also appears in isospin-limit $\pi N$ scattering and features in the chiral expansion of $\sigma_{\pi N}$, see Sect.~\ref{sec:ChPT}, $f_1$ parameterizes an electromagnetic mass shift that contributes equally to $\mpp$ and $\mn$, while $f_2$ follows from the electromagnetic part of the proton--neutron mass difference. For the numerical analysis we will later use the estimate $|f_1|\leq 1.4\GeV^{-1}$~\cite{Gasser:2002am,Fettes:2000vm} and $f_2=(-0.97\pm 0.38)\GeV^{-1}$ extracted from~\cite{Gasser:1974wd}.\footnote{The analysis~\cite{Gasser:1974wd} based on the Cottingham formula~\cite{Cottingham:1963zz} was criticized in~\cite{WalkerLoud:2012bg} regarding the treatment of the subtraction function. However, 
the underlying misconceptions have recently been rectified in~\cite{Gasser:2015dwa}, where in addition the consequences for the nucleon polarizabilities have been worked out and found to be in line with phenomenology~\cite{McGovern:2012ew,Myers:2014ace}. Moreover, the corresponding number for $f_2$ is consistent with extractions from $pn\to d\pi^0$~\cite{Filin:2009yh},  $f_2=(-0.3\pm 1.2)\GeV^{-1}$, and the latest determination on the lattice in full QCD+QED~\cite{Borsanyi:2014jba}, 
$f_2=(-1.28\pm 0.20)\GeV^{-1}$.  The latter is more precise than the estimate from the Cottingham formula, but given that the error on $f_2$ is already insignificant for the final uncertainty estimates, we stick to the input used in~\cite{Baru:2010xn,Baru:2011bw}.}
In order to minimize bias from theory, we do not insert these estimates yet, but perform the analysis in terms of $\tilde a^+$ and $a^-$, as well as the corrections
\beq
\Delta \tilde a_{\pi^-p}=a_{\pi^-p}-(\tilde a^++a^-),\qquad \Delta \tilde a_{\pi^-n}=a_{\pi^-n}-(\tilde a^+-a^-).
\eeq
Numerically, the NLO calculation of~\cite{Hoferichter:2009ez,Hoferichter:2009gn,Gasser:2002am} yields
\begin{align}
 \Delta \tilde a^+=\frac{1}{2}\big(\Delta \tilde a_{\pi^-p}+\Delta \tilde a_{\pi^-n}\big)&=(-3.3\pm 0.3)\times 10^{-3}\mpi^{-1}, & 
\Delta a^-&=(1.4\pm 1.3)\times 10^{-3}\mpi^{-1},\notag\\
\Delta \tilde a_{\pi^- p}&=(-2.0\pm 1.3)\times 10^{-3}\mpi^{-1},& 
\Delta a^\text{cex}_{\pi^-p}&=(0.4\pm 0.9)\times 10^{-3}\mpi^{-1},
\end{align}
where the error is dominated by NLO LECs.

\subsubsection{Three-body effects}

\begin{figure}[t!]
\centering
\includegraphics[width=0.8\linewidth,clip]{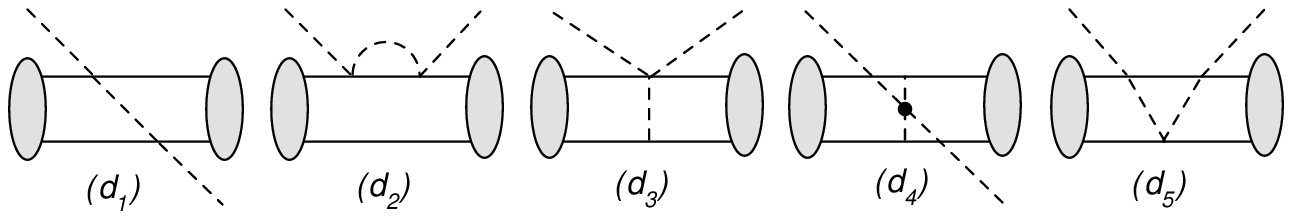}\\[0.1cm]
\includegraphics[width=0.8\linewidth,clip]{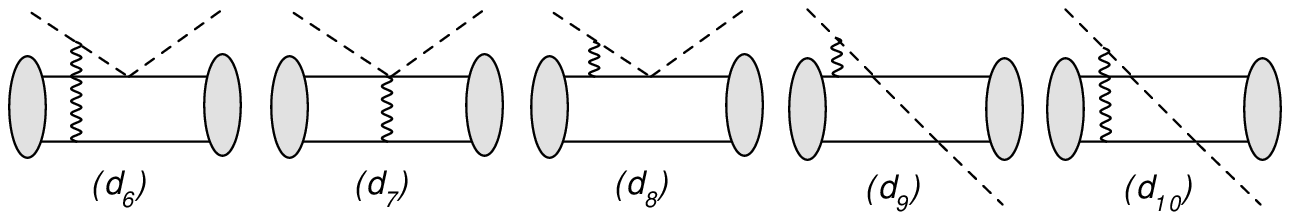}\\[0.1cm]
\raisebox{0.05cm}{\includegraphics[width=0.45\linewidth,clip]{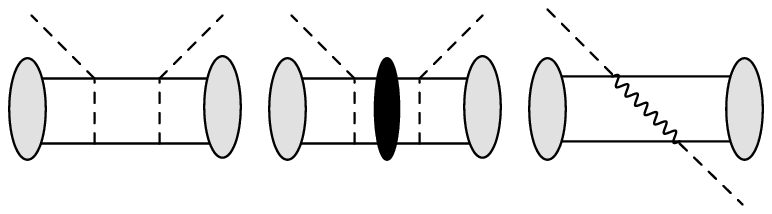}}\qquad
\includegraphics[width=0.45\linewidth,clip]{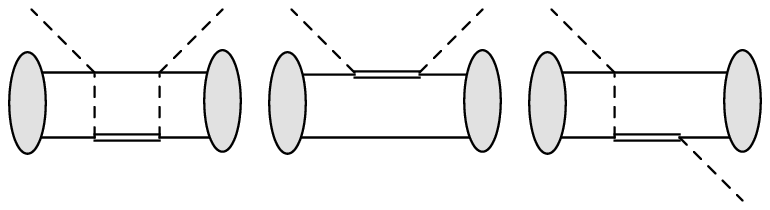}
\caption{Three-body diagrams in $\pi^-d$ scattering. Solid, dashed, wiggly, and double lines denote nucleons, pions, photons, and $\Delta(1232)$ degrees of freedom, respectively, the gray blobs the deuteron wave functions, and the black ellipse intermediate-state $NN$ interactions. The first row shows diagrams that enter $a^\text{str}$, the second diagrams relevant for $a^\text{EM}$, and the third representatives of $a^{\text{disp}+\Delta}$, see main text for details.}
\label{fig:three_body}
\end{figure}

The calculation of the three-body contributions is organized in ChEFT~\cite{Weinberg:1992yk}, see~\cite{Baru:2011bw} for a detailed account. The numerically by far dominant double-scattering diagram, $(d_1)$ in Fig.~\ref{fig:three_body}, had already been identified early on~\cite{Weinberg:1992yk,Baru:1997xf,Beane:1997yg,Ericson:2000md}, and all other contributions will be given compared relative to this diagram. Further LO diagrams are of the type $(d_3)$ and $(d_4)$, while NLO corrections to these topologies were shown to vanish in~\cite{Beane:2002wk}. A special role is played by diagram $(d_5)$, which formally enters at N$^2$LO, but is enhanced by a large numerical factor $\pi^2$, and therefore has to be included in the calculation~\cite{Liebig:2010ki}. 
The strategy of~\cite{Baru:2011bw} was to include all contributions below the order of the first $(N^\dagger N)^2\pi\pi$ contact term, which enters at N$^2$LO. In addition to the diagrams mentioned so far, there are contributions from nucleon recoil~\cite{Kolybasov:1976mm,Faldt:1974sm,Baru:2004kw,Lensky:2005hb,Baru:2009tx}, which originate from the interplay between diagrams $(d_1)$ and $(d_2)$, start at NLO, and can also contribute at fractional orders in the chiral expansion. The sum $(d_1)$ to $(d_5)$ is represented by $a^\text{str}$ in~\eqref{eq:apid3}.
The special role of $(d_1)$ and $(d_5)$, the first two terms in the multiple-scattering series~\cite{Brueckner:1953zz,Kolybasov:1972bn,Kamalov:2000iy,Meissner:2006gx,Baru:2012iv,Kudryavtsev:2012ap}, could suggest that even higher-order terms might be important. However, the full series can be resummed in configuration space, with the result that multiple-scattering topologies beyond $(d_5)$ are numerically much less important than the parametric estimate for the N$^2$LO contact term or the explicit wave-function dependence observed in the integrals, both of which point to an accuracy of a few percent. 
The final result quoted in~\cite{Baru:2011bw} (for a fixed $a^-=86.1\times 10^{-3}\mpi^{-1}$), based on AV18~\cite{Wiringa:1994wb}, CD-Bonn~\cite{Machleidt:2000ge}, and N$^2$LO chiral interactions~\cite{Epelbaum:2004fk}, 
becomes
\beq  
a^\text{str} = (-22.6\pm 1.1\pm 0.4)\times10^{-3}\mpi^{-1},
\eeq
where the first uncertainty arises from the different short-distance 
physics of the deuteron wave functions,\footnote{In fact, it has been demonstrated that, for deuteron wave functions based on the one-pion-exchange
interaction, the results for the individual diagrams at LO and NLO
become cutoff-independent in the limit of a large cutoff~\cite{Platter:2006pt,PavonValderrama:2006np,Nogga:2005fv,Liebig:2010ki}.} and the second from the isospin-breaking shifts in the $\pi N$ scattering lengths that enter at the vertices in the double-scattering diagram.
Finally, there are isospin-conserving diagrams that generate an absorptive part~\cite{Lensky:2006wd} as well as those where the $\Delta(1232)$ becomes important~\cite{Baru:2007wf}, see the last row in Fig.~\ref{fig:three_body}. Both contributions enter at the fractional order $\Order(p^{3/2})$, their combined effect is~\cite{Lensky:2006wd,Baru:2007wf}
\beq
 a^{\text{disp}+\Delta} = (-0.6\pm 1.5)\times 10^{-3} \mpi^{-1}.
\eeq

In addition to these isospin-conserving contributions, the required few-percent-level accuracy makes the consideration of virtual-photon effects mandatory. In fact, the scale separation in the integrals involving virtual photons amounts to a definition of $\Re a_{\pi^-d}$: photons with momenta of the order of the atomic-physics scale $p\sim \alpha \mpi$ are included in the derivation of the Deser formula~\eqref{pid_Deser}, so that contributions above that scale need to be separated. The corresponding diagrams shown in the second line of Fig.~\ref{fig:three_body} are infrared enhanced due to the photon and pion propagators, and, therefore, albeit higher order the virtual-photon corrections to the numerically dominant double-scattering topologies might have become relevant. However, as shown in~\cite{Baru:2011bw}, in the end it suffices
to consider the diagrams $(d_6)$, $(d_7)$, and $(d_8)$, with the result
\beq 
a^\text{EM} =(0.94\pm 0.01)\times  10^{-3} \mpi^{-1}.
\eeq
As pointed out in~\cite{Baru:2011bw}, there is a remarkable cancellation amongst various contributions that go beyond the static approximation for the double- and triple-scattering diagrams. 
In consequence, the main impact of all the additional corrections discussed in this section on the extraction of the scattering lengths
actually originates from the large isospin-breaking shift in the two-body sector $\Delta \tilde a^+=(-3.3\pm 0.3)\times 10^{-3}\mpi^{-1}$~\cite{Hoferichter:2009cq}.

\subsection{Results}

\begin{figure}
\centering
\includegraphics[width=0.7\linewidth,clip]{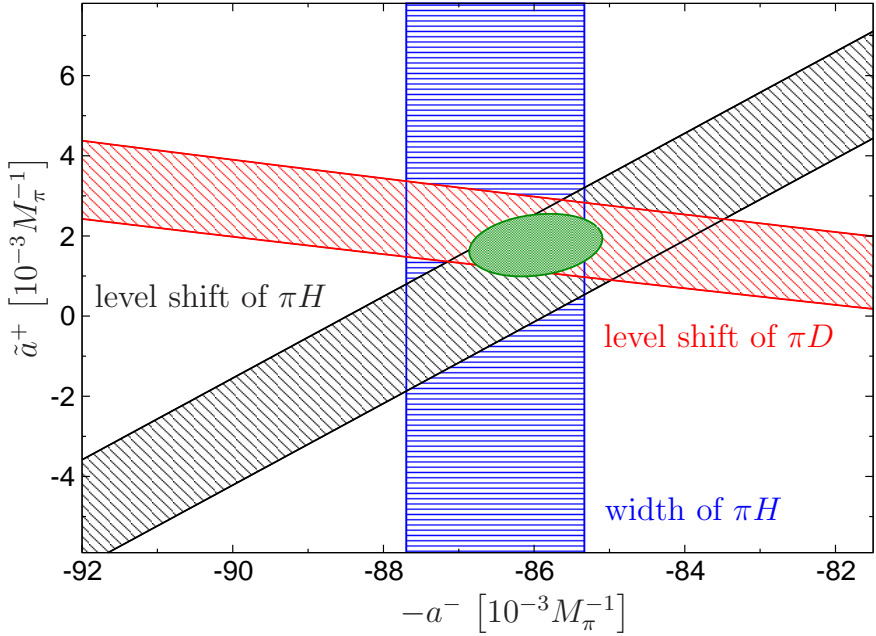}
\caption{Combined constraints on $\tilde a^+$ and $a^-$ from data on the width and level
shift of $\pi H$, as well as the $\pi D$ level shift. The figure is an updated version of the ones from~\cite{Baru:2010xn,Baru:2011bw} to account for the new value of the $\pi H$ level shift~\cite{Hennebach:2014lsa}.}
\label{fig:bands}
\end{figure}

Based on the theory input reviewed in the previous section, one obtains the constraints on $\tilde a^+$ and $a^-$ shown in Fig.~\ref{fig:bands}. The figure is slightly updated compared to~\cite{Baru:2010xn,Baru:2011bw} to account for the new value for the $\pi H$ level shift~\cite{Hennebach:2014lsa} that has become available in the meantime. In particular, one obtains from the $1\sigma$ error ellipse  
\beq
\tilde{a}^+=(1.8\pm 0.8)\times 10^{-3} \mpi^{-1},\qquad a^-=(86.0\pm 0.9)\times 10^{-3}\mpi^{-1},
\label{atilde}
\eeq
with a correlation coefficient $\rho_{a^-\tilde a^+}=-0.21$. If the error on the $\pi H$ width is increased, keeping in mind that~\eqref{width} is preliminary, the error on $a^-$ will increase as well, while $\tilde a^+$ is hardly affected. In the extreme case that the band from the $\pi H$ width is discarded altogether, $a^-$ reduces to $(85.3\pm 1.3)\times 10^{-3}\mpi^{-1}$, while $\tilde a^+=(1.9\pm 0.8)\times 10^{-3} \mpi^{-1}$ remains nearly unchanged. For the application in the RS equations we will actually use $\tilde a^+$ and $a_{\pi^-p}$, the latter extracted directly from~\eqref{level_shifts} by means of~\eqref{eq:eps1s}, so that the results will be largely immune to a potential shift in the $\pi H$ width.

Using the appropriate $c_1=(-1.07\pm 0.02)\GeV^{-1}$ from Sect.~\ref{sec:LECs} as well as the estimate $|f_1|\leq 1.4\GeV^{-1}$~\cite{Gasser:2002am,Fettes:2000vm}, the isoscalar scattering length becomes
\beq
a^+=(7.9\pm 2.6)\times 10^{-3}\mpi^{-1},
\eeq
where the error is dominated by the uncertainty in $f_1$, which is so large that it should safely cover higher-order corrections.
Combining the results for $\tilde a^+$, $a^-$, and $a^+$ with the isospin-breaking corrections from~\cite{Hoferichter:2009ez} one arrives at the scattering lengths for the physical channels summarized in the upper panel of Table~\ref{table:physical_channels}. As mentioned above, a more precise result for $a_{\pi^-p}$ follows directly from~\eqref{eq:eps1s}, and a similar strategy can be used to improve the scattering lengths in the channels that in the isospin limit collapse to $a^+\pm a^-$. Rewriting
\begin{align}
 a_{\pi^+p}&=2\tilde a^+-a_{\pi^-p}-\frac{1}{4\pi\xip}\Bigg\{e^2 f_2+\frac{\ga^2\mpi}{16\pi\Fpi^2}\bigg(\frac{33\Delta_\pi}{4\Fpi^2}+e^2\bigg)\Bigg\},\notag\\
 a_{\pi^-n}&=2\tilde a^+-a_{\pi^-p}+\frac{1}{4\pi\xip}\Bigg\{-\frac{\ga^2\mpi}{16\pi\Fpi^2}\bigg(\frac{33\Delta_\pi}{4\Fpi^2}+e^2\bigg)+2e^2\mpi\big(2g_6^\text{r}+g_8^\text{r}\big)\Bigg\},\notag\\
 a_{\pi^+n}&=a_{\pi^-p}+\frac{e^2}{4\pi\xi_p}\Big\{f_2-2\mpi\big(2g_6^\text{r}+g_8^\text{r}\big)\Big\},
\end{align}
with $\Order(e^2p)$ LECs $g_i^\text{r}$~\cite{Gasser:2002am}, produces the results in the lower panel of Table~\ref{table:physical_channels}: a significant reduction in the uncertainty for $a_{\pi^-p}$ and $a_{\pi^+n}$, and still a slight improvement for the other two channels. 
As argued in Sect.~\ref{sec:schannel_sol}, for the RS solution we are interested in $a_{\pi^\pm p}$, or, equivalently, the $I_s=1/2,3/2$ combinations constructed from these quantities. However, the numbers given in Table~\ref{table:physical_channels} still contain effects from virtual photons, which, for the application in dispersion relations, should be removed as far as possible, as will be discussed in the next section.

\begin{table}
\centering
\renewcommand{\arraystretch}{1.3}
\begin{tabular}{ccrcr}
\toprule
isospin limit & channel & scattering length & channel & scattering length \\\midrule
$a^++a^-$ & $\pi^-p\rightarrow \pi^-p$ & $85.8\pm 1.8$ & $\pi^+n\rightarrow \pi^+n$ & $84.9\pm 1.8$ \\
$a^+-a^-$ & $\pi^+p\rightarrow \pi^+p$ & $-88.1\pm 1.8$ & 	$\pi^-n\rightarrow \pi^-n$  & $-89.0\pm 1.8$ \\
$-\sqrt{2}\,a^-$ & $\pi^-p\rightarrow \pi^0n$ & $-121.2\pm 1.6$ &  	$\pi^+n\rightarrow \pi^0 p$ & $-119.3\pm 1.6$\\
$a^+$ & $\pi^0p\rightarrow \pi^0p$ & $2.1\pm 3.1$ & 		$\pi^0n\rightarrow \pi^0 n$ &  $5.5\pm 3.1$\\\midrule
$a^++a^-$ & $\pi^-p\rightarrow \pi^-p$ & $85.25\pm 0.11$ & $\pi^+n\rightarrow \pi^+n$ & $84.4\pm 0.7$ \\
$a^+-a^-$ & $\pi^+p\rightarrow \pi^+p$ & $-87.6\pm 1.6$ & 	$\pi^-n\rightarrow \pi^-n$  & $-88.4\pm 1.6$ \\\bottomrule
\end{tabular}
\caption{$\pi N$ scattering lengths for the physical channels in units of $10^{-3}\mpi^{-1}$, including virtual photons. The upper panel refers to the results based on $\tilde a^+$, $a^-$, and $a^+$, while in the lower panel $a^-$ is eliminated in favor of $a_{\pi^-p}$.}
\label{table:physical_channels}
\renewcommand{\arraystretch}{1.0}
\end{table}

\subsection{Virtual photons}

The complications that arise if one aims for a definition of a purely strong scattering length, with all electromagnetic effects removed, is well known in proton--proton scattering, which we will discuss first to illustrate the corresponding subtleties and later compare to the case of $\pi N$~\cite{Hoferichter:2012bz}.
As a first step, the pure Coulomb phase shift $\sigma^C$ is removed, so that the remainder $\delta^C_{pp}$ is related to the strong amplitude $T_{pp}(q)$ according to
\beq
q\big(\cot \delta^C_{pp}-i\big)=-\frac{4\pi}{\mpp}\frac{e^{2i\sigma^C}}{T_{pp}(q)},\qquad q=|\qq|,
\eeq
 and obeys the modified effective range expansion~\cite{Bethe:1949yr}
\begin{align}
\label{mERE}
q\Big[C_\eta^2\big(\cot \delta^C_{pp}-i\big)+2\eta H(\eta)\Big]&=-\frac{1}{a_{pp}^C}+\frac{1}{2}r_0q^2+\Order(q^4),\notag\\
C_\eta^2=\frac{2\pi\eta}{e^{2\pi\eta}-1},\qquad \eta&=\frac{\alpha \mpp}{2q},\qquad
 H(\eta)=\psi(i\eta)+\frac{1}{2i\eta}-\log(i\eta),\qquad \psi(x)=\frac{\Gamma'(x)}{\Gamma(x)},
\end{align}
where $\qq$ is the CMS momentum and $\Gamma(x)$ denotes the conventional Gamma function.
While the parameters $a_{pp}^C$ and $r_0$ appearing in~\eqref{mERE} are scale-independent, the separation of residual 
Coulomb interactions to define a purely strong scattering length $a_{pp}$ introduces a scale~\cite{Jackson:1950zz,Kong:1999sf}
\beq
\frac{1}{a_{pp}}=\frac{1}{a_{pp}^C}+\alpha \mpp\bigg[\log\frac{\mu\sqrt{\pi}}{\alpha \mpp}+1-\frac{3}{2}\gamma_E\bigg],
\eeq
since the Coulomb--nuclear interference depends on the short-distance part of the nuclear force. 
The origin of this scale dependence can be traced back to the fact that for 
the subtraction of virtual photons the electromagnetic coupling has to be switched off in the running of operators as well, 
which can only be done fully consistently if the underlying theory is known~\cite{Gegelia:2003ta,Gasser:2003hk}. For $pp$ scattering such residual Coulomb effects are by no means negligible, they induce a sizable difference between $a_{pp}$ and $a_{pp}^C$~\cite{Bergervoet:1988zz,Miller:1990iz}
\beq
a_{pp}^C=(-7.8063\pm 0.0026)\,\text{fm},\qquad a_{pp}=(-17.3\pm 0.4)\,\text{fm}.
\eeq

For $\pi N$ scattering the standard ChPT definition for the scattering lengths~\cite{Gasser:2007zt}
\beq
e^{-2i\sigma^C}T_{\pi^-p}=\frac{\pi\alpha\mu_H a_{\pi^-p}}{q}-2\alpha\mu_H\big(a_{\pi^-p}\big)^2\log\frac{q}{\mu_H}+a_{\pi^-p}+\Order(q,\alpha^2),
\eeq
with the Coulomb pole $\propto 1/q$ and the term $\propto \log q/\mu_H$ first generated at one- and two-loop level,
matches onto the modified effective range expansion~\eqref{mERE} as follows
\beq
e^{-2i\sigma^C}T_{\pi^-p}=\frac{\pi\alpha\mu_Ha_{\pi^-p}^C}{q}-2\alpha\mu_H\big(a_{\pi^-p}^C\big)^2\bigg(\gamma_E+\log\frac{q}{\alpha\mu_H}\bigg)+a_{\pi^-p}^C+\Order(q,\alpha^2).
\eeq
This leads to the identification
\beq
a_{\pi^-p}=a_{\pi^-p}^C+2\alpha\mu_H\big(a_{\pi^-p}^C)^2(\log\alpha-\gamma_E)+\Order(\alpha^2),
\eeq
so that $a_{\pi^-p}$ (nearly) corresponds to the concept of a Coulomb-subtracted scattering length as defined in terms of the modified effective range expansion, up to higher orders in $\alpha$. In particular, it is a scale-independent quantity as well. 

The main difference between the $\pi N$ and the $NN$ system is that the former is perturbative, while the fine-tuning in the $NN$ potential will enhance any residual virtual-photon effects. In ChPT the necessity for a choice of scale when removing virtual photons manifests itself in the need for the regularization of UV divergent virtual-photon diagrams, where the separation between mass-difference ($\Delta_\pi$) and virtual-photon contributions to LECs requires a scale. However, due to the $\pi N$ system being perturbative, these virtual-photon contributions are numerically of similar size as the uncertainties in the scattering lengths themselves, so that any induced scale dependence of the correction will be tiny and becomes completely negligible in practice. 

Following~\cite{Baru:2011bw}, we perform this separation of the LECs based on their $\beta$-functions. 
The scattering lengths with virtual-photon effects removed can then be written as
\begin{align}
\label{scatt_pionic_atoms_final}
 a^{1/2}_{0+}&=2a_{\pi^-p}-\tilde a^++\frac{1}{8\pi\xip}\Bigg\{e^2 f_2+\frac{\ga^2\mpi}{16\pi\Fpi^2}\bigg(\frac{33\Delta_\pi}{4\Fpi^2}+2e^2\bigg)\Bigg\}
 +\frac{\mpi}{2\pi\xip}\Bigg\{\frac{e^2\ga^2}{16\pi^2\Fpi^2}\bigg(1+4\log 2+3\log\frac{\mpi^2}{\mu^2}\bigg)-2e^2C_{gk}\Bigg\}\notag\\
&=(169.8\pm 2.0)\times 10^{-3}\mpi^{-1},\notag\\
a^{3/2}_{0+}&=-a_{\pi^-p}+2\tilde a^+-\frac{1}{4\pi\xip}\Bigg\{e^2 f_2+\frac{\ga^2\mpi}{32\pi\Fpi^2}\bigg(\frac{33\Delta_\pi}{2\Fpi^2}+e^2\bigg)\Bigg\}
-\frac{\mpi}{4\pi\xip}\Bigg\{\frac{e^2\ga^2}{16\pi^2\Fpi^2}\bigg(1+4\log 2+3\log\frac{\mpi^2}{\mu^2}\bigg)-2e^2C_{gk}\Bigg\}\notag\\
&=(-86.3\pm 1.8)\times 10^{-3}\mpi^{-1},
\end{align}
where $C_{gk}=\tilde g^\text{r}_6+\tilde g^\text{r}_8-\frac{5}{9\Fpi^2}\tilde k_1^\text{r}$ (see~\cite{Baru:2011bw} for precise definitions) is exactly the combination of LECs whose virtual-photon contribution induces the scale dependence in the one-loop calculation. However, given that 
little is known about the numerical values particularly for the $g_i^\text{r}$ anyway, this effect can be safely ignored in practice.

In the remainder of this paper we will adopt the results quoted in~\eqref{scatt_pionic_atoms_final} as our central values.
Moreover, since the uncertainty in $a^{1/2}_{0+}$ is dominated by the LECs, while the main source of uncertainty in the case of $a^{3/2}_{0+}$ is $\tilde a^+$, their errors can be considered uncorrelated to very good approximation.

\section{Results for low-energy phase shifts and subthreshold parameters}
\label{sec:results}

\begin{figure}[t!]
\includegraphics[width=0.45\linewidth,clip]{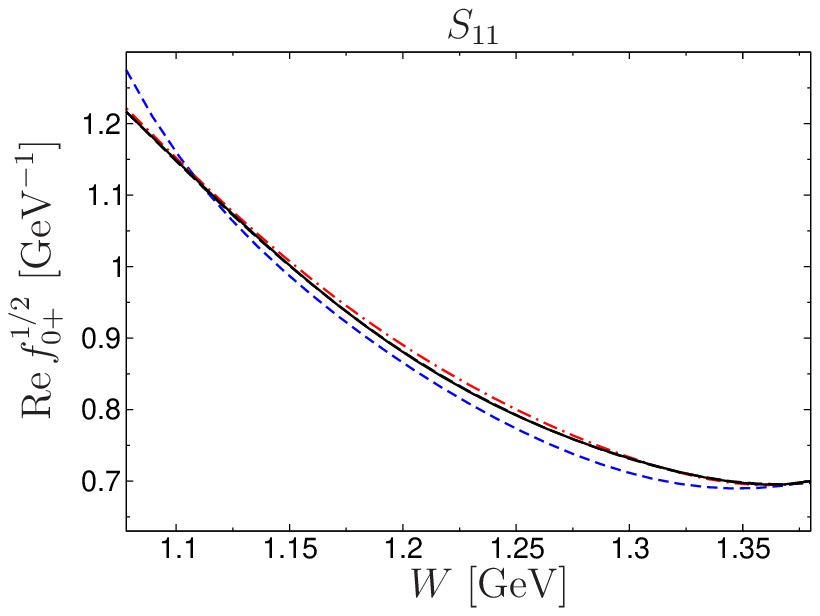}\quad
\includegraphics[width=0.45\linewidth,clip]{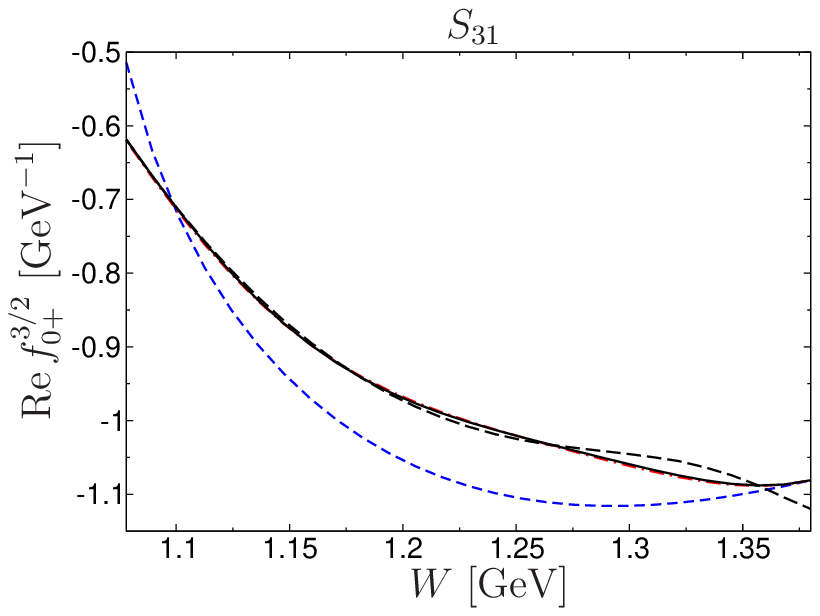}\\[0.1cm]
\includegraphics[width=0.45\linewidth,clip]{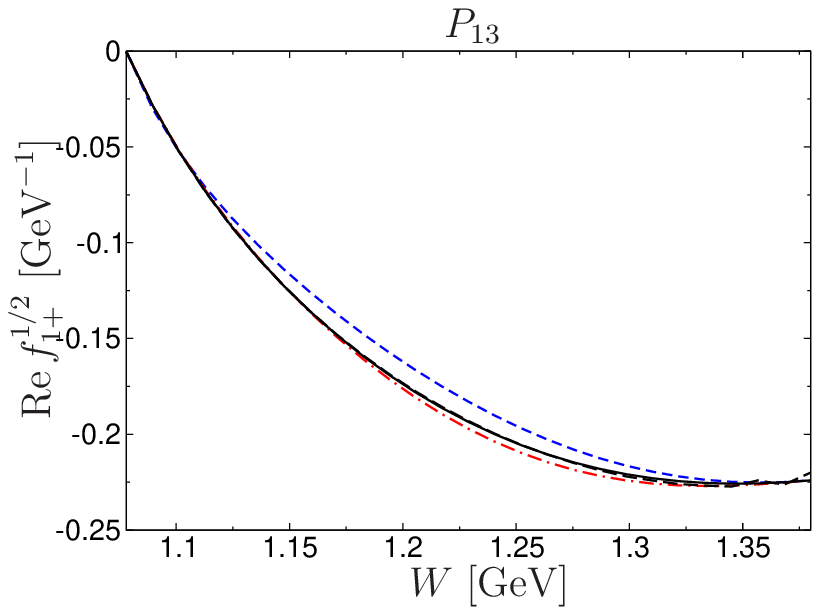}\quad
\includegraphics[width=0.45\linewidth,clip]{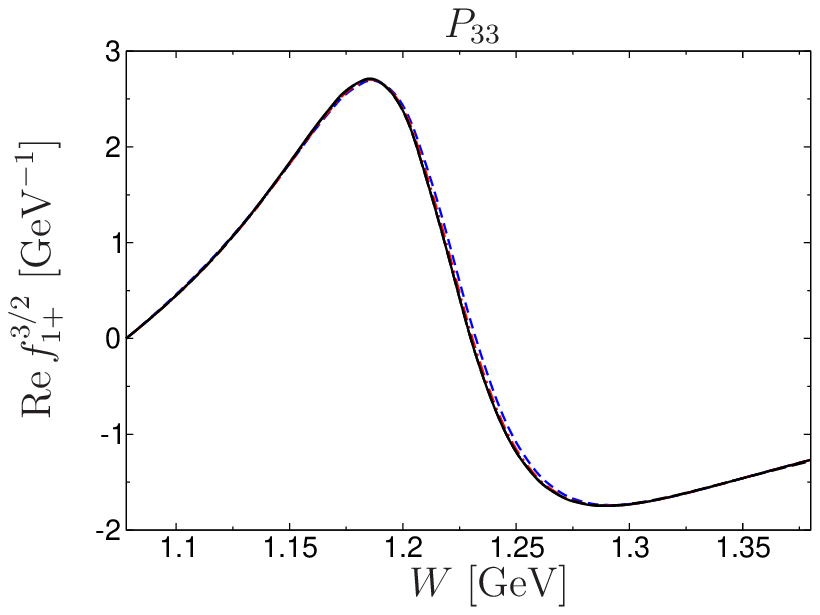}\\[0.1cm]
\includegraphics[width=0.45\linewidth,clip]{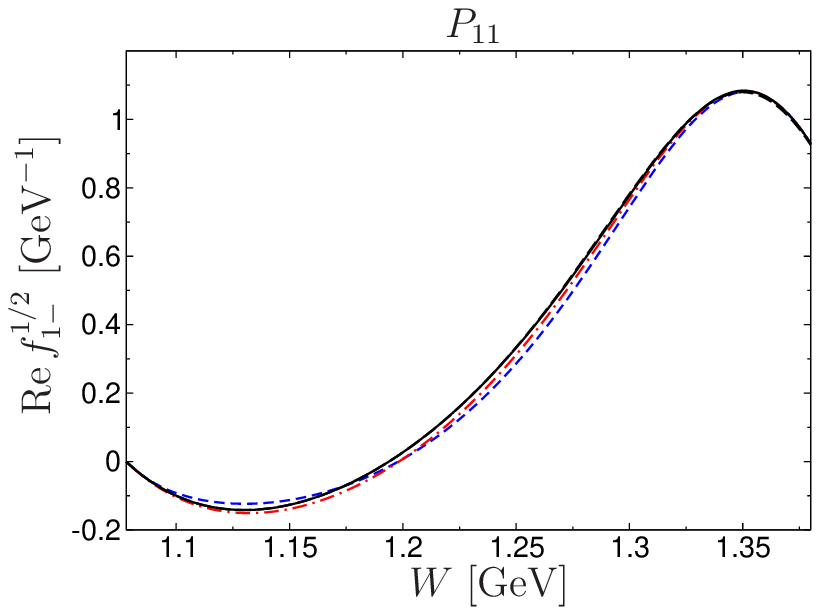}\quad
\includegraphics[width=0.45\linewidth,clip]{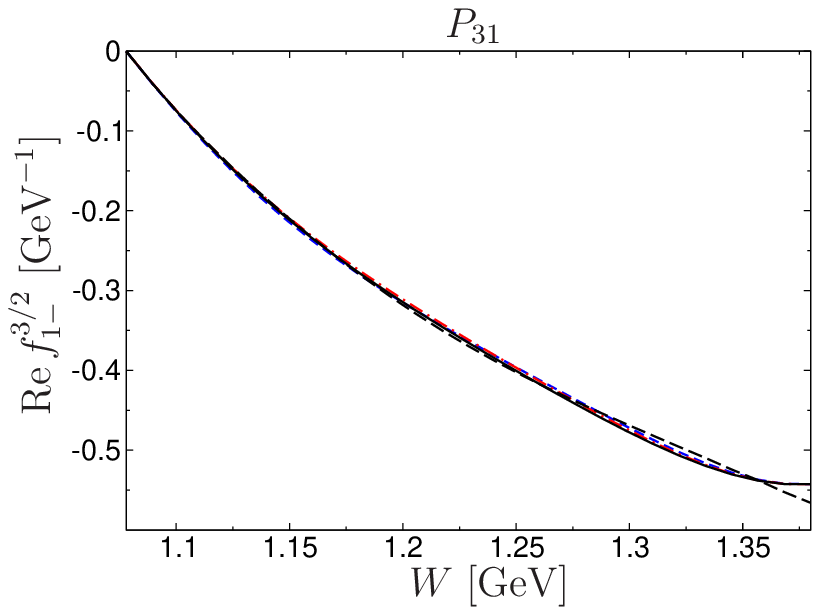}
  \caption{$\Re f_{l\pm}^{I_s}$ including the $t$-channel recoupling via the subthreshold parameters: the black solid and dashed lines refer to the LHS and RHS of~\eqref{spwhdr}, respectively.
  For comparison, the blue short-dashed line shows the SAID results~\cite{Arndt:2006bf,Workman:2012hx}, and the red dot-dashed line
   the LHS of the solution from Sect.~\ref{sec:schannel_sol} without $t$-channel recoupling.}
  \label{fig:fit-tsub}
\end{figure} 

\begin{figure}[t!]
\center
 \includegraphics[width=0.447\linewidth,clip]{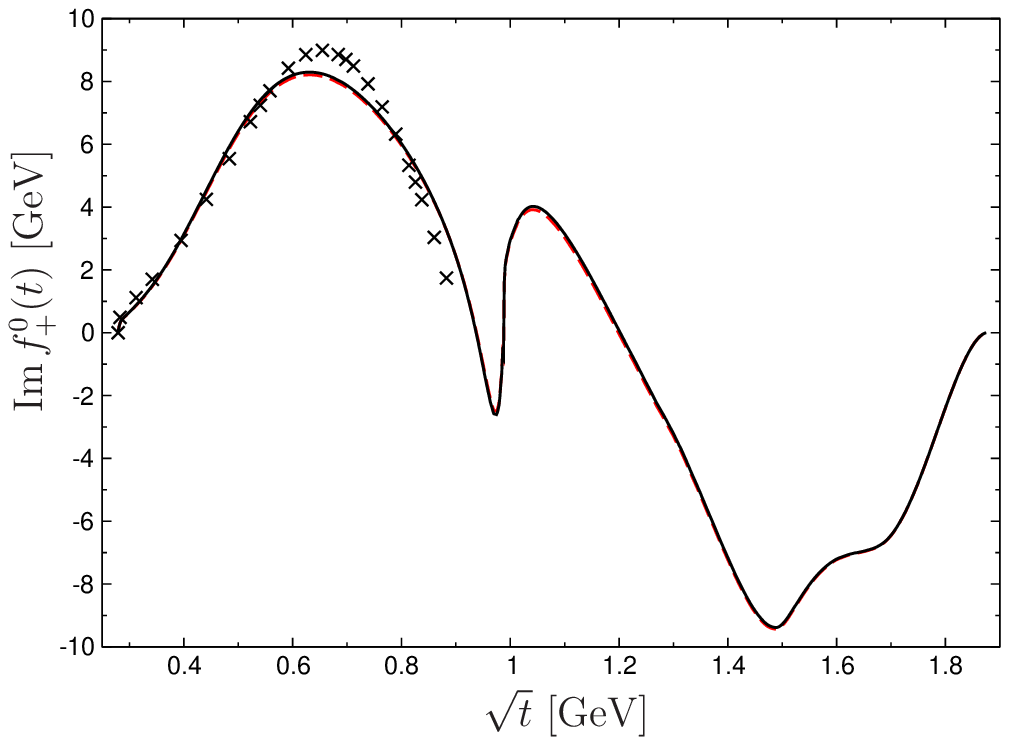}\\[1mm]
\includegraphics[width=0.447\linewidth,clip]{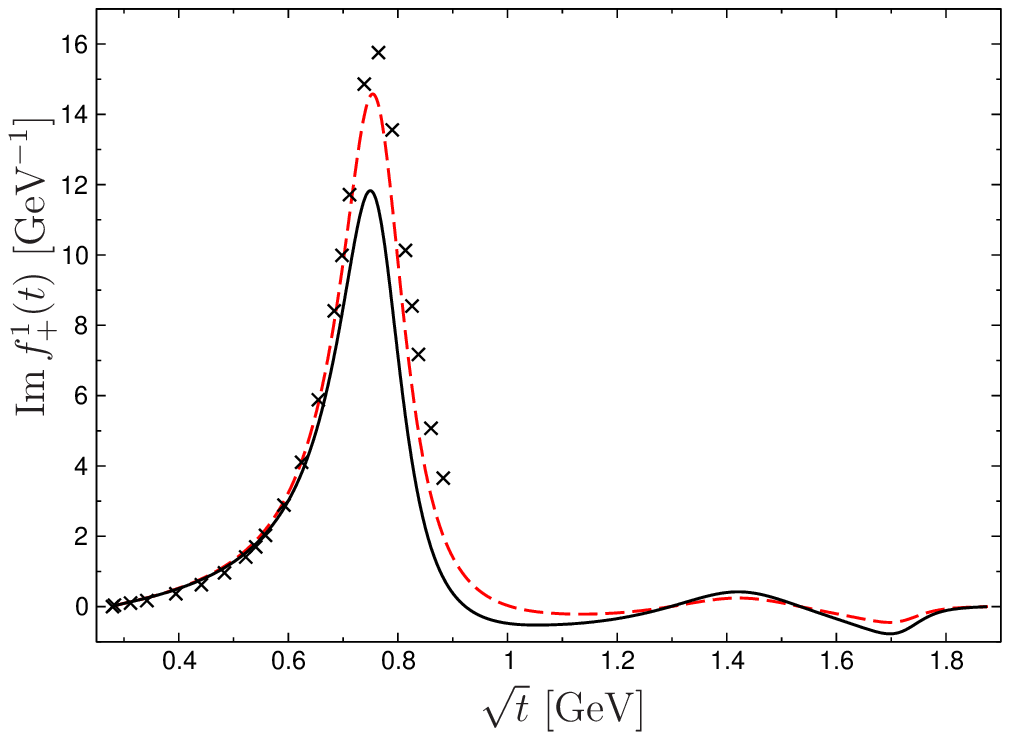}\quad
\includegraphics[width=0.447\linewidth,clip]{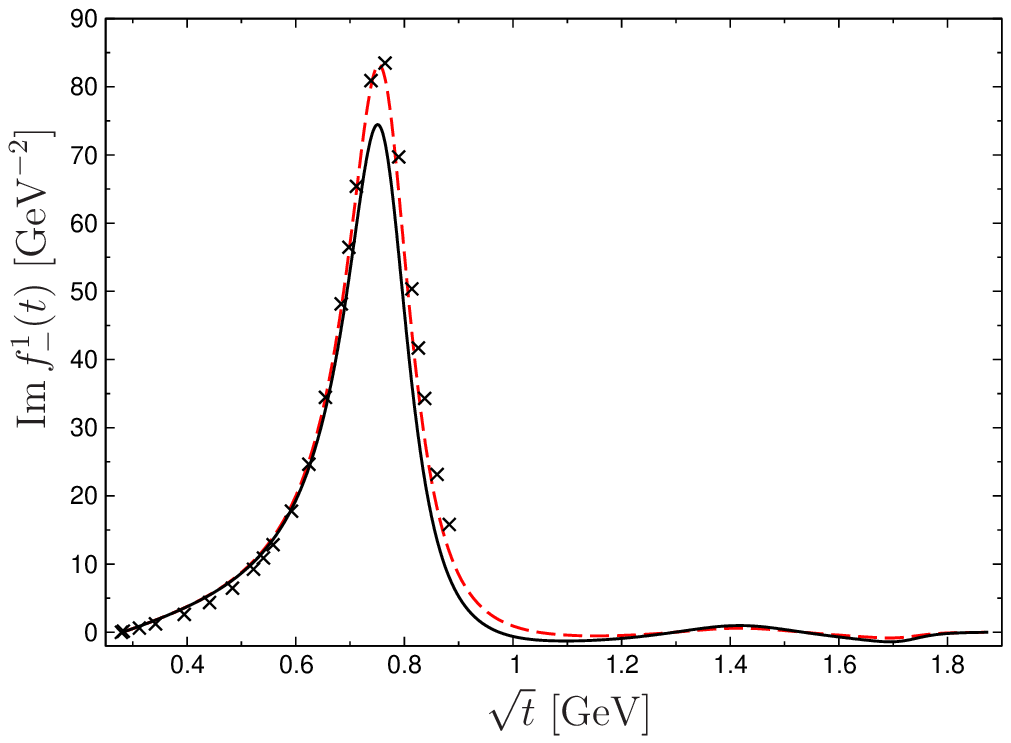}\\[1mm]
\includegraphics[width=0.447\linewidth,clip]{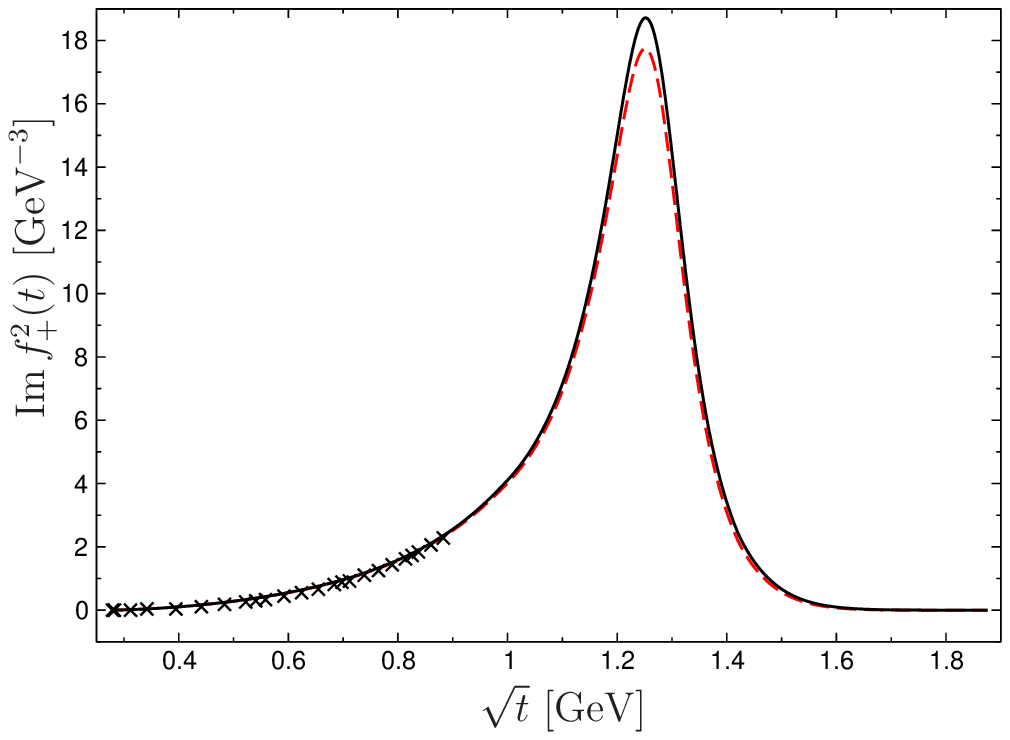}\quad
\includegraphics[width=0.447\linewidth,clip]{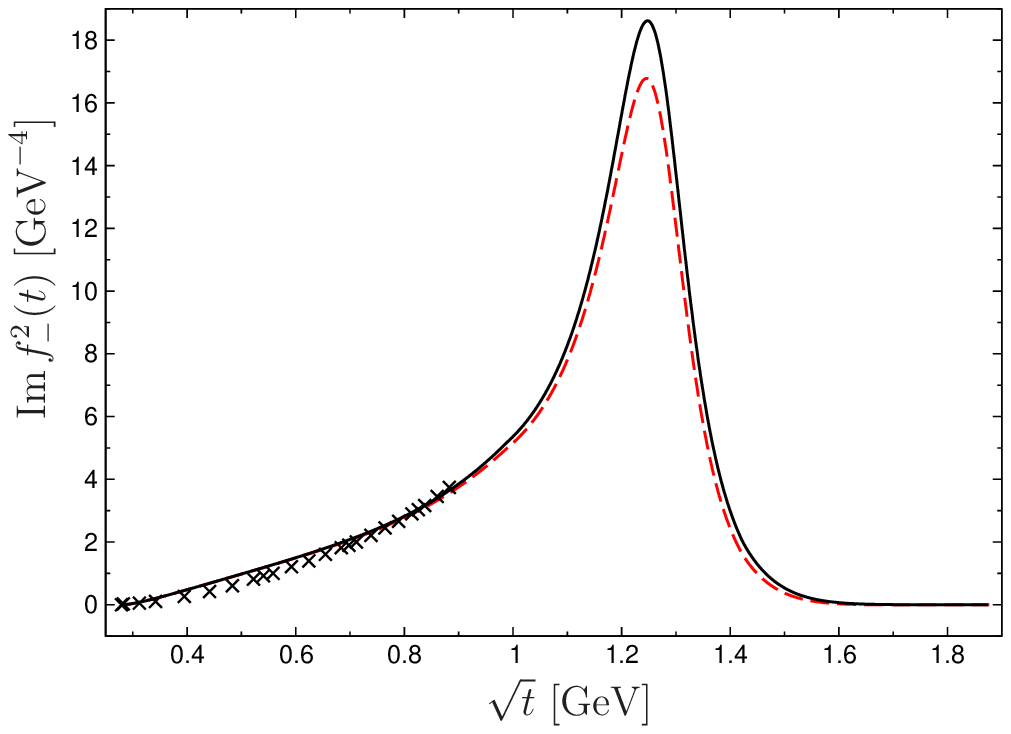}
  \caption{Imaginary parts of the $t$-channel partial waves with KH80 subthreshold parameters (red dashed lines), and with the subthreshold parameters obtained from the full RS solution (black solid lines).
 The black crosses indicate the results from~\cite{Hoehler:1983}.}
  \label{fig:Tsub}
\end{figure} 

In Sects.~\ref{sec:tchannel_sol} and \ref{sec:schannel_sol} we have reviewed how to obtain a separate solution of the $t$- and $s$-channel subsystems, respectively, while 
of course a full solution of the RS system of equations requires a consistent set of $t$- and $s$-channel phase shifts and subthreshold parameters.
Eventually, such a solution could be reached by iterating this procedure until all partial waves and parameters are determined self-consistently,
however, as already explained in detail in Sect.~\ref{sec:num_sol} there are two factors which allow for a considerable simplification. 
On the one hand, the dependence of the $t$-channel RS equations on the $s$-channel phase shifts is very weak,
so in practice, the interdependence between both proceeds via the subthreshold parameters. 
On the other hand, the results for the $s$-channel phase shifts change little with respect to the GWU/SAID solutions~\cite{Arndt:2006bf,Workman:2012hx},
which were used as input for the $t$-channel problem in the first place. 
Therefore, a full solution can be obtained by including the dependence of the $t$-channel partial-wave solutions on the subthreshold parameters explicitly in the $s$-channel fit,
which was already identified in~\eqref{P-wave-subth} and~\eqref{S-wave-subth}.
In this way, we modify the $t$-channel kernels $K_tT^{I_s}_{l\pm}$ in~\eqref{decomp} by
\beq
\widehat{K_tT}(W)^{I_s}_{l\pm}=\frac{1}{\pi}\int\limits^\infty_{\tpi}\diff t'\;\sum\limits_J
\left\{G_{lJ}(W,t')\,\left(\Im f^J_+(t')+ \Im \Delta f^J_+(t)\right)+H_{lJ}(W,t')\,\left(\Im f^J_-(t')+\Im\Delta f^J_-(t)\right)\right\}, 
\eeq
where $\Im f^J_\pm(t')$ are the solutions of the MO problem when the subthreshold parameters are fixed at KH80 values, see Fig.~\ref{fig:MO_KH80},
and $\Im \Delta f^J_{\pm}(t)$, as given in~\eqref{P-wave-subth} and \eqref{S-wave-subth}, collect the shifts in the $t$-channel amplitudes due to the subthreshold-parameter correction with respect to the KH80 reference point. Thus, replacing $K_tT^{I_s}_{l\pm}$ by $\widehat{K_tT}^{I_s}_{l\pm}$ in~\eqref{decomp}, the minimization of~\eqref{chitot} provides a new set of consistent $s$- and $t$-channel phase shifts and subthreshold parameters as the solution.   

The corresponding results for the real part of the $s$-channel partial waves are plotted in Fig.~\ref{fig:fit-tsub}. Black solid and dashed lines correspond to the LHS and RHS of~\eqref{spwhdr}. As in Fig.~\ref{fig:fit-notsub}, the differences between them are imperceptible but for the $S_{31}$-wave, where deviations emerge close to $\Wm$. These deviations can be cured allowing for more parameters in the Schenk phase-shift parameterization in~\eqref{eq:schenk-par-Sw}, 
but the differences are sufficiently small to be included in the uncertainty estimate. In addition, for better comparison, we also include the GWU/SAID solutions~\cite{Arndt:2006bf,Arndt:2008zz,Workman:2012hx} in Fig.~\ref{fig:fit-tsub} (blue short-dashed lines),
and the results of Sect.~\ref{sec:num_sol}, i.e.\ the results of the $s$-channel problem when the dependence of the $t$-channel solution on the subthreshold parameters is neglected (red dot-dashed curves).
The differences with respect to SAID results are only sizable for the $S_{31}$, whereas the comparison between the $s$-channel and the full RS solution is small in all cases, 
reflective of the fact that the variation of the subthreshold parameters with respect to the KH80 values~\cite{Koch:1980ay,Hoehler:1983} remains small. 
The resulting subthreshold parameters are given in Table~\ref{tab:RS_subthr}, where, for completeness, we also show the KH80 values.
The comparison between them reveals fair agreement, all of them lie within $2\sigma$, with the only exception of $d_{00}^-$.\footnote{The sum rules given in~\ref{app:subthr_sum_rules} imply that, everything else being equal, the values for $d_{00}^-$ should differ already because of the different $\pi N$ coupling constant. Indeed, if we used the KH80 value thereof in the pseudovector definition of the subthreshold parameters~\eqref{def_PV_pole}, we would obtain $d_{00}^-=1.49\mpi^{-2}$ instead, much closer to the KH80 expectation. The same argument deteriorates agreement slightly for $b_{00}^-$, but fails completely for $d_{00}^+$, which is most likely related to the fact that the corresponding sum rule for $A^+$ does not appear to converge.}    
In addition, the impact of the shifts in the subthreshold parameters on the $t$-channel imaginary parts is illustrated in Fig.~\ref{fig:Tsub}.
Black solid lines correspond to the full RS values, whereas the dashed red lines denote the results of the $t$-channel subsystem with subthreshold parameters fixed to the KH80 values as described in Sect.~\ref{sec:tchannel_sol}.
Note that the differences between both curves are only due to the subthreshold parameter values. Finally, the black crosses indicate again the results from~\cite{Hoehler:1983}, which, as expected, lie closer to the red dashed lines than to the full RS results.

\begin{table}[t]
\renewcommand{\arraystretch}{1.3}
\centering
\begin{tabular}{crr}
\toprule
& RS & KH80\\
\midrule
$d_{00}^+\,[\mpi^{-1}]$ & $-1.361\pm 0.032$ & $-1.46\pm0.10$\\
$d_{10}^+\,[\mpi^{-3}]$ & $1.156\pm 0.019$ &$1.12\pm0.02$\\
$d_{01}^+\,[\mpi^{-3}]$ & $1.155\pm 0.016$ &$1.14\pm0.02$\\
$d_{20}^+\,[\mpi^{-5}]$ & $0.196\pm 0.003$ &$0.200\pm0.005$\\
$d_{11}^+\,[\mpi^{-5}]$ & $0.185\pm 0.003$ &$0.17\pm0.01$\\
$d_{02}^+\,[\mpi^{-5}]$ & $0.0336\pm 0.0006$ &$0.036\pm0.003$\\\midrule
$d_{00}^-\,[\mpi^{-2}]$ & $1.411\pm 0.015$ &$1.53\pm0.02$\\
$d_{10}^-\,[\mpi^{-4}]$ & $-0.159\pm 0.004$ &$-0.167\pm0.005$\\
$d_{01}^-\,[\mpi^{-4}]$ & $-0.141\pm 0.005$ &$-0.134\pm0.005$\\\midrule
$b_{00}^+\,[\mpi^{-3}]$ & $-3.455\pm 0.072$ &$-3.54\pm0.06$\\\midrule
$b_{00}^-\,[\mpi^{-2}]$ & $10.49\pm 0.11$ &$10.36\pm0.10$\\
$b_{10}^-\,[\mpi^{-4}]$ & $1.000\pm 0.029$ &$1.08\pm0.05$\\
$b_{01}^-\,[\mpi^{-4}]$ & $0.208\pm 0.020$ &$0.24\pm0.01$\\
\bottomrule
\end{tabular}
\caption{Subthreshold parameters from the RS analysis in comparison with the KH80 values~\cite{Koch:1980ay,Hoehler:1983}.}
\label{tab:RS_subthr}
\renewcommand{\arraystretch}{1.0}
\end{table}

\subsection{Error analysis}
\label{sec:error_analysis}

The $s$- and $t$-channel partial-wave results depicted in Figs.~\ref{fig:fit-tsub} and \ref{fig:Tsub}, as well as the subthreshold parameters collected in Table~\ref{tab:RS_subthr} 
correspond to our RS central solution, which, as detailed in Sect.~\ref{sec:schannel_sol}, depend on several input quantities. 
The imaginary parts of the $s$-channel $S$- and $P$-waves are required above $\Wm$, those for the partial waves with $l>1$ between $W_+$ and $\Wa$. 
In our central solution this information is taken from the GWU/SAID~\cite{Arndt:2006bf,Workman:2012hx} PWA.
Particularly relevant are the values of the $S$- and $P$-wave phase shifts and their derivatives at $\Wm$,
which fix the matching conditions imposed in the $S$- and $P$-wave parameterizations of~\eqref{eq:schenk-par-Sw}, \eqref{eq:schenk-par-Pw}, and \eqref{eq:schenk-par-Pw-conf}.
Above $\Wa$ we have used the Regge model from~\cite{Huang:2009pv}, although the contribution from that energy range is completely negligible.
In addition, we also need the $S$- and $P$-wave inelasticities below $\Wm$, which for our central value solutions were taken from the fit to the SAID PWA as described in Sect.~\ref{sec:schannel_sol}. 
Another essential piece of information is the value of the $S$-wave scattering lengths extracted from pionic atoms that we have reviewed in Sect.~\ref{sec:pionic_atom}, 
which play a crucial role in stabilizing the RS solution (and in the determination of the $\pi N$ coupling constant by means of the GMO sum rule).
Finally, we have also included the elastic $\pi\pi$ phase shifts from~\cite{Caprini:2011ky,GarciaMartin:2011cn} as input, which enter in the $t$-channel MO problem.
Above roughly $\sqrt{t}\sim1\GeV$ for the $P$- and $D$-waves and $\sqrt{t}\sim1.3\GeV$ for the $S$-wave, the single- and two-channel approximations considered respectively in Sect.~\ref{sec:tchannel_sol} break down, and further inelastic input would be required.
All this information has been included in our solution as input, and, thus, the corresponding uncertainties propagate directly to our results as a systematic error.   
In order to quantify them, we will study the sensitivity to all the input quantities next. 

Let us first start with the effects of the $s$-channel PWA input. 
In this case there are two main sources of uncertainty: the errors associated with the $S$- and $P$-wave inelasticities and imaginary parts of the SAID PWA, and the effect of truncating the $s$-channel partial-wave expansion. 
In the first case, unfortunately, the GWU/SAID collaboration does not provide an error estimate, so that, instead of fixing the errors by hand, we use the results of the KH80 analysis~\cite{Koch:1980ay,Hoehler:1983} as an alternative, 
and interpret the differences between them as uncertainty estimate. 
In the second case, we will check the effect of truncating the partial-wave expansion at $l_\text{max}=4$ or $5$.

\begin{figure}[t!]
\includegraphics[width=0.45\linewidth,clip]{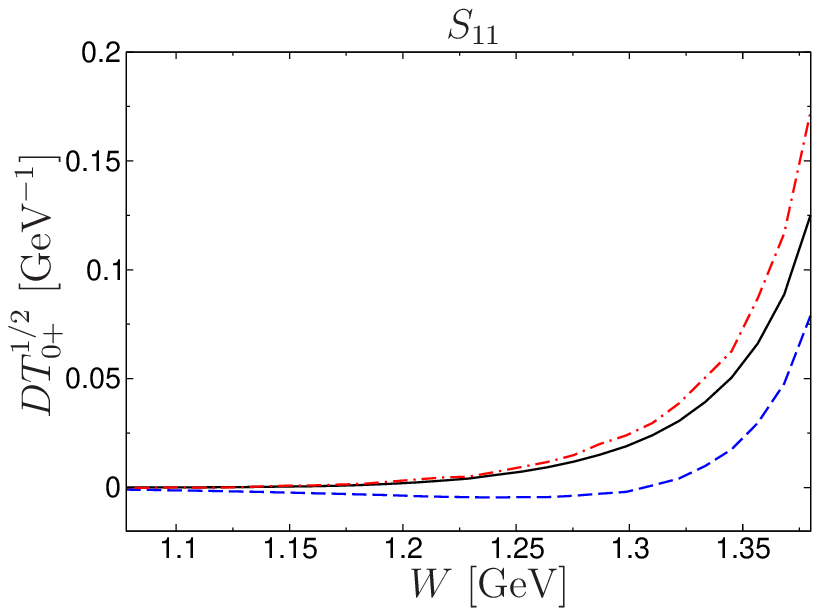}\quad
\includegraphics[width=0.45\linewidth,clip]{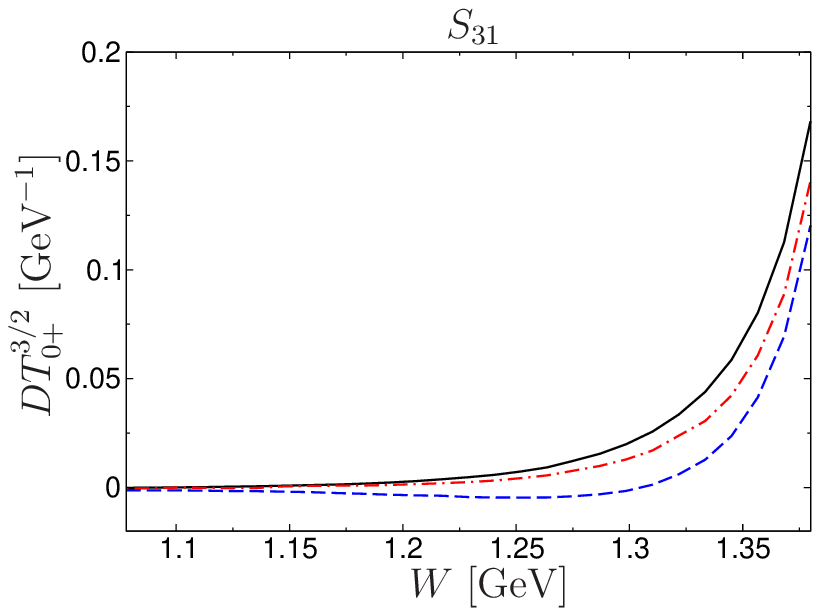}\\[0.1cm]
\includegraphics[width=0.45\linewidth,clip]{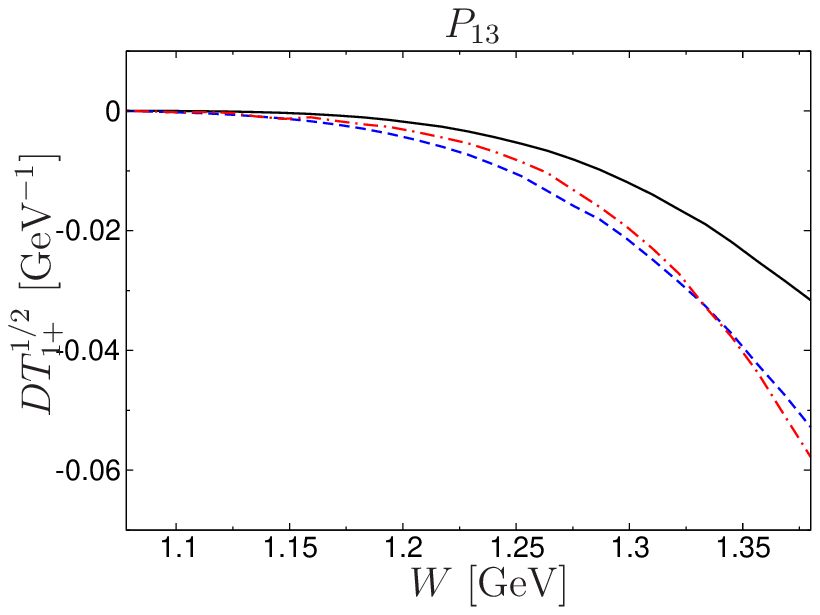}\quad
\includegraphics[width=0.45\linewidth,clip]{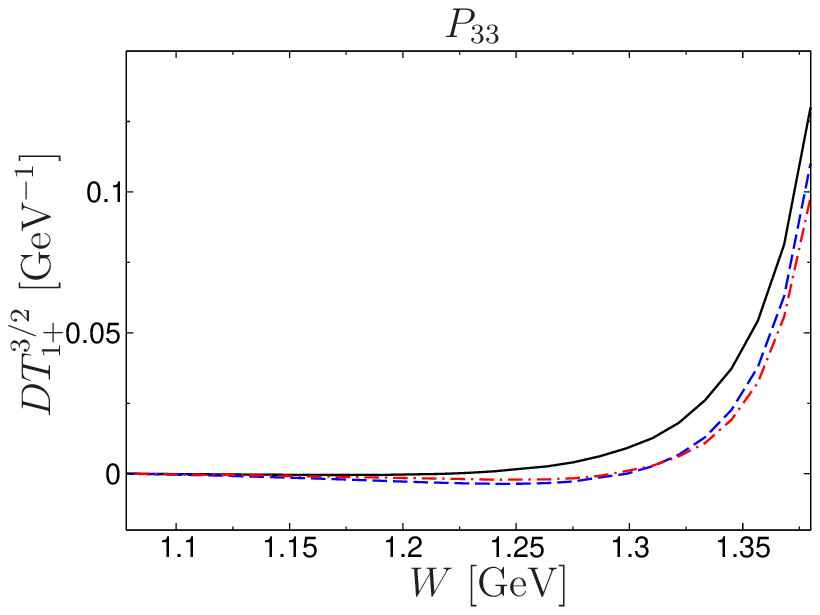}\\[0.1cm]
\includegraphics[width=0.45\linewidth,clip]{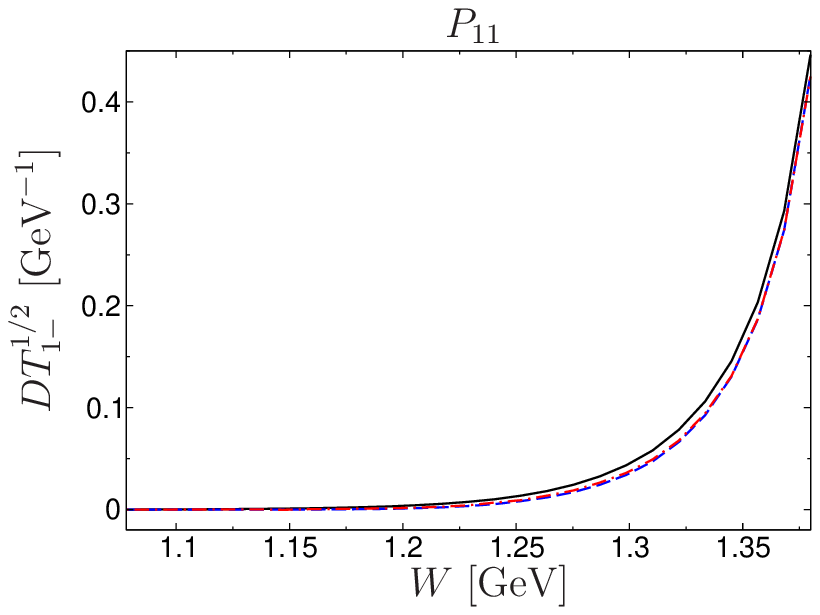}\quad
\includegraphics[width=0.45\linewidth,clip]{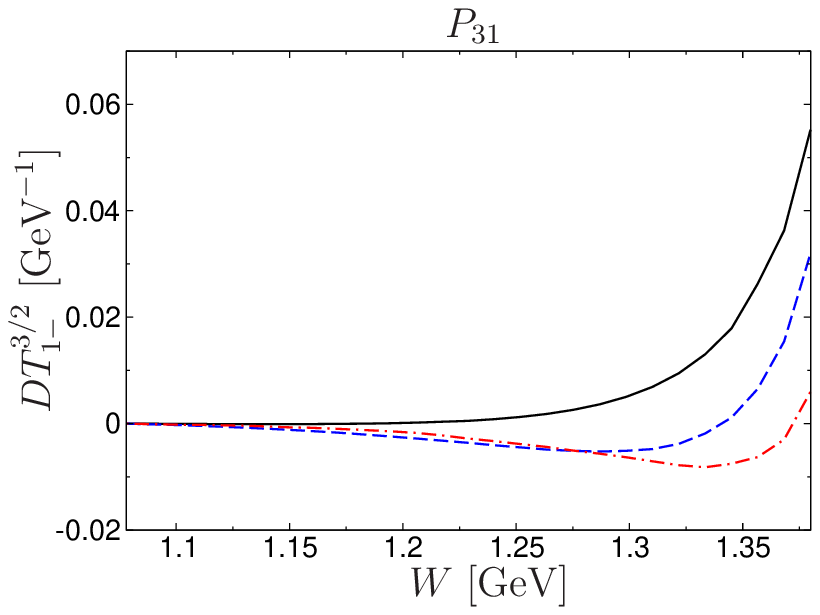}
  \caption{$DT(W)^{I_s}_{l\pm}$ for the central solution (black solid lines), the version with KH80 input instead of SAID (red dot-dashed lines), and with $l_\text{max}=5$ (blue dashed lines).}
  \label{fig:schannel-input}
\end{figure} 

As anticipated in Sect.~\ref{sec:decomp},
the variation of the $s$-channel input only affects the driving terms $DT(W)$, which are already small in comparison to the other contributions. 
The values of $DT(W)^{I_s}_{l\pm}$ for both cases and for each partial wave are plotted in Fig.~\ref{fig:schannel-input}.
Black solid lines denote the driving terms for the central solution, red dot-dashed lines the results obtained when the $s$-channel partial waves and inelasticities are taken from KH80, and the blue dashed lines correspond to $l_\text{max}=5$. From the scales in Fig.~\ref{fig:schannel-input} we can conclude that the effect of considering higher partial waves is indeed small, given that all contributions are almost one order of magnitude smaller than in Fig.~\ref{fig:decomp}. The inclusion of KH80 partial waves instead of SAID for the $s$-channel input produces effects of similar size, 
which, given the overall magnitude of the driving terms, hardly propagate to the full RS solution for the subthreshold parameters at all, and remain very small for the $s$-channel partial waves (the biggest change occurs for the $P_{13}$-wave and is still less than $1\%$).

\begin{figure}[t!]
\center
\includegraphics[width=0.45\linewidth,clip]{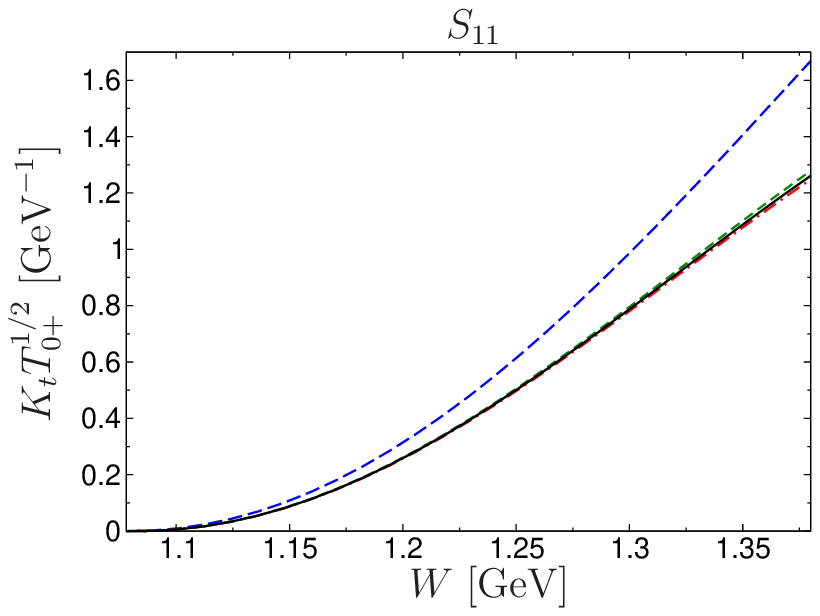}\quad
\includegraphics[width=0.45\linewidth,clip]{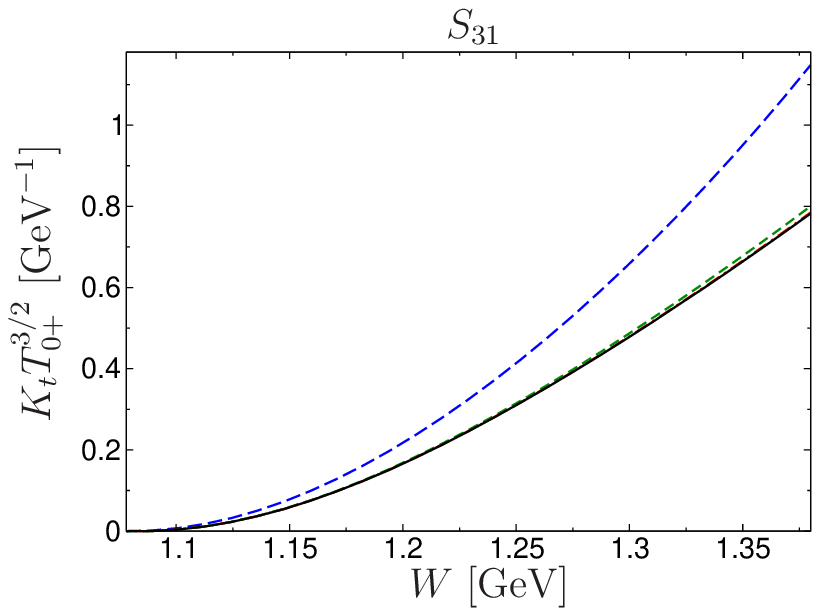}\\[0.1cm]
\includegraphics[width=0.45\linewidth,clip]{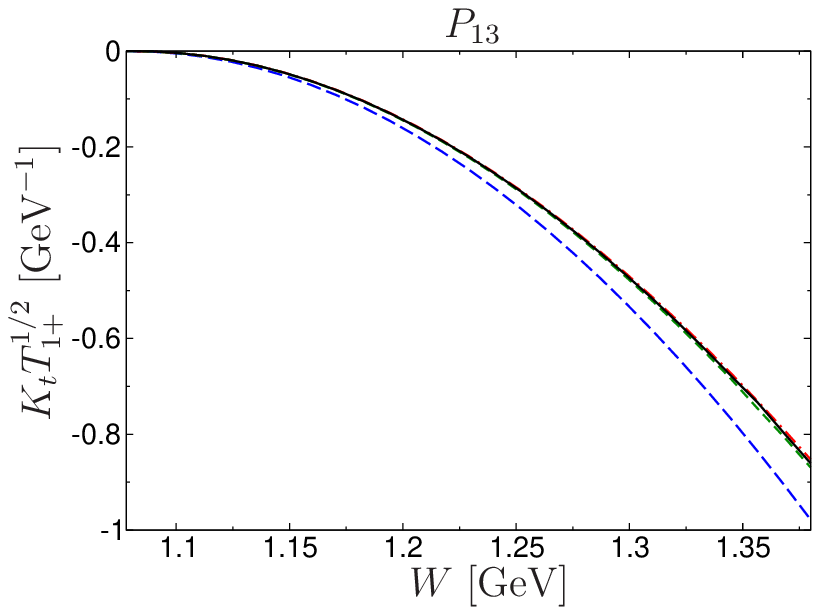}\quad
\includegraphics[width=0.45\linewidth,clip]{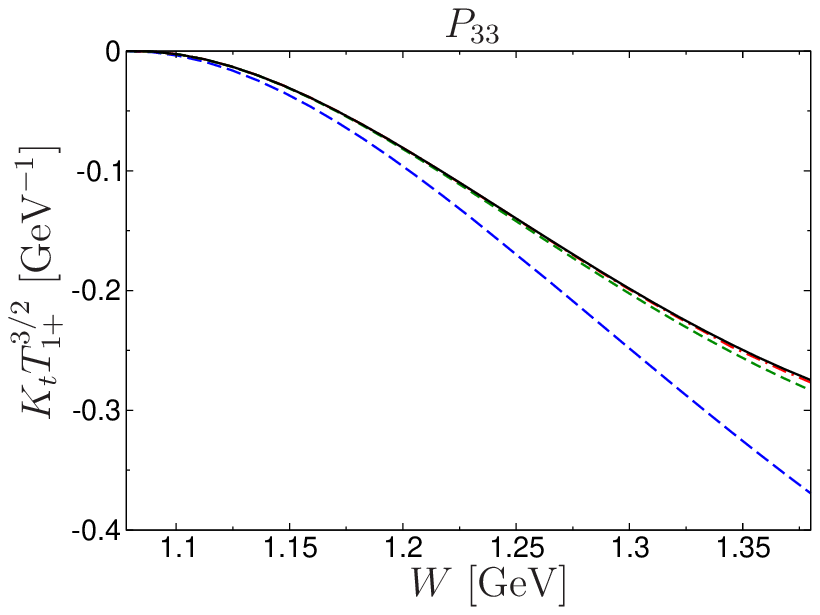}\\[0.1cm]
\includegraphics[width=0.45\linewidth,clip]{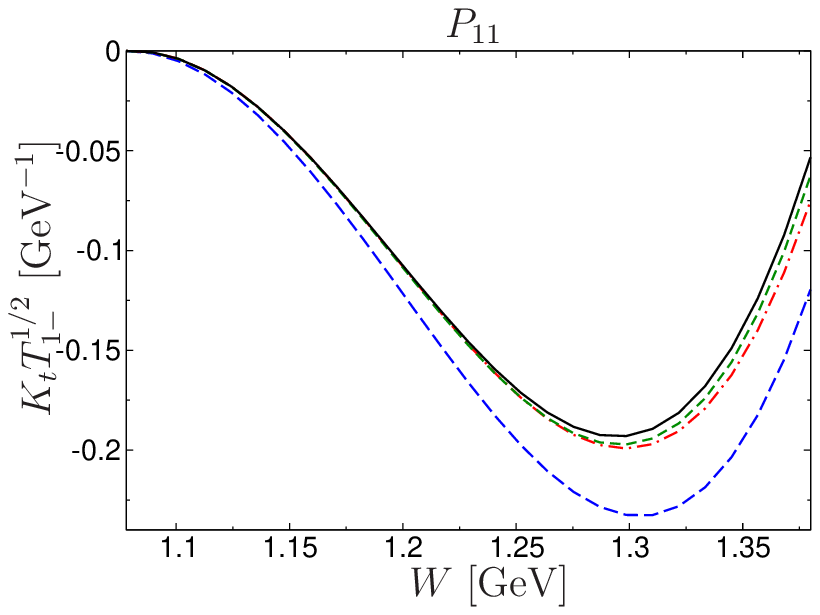}\quad
\includegraphics[width=0.45\linewidth,clip]{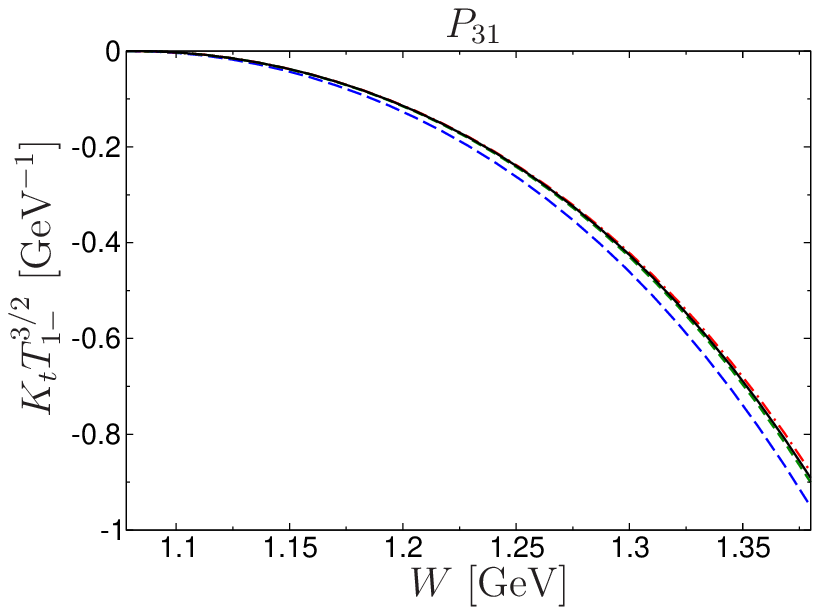}
  \caption{$K_tT(W)^{I_s}_{l\pm}$ for each partial wave. Black solid lines correspond to our central solution, which incorporate the effect of the $f_2(1270)$, $\rho'$, and $\rho''$ resonances, blue long-dashed lines to the solution without the $f_2(1270)$ resonance. The effect of the $F$-waves is represented by the red dot-dashed lines. Finally the green short-dashed lines include the version of the $S$-wave where the phase shifts above $1.3 \GeV$ are kept constant. All other effects described in the text are negligible and cannot be distinguished from the central solution curves.}
  \label{fig:tchannel-input}
\end{figure}

The error propagation of the $t$-channel information is much more involved, since, in addition to the $\pi\pi$ input uncertainties, we have to estimate the effects of the inelasticities that could not be accounted for explicitly. 
In fact, the uncertainties in the elastic $\pi\pi$ phase shifts are in general very small, as can be either seen from their respective error bands or by comparing the solutions from the dispersive analyses of~\cite{GarciaMartin:2011cn,Caprini:2011ky}.
In the RS system the variation of the $t$-channel input affects the $K_tT^{I_s}_{l\pm}$ kernels, which, in contrast to the $s$-channel driving terms, do play an important role. Their values for the different cases described above are depicted in Fig.~\ref{fig:tchannel-input}.
Our central solution (black solid lines) incorporates $S$-, $P$-, and $D$-waves from the $\pi\pi$ threshold to the $\bar NN$ threshold, with $\pi\pi$ phase shifts from~\cite{Caprini:2011ky} ($S$- and $P$-waves) and~\cite{GarciaMartin:2011cn} ($D$-waves). 

For the $S$-wave, the main uncertainty is generated by $4\pi$ intermediate states, which start to become relevant around $1.3\GeV$. 
Above this energy, the uncertainties were estimated by continuing the phase shifts in two extreme cases, cf.\ the first panel of Fig.~\ref{fig:MO_KH80}: 
as our central solution, we guide the phase shifts smoothly to $2\pi$; alternatively, we keep them constant above $1.3\GeV$ (green short-dashed lines). 
The difference suggests a small impact and allows us to include all the complicated inelastic effects above that energy as a (moderate) uncertainty.
For the $P$-waves, we estimate the imprint of the dominant $4\pi$ inelasticities by including the contributions of the $\rho'$ and $\rho''$ resonances
as part of our central solution, albeit in an approximate, elastic way, compare the second row of Fig.~\ref{fig:MO_KH80};
their impact is however negligible compared to the other variations studied here.
The $D$-waves are dominated by the $f_2(1270)$ resonance, which is $85\%$ elastic. The potential impact of the $15\%$ inelasticity is estimated by replacing the $\pi\pi$ phase shift in the MO solution by the phase of the $\pi\pi$ partial wave, third row of Fig.~\ref{fig:MO_KH80}; its effect on the final solution turns out to be entirely negligible. However, the effect of neglecting the $f_2(1270)$ resonance altogether corresponds to the blue long-dashed lines in Fig.~\ref{fig:tchannel-input}: the differences to the black solid lines highlight the large impact of this resonance, which we even found to be necessary to obtain an acceptable solution to the full RS problem at all. 
Finally, the contribution of higher partial waves (not part of the central solution) is estimated by considering $F$-waves, which are constructed by matching the $\pi\pi$ parameterization from~\cite{GarciaMartin:2011cn} to a Breit--Wigner description of the $\rho_3(1690)$. This is a rather crude model, as the $\rho_3(1690)$ is predominantly inelastic.  Its inclusion is shown by the red dot-dashed line of Fig.~\ref{fig:tchannel-input}: the comparison with the central solution shows that despite the prominent $D$-wave effect, $F$-waves are already very small, which is related to the fact that the first resonance occurs at much higher energies and justifies the treatment as a part of the uncertainty estimate. 
Above the $\bar NN$ threshold, one could incorporate experimental results from~\cite{Anisovich:2011bw}, but the contribution from this high-energy tail is negligible compared to the uncertainties at intermediate energies.

As discussed in Sect.~\ref{sec:schannel_sol}, the matching conditions have been imposed in order to ensure a unique solution to the RS equations. 
They have been fixed by constraining the family of allowed $S$- and $P$-wave phase-shift parameterizations directly, so that the values of the phases and their derivatives at $\Wm$ satisfy the input values for any set of parameters explicitly, see~\eqref{eq:schenk-par-Sw}, \eqref{eq:schenk-par-Pw}, and \eqref{eq:schenk-par-Pw-conf}. 
Consequently, the variation of the matching conditions constitutes an important source of uncertainty that we have to take into account.
Since we have defined our central solution in terms of the  GWU/SAID PWA~\cite{Arndt:2006bf,Arndt:2008zz,Workman:2012hx}, the matching conditions are given by the SAID phase-shift values depicted in Fig.~\ref{fig:phases-PWAs}. 
However, in view of lack of errors of the SAID solutions we assess the uncertainties of the matching conditions by varying the phase shifts at the matching point as suggested by the KH80 solutions, which were also plotted in Fig.~\ref{fig:phases-PWAs}. Due to the fact that the latter is not sufficiently smooth to define a complete matching condition, it is not possible to extract meaningful values for the derivatives. Alternatively, we deduce them from fits to the KH80 phase shifts in the region around $\Wm$, so that the unphysical oscillations are considerably smoothed. 
In addition, in the case of the $S_{11}$-wave even extracting an unambiguous value for the phase shift at the matching point was difficult, so that in this case we varied its value in the direction suggested by KH80, but by the (relative) amount as seen in the $S_{31}$-wave.
The effect of varying the matching conditions has considerable impact on the RS equations and dominates the RS error close to $\Wm$.  

The propagation of the scattering-length uncertainties also plays a key role, since they fix the value of the $S$-wave phase shifts at threshold. In this case, the precise analysis carried out in Sect.~\ref{sec:pionic_atom} allows for a rigorous determination of the scattering-length errors, which we propagate into subtraction constants and phase shifts. The variation of the scattering lengths dominate the RS error for the subthreshold parameters and the $S$-wave phase shifts close to $W_+$. 
Similarly, the uncertainty in the $\pi N$ coupling constant is propagated by replacing $g^2/(4\pi) = 13.7\pm 0.2$ by its upwards and downwards shift in the central-value fit. 

The variation of the input discussed might be understood as a systematic uncertainty. However, our results are based on numerical solutions of the RS equations, and thus, they also suffer from statistical errors. 
The first statistical effect is related to the numerical minimization of~\eqref{chitot}, which led to the results depicted in Fig.~\ref{fig:fit-tsub}. The agreement between the LHS and RHS of the RS equations in the central solution is good enough for each partial wave to neglect the numerical error with the exception of the $S_{31}$-wave, where deviations become meaningful close to $\Wm$. These differences can be cured by allowing for more free parameters in the $S_{31}$ phase-shift parameterization. Given that the LHS changes only slightly, we continue to use the same number of free parameters for all partial waves and only include the differences found when using an extra parameter in the $S_{31}$-wave as uncertainties.  

Finally, the second effect is related to important correlations in the phase space of parameters. We observed that in some cases fit minima with a similar $\Delta_\text{RS}$, but significantly different values for the subthreshold parameters could be found, which signals the existence of flat directions in the $10$-dimensional space of subthreshold parameters to which the RS constraints are only weakly sensitive. However, this ambiguity can be removed by recalling that the number of subtractions we are using is motivated by the number of degrees of freedom of the RS system, not by necessity since the dispersive integrals would not converge otherwise. For this reason, the physical space of solutions can be restricted to sets of parameters that do not grossly violate the sum rules for the higher subthreshold parameters (see~\ref{app:subthr_sum_rules}). Indeed, we find that fit minima with significantly different subthreshold parameters violate these sum rules at an unacceptable level and can therefore be discarded.

In more detail, we first studied the sum rules in the context of the KH80 PWA. We found that the sum rule for $d_{00}^+$ does not show any sign of convergence, which might be related to the finding in fixed-$t$ dispersion relations that the amplitude $D^+$ requires a subtraction.\footnote{From the point of view of ChPT the latter is related to the fact that $D^+$ has an Adler zero close to the Cheng--Dashen point~\cite{Becher:1999he}.} The sum rules for $d_{00}^-$, $b_{00}^-$, and $d_{01}^+$, evaluated with the same input as used for the RS equations, are fulfilled at the $10$--$20\%$ level, and therefore not suited as constraints for a precision analysis. In contrast, the sum rules for the remaining subthreshold parameters are fulfilled very accurately, in most cases the KH80 value is even included in the range obtained by varying the input for the evaluation of the dispersive integrals as discussed above. In practice, we implemented these sum rules in the estimate of the impact of the flat directions according to the following recipe: starting from our central solution, we defined another set of fits by forcing one of the subthreshold parameters closer to its KH80 value by means of an additional penalty function in $\Delta_\text{RS}$, and, if a fit minimum with $\Delta_\text{RS}$ comparable to the central solution could be found, kept the corresponding set of parameters as an alternative solution. To cover the space of possible flat directions reasonably comprehensively, we repeated the same exercise by forcing pairs of subthreshold parameters simultaneously closer to their KH80 values. Out of the $22$ fit minima generated in this way, several could be discarded because $\Delta_\text{RS}$ was significantly worse than for the central solution. In addition, we demanded that the sum rules be violated by less than $2\%$ for $b_{00}^+$ and $a_{10}^+$ and less than $5\%$ for $b_{01}^-$, $a_{01}^-$, $b_{10}^-$, $a_{10}^-$, a rather generous requirement based on the study of the KH80 starting point.\footnote{We note that the KH80 analysis was based on fixed-$t$ dispersion relations, so that the fact that the HDRs yield sum rules for the subthreshold parameters that are fulfilled at the percent level is a highly non-trivial consistency check.}
In the end, $N_\text{FD}=16$ minima survive both the constraints on $\Delta_\text{RS}$ and the sum rules, and their distribution is considered as a statistical ensemble.

\subsection{Uncertainty estimates for the subthreshold parameters}
\label{sec:subthreshold_parameters}

The final sources of uncertainty that remain after the detailed study of all input quantities discussed in the previous section fall into the following categories:
\begin{enumerate}
 \item uncertainty in the scattering lengths,
 \item systematic effects in the PWA input for $s$- and $t$-channel,
 \item uncertainties in the matching condition,
 \item input for the $\pi N$ coupling constant,
 \item flat directions in the space of subthreshold parameters.
\end{enumerate}

In the following, we will describe how each of these sources of uncertainty is propagated to final uncertainty estimates for the subthreshold parameters, while the more complicated case of the phase shifts is postponed to Sect.~\ref{sec:err_phase_shift}. 
First, to propagate the uncertainty in the scattering lengths we repeated the fit on a grid around the central values~\eqref{scatt_pionic_atoms_final}, with a grid spacing given by the $1\sigma$ errors in each direction. The changes in the subthreshold parameters are sufficiently small that they can be fit by a linear ansatz of the form
\beq
X_i=X_i^0+\Delta X_i^1\Big(a_{0+}^{1/2}-\bar a^{1/2}_{0+}\Big)+\Delta X_i^2\Big(a_{0+}^{3/2}-\bar a^{3/2}_{0+}\Big),
\eeq
where $X_i\in\{d_{00}^+,\ldots\}$ and $\bar a^{I_s}_{0+}$ refers to the central values~\eqref{scatt_pionic_atoms_final}. The covariance matrix in the $10$-dimensional space of subthreshold parameters from this source of uncertainty becomes
\beq
\Sigma^\text{SL}_{ij}=\Delta X_i^1\Delta X_j^1 \big(\Delta a_{0+}^{1/2}\big)^2+\Delta X_i^2\Delta X_j^2 \big(\Delta a_{0+}^{3/2}\big)^2,
\eeq
where we have neglected the correlations between $\Delta a_{0+}^{1/2}$ and $\Delta a_{0+}^{3/2}$ as discussed in Sect.~\ref{sec:pionic_atom}.

\begin{table}[t]
\footnotesize
\renewcommand{\arraystretch}{1.3}
\centering
\begin{tabular}{crrrrrrrrrrrrr}
\toprule
& $d_{00}^+$ & $d_{10}^+$ & $d_{01}^+$ & $d_{20}^+$ & $d_{11}^+$ & $d_{02}^+$ & $d_{00}^-$ & $d_{10}^-$ & $d_{01}^-$ & $b_{00}^+$ & $b_{00}^-$ & $b_{10}^-$ & $b_{01}^-$ \\
$d_{00}^+$ & $1$ & $-0.77$ & $-0.51$ & $-0.43$ & $-0.39$ & $-0.30$ & $-0.34$ & $0.43$ & $0.46$ & $0.37$ & $-0.08$ & $-0.39$ & $0.14$\\
$d_{10}^+$ & & $1$ & $0.85$ & $0.48$ & $0.53$ & $0.58$ & $0.16$ & $-0.40$ & $-0.64$ & $-0.48$ & $0.06$ & $0.56$ & $-0.21$\\
$d_{01}^+$ &&& $1$ & $0.59$ & $0.68$ & $0.90$ & $0.08$ & $-0.55$ & $-0.71$ & $-0.67$ & $0.04$ & $0.58$ & $-0.24$\\
$d_{20}^+$ &&&& $1$ & $0.97$ & $0.64$ & $0.14$ & $-0.35$ & $-0.79$ & $-0.63$ & $0.01$ & $0.72$ & $-0.29$\\
$d_{11}^+$ &&&&& $1$ & $0.67$ & $-0.04$ & $-0.26$ & $-0.79$ & $-0.60$ & $0.01$ & $0.78$ & $-0.28$\\
$d_{02}^+$ &&&&&& $1$ & $0.13$ & $-0.73$ & $-0.75$ & $-0.83$ & $0.06$ & $0.47$ & $-0.22$\\\midrule
$d_{00}^-$ &&&&&&& $1$ & $-0.50$ & $0.01$ & $-0.23$ & $0.04$ & $-0.11$ & $-0.04$\\
$d_{10}^-$ &&&&&&&& $1$ & $0.61$ & $0.86$ & $-0.15$ & $-0.05$ & $0.21$\\
$d_{01}^-$ &&&&&&&&& $1$ & $0.86$ & $-0.09$ & $-0.55$ & $0.41$\\\midrule
$b_{00}^+$ &&&&&&&&&& $1$ & $-0.10$ & $-0.42$ & $0.20$\\\midrule
$b_{00}^-$ &&&&&&&&&&& $1$ & $-0.20$ & $0.34$\\
$b_{10}^-$ &&&&&&&&&&&& $1$ & $0.18$\\
$b_{01}^-$ &&&&&&&&&&&&& $1$\\
\bottomrule
\end{tabular}
\caption{Correlation coefficients for the subthreshold parameters from the RS analysis.}
\label{tab:corr_RS_subthr}
\renewcommand{\arraystretch}{1.0}
\end{table}

To assess the sensitivity to the input quantities we retained the following effects that had a non-negligible impact on the solution:
\begin{enumerate}
 \item $s$-channel partial waves and inelasticities from KH80 instead of SAID,
 \item $l_\text{max}=5$ instead of $l_\text{max}=4$,
 \item $J_\text{max}=3$ instead of $J_\text{max}=2$ ($t$-channel $F$-waves),
 \item different continuation of the $t$-channel $S$-wave phase above $\sqrt{t_0}=1.3\GeV$,
 \item additional parameter for $S_{31}$-wave.
\end{enumerate}
Each of these fits $k=1,\ldots,5$ produces another set of subthreshold parameters $X_i^k$.
Assuming that the fluctuations are equally likely in either direction around the central solution, this translates to a covariance matrix
\beq
\label{Sigma_ij}
\Sigma^\text{syst}_{ij}=\frac{1}{4}\sum_{k=1}^5\big(X_i^0-X_i^k\big)\big(X_j^0-X_j^k\big).
\eeq

The uncertainties in the matching conditions and in the $\pi N$ coupling constant are treated similarly. However, in view of the deficiencies in constructing another completely independent matching condition discussed above, 
we interpreted the corresponding variation as a full $1\sigma$ effect in $\Sigma^\text{match}_{ij}$, equivalent to dropping the factor $1/4$ in~\eqref{Sigma_ij}, to remain on the conservative side.
Similarly, we found that the upwards and downwards shifts in $g^2/(4\pi) = 13.7\pm 0.2$ did not always produce symmetric errors in the subthreshold parameters, so that we decided to interpret both variations independently as a $1\sigma$ effect, i.e.\
\beq
\Sigma^\text{g}_{ij}=\big(X_i^0-X_i^{g \,\text{up}}\big)\big(X_j^0-X_j^{g \,\text{up}}\big)+\big(X_i^0-X_i^{g \,\text{down}}\big)\big(X_j^0-X_j^{g \,\text{down}}\big).
\eeq

Finally, in the case of the flat minima, we interpret the distribution of allowed solutions as a statistical ensemble and thereby estimate the uncertainty from the flat directions as
\beq
\Sigma^\text{FD}_{ij}=\frac{1}{N_\text{FD}}\sum_{k=1}^{N_\text{FD}}\big(X_i^k-\bar X_i\big)\big(X_j^k-\bar X_j\big),\qquad \bar X_i=\frac{1}{N_\text{FD}}\sum_{k=1}^{N_\text{FD}}X_i^k.
\eeq
The results for the subthreshold parameters with full uncertainty estimates derived from the covariance matrix
\beq
\Sigma=\Sigma^\text{SL}+\Sigma^\text{syst}+\Sigma^\text{match}+\Sigma^\text{g}+\Sigma^\text{FD}
\eeq
are given in Table~\ref{tab:RS_subthr}. In addition, we show the results for $d_{20}^+$, $d_{11}^+$, $d_{02}^+$ calculated from the sum rules in~\ref{app:ChPT_subthreshold}, with errors propagated from all the sources of uncertainty discussed above. In particular, we checked that for these parameters the sum rules converge so rapidly that the uncertainties from the truncation of the dispersive integrals are completely negligible in comparison. In general, we find reasonable agreement with the KH80 values~\cite{Koch:1980ay,Hoehler:1983} within uncertainties. The correlation coefficients $\rho_{ij}=\Sigma_{ij}/\sqrt{\Sigma_{ii}\Sigma_{jj}}$ are listed in Table~\ref{tab:corr_RS_subthr}.

\subsection{Uncertainty estimates for the $\pi N$ phase shifts}
\label{sec:err_phase_shift}

\begin{figure}[t!]
\centering
\includegraphics[width=0.45\linewidth,clip]{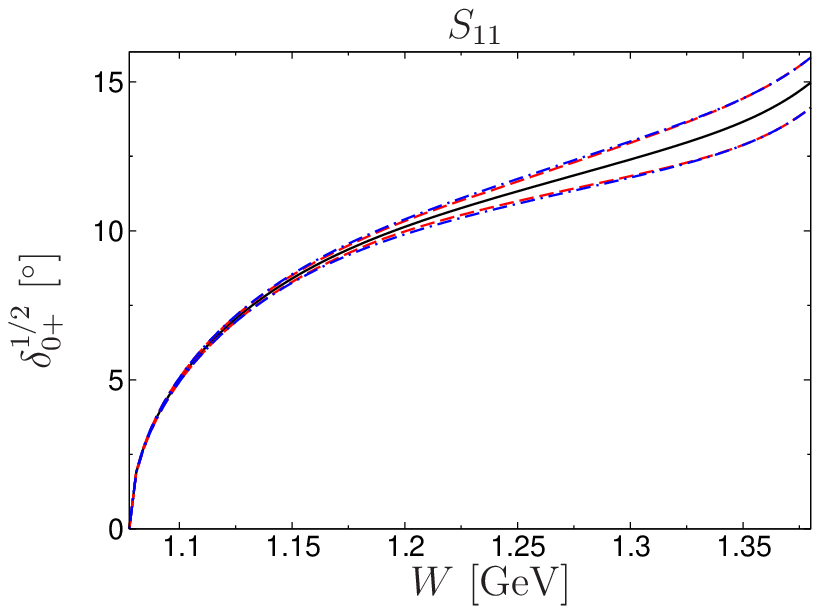}\quad
\includegraphics[width=0.45\linewidth,clip]{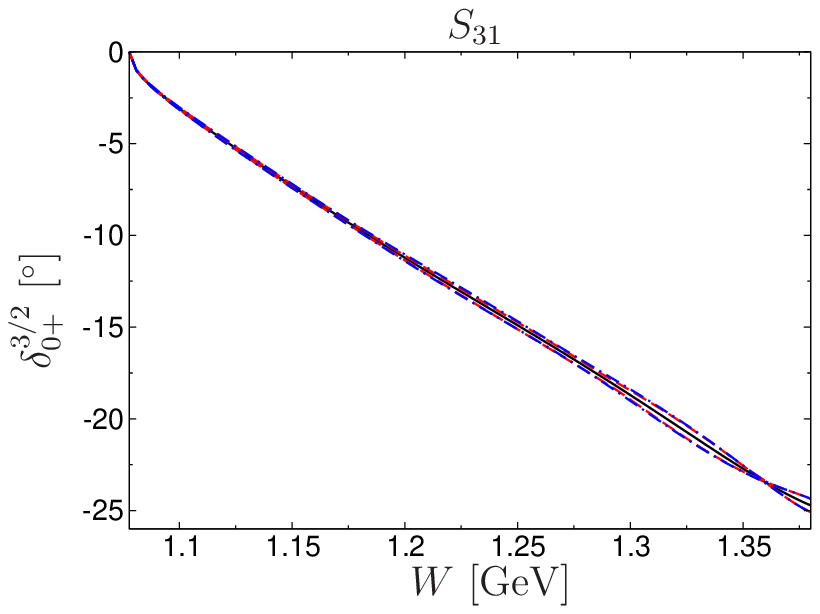}\\[0.1cm]
\includegraphics[width=0.45\linewidth,clip]{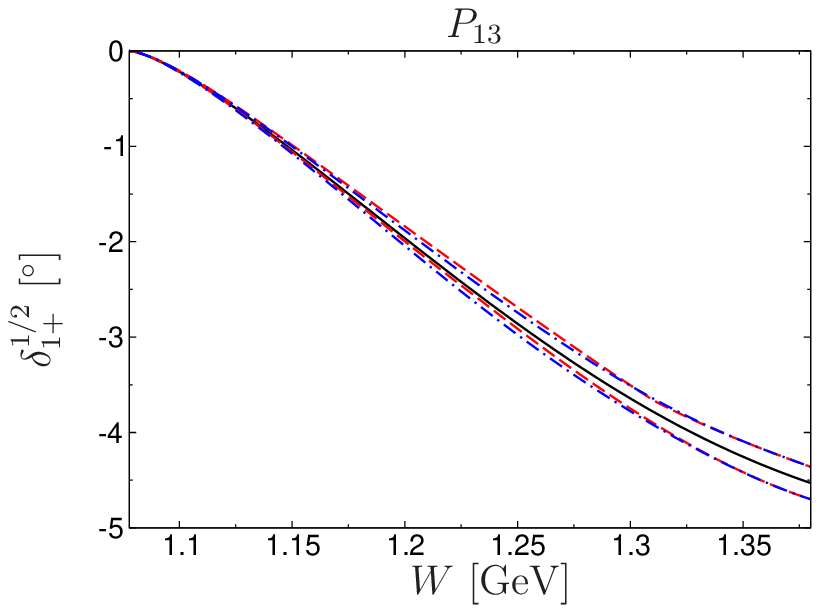}\quad
\includegraphics[width=0.45\linewidth,clip]{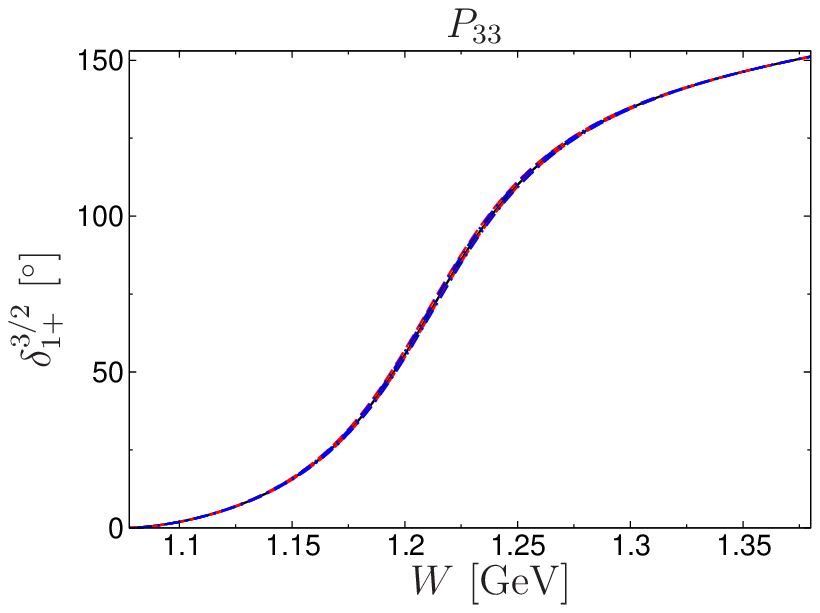}\\[0.1cm]
\includegraphics[width=0.45\linewidth,clip]{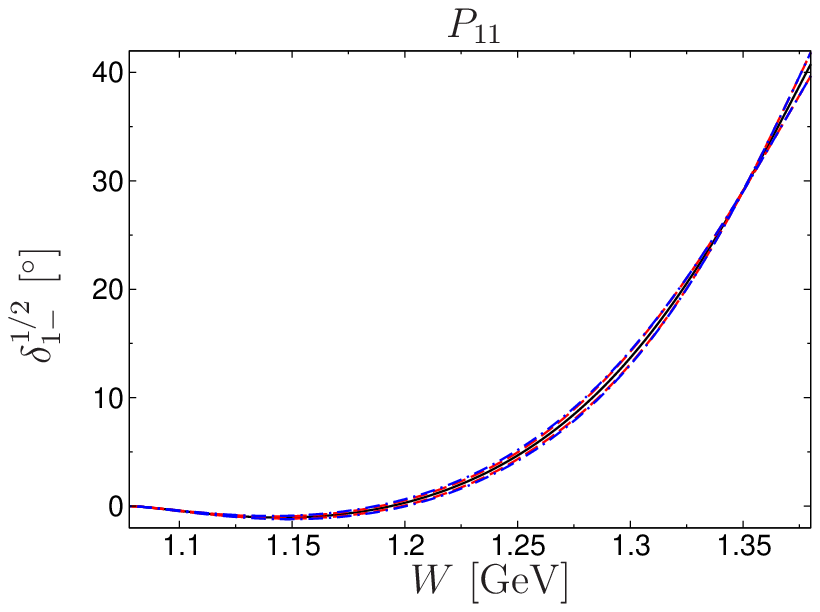}\quad
\includegraphics[width=0.45\linewidth,clip]{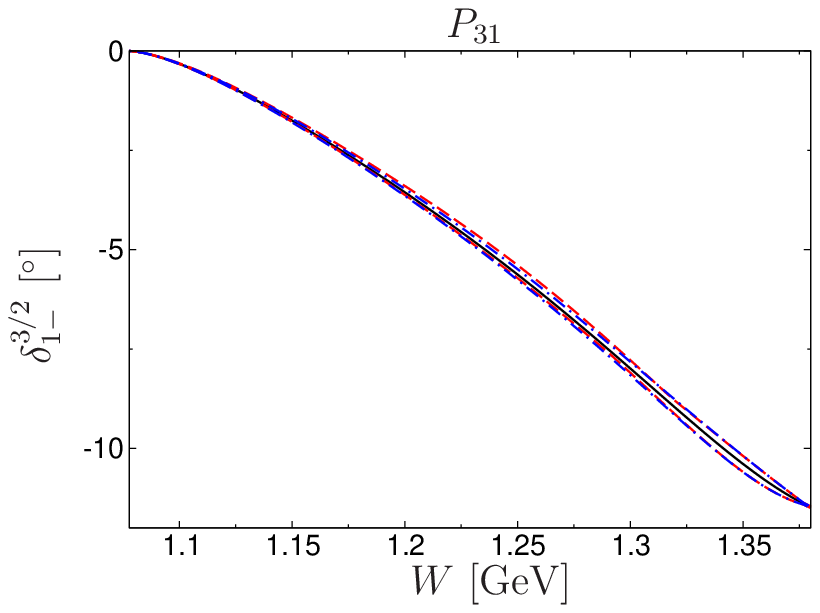}
\caption{Envelope (red dashed) and quadrature (blue dot-dashed) error bands around the central solution (black solid line). A close-up on the error bands when the central value is subtracted is shown in Fig.~\ref{bands-sub}.}
\label{bands}
\end{figure}

The procedure adopted in the previous section cannot be straightforwardly generalized to the phase shifts, since
the parameterizations are by no means unique: it is possible to describe the same curve with two different sets of parameters equally well, 
so that even when two fits produce similar curves, they can differ substantially in the parameter values, 
which would artificially increase the errors of the parameters.  
Therefore, instead of computing the errors of the phase-shift parameters directly from the values of the different solutions, 
we will first identify the phase-shift error bands, 
and subsequently translate them into parameter errors.  

We consider two different ways to define the phase-shift errors bands:
\begin{enumerate}
\item taking the envelope of the full set of fits as the error,
\item calculating the error of each source of uncertainty identified above separately, i.e.\ scattering lengths, matching point, input variation, 
and statistical errors from flat directions point by point, and adding them in quadrature.
\end{enumerate}

In the first case, for each of the energy points $W_i$ where we have solved the RS equations, 
we take the biggest difference to the central value as the error, i.e.\ 
\beq
\Delta_{E}\delta^{I_s}_{l\pm}(W_i)=\smash{\displaystyle\max_{i\,\in\, G}}\Big\{\delta^{I_s}_{l\pm}(W_i) -\delta^{I_s}_{i\;l\pm}(W_i) \Big\},
\eeq
where $G$ denotes the whole set of fits. In the case of the scattering-length error the points at the corner of the grid have been discarded, since they are more than one sigma away from the center.
In general, this method of taking the envelope of all fits defines asymmetric errors. 

\begin{figure}[t!]
\centering
\includegraphics[width=0.45\linewidth,clip]{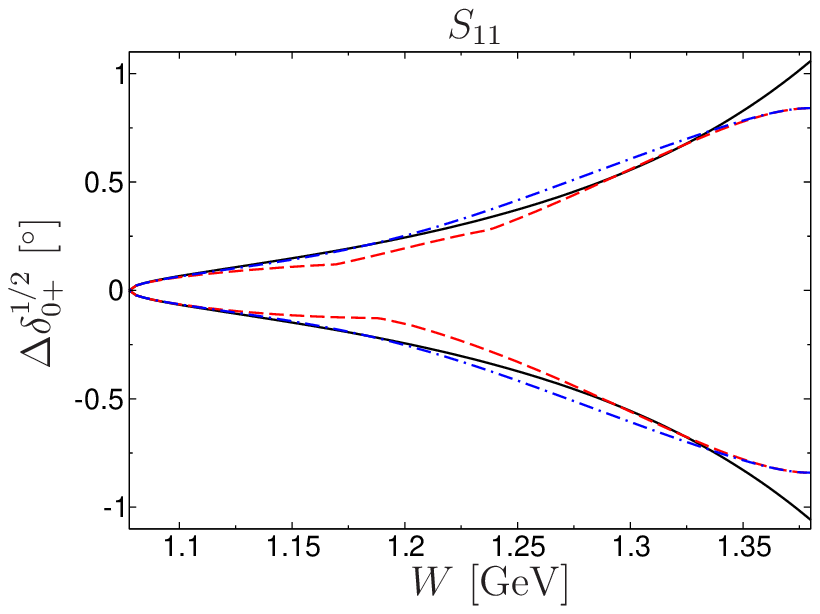}\quad
\includegraphics[width=0.45\linewidth,clip]{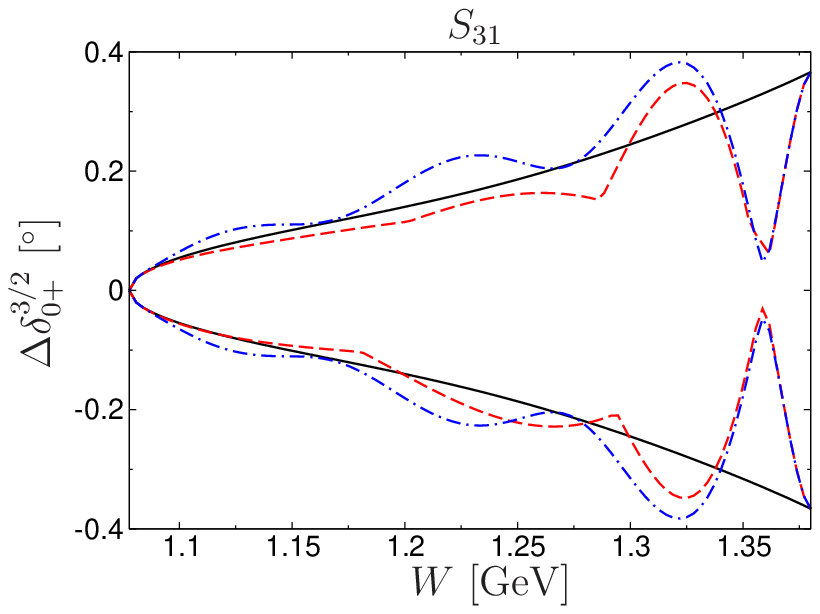}\\[0.1cm]
\includegraphics[width=0.45\linewidth,clip]{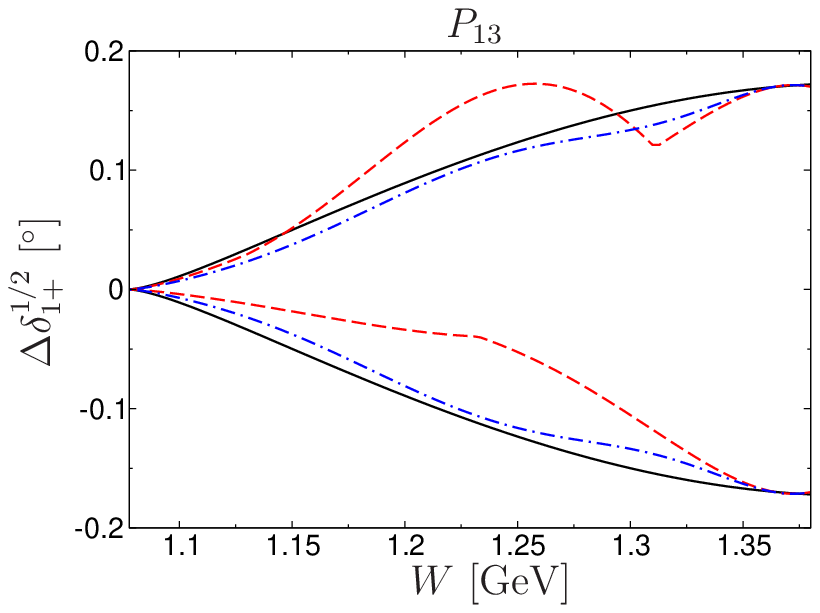}\quad
\includegraphics[width=0.45\linewidth,clip]{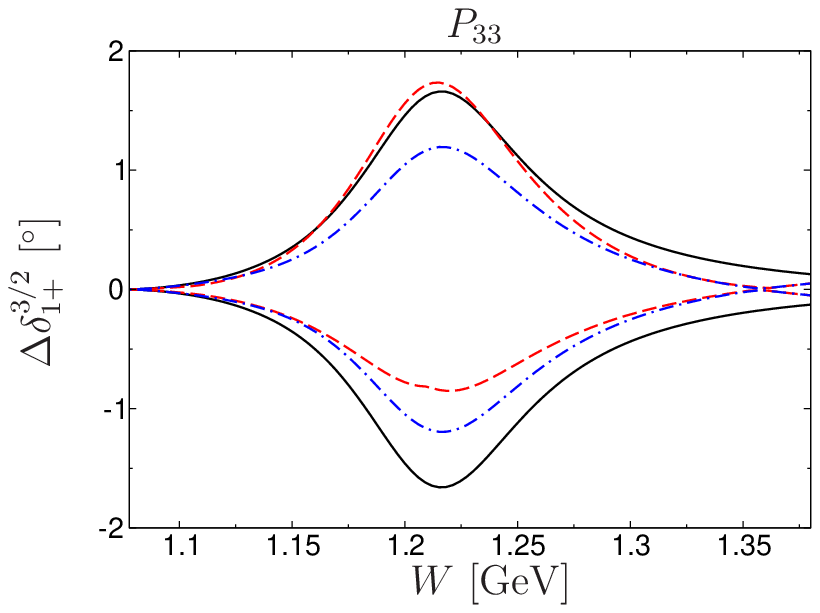}\\[0.1cm]
\includegraphics[width=0.45\linewidth,clip]{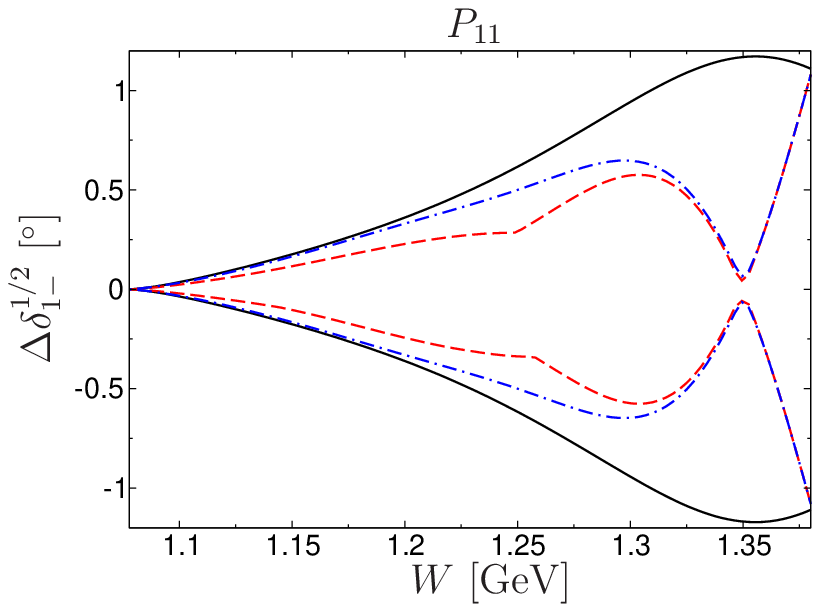}\quad
\includegraphics[width=0.45\linewidth,clip]{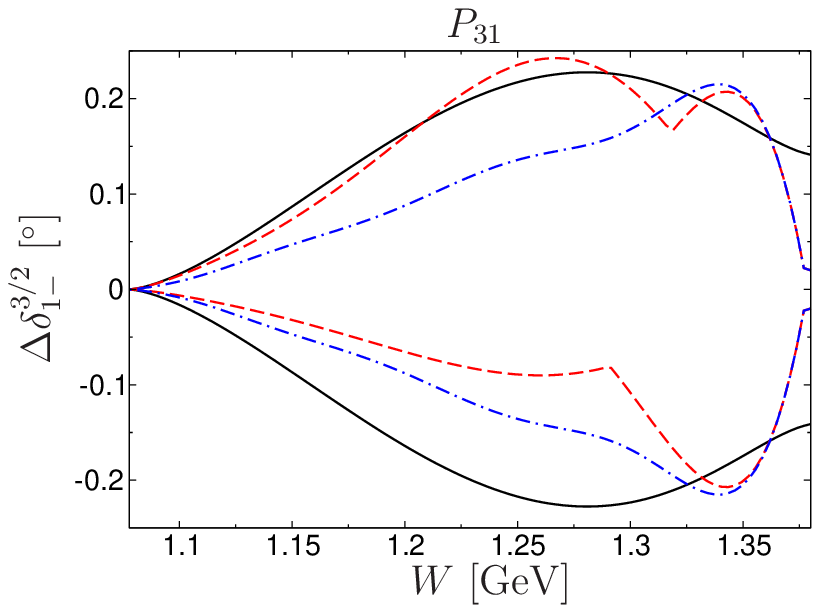}
\caption{Envelope (red dashed) and quadrature (blue dot-dashed) error bands calculated for each phase shift when the central value is subtracted. The black solid lines correspond to the fitted lines obtained from the propagation of the uncertainties of the parameters in the phase shifts, as described in the main text.}
\label{bands-sub}
\end{figure}

In the second case a separate analysis for each of the different sources of uncertainty has been performed in the same way as for the subthreshold parameters. 
Therefore, the full error band can be expressed as
\beq
\left(\Delta\Re f^{I_s}_{l\pm}(W_i)\right)_\text{full}=\sqrt{\left(\Delta\Re f^{I_s}_{l\pm}(W_i)\right)_\text{SL}^2+\left(\Delta\Re f^{I_s}_{l\pm}(W_i)\right)_\text{syst}^2+\left(\Delta\Re f^{I_s}_{l\pm}(W_i)\right)_\text{match}^2+\left(\Delta\Re f^{I_s}_{l\pm}(W_i)\right)_\text{g}^2+\left(\Delta\Re f^{I_s}_{l\pm}(W_i)\right)_\text{FD}^2}.
\eeq

In Fig.~\ref{bands} we compare the error bands obtained from these two methods. 
In general, the results are quite similar, so that the difference between them can only be highlighted if the central value is subtracted, 
as is done in Fig.~\ref{bands-sub}. 
The main difference between these two definitions is the non-symmetric treatment of the envelope error. 
However, both ways to estimate the errors provide similar results for all the partial waves, except for the $P_{13}$ and $P_{31}$, 
for which the flat-direction mean $\Re \bar{f}^{I_s}_{l\pm}$ differs noticeably from the central solution.

The analysis leading to the bands in Figs.~\ref{bands} and~\ref{bands-sub} provides us with two different sets of error bands. 
However, for practical reasons it is more convenient to translate them into errors for the parameters in the description of the phase shifts, 
so that the error bands can be reproduced directly from the parameterizations.  

\begin{figure}[t!]
\centering
\includegraphics[width=0.45\linewidth,clip]{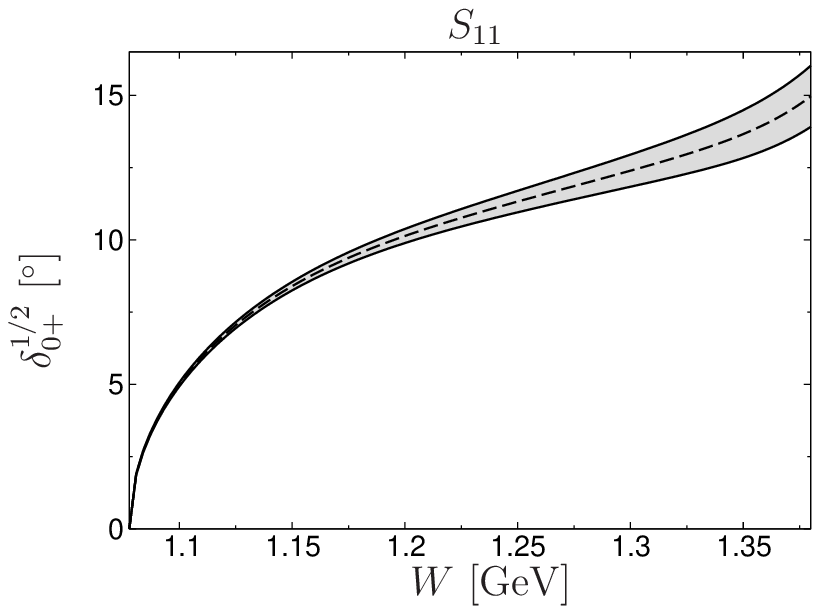}\quad
\includegraphics[width=0.45\linewidth,clip]{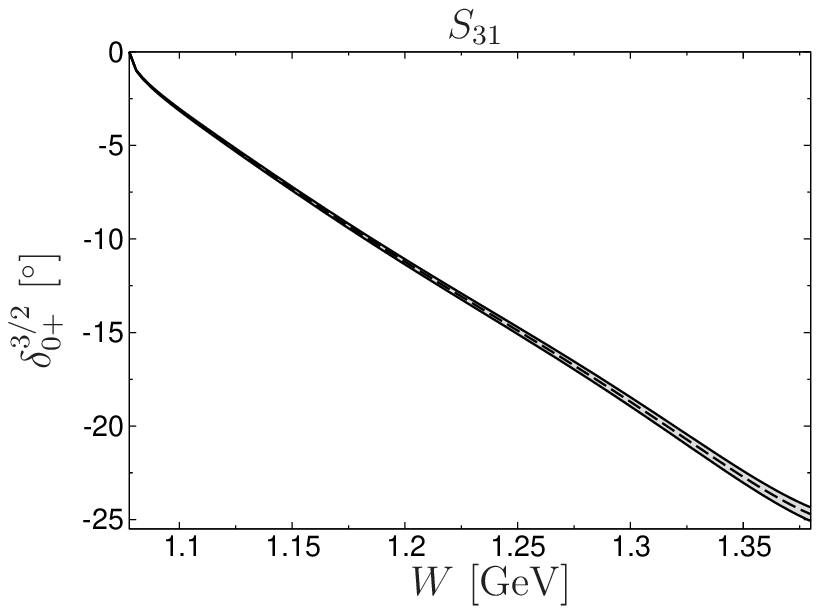}\\[0.1cm]
\includegraphics[width=0.45\linewidth,clip]{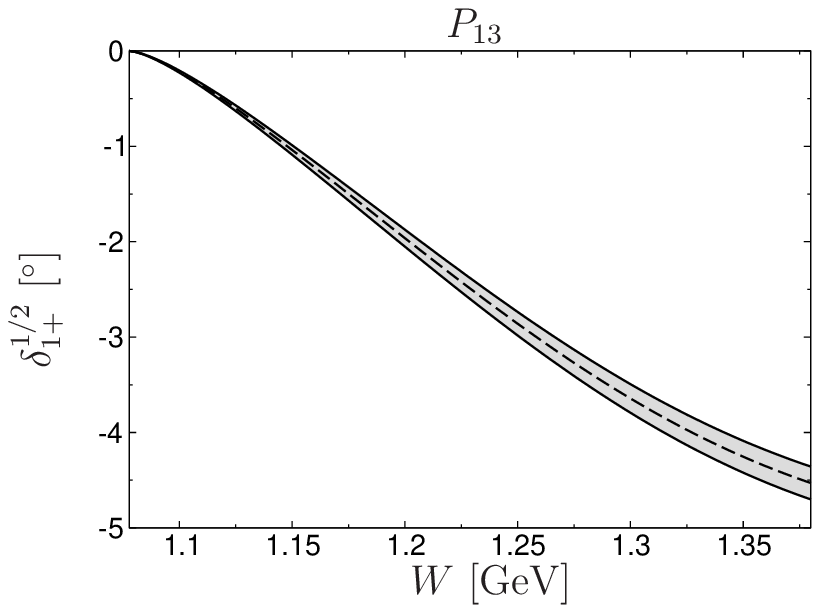}\quad
\includegraphics[width=0.45\linewidth,clip]{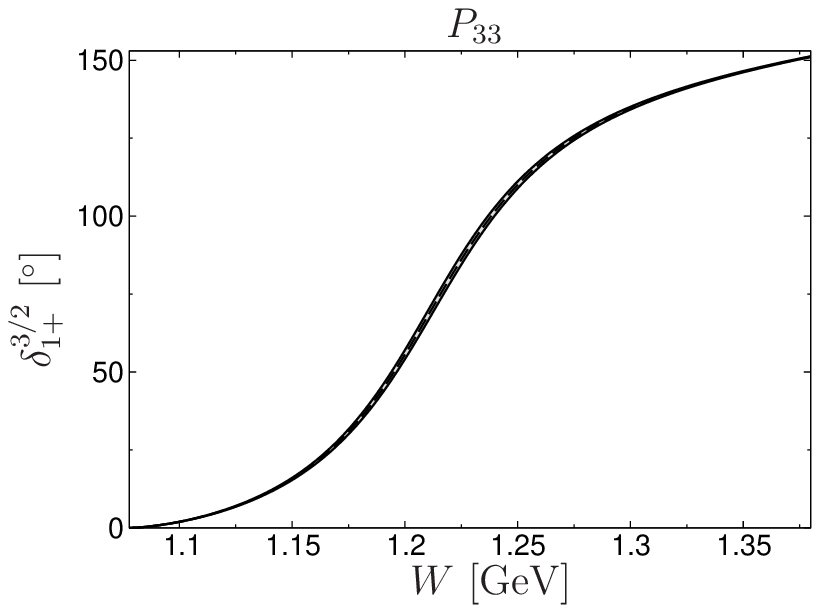}\\[0.1cm]
\includegraphics[width=0.45\linewidth,clip]{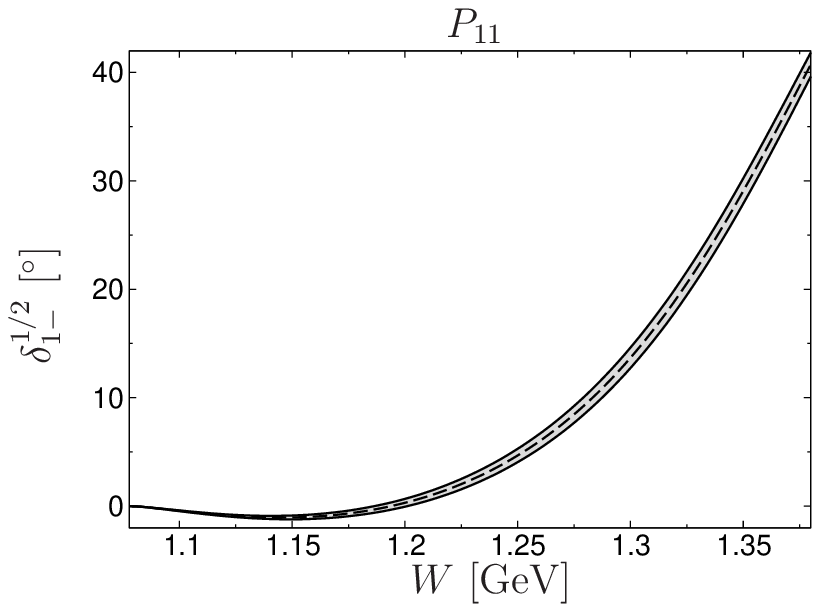}\quad
\includegraphics[width=0.45\linewidth,clip]{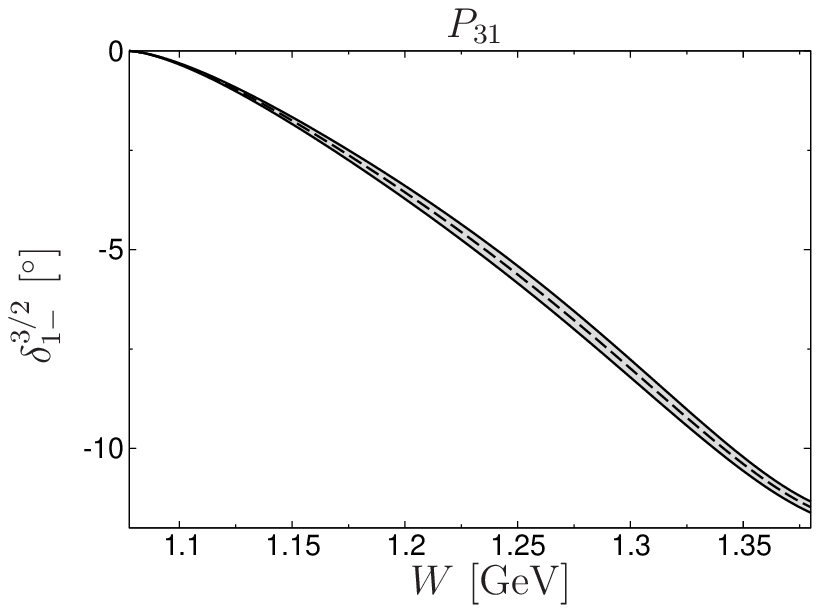}
\caption{Final errors bands for the $\pi N$ phase shifts. The dashed lines indicate the central curves.}
\label{final-bands}
\end{figure}

As a first step we define a common error band, choosing for each wave the biggest uncertainty to stay conservative.  
Accordingly, we take the quadratic approach for the $S_{11}$-, $S_{31}$-, and $P_{11}$-waves, and the upwards shift of the envelope band for the $P_{13}$-, $P_{33}$-, and $P_{31}$-waves 
(we assume a symmetric error, so we adopt the biggest shift for all the cases).  
Subsequently, we follow the strategy to fit the errors obtained from the propagation of the uncertainty of the parameters to the corresponding errors bands. 
In all the cases we checked that it can be done easily using only the first two parameters of the Schenk or conformal parameterization and their corresponding correlation.
This observation is consistent with the fact that the parameterizations are defined as an expansion in the CMS momentum, so higher parameters become less and less relevant in the determination of the error. 
Therefore, the expression we have considered to fit the errors bands is:
\beq
\Delta\delta_{l{\pm}}^{I_s}=\sqrt{\left(\frac{\partial \delta_{l{\pm}}^{I_s}}{\partial A_{l\pm}^{I_s}}\right)^2\left(\Delta A_{l\pm}^{I_s}\right)^2+\left(\frac{\partial \delta_{l{\pm}}^{I_s}}{\partial B_{l\pm}^{I_s}}\right)^2\left(\Delta B_{l\pm}^{I_s}\right)^2+2\left(\frac{\partial \delta_{l{\pm}}^{I_s}}{\partial A_{l\pm}^{I_s}}\right)\left(\frac{\partial \delta_{l{\pm}}^{I_s}}{\partial B_{l\pm}^{I_s}}\right)\rho_{AB}\Delta A_{l\pm}^{I_s}\Delta B_{l\pm}^{I_s}},
\eeq
where $\Delta A_{l\pm}^{I_s}$, $\Delta B_{l\pm}^{I_s}$, and $\rho_{AB}$ refer to the parameter errors and their correlation coefficients, respectively (and analogously for $\tilde A_{l\pm}^{I_s}$ and $\tilde B_{l\pm}^{I_s}$ in the case of the $P_{33}$).
In addition, we impose that $\Delta A_{l\pm}^{I_s}$ is fixed to the pionic-atom scattering-length error for the $S$-waves, 
so that the corresponding error is reproduced exactly at threshold. 
Besides, we also impose that the fitted error has to be at least as large as the error bands at the matching point, 
since it corresponds directly to the difference between SAID and KH80.
The curves obtained in this way are reproduced in Fig.~\ref{bands-sub}, whereas the final error bands are depicted in Fig.~\ref{final-bands}. Explicit numerical solutions for the phase shifts are provided in~\ref{app:numerical_sol}.

\begin{figure}[t!]
\centering
\includegraphics[width=0.45\linewidth,clip]{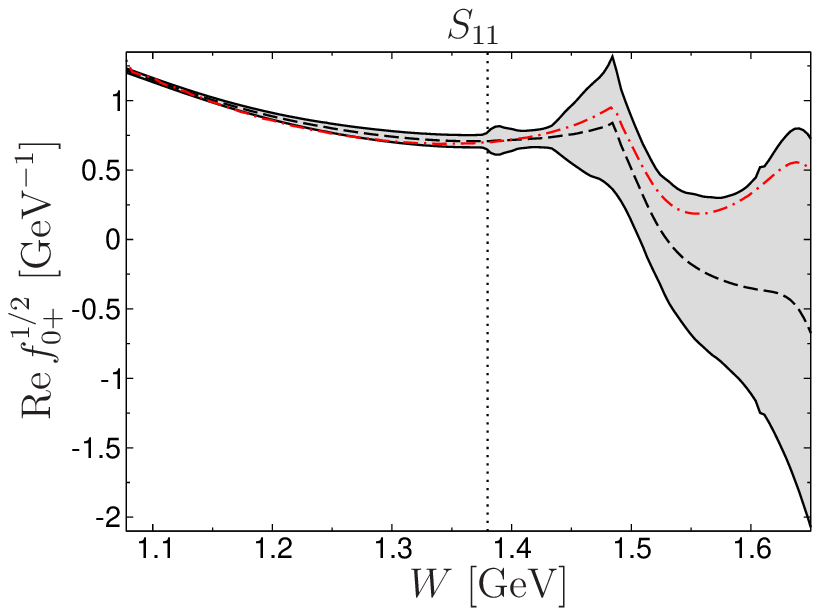}\quad
\includegraphics[width=0.45\linewidth,clip]{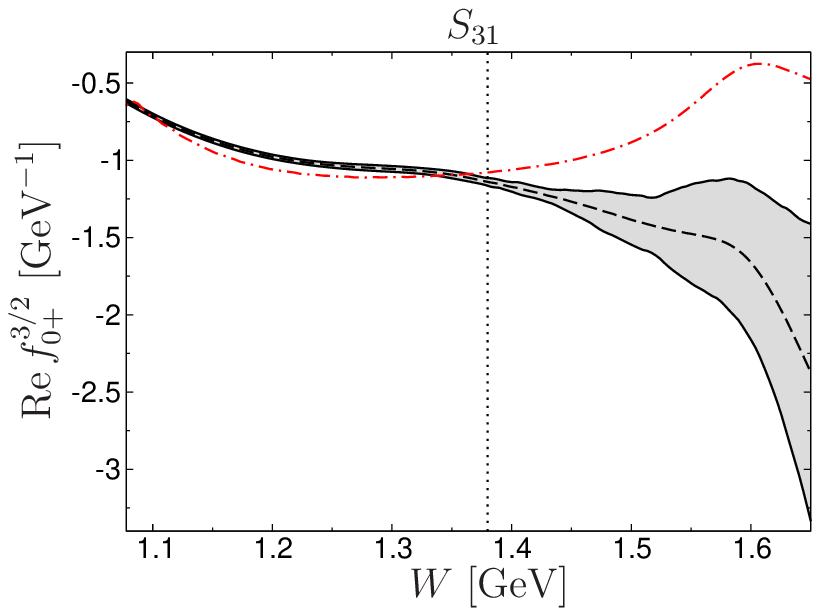}\\[0.1cm]
\includegraphics[width=0.45\linewidth,clip]{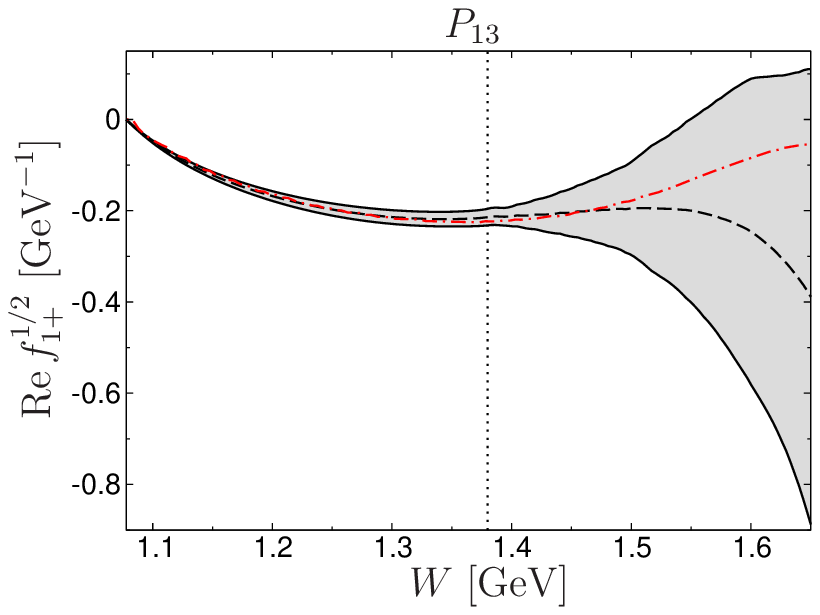}\quad
\includegraphics[width=0.45\linewidth,clip]{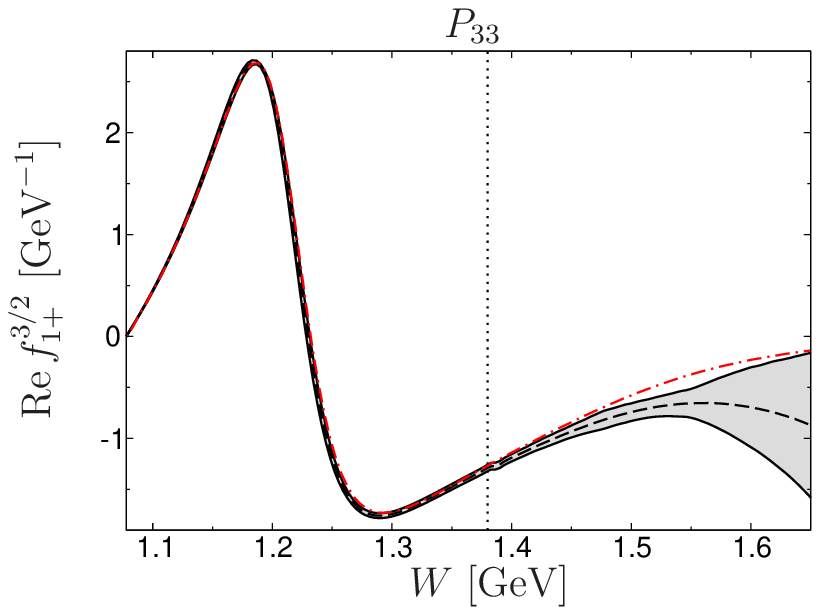}\\[0.1cm]
\includegraphics[width=0.45\linewidth,clip]{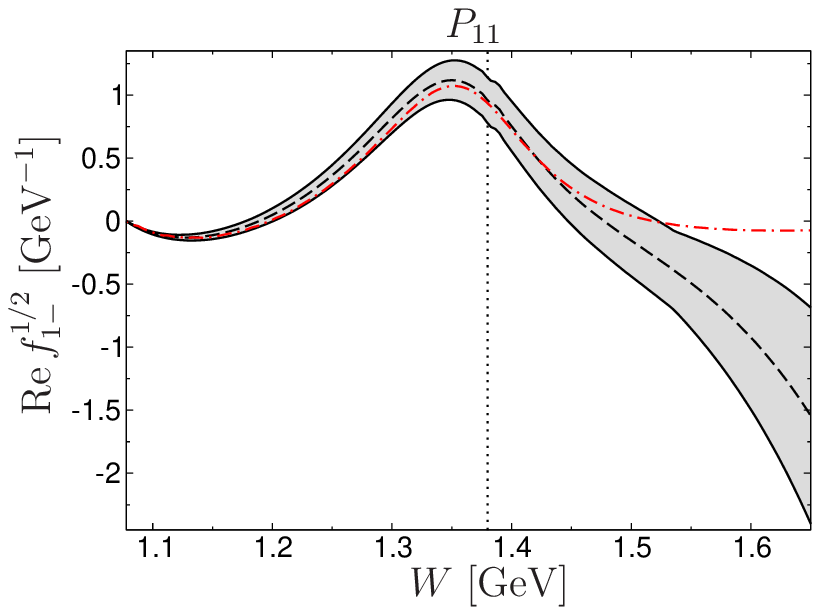}\quad
\includegraphics[width=0.45\linewidth,clip]{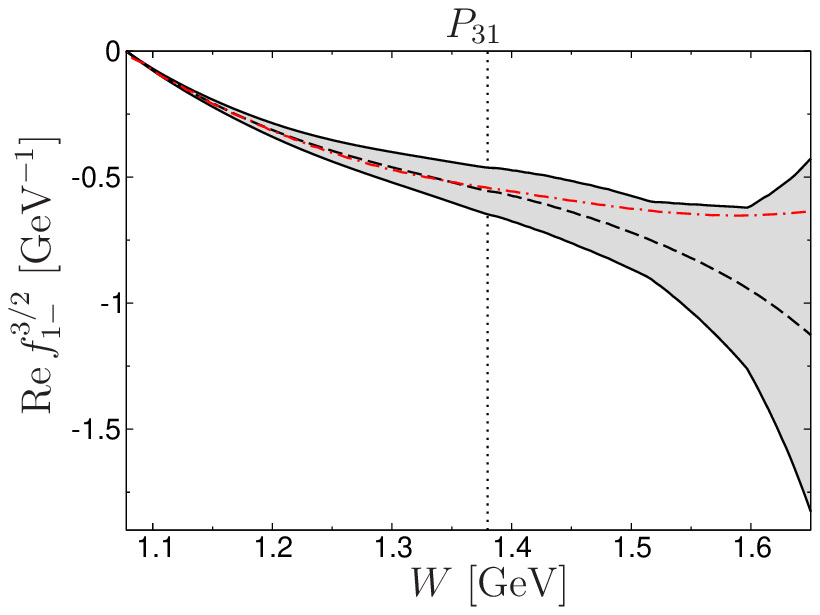}
\caption{Real parts of the $s$-channel partial waves calculated from the RHS of the RS equations (dashed lines) in comparison to the SAID results (red dot-dashed lines). The dashed lines indicate the position of the matching point $\sqrt{\sm}=1.38\GeV$.}
\label{fig:real}
\end{figure}

\begin{figure}[t!]
\center
 \includegraphics[width=0.447\linewidth,clip]{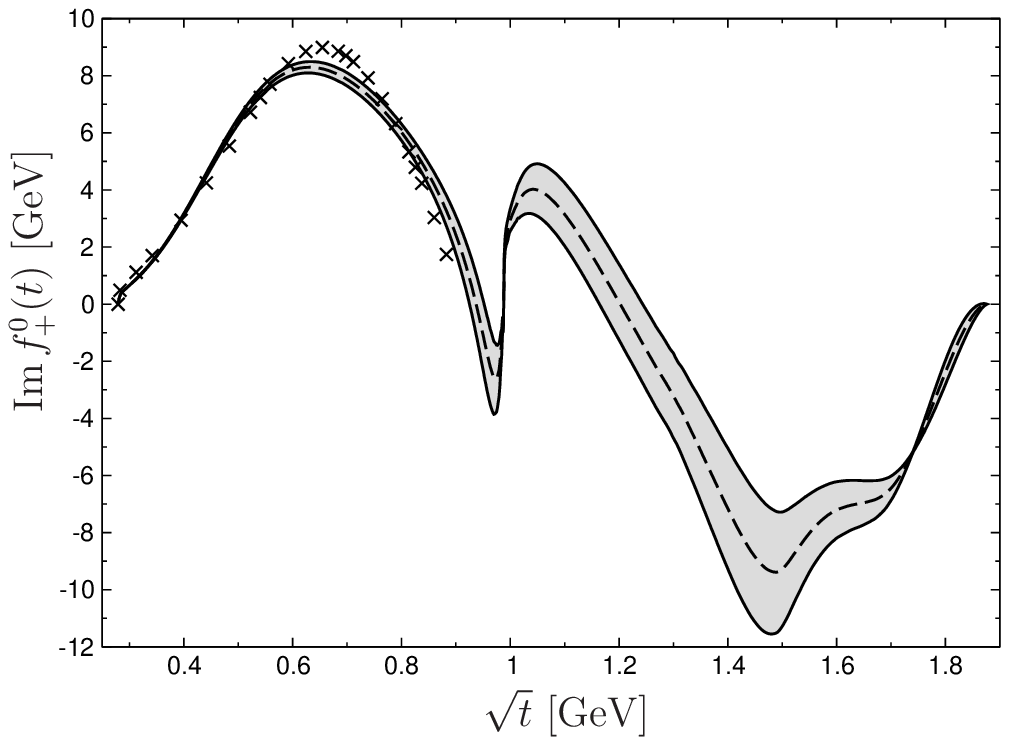}\\[1mm]
\includegraphics[width=0.447\linewidth,clip]{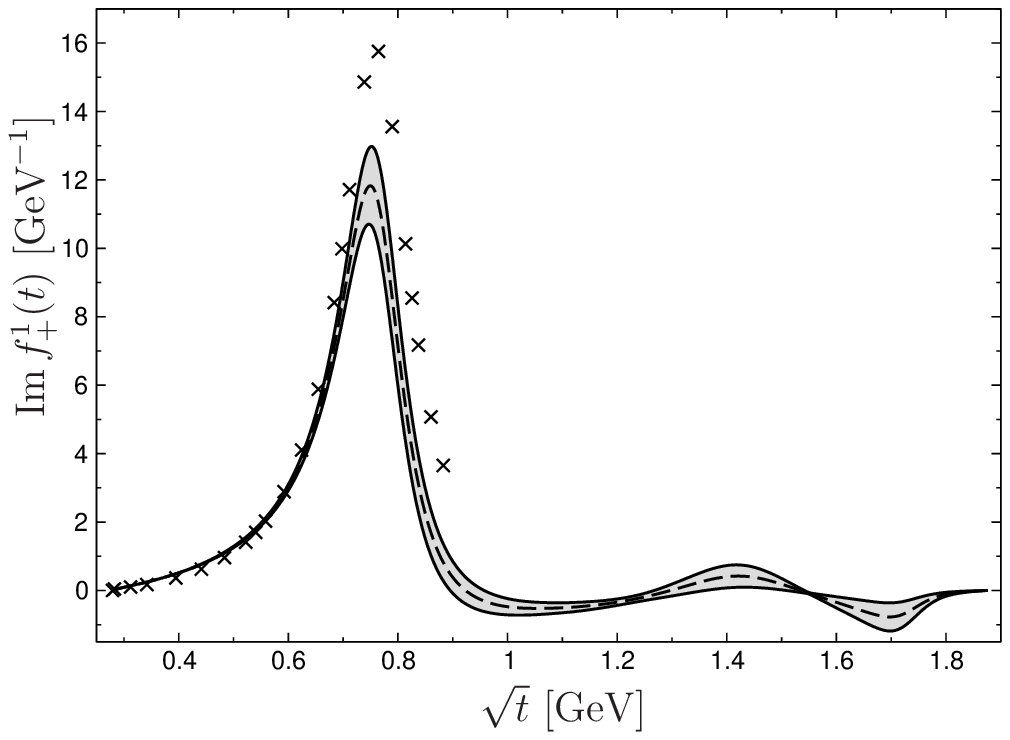}\quad
\includegraphics[width=0.447\linewidth,clip]{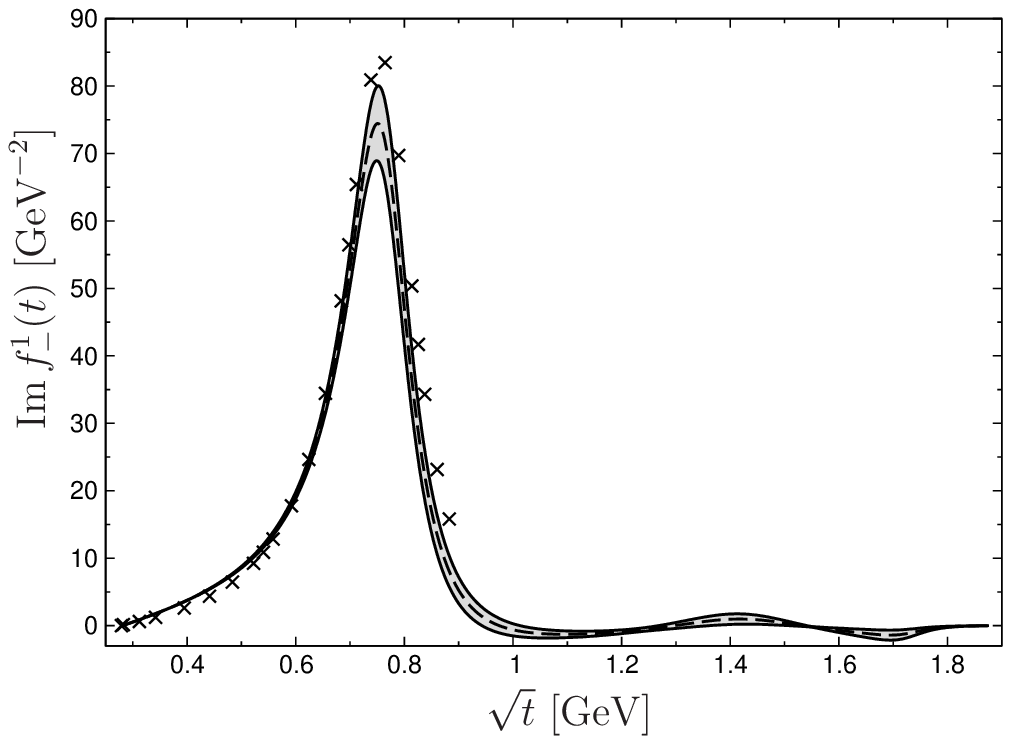}\\[1mm]
\includegraphics[width=0.447\linewidth,clip]{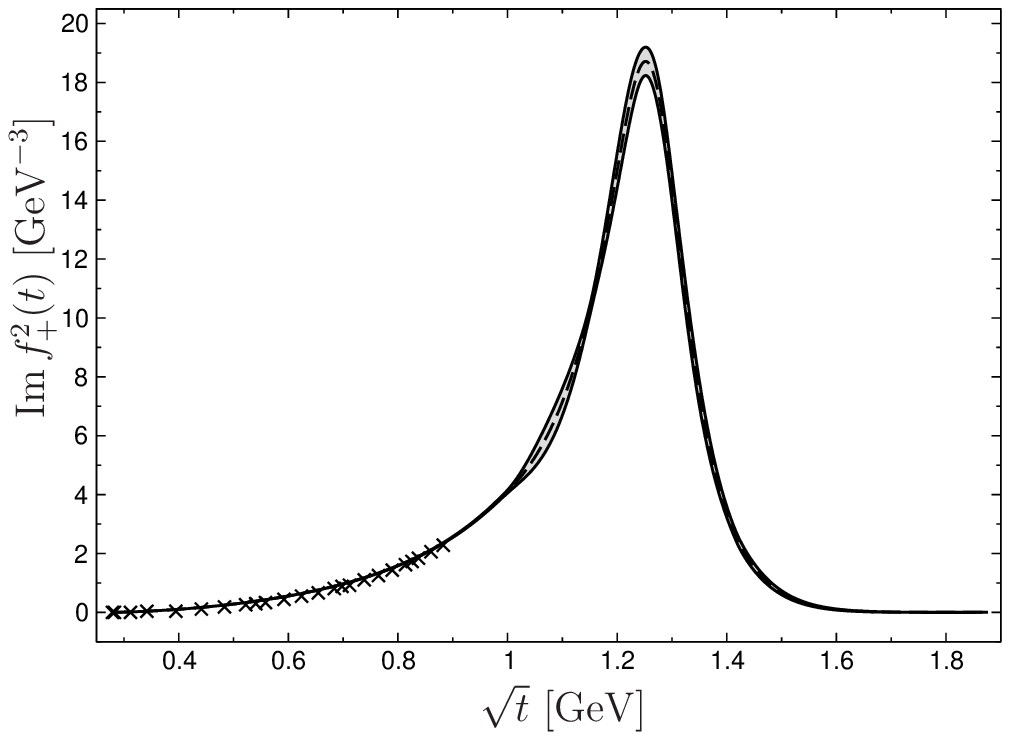}\quad
\includegraphics[width=0.447\linewidth,clip]{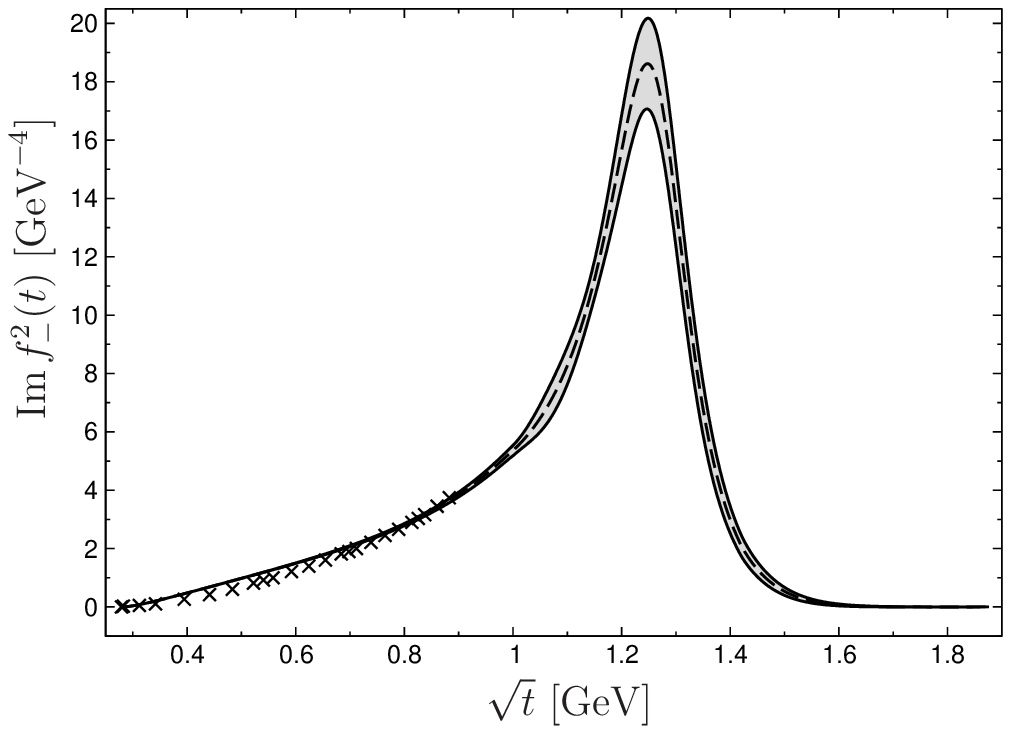}
  \caption{Final results for the imaginary parts of the $t$-channel partial waves, with error bands including both uncertainties in the subthreshold parameters and the MO input.
 The black crosses refer to the results from~\cite{Hoehler:1983}.}
  \label{fig:ImfJ_band}
\end{figure} 

Finally, although the RS equations are strictly valid only up to the matching point, the real parts can be formally evaluated even above that energy from the RHS of the RS equations in terms of a principal-value integral, while the imaginary parts agree by construction with the imaginary parts used as input. The results for the real parts obtained in this way are shown in Fig.~\ref{fig:real} up to energies well beyond $\sqrt{\sm}$. The error bands are derived in the same way as before: we first study the quadratic and (symmetrized) envelope error bands separately and then take the larger one of the two at a given energy. If the RS equations at the matching point were fulfilled perfectly, the central curves (black dashed lines) would agree there with the SAID input (red dot-dashed lines). The fact that this is not exactly the case reflects the remaining small differences between LHS and RHS in Fig.~\ref{fig:fit-tsub}. As expected, the uncertainties grow quickly above the matching point, but within those uncertainties the outcome is consistent with the SAID real parts up to energies as high as $1.6\GeV$ in most partial waves, apart from the $S_{31}$, where differences emerge already below the matching point. Similarly to the case of $\pi\pi$ Roy equations~\cite{Ananthanarayan:2000ht}, we conclude that the RS equations appear not to break down abruptly above the boundary of their strict domain of validity, but to remain approximately valid even above that energy.   

\subsection{Uncertainty estimates for the $t$-channel partial waves}
\label{sec:err_tchannel_waves}

\begin{figure}[t!]
\center
 \includegraphics[width=0.447\linewidth,clip]{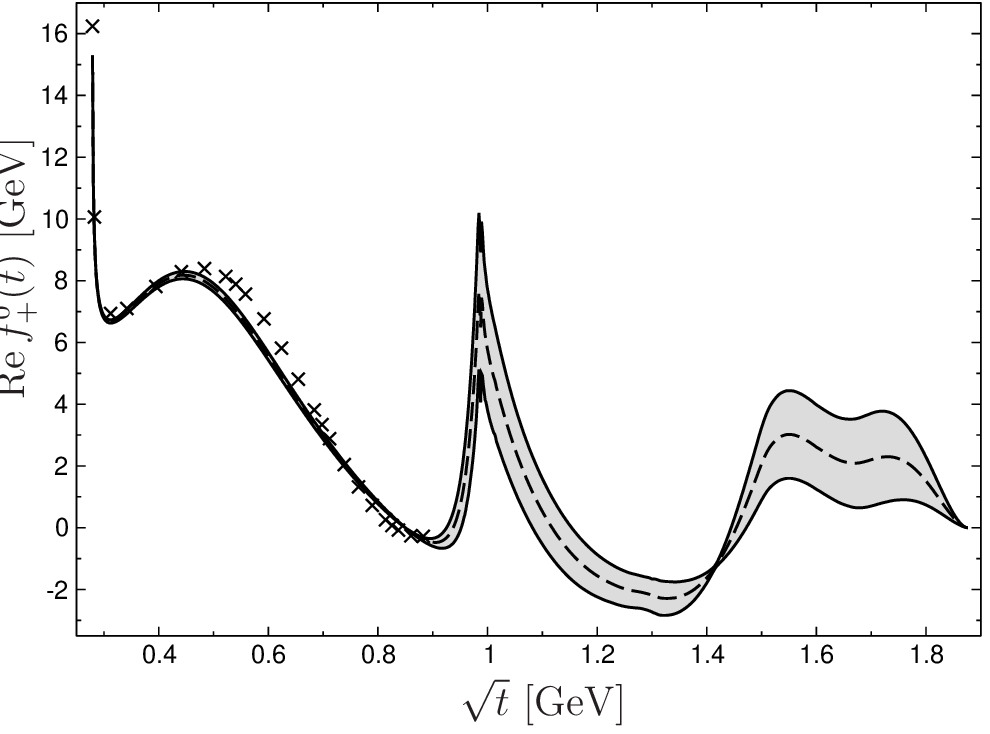}\\[1mm]
\includegraphics[width=0.447\linewidth,clip]{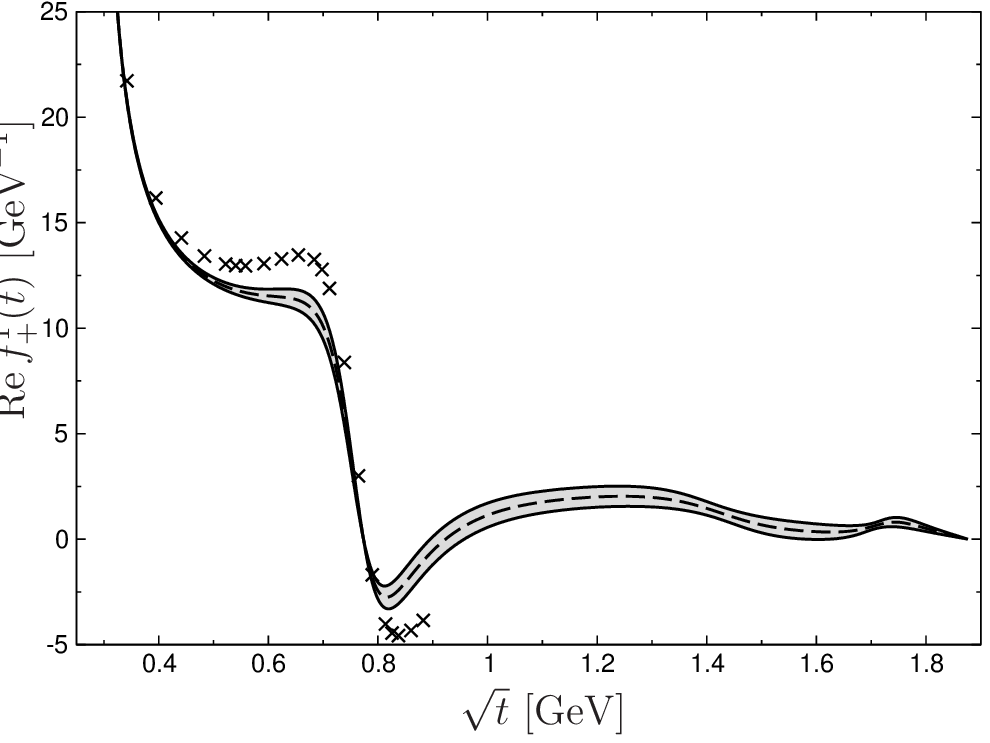}\quad
\includegraphics[width=0.447\linewidth,clip]{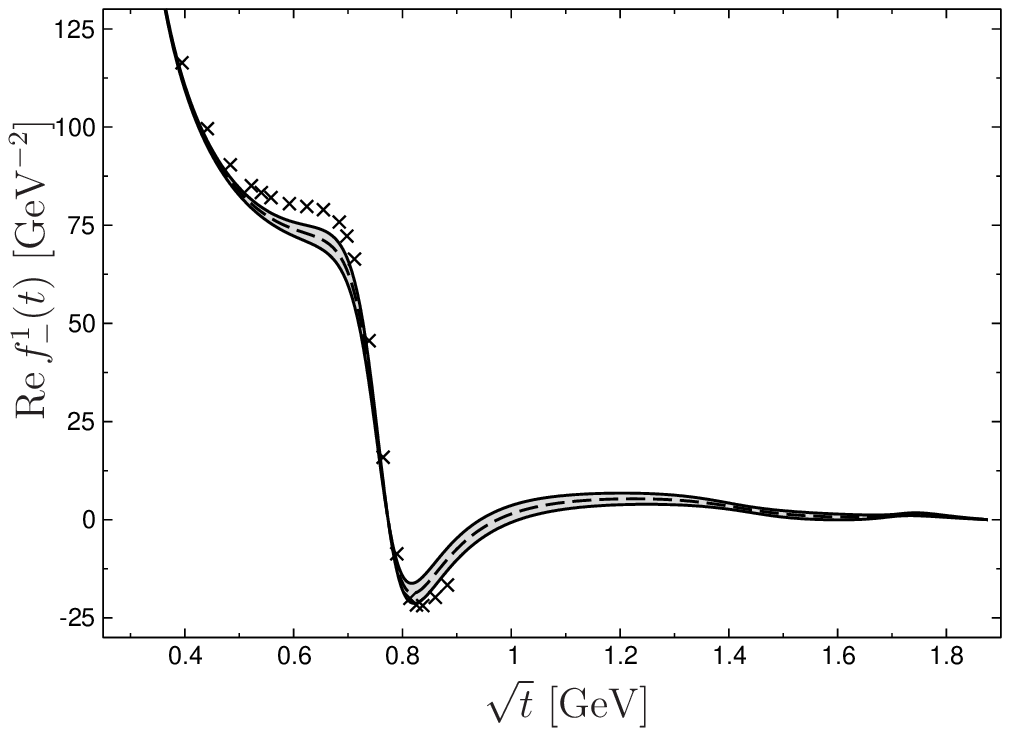}\\[1mm]
\includegraphics[width=0.447\linewidth,clip]{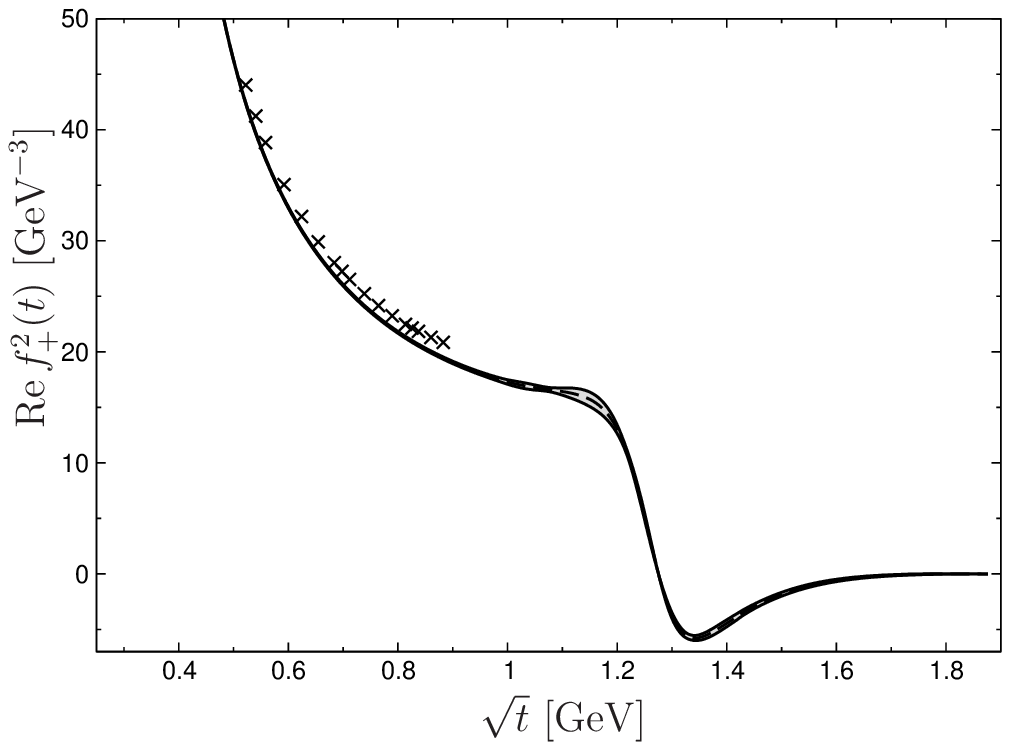}\quad
\includegraphics[width=0.447\linewidth,clip]{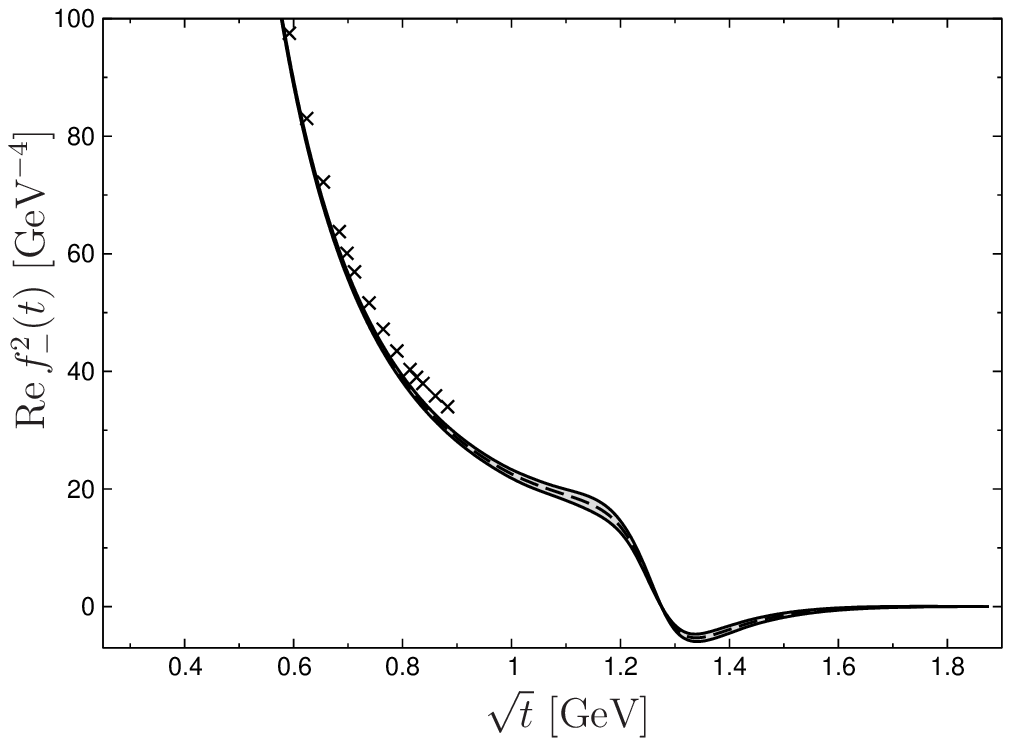}
  \caption{Final results for the real parts of the $t$-channel partial waves, with error bands including both uncertainties in the subthreshold parameters and the MO input.
 The black crosses refer to the results from~\cite{Hoehler:1983}.}
  \label{fig:RefJ_band}
\end{figure}

Once the subthreshold parameters and their covariance matrix are finalized, so can be the uncertainties of the $t$-channel partial waves. 
They cover both the systematic errors associated with the inelastic input and the $\pi\pi$ phase shifts, as well as the subthreshold-parameter errors propagated through~\eqref{P-wave-subth} and \eqref{S-wave-subth}.
Following the same conventions as adopted in Sect.~\ref{sec:subthreshold_parameters}, the systematic errors are deduced from the curves depicted in Fig.~\ref{fig:MO_KH80}, whose spread is interpreted as a full $1\sigma$ band, to be attached around the central solution. 
For the propagation of the subthreshold-parameter errors we also take into account their correlations, which in particular play a key role for $f^1_+$ and $f^2_+$.   
We combine both effects by adding them in quadrature, leading to the results for the imaginary parts plotted in Fig.~\ref{fig:ImfJ_band}.
The systematic uncertainties from inelasticities are most relevant for the $S$-wave, the high-energy tail of the $P$-waves and, to a certain extent, the $D$-waves, although for the latter 
 the error from the subthreshold parameters is still a factor of three larger. As expected, these effects set in roughly around energies where inelastic channels open, while
 the errors of the subthreshold parameters fully dominate in the low-energy region, in particular around the peak of the $\rho(770)$ resonance. 

For completeness we also show the results for the real parts, see Fig.~\ref{fig:RefJ_band}. Apart from the $S$-wave all partial waves are strongly dominated by the Born terms close to threshold, where they take a large (but finite) value that would overshadow any structure in the remainder of the amplitude if included in the plot. For this reason, the scale is cut off much earlier, focusing on the part of the partial waves where the respective resonances occur. In general, we find that deviations from the KH80 results are at a similar level as already observed for the imaginary parts, with error analysis performed in the same way as in Fig.~\ref{fig:ImfJ_band}. Due to the Born-term dominance at threshold there is in principle one additional source of uncertainty originating from the $\pi N$ coupling constant, whose effect, however, is concentrated to the region very close to threshold and would therefore not even be visible in Fig.~\ref{fig:RefJ_band}. Moreover, in spectral functions that involve $\pi\pi\to\bar N N$ amplitudes the corresponding contribution will be suppressed by phase-space factors, so that the residual uncertainty from the $\pi N$ coupling constant becomes irrelevant in practice.

\section{Threshold parameters}
\label{sec:threshold}

In order to implement the constraints from the $S$-wave scattering lengths from pionic atoms, we demand that the RHS of the RS equations for the $S$-waves at threshold reproduce these values, see Sect.~\ref{sec:schannel_sol}. In fact, this relation between the RHS of the RS equations and the $S$-wave scattering lengths is a special case of sum rules that express the threshold parameters in terms of HDRs and derivatives thereof.  
The threshold parameters are defined as the expansion coefficients in
\beq
\label{threshold_expansion}
\Re f_{l\pm}^I(s)=\qq^{2l}\Big\{\athr+\bthr \qq^2+\cthr \qq^4+\dthr \qq^6+\Order\big(\qq^8\big)\Big\}.
\eeq
The leading terms are the scattering lengths (for the $P$-waves also referred to as scattering volumes), 
while the first correction is determined by the effective ranges $\bthr$ and even higher terms are referred to as shape parameters.

As a direct calculation of these parameters from derivatives of the partial waves is numerically rather delicate, the most promising framework for a stable evaluation is based on sum rules involving dispersive integrals over the pertinent amplitudes, see~\cite{Ananthanarayan:2000ht,GarciaMartin:2011cn,Wanders:1966,Palou:1974ma} for the case of $\pi\pi$ scattering. Such sum rules could be derived directly from~\eqref{sRSpwhdr} by taking derivatives with respect to $\qq^2$ and identifying the results with the coefficients in~\eqref{threshold_expansion}. However, this procedure is unfavorable from a technical point of view since a substantial part of the effort in calculating the derivatives is wasted on reproducing the kinematic structure of the partial-wave expansion~\eqref{sprojform}, i.e.\ its decomposition into invariant amplitudes $A^I$ and $B^I$ with known, $\qq$-dependent prefactors.

For this reason, we will consider an approach that is directly based on the threshold expansion of the projections~\eqref{sprojinvampl}. Suppressing isospin indices for the time being, we have
\beq
\label{threshold_expansion_Al}
X_l(s)=X_l^{(l)}\qq^{2l}+X_l^{(l+1)}\qq^{2l+2}+X_l^{(l+2)}\qq^{2l+4}+\Order\big(\qq^{2l+6}\big),\qquad X\in\{A,B\},
\eeq
and the inversion
\beq
X_l^{(n)}=\frac{1}{n!}\zeq{\partial_{\qq^2}^nX_l(s)}.
\eeq
By means of the expansion
\beq
X(s,t)=X(s,0)+t\zet{\partial_tX(s,t)}+\frac{t^2}{2}\zet{\partial_t^2X(s,t)}+\cdots,
\eeq
the lowest coefficients in~\eqref{threshold_expansion_Al} are given by
\begin{align}
 X_0^{(0)}&=2\zeq{X(s,0)}, & X_0^{(1)}&=2\zeq{\partial_{\qq^2}X(s,0)}-4\zetq{\partial_tX(s,t)},\notag\\
X_1^{(1)}&=\frac{4}{3}\zetq{\partial_tX(s,t)}, & 
X_1^{(2)}&=\frac{4}{3}\zetq{\partial_{\qq^2}\partial_t X(s,t)}-\frac{8}{3}\zetq{\partial_t^2X(s,t)},
\end{align}
and expanding~\eqref{sprojform} in $\qq^2$ then shows
\begin{align}
\label{theshold_par}
 a_{0+}&=\frac{\mN}{8\pi W_+}\Big(A_0^{(0)}+\mpi B_0^{(0)}\Big),\notag\\
a_{1+}&=\frac{\mN}{8\pi W_+}\Big(A_1^{(1)}+\mpi B_1^{(1)}\Big),\notag\\
a_{1-}&=a_{1+}+\frac{1}{32\pi \mN W_+}\Big(-A_0^{(0)}+B_0^{(0)}(2\mN+\mpi)\Big),\notag\\
b_{0+}&=\frac{1}{32\pi \mN\mpi W_+}\Big(-A_0^{(0)}(2\mN-\mpi)+B_0^{(0)}(2\mN^2+\mpi^2)\Big)+\frac{\mN}{8\pi W_+}\Big(A_0^{(1)}+\mpi B_0^{(1)}\Big).
\end{align}
In combination with the HDRs~\eqref{hdr} as well as the subtracted versions thereof, these equations lead to sum rules for the threshold parameters, explicitly for the $S$- and $P$-wave scattering lengths 
\begin{align}
\label{sum_rule_sl}
a_{0+}^\pm&=\frac{\mN}{4\pi W_+}\Big(\zeq{A^\pm(s,0)}+\mpi \zeq{B^\pm(s,0)}\Big),\notag\\
a_{1+}^\pm&=\frac{\mN}{6\pi W_+}\Big(\zetq{\partial_tA^\pm(s,t)}+\mpi \zetq{\partial_t B^\pm(s,t)}\Big),\notag\\
a_{1-}^\pm&=a_{1+}^\pm+\frac{1}{16\pi \mN W_+}\Big(-\zeq{A^\pm(s,0)}+(2\mN+\mpi) \zeq{B^\pm(s,0)}\Big),
\end{align}
and the $S$-wave effective ranges
\begin{align}
\label{sum_rule_er}
 b_{0+}^\pm&=\frac{1}{16\pi \mN\mpi W_+}\Big(-\zeq{A^\pm(s,0)}(2\mN-\mpi)+ \zeq{B^\pm(s,0)}(2\mN^2+\mpi^2)\Big)\notag\\
&+\frac{\mN}{4\pi W_+}\Big(\zeq{\partial_{\qq^2}A^\pm(s,0)}-2\zetq{\partial_tA^\pm(s,t)}
+\mpi\Big[\zeq{\partial_{\qq^2}B^\pm(s,0)}-2\zetq{\partial_tB^\pm(s,t)}\Big]\Big).
\end{align}
Thus, the equations for the effective ranges involve, in addition, the derivatives with respect to $\qq^2$. However, integration and differentiation may only be exchanged after the threshold singularity has been removed, since otherwise the integral over
\beq
\zeq{\partial_{\qq^2}\frac{\Im f_{0+}^{I_s}(W')}{s'-s}}\sim\frac{W_+}{\sqrt{\mN\mpi}}\frac{\big(a_{0+}^{I_s}\big)^2}{(s'-s_+)^{3/2}}\qquad\text{for}\qquad s'\to s_+
\eeq
would diverge at threshold. This divergence can be removed by adding a suitable term proportional to
\beq
\dashint[0.5pt]\limits_{s_+}^{\infty}\frac{\diff s'}{(s'-s)\sqrt{s'-s_+}}=0 \qquad \text{for} \qquad s>s_+
\eeq
before taking the derivative (the dash indicates the principal value of the integral). 
The sum rules for the covariant amplitudes and their derivatives required for the explicit evaluation of~\eqref{sum_rule_sl} and~\eqref{sum_rule_er} are summarized in~\ref{app:thr_sum_rules}. The error estimate for their evaluation proceeds in analogy to the higher subthreshold parameters calculated from similar sum rules in Sect.~\ref{sec:results}, i.e.\ by propagating the errors from the various uncertainties in the RS solution, and it leads to the results shown in Table~\ref{tab:threshold_parameters}. 
Note that by construction $a_{0+}^{1/2}$ and $a_{0+}^{3/2}$ coincide with~\eqref{scatt_pionic_atoms_final}, since these sum rules were imposed as constraints.
The comparison to KH80 shows that the discrepancy 
in the $S$-wave scattering lengths originates from the isospin-$3/2$ channel alone, while the isospin-$1/2$ analog agrees within uncertainties. In general, apart from some tension in $a_{1-}^{1/2}$, we find reasonable agreement for the remaining threshold parameters.

\begin{table}[t]
\renewcommand{\arraystretch}{1.3}
\centering
\begin{tabular}{crr}\toprule
& RS & KH80\\\midrule
$a_{0+}^{1/2} \ [10^{-3}\mpi^{-1}]$ & $169.8\pm 2.0$ & $173\pm 3$\\
$a_{0+}^{3/2} \ [10^{-3}\mpi^{-1}]$ & $-86.3\pm 1.8$ & $-101\pm 4$\\
$a_{1+}^{1/2} \ [10^{-3}\mpi^{-3}]$ &  $-29.4\pm 1.0$ & $-30\pm 2$\\
$a_{1+}^{3/2} \ [10^{-3}\mpi^{-3}]$ & $211.5\pm 2.8$ & $214\pm 2$\\
$a_{1-}^{1/2} \ [10^{-3}\mpi^{-3}]$ &  $-70.7\pm 4.1$ & $-81\pm 2$\\
$a_{1-}^{3/2} \ [10^{-3}\mpi^{-3}]$ & $-41.0\pm 1.1$ & $-45\pm 2$\\
$b_{0+}^{1/2} \ [10^{-3}\mpi^{-3}]$ &  $-35.2 \pm 2.2$ & $-18\pm 12$\\
$b_{0+}^{3/2} \ [10^{-3}\mpi^{-3}]$ & $-49.8\pm 1.1$ & $-58 \pm 9$\\
\bottomrule
\end{tabular}
\qquad
\begin{tabular}{crr}\toprule
& RS & KH80\\\midrule
$a_{0+}^+ \ [10^{-3}\mpi^{-1}]$ & $-0.9\pm 1.4$ & $-9.7\pm 1.7$\\
$a_{0+}^- \ [10^{-3}\mpi^{-1}]$ & $85.4\pm 0.9$ & $91.3\pm 1.7$\\
$a_{1+}^+ \ [10^{-3}\mpi^{-3}]$ &  $131.2\pm 1.7$ & $132.7\pm 1.3$\\
$a_{1+}^- \ [10^{-3}\mpi^{-3}]$ & $-80.3\pm 1.1$ & $-81.3\pm 1.0$\\
$a_{1-}^+ \ [10^{-3}\mpi^{-3}]$ &  $-50.9\pm 1.9$ & $-56.7\pm 1.3$\\
$a_{1-}^- \ [10^{-3}\mpi^{-3}]$ & $-9.9\pm 1.2$ & $-11.7\pm 1.0$\\
$b_{0+}^+ \ [10^{-3}\mpi^{-3}]$ &  $-45.0 \pm 1.0$ & $-44.3\pm 6.7$\\
$b_{0+}^- \ [10^{-3}\mpi^{-3}]$ & $4.9\pm 0.8$ & $13.3\pm 6.0$\\
\bottomrule
\end{tabular}
\caption{Predictions for the threshold parameters from the RS solution, compared to KH80~\cite{Hoehler:1983}.}
\label{tab:threshold_parameters}
\renewcommand{\arraystretch}{1.0}
\end{table}

\section{Consequences for the $\boldsymbol{\pi N}$ $\boldsymbol{\sigma}$-term}
\label{sec:sigma_term}

The Cheng--Dashen LET~\cite{Cheng:1970mx,Brown:1971pn} relates the Born-term-subtracted isoscalar amplitude  $\bar D^+(\nu,t)$ (see Sect.~\ref{sec:piN_subtractions}) evaluated at the Cheng--Dashen point $(\nu=0,t=2\mpi^2)$ to 
the scalar form factor of the nucleon
\beq
\sigma(t)=\frac{1}{2\mN}\langle N(p')|\hat m(\bar u u+\bar d d)|N(p)\rangle,\qquad \hat m=\frac{\muu+\md}{2},
\eeq
evaluated at momentum transfer $t=(p'-p)^2=2\mpi^2$,
\beq
\label{LET}
\bar D^+(0,2\mpi^2)=\sigma(2\mpi^2)+\Delta_R,
\eeq
where $\Delta_R$ represents higher-order corrections in the chiral expansion. These corrections are expected to be very small: the non-analytic terms agree at full one-loop order~\cite{Bernard:1996nu,Becher:2001hv}, so that, based on the $SU(2)$ expansion parameter, the remaining effect would scale as $(\mpi^2/\mN^2)\sigma_{\pi N}\sim 1\MeV$. In this paper, we will use the estimate~\cite{Bernard:1996nu}
\beq
|\Delta_R|\lesssim 2\MeV,
\eeq
derived from resonance saturation for the $\Order(p^4)$ LECs.

In practice, the relation~\eqref{LET} is often rewritten as
\beq
\sigma_{\pi N}=\sigma(0)=\Sigma_d+\Delta_D-\Delta_\sigma-\Delta_R,
\eeq
with correction terms 
\beq
 \Delta_\sigma=\sigma(2\mpi^2)-\sigma_{\pi N},\qquad \Delta_D=\bar D^+(0,2\mpi^2)-\Sigma_d,\qquad
\Sigma_d=\Fpi^2\big(d_{00}^++2\mpi^2d_{01}^+\big).
\eeq
$\Delta_\sigma$ measures the curvature in the scalar form factor, while $\Delta_D$ parameterizes contributions to the $\pi N$ amplitude beyond the first two terms in the subthreshold expansion. 
As shown in~\cite{Gasser:1990ap}, although these corrections are large individually due to strong rescattering in the isospin-$0$ $\pi\pi$ $S$-wave, they cancel to a large extent in the difference. For the numerical analysis we will use~\cite{Hoferichter:2012tu,Hoferichter:2012wf}
\beq
\label{DeltaDDeltasigma}
\Delta_D-\Delta_\sigma=(-1.8\pm 0.2)\MeV.
\eeq 
The crucial remaining challenge thus consists of determining the subthreshold parameters to sufficient accuracy.
As discussed in Sect.~\ref{sec:cheng_dashen_theorem}, also isospin-breaking corrections to the LET become relevant when it comes to a precision determination of $\sigma_{\pi N}$.

\subsection{Previous extractions}
\label{sec:sigma_term_history}

The determination of the subthreshold parameters $d_{00}^+$ and $d_{01}^+$ from $\pi N$ phenomenology requires an analytic continuation from the physical region, where scattering data and PWAs are available, to the unphysical region of the Mandelstam plane.
The KH80 values translate to $\Sigma_d=(50\pm 7)\MeV$, and other early determinations led to similar results $\Sigma=\Sigma_d+\Delta_D=(64\pm 8)\MeV$~\cite{Koch:1982pu}.
The relation between the input from the physical region and the analytic continuation to the Cheng--Dashen point was made more transparent by the formalism developed in~\cite{Gasser:1988jt,Gasser:1990ce}, where the subthreshold parameters are eliminated in favor of the threshold parameters $a_{0+}^+$ and $a_{1+}^+$, with corrections expressed in terms of dispersive integrals. Using input for the threshold parameters from~\cite{Koch:1980ay}, this analysis led to $\Sigma_d=(49\pm 8)\MeV$~\cite{Gasser:1990ce}, in good agreement with the original KH80 extraction. The corresponding $\sigma$-term $\sigma_{\pi N}\sim 45\MeV$ has often been referred to as its ``canonical value.''

This result was contested in~\cite{Pavan:2001wz}, where a new PWA indicated a much larger $\Sigma_d=(67\pm 6)\MeV$, although based on the same formalism~\cite{Gasser:1988jt,Gasser:1990ce} for the extraction from the physical region. In more detail, the discrepancy could be attributed to about equal parts to the $\pi N$ coupling constant, $a_{0+}^+$, and the dispersive integrals.\footnote{The shift related to the coupling constant reflects the fact that the old standard value $g^2/(4\pi)=14.3\pm 0.2$~\cite{Bugg:1973rv} is inconsistent with several recent, independent extractions, such as from $NN$ scattering, $g^2/(4\pi)=13.54\pm 0.05$~\cite{deSwart:1997ep}, a different PWA, $g^2/(4\pi)=13.75\pm 0.10$~\cite{Arndt:2003if}, and the GMO sum rule, $g^2/(4\pi)=13.69\pm 0.19$~\cite{Baru:2011bw}.
The range $g^2/(4\pi)=13.7\pm 0.2$ adopted in the present paper safely covers all these determinations. Note that to get significantly below this uncertainty the analysis of radiative corrections becomes critical, see~\cite{Baru:2011bw}.}
A similar picture emerged from the subthreshold parameters extracted with dispersive techniques from PWAs in~\cite{Stahov:1999,Stahov:2002,Hite:2005tg,Kaufmann:1999dd,Oades:1999ap,Martin:2002mz}, where the input from the GWU group~\cite{Arndt:1995bj,Arndt:1998zd,Pavan:1999cr,Arndt:2003if} consistently produced a larger $\sigma$-term than the PWAs from~\cite{Koch:1980ay,Hoehler:1983,Koch:1985bp} (see also~\cite{Olsson:1999jt,Stahov:2012ca,Matsinos:2013jda,Matsinos:2013era,Matsinos:2015wxa} in this context).
Similarly, extractions of the $\sigma$-term from ChPT~\cite{Fettes:2000xg,Alarcon:2011zs} tend to reproduce the value corresponding to the PWA that is used to determine the LECs in the physical region.

In recent years, the $\sigma$-term has also become the subject of many lattice-QCD calculations~\cite{Fukugita:1994ba,Dong:1995ec,Gusken:1998wy,Leinweber:2000sa,Leinweber:2003dg,Procura:2003ig,Procura:2006bj,Alexandrou:2008tn,WalkerLoud:2008bp,Ohki:2008ff,Young:2009zb,Ishikawa:2009vc,Alexandrou:2009qu,MartinCamalich:2010fp,Durr:2011mp,Horsley:2011wr,Bali:2011ks,Dinter:2012tt,Semke:2012gs,Shanahan:2012wh,Bali:2012qs,Ren:2012aj,Young:2013nn,Alvarez-Ruso:2013fza,Lutz:2014oxa,Ren:2014vea,Alexandrou:2014sha,Torrero:2014pxa}. However, while the strangeness coupling has been determined rather accurately on the lattice~\cite{Junnarkar:2013ac}, the coupling to the light quarks remains challenging, in particular due to the presence of disconnected diagrams. The direct calculation of the three-point function becomes even more demanding, so that most lattice results rely on the Feynman--Hellmann theorem~\cite{Hellmann:1937,Feynman:1939zza,Gasser:1979hf} to extract the $\sigma$-term from the quark-mass derivative of the nucleon mass.
At this point in time there is no global average of lattice results for $\sigma_{\pi N}$ available, similar to the FLAG reviews~\cite{Colangelo:2010et,Aoki:2013ldr} in meson physics or~\cite{Junnarkar:2013ac} for the strangeness coupling. According to the criteria suggested in~\cite{Junnarkar:2013ac} (in analogy to~\cite{Colangelo:2010et}), the only calculation that would have passed all requirements would have been~\cite{Durr:2011mp}, which quotes $\sigma_{\pi N}=(39\pm 4^{+18}_{-7})\MeV$. The error is dominated by the chiral extrapolation, e.g.\ it can make a huge difference whether the $\Delta$ is included in the extrapolation formula or not. For this reason, the most recent published calculation~\cite{Alexandrou:2014sha} quotes $\sigma_{\pi N}=(64.9\pm 1.5\pm 19.6)\MeV$, where the second uncertainty refers to the difference between $\Delta$-full and $\Delta$-less extrapolations. Computations at the physical point will remove this particular systematic effect~\cite{Torrero:2014pxa}.

\subsection{Cheng--Dashen theorem}
\label{sec:cheng_dashen_theorem}

The original evaluation of the corrections $\Delta_D-\Delta_\sigma=(-3.3\pm 0.2)\MeV$~\cite{Gasser:1990ap} to the Cheng--Dashen theorem demonstrated that the net effect of deviations from $\Sigma_d$ is quite moderate. However, the quoted uncertainty only refers to the $\pi\pi$ phase shifts, effects from $\bar K K$ intermediate states and uncertainties from $\pi N$ input are not included in the error estimate. Those effects can be captured in a first application of the RS solution for the $t$-channel partial waves~\cite{Hoferichter:2012wf,Hoferichter:2012tu}.
The scalar form factor fulfills a once-subtracted dispersion relation
\beq
\sigma(t)=\sigma_{\pi N}+\frac{t}{\pi}\int\limits_{\tpi}^\infty\diff t'\frac{\Im\sigma(t')}{t'(t'-t)},
\eeq
which, evaluated at $t=2\mpi^2$, determines $\Delta_\sigma$. Similarly, if one assumed even an unsubtracted dispersion relation to converge, one would obtain an estimate for the $\sigma$-term itself. 
The imaginary part generated by $\pi\pi$ and $\bar K K$ intermediate states reads
\beq
\label{unitarity_sigma_term}
\Im\sigma(t)=-\frac{1}{p_t^2\sqrt{t}}\bigg\{\frac{3}{4}q_t\big(F^S_\pi(t)\big)^*f^0_+(t)\,\theta\big(t-\tpi\big)+k_t\big(F^S_K(t)\big)^*h^0_+(t)\,\theta\big(t-\tK\big)\bigg\},
\eeq
where $f^0_+(t)$ and $h^0_+(t)$ are the $S$-waves for $\pi\pi\to\bar N N$ and $\bar K K\to\bar N N$, respectively, see~\ref{app:tchannel_unitarity}, and the meson scalar form factors $F^S_{\pi,K}(t)$ are defined by
\beq
F^S_\pi(t)=\langle \pi(p')|\hat m(\bar u u+\bar d d)|\pi(p)\rangle,\qquad F^S_K(t)=\langle K(p')|\hat m(\bar u u+\bar d d)|K(p)\rangle.
\eeq
Once the input for the $T$-matrix is specified, they can be calculated with MO techniques as well.  

\begin{figure}[t]
\centering
\includegraphics[width=0.49\linewidth,clip]{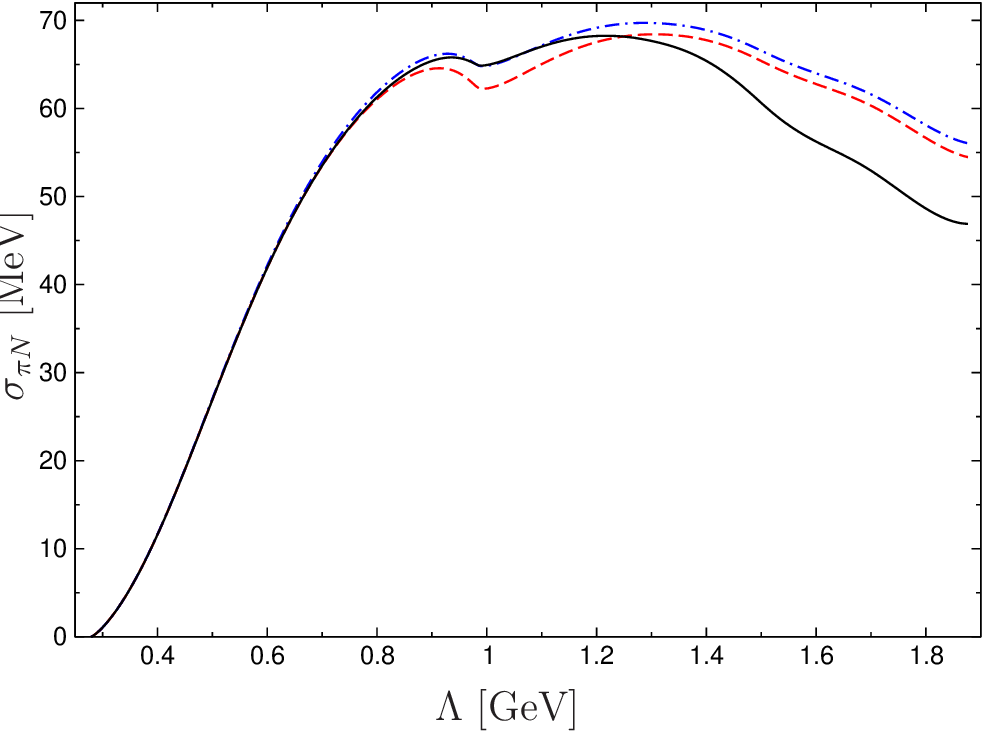}
\includegraphics[width=0.49\linewidth,clip]{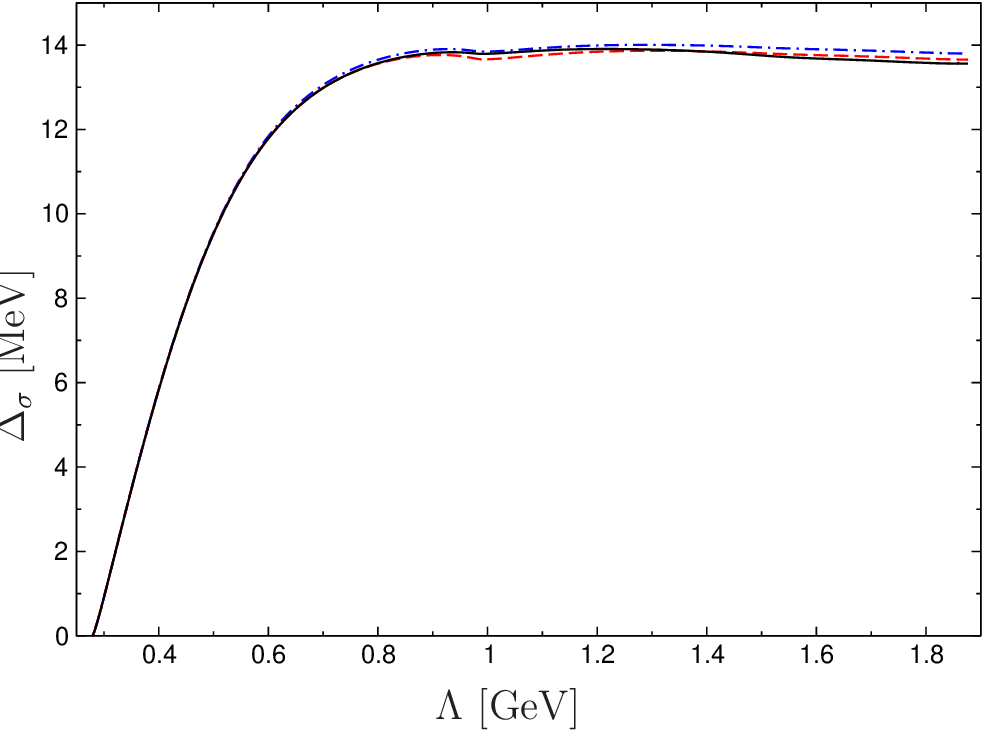}
\caption{$\sigma_{\pi N}$ and $\Delta_\sigma$ as a function of the integral cutoff $\Lambda$ for different input in the RS solution. The black solid and red dashed lines refer to the respective curves in Fig.~\ref{fig:MO_KH80}, while the blue dot-dashed curve indicates a variant of the red dashed one where the kaon input in the MO inhomogeneity is switched off.}
\label{fig:Delta_sigma}
\end{figure}

The results for $\sigma_{\pi N}$ and $\Delta_\sigma$ as a function of the cutoff $\Lambda$ of the dispersive integral are shown in Fig.~\ref{fig:Delta_sigma}. Although the unsubtracted integral for $\sigma_{\pi N}$ does not converge sufficiently fast to be taken seriously as another independent constraint, the once-subtracted result for $\Delta_\sigma$ proves already very stable both regarding the extrapolation $\Lambda\to\infty$ and variation of the input.
Since the RS solution for the $t$-channel partial waves is linear in the subtraction constants, the result can be represented in the form
\begin{align}
\label{DelSigRes}
 \Delta_\sigma&=(13.9\pm 0.3)\MeV
 + Z_1 \bigg(\frac{g^2}{4\pi}-13.7\bigg)+ Z_2\Big(d_{00}^+\,\mpi+1.36\Big)+
Z_3\Big(d_{01}^+\,\mpi^{3}-1.16\Big)+Z_4\Big(b_{00}^+\,\mpi^{3}+3.46\Big),\notag\\
Z_1&=0.36\MeV,\qquad Z_2= 0.57\MeV,\qquad Z_3= 12.0\MeV,\qquad Z_4=-0.81\MeV,
\end{align}
where the $Z_i$ parameterize the sensitivity to the $\pi N$ input. 

In a similar way, the solution of the $t$-channel equations determines $\Delta_D$.
Starting from the expansion
\beq
\bar D^+(\nu=0,t)=4\pi\bigg\{-\frac{1}{p_t^2}\bar f_+^0(t)+\frac{5}{2}q_t^2\bar f_+^2(t)-\frac{27}{8}p_t^2q_t^4\bar f_+^4(t)+\frac{65}{16}p_t^4q_t^6\bar f_+^6(t)+\cdots\bigg\},
\eeq
where $\bar f^J_+(t)$ denotes the Born-term-subtracted part of $f^J_+(t)$, and neglecting the imaginary parts of $f^J_\pm(t)$ for $J\geq 4$, the RS equations for $f^J_+(t)$ yield
\begin{align}
\label{barD}
\bar D^+(\nu=0,t)&=d_{00}^++d_{01}^+ t-16 t^2\int\limits_{\tpi}^\infty\diff t'\frac{\Im f_+^0(t')}{t'^2(t'-4m^2)(t'-t)}\notag\\
&-\frac{4}{p_t^2}\big(I_0^t(t)+I_0^s(t)\big)+10q_t^2\big(I_2^t(t)+I_2^s(t)\big)-\frac{27}{2}p_t^2q_t^4I_4^s(t)+\frac{65}{4}p_t^4q_t^6I_6^s(t)+\cdots,\notag\\
3I_J^s(t)&=\int\limits_{W_+}^\infty\diff W'\sum\limits_{l=0}^\infty\bigg\{\tilde G_{Jl}(t,W')\Big(\Im f_{l+}^{1/2}(W')+2\Im f_{l+}^{3/2}(W')\Big)
+\tilde G_{Jl}(t,-W')\Big(\Im f_{(l+1)-}^{1/2}(W')+2\Im f_{(l+1)-}^{3/2}(W')\Big)\bigg\},\notag\\
I_J^t(t)&=\int\limits_{\tpi}^\infty\diff t'\bigg\{\tilde K^1_{J2}(t,t')\Im f^2_+(t')+\tilde K^2_{J2}(t,t')\Im f^2_-(t')\bigg\}.
\end{align}
Evaluation of this formula at $t=2\mpi^2$ provides the desired expression for $\Delta_D$, whose
numerical analysis in close analogy to $\Delta_\sigma$ leads to
\begin{align}
\label{DelD}
 \Delta_D&=(12.1\pm 0.3)\MeV
 + \tilde Z_1 \bigg(\frac{g^2}{4\pi}-13.7\bigg)+ \tilde Z_2\Big(d_{00}^+\,\mpi+1.36\Big)+
\tilde Z_3\Big(d_{01}^+\,\mpi^{3}-1.16\Big)+\tilde Z_4\Big(b_{00}^+\,\mpi^{3}+3.46\Big),\notag\\
\tilde Z_1&=0.42\MeV,\qquad \tilde Z_2= 0.67\MeV,\qquad \tilde Z_3= 12.0\MeV,\qquad \tilde Z_4=-0.77\MeV.
\end{align}
The comparison to~\eqref{DelSigRes} shows that, since $Z_i\sim\tilde Z_i$, most of the dependence on the $\pi N$ parameters cancels in the difference. Similarly, part of the uncertainty from the $\Lambda\to\infty$ limit and the $\pi\pi/\bar K K$ $T$-matrix cancels as well, which in the end produces the small uncertainty quoted in~\eqref{DeltaDDeltasigma}.

Finally, the LET itself as well as $\Delta_D-\Delta_\sigma$ are defined in the isospin limit, where all masses are identified with the charged-particle ones. For the isospin conventions specified in Sect.~\ref{sec:schannel_sol}, we therefore need a version of the LET that accounts for possible effects from mass differences, in particular as similar corrections are known to be important in the case of the scattering lengths. First of all, we define the $\sigma$-term as the average value of proton and neutron scalar-current matrix elements ($N\in\{p,n\}$)
\beq
\label{sigma_term_def}
\sigma_{\pi N}=\frac{1}{2}\big(\sigma_p+\sigma_n\big),\qquad \sigma_N=\frac{1}{2\mN}\langle N|\hat m(\bar u u+\bar d d)|N\rangle,
\eeq
although $\sigma_p$ and $\sigma_n$ agree up to third order in the chiral expansion~\cite{Meissner:1997ii}. Next, the curvature in the scalar form factor indeed receives a contribution from neutral-pion loops as well, so that~\cite{Meissner:1997ii}
\beq
\Delta_{\sigma}^p\equiv\sigma_p\big(2\mpi^2\big)-\sigma_p
=\frac{3\ga^2\mpi^3}{64\pi\Fpi^2}+\frac{\ga^2\mpi\Delta_\pi}{128\pi\Fpi^2}\Big(-7+2\sqrt{2}\log\big(1+\sqrt{2}\big)\Big)
\eeq
is shifted accordingly from its isospin-limit value.
Similarly, mass effects in $\Delta_D$ can be extracted from~\cite{Hoferichter:2009gn}
\beq
 \Delta_D^p\equiv\Fpi^2\Big\{D_p\big(0,2\mpi^2\big)-d_{00}^p-2\mpi^2d_{01}^p\Big\}=\frac{23\ga^2\mpi^3}{384\pi\Fpi^2}+\frac{\ga^2\mpi\Delta_\pi}{256\pi\Fpi^2}\Big(3+4\sqrt{2}\log\big(1+\sqrt{2}\big)\Big),
\eeq
where we have used the notation 
\beq
X^+\to X^p=\frac{1}{2}\big(X_{\pi^+p\to\pi^+ p}+X_{\pi^-p\to\pi^- p}\big),\qquad X\in\{D,d_{00},d_{01}\},
\eeq
to differentiate between isospin limit and the amplitudes used in the RS solution.
In principle, there are also isospin-breaking corrections from virtual photons, which, at LO, are encoded in the LECs $f_1$ and $f_2$, see Sect.~\ref{sec:isospin_breaking}. At loop level, such corrections are more difficult to account for conceptually, since, unlike in the case of the scattering lengths, where threshold kinematics forbids the emission of bremsstrahlung, the amplitudes would display infrared singularities, whose removal would require a full treatment of radiative corrections including real bremsstrahlung. However, given that virtual-photon effects were found to be much smaller than the contribution from the pion-mass difference for the scattering lengths, and in view of the substantial uncertainty in $f_1$, such effects can be safely neglected here.
Taking everything together, our version of the LET that includes the leading isospin-breaking effects becomes
\begin{align}
\label{LET_IV_corr}
\sigma_{\pi N}&=\Fpi^2\big(d_{00}^p+2\mpi^2d_{01}^p\big)+\Delta_D-\Delta_\sigma-\Delta_R
+\frac{81\ga^2\mpi\Delta_\pi}{256\pi\Fpi^2}+\frac{e^2}{2}\Fpi^2\big(4f_1+f_2\big)\notag\\
 &=\Fpi^2\big(d_{00}^p+2\mpi^2d_{01}^p\big)+(1.2\pm 3.0)\MeV.
\end{align}
As expected, the single largest correction is generated by $\Delta_\pi$, an upward shift of $3.4\MeV$.

\subsection{Determination from the Roy--Steiner solution}
\label{sec:sigma_term_RS}

Based on~\eqref{LET_IV_corr} the RS results for the subthreshold parameters translate immediately to a corresponding value of $\sigma_{\pi N}$. To illustrate the dependence of the $\sigma$-term on the scattering lengths used as input to the solution, we expand $\Sigma_d$ linearly around the central values~\eqref{scatt_pionic_atoms_final} and find
\beq
\label{Sigma_d_lin}
\Sigma_d=(57.9\pm 0.9)\MeV + \sum_{I_s}c_{I_s}\Delta a^{I_s}_{0+},\qquad
c_{1/2}=0.24\MeV,\qquad c_{3/2}=0.89\MeV,
\eeq
where $\Delta a^{I_s}_{0+}$ gives the deviation from~\eqref{scatt_pionic_atoms_final} in units of $10^{-3}\mpi^{-1}$.
Already this linearized version produces $\Sigma_d=(46\pm 4)\MeV$ if the KH80 scattering lengths given in Table~\ref{tab:threshold_parameters} are used, and the agreement with the original KH80 value $\Sigma_d=(50\pm 7)\MeV$ improves further in a full solution.
In contrast, our central solution~\eqref{scatt_pionic_atoms_final} corresponds to 
\beq
\label{Sigma_d}
\Sigma_d=(57.9\pm 1.9)\MeV, 
\eeq
and thus to a significant increase compared to the early estimates. 

However, we do not find as large a value as $\Sigma_d=(67\pm 6)\MeV$~\cite{Pavan:2001wz}. To isolate the origin of this discrepancy, let us analyze $\Sigma_d$ in the formalism from~\cite{Gasser:1988jt,Gasser:1990ce}
\begin{align}
\label{sum_rule_Sigma_d}
 \Sigma_d&=\Fpi^2\Bigg\{4\pi\bigg(1+\frac{\mpi}{\mN}\bigg)a_{0+}^++\frac{g^2\mpi^2}{\mN\big(4\mN^2-\mpi^2\big)}-J^+
 +2\mpi^2\bigg[6\pi\bigg(1+\frac{\mpi}{\mN}\bigg) a_{1+}^+-\frac{g^2}{\mpi(2\mN-\mpi)^2}-\tilde J^+\bigg]\Bigg\},\notag\\ 
 J^+&=\frac{\mpi^2}{\pi}\int\limits_0^\infty\diff |\qq'|\frac{\sigma_{\pi^-p}^\text{tot}(|\qq'|)+\sigma_{\pi^+p}^\text{tot}(|\qq'|)}{\mpi^2+\qq'^2},\notag\\
\tilde J^+&=\frac{2\mpi^2}{\pi}\int\limits_{\mpi}^\infty\diff\omega'\frac{\Im E^+(\omega')}{\omega'(\omega'^2-\mpi^2)}
+\frac{1}{2\pi \mN}\int\limits_{\mpi}^\infty\diff\omega'\Im D^+(\omega')\bigg(\frac{1}{\omega'^2}-\frac{1}{(\omega'+\mpi)^2}\bigg)\notag\\
&-\frac{1}{2\pi \mN}\int\limits_{\mpi}^\infty\diff\omega'\Im B^+(\omega')\bigg(\frac{1}{\omega'}-\frac{1}{\omega'+\mpi}\bigg),
\end{align}
where $\omega=\sqrt{\qq^2+\mpi^2}$, $s=\mN^2+\mpi^2+2\mN\omega$, and
\begin{align}
D^+(\omega)&=\swt{A^++\omega B^+},\qquad
 E^+(\omega)=\swt{\frac{\partial}{\partial t}\big(A^+(s,t)+\omega B^+(s,t)\big)}.
\end{align}
Accordingly, there is a direct connection between $J^+$ and cross-section data,\footnote{The analogous integral $J^-$ of the cross-section difference becomes relevant for the evaluation of the GMO sum rule~\cite{Ericson:2000md,Abaev:2007nq}.} whereas $\tilde J^+$ requires a PWA. 
Using $J^+=(1.459\pm 0.005)\mpi^{-1}$ from~\cite{Metsa:2007gn}, $a_{0+}^+=(-0.9\pm 1.4)\times 10^{-3}\mpi^{-1}$ from~\eqref{scatt_pionic_atoms_final}, $a_{1+}^+=(131.2\pm 1.7)\times 10^{-3}\mpi^{-3}$ from Table~\ref{tab:threshold_parameters}, and $g^2/(4\pi)=13.7\pm 0.2$ as before, the only missing ingredient is $\tilde J^+$. Adopting a value of $(-70.5\pm 1.5)\MeV$ for its contribution to $\Sigma_d$ (to cover the two evaluations given in~\cite{Pavan:2001wz}), we find $\Sigma_d=(59.2\pm 5.2)\MeV$, in excellent agreement with the RS result, but considerably less precise. 
In the end, the difference originates almost exclusively from $a_{1+}^+=133\times 10^{-3}\mpi^{-3}$ as used in~\cite{Pavan:2001wz}. Keeping the rest of the input fixed but increasing $a_{1+}^+$ accordingly, the central value indeed increases to $\Sigma_d=64\MeV$. 
This comparison shows that the decomposition~\eqref{sum_rule_Sigma_d} is much more sensitive to $a_{1+}^+$ than previously appreciated, already the rather precise prediction from the RS solution amounts to an uncertainty of $5\MeV$ in the $\sigma$-term. To obtain a result comparable to~\eqref{Sigma_d}, $a_{1+}^+$ would need to be known at sub-percent accuracy. The elimination of the need for independent input for $a_{1+}^+$ thus constitutes the main advantage of the RS approach. 

In conclusion, the final result~\cite{Hoferichter:2015dsa}
\beq
\label{sigma_term_result}
\sigma_{\pi N}=(59.1\pm 3.5)\MeV
\eeq
does amount to a significant increase compared to the ``canonical value'' of $\sigma_{\pi N}\sim 45\MeV$, although 
already $4.2\MeV$ are due to new corrections to the LET (and thereof $3.0\MeV$ from isospin breaking).
The remaining increase of nearly $10\MeV$ is dictated by experiment: the new scattering lengths from pionic atoms 
determine the position of the $\sigma$-term on the curve approximately described by~\eqref{Sigma_d_lin}.

The $\sigma$-term has also been extracted from $\pi N$ scattering phase shifts using ChPT at one loop~\cite{Fettes:2000xg,Alarcon:2011zs,Chen:2012nx}, partly finding central values that are well compatible with~\eqref{sigma_term_result}.  In such analyses, the LECs of the chiral representation are fixed from fits to various PWAs, and chiral LETs used subsequently to determine the $\sigma$-term.  All the (dispersive) relations that constitute the Cheng--Dashen LET used in the extraction from the RS solution are fulfilled by the chiral representation, too, albeit only in a perturbative way.  In particular, one implicitly needs to extrapolate from the physical $s$-channel to the subthreshold region; we will comment on this relation in Sect.~\ref{sec:threshold_ChPT}.  Based on the analysis performed up to here, we point out that the chiral one-loop representation is likely problematic for a precision determination of the $\sigma$-term.  
It is well-known that it does not provide sufficient curvature to the scalar form factor of the nucleon~\cite{Gasser:1990ap}; similarly, the quantity $\Delta_D$ is severely underestimated~\cite{Alarcon:2012kn}. Therefore, the one-loop representation of the $\pi N$ scattering amplitude does not describe the subthreshold region very accurately: the extraction of the $\sigma$-term is enabled only by the large cancellation in $\Delta_D-\Delta_\sigma$ as described above.
Furthermore, we have explained in Sect.~\ref{sec:error_analysis} how $t$-channel $D$-waves including the $f_2(1270)$ resonance are an essential ingredient to a consistent solution of the RS system---omitting its contribution leads to a significantly larger $\sigma$-term.  The $t$-channel $D$-waves of the chiral one-loop representation, however, are real: imaginary parts will only begin to contribute at two-loop order.  Hence, the large modifications induced by the $f_2(1270)$ are part of the uncertainties ignored at one loop.  The solution to this problem lies in the use of the RS equations for the momentum dependence of the $\pi N$ amplitude.  The convergence of the chiral series as an expansion in powers of the light quark masses can be studied subsequently, as we will see in Sect.~\ref{sec:ChPT}.

\subsection{On the strangeness content of the nucleon}
\label{sec:y}

The $\pi N$ $\sigma$-term can also be related to the mass shift in the nucleon due to strange quarks,  $m_s \langle N| \bar s s|N\rangle$. 
For that, one usually considers the so-called strangeness fraction $y$,  given by 
\beq
\label{eq:strangeness}
\sigma_{\pi N} = \frac{\hat m}{2\mN}  \frac{\langle N|\bar u u + \bar dd -2 \bar ss |N\rangle }{1-y}
=  \frac{\sigma_0}{1-y}, \qquad y \equiv\frac{2\langle N | \bar ss |N\rangle}{\langle N | \bar uu + \bar dd |N\rangle}.
\eeq
The leading $SU(3)$ breaking is generated by the operator $(m_s- \hat m)(\bar uu + \bar dd - 2\bar ss)$ so that $\sigma_0$ can 
be expressed through baryon mass splittings
\beq
\sigma_0 = \frac{\hat m}{m_s - \hat m}\left( m_\Xi + m_\Sigma - 2\mN  \right) \sim 26\MeV.
\eeq
The first calculation of the higher-order corrections to this relation were performed in the pioneering work by Gasser,
leading to $\sigma_0 = (35\pm 5)\MeV$~\cite{Gasser:1980sb}. This was updated in a modern version of three-flavor baryon ChPT in~\cite{Borasoy:1996bx},
giving a similar value,  $\sigma_0 = (36\pm 7)\MeV$. Combining this with our value for $\sigma_{\pi N}$~\eqref{sigma_term_result}  would
lead to unrealistically large values of the strangeness fraction, $y = 0.4\pm 0.1$. We note that by now it is established that the expectation
values of other strange operators such as the vector current $\bar  s\gamma_\mu s$ are small in the nucleon~\cite{Armstrong:2012bi}.
However, more recent calculations using covariant baryon ChPT and/or including the effects from the baryon decuplet~\cite{Jenkins:1991bs} give sizably 
larger values of $\sigma_0$, for example the covariant calculation of~\cite{Alarcon:2012nr} results in $\sigma_0 = (58\pm 8)\MeV$.
Such values for $\sigma_0$  lead to very small or even vanishing strangeness fractions. 
Clearly, in such a scenario our value for $\sigma_{\pi N}$ is not incompatible with a small strangeness fraction, but one also has to realize that
the chiral convergence of  $\sigma_0$ and thus of  $m_s \langle N| \bar s s|N\rangle$ is very doubtful. Therefore, at present one cannot draw a firm conclusion on the size of $y$ based on~\eqref{eq:strangeness}. 
Calculations that combine covariant baryon ChPT with recent lattice data also do not lead to a clear picture: 
whereas~\cite{Lutz:2014oxa} suggests a sizable mass shift
from strange quarks, in~\cite{Ren:2014vea} a small shift is found.
Most recent lattice calculations based on the application
of the Feynman--Hellmann theorem or performing direct calculations of the matrix element $ \langle N| \bar s s|N\rangle$ (including also the
disconnected contributions) give small
values of $y$, see~\cite{Young:2009zb,Toussaint:2009pz,Takeda:2010cw,Durr:2011mp,Bali:2011ks,Horsley:2011wr,Dinter:2012tt,Semke:2012gs,Freeman:2012ry,Oksuzian:2012rzb,Engelhardt:2012gd,Junnarkar:2013ac,Gong:2013vja,Alexandrou:2013nda}.

\section{Matching to chiral perturbation theory}
\label{sec:ChPT}

The matching to ChPT is one of the most fundamental applications of the RS solution, since it offers a unique opportunity for a systematic determination of $\pi N$ LECs.
However, the detailed procedure of the matching plays an important role when it comes to minimizing higher-order effects in the chiral expansion. Even in the $\pi\pi$ case, where 
in general the chiral series converges rapidly, the uncertainties can be reduced substantially with a prudent choice of the matching procedure. 
For instance, to determine the $\pi\pi$ scattering lengths in~\cite{Colangelo:2000jc}, the chiral and the Roy-equation representation were first brought into such a form that the singularities in the physical region can be identified up to higher orders, so that the remaining matching condition amounts to equating a polynomial in the Mandelstam variables. Indeed, this procedure improves convergence significantly compared to naive matching at threshold, where the branch cut required by unitarity sets in. The onset of the square-root singularity, which the chiral series only generates perturbatively, is precisely the reason why truncations of the chiral expansion become more severe near threshold. 
For this reason, one would expect the chiral expansion to work best in a kinematic region where no singularities occur, i.e.\ where the amplitude can be described solely by a polynomial in the Mandelstam variables. This is precisely the situation encountered in the subthreshold region: the amplitude is purely real, and characterized by its expansion coefficients around $(\nu=0,t=0)$.
The matching is thus most conveniently performed by equating the chiral expansion for the subthreshold parameters, see~\ref{app:ChPT_subthreshold}, to the RS results given in Table~\ref{tab:RS_subthr}. The error propagation will be based on the correlation coefficients listed in Table~\ref{tab:corr_RS_subthr}. We stress that the errors derived in this way pertain to a given chiral order, we do not attempt to attach an additional uncertainty that reflects the potential impact of higher-order terms in the chiral expansion. In practice, this implies that those uncertainties need to be studied for each application separately based on the respective convergence pattern, a strategy that is indeed becoming more and more common in ChEFT calculations~\cite{Epelbaum:2014efa,Furnstahl:2014xsa,Furnstahl:2015rha}. 

Another motivation for performing the matching at the subthreshold point is provided by the application to $NN$ scattering. The $\pi N$ LECs enter the $NN$ potential in $2\pi$-exchange diagrams, so that a one-loop $\pi N$ amplitude corresponds to a two-loop contribution to the chiral potential. Such diagrams can be elegantly represented by means of Cutkosky rules~\cite{Kaiser:2001pc}, based on methods developed earlier for the calculation of the $3\pi$ contribution to the spectral function of nuclear form factors~\cite{Bernard:1996cc,Kaiser:2003dr} as well as $3\pi$-exchange potentials~\cite{Kaiser:1999ff,Kaiser:1999jg,Kaiser:2001dm}, and recently applied to even higher chiral orders~\cite{Entem:2014msa,Epelbaum:2014sza,Entem:2015xwa}.
The explicit expressions from~\cite{Kaiser:2001pc} show that the $\pi N$ amplitude that enters the dispersive integrals is weighted towards (or even evaluated at) zero pion CMS energy, i.e.\
 $s=\mN^2-\mpi^2$. Similarly, the Cauchy integrals are most sensitive to physical momentum transfer $t=0$. The corresponding combination $(\nu,t)=(-\mpi^2/\mN,0)$ is much closer to subthreshold kinematics $(0,0)$ than for instance to the threshold point $(\mpi,0)$.

\subsection{Low-energy constants}
\label{sec:LECs}

\begin{table}[t]
\renewcommand{\arraystretch}{1.3}
\centering
\begin{tabular}{crrrrrr}
\toprule
& NLO & N$^2$LO & N$^3$LO & N$^3$LO$^{NN}$\\
\midrule
$c_1\, [\GeV^{-1}]$ & $-0.74\pm 0.02$ & $-1.07\pm 0.02$ & $-1.11\pm 0.03$ & $-1.10\pm 0.03$ \\
$c_2\, [\GeV^{-1}]$ & $1.81\pm 0.03$ & $3.20\pm 0.03$ & $3.13\pm 0.03$ & $3.57\pm 0.04$ \\
$c_3\, [\GeV^{-1}]$ & $-3.61\pm 0.05$ & $-5.32\pm 0.05$ & $-5.61\pm 0.06$ & $-5.54\pm 0.06$ \\
$c_4\, [\GeV^{-1}]$ & $2.17\pm 0.03$ & $3.56\pm 0.03$ & $4.26\pm 0.04$ & $4.17\pm 0.04$ \\
$\bar d_1+\bar d_2\, [\GeV^{-2}]$ & --- & $1.04\pm 0.06$ & $7.42\pm 0.08$ & $6.18\pm 0.08$ \\
$\bar d_3\, [\GeV^{-2}]$ & --- & $-0.48\pm 0.02$ & $-10.46\pm 0.10$ & $-8.91\pm 0.09$ \\
$\bar d_5\, [\GeV^{-2}]$ & --- & $0.14\pm 0.05$ & $0.59\pm 0.05$ & $0.86\pm 0.05$ \\
$\bar d_{14}-\bar d_{15}\, [\GeV^{-2}]$ & --- & $-1.90\pm 0.06$ & $-13.02\pm 0.12$ & $-12.18\pm 0.12$ \\
$\bar e_{14}\, [\GeV^{-3}]$ & --- & --- & $0.89\pm 0.04$ & $1.18\pm 0.04$\\
$\bar e_{15}\, [\GeV^{-3}]$ & --- & --- & $-0.97\pm 0.06$ & $-2.33\pm 0.06$\\
$\bar e_{16}\, [\GeV^{-3}]$ & --- & --- & $-2.61\pm 0.03$ & $-0.23\pm 0.03$ \\
$\bar e_{17}\, [\GeV^{-3}]$ & --- & --- & $0.01\pm 0.06$ & $-0.18\pm 0.06$ \\
$\bar e_{18}\, [\GeV^{-3}]$ & --- & --- & $-4.20\pm 0.05$ & $-3.24\pm 0.05$\\
\bottomrule
\end{tabular}
\caption{Results for the $\pi N$ LECs at different orders in the chiral expansion. In most cases, standard and $NN$ counting coincide up to N$^2$LO, except for NLO in $c_4$, which in the $NN$ scheme becomes $(2.44\pm 0.03)\GeV^{-1}$.}
\label{tab:LECs}
\renewcommand{\arraystretch}{1.0}
\end{table}

The $\pi N$ amplitude at N$^3$LO, $\Order(p^4)$, involves four NLO LECs, $c_i$, four (combinations of) N$^2$LO LECs, $\bar d_i$, and five N$^3$LO LECs, $\bar e_i$, see~\cite{Fettes:2000gb} and~\ref{app:ChPT}.
These $13$ LECs correspond to the $13$ subthreshold parameters that receive contributions from LECs in a fourth-order calculation (all higher parameters are given by LETs at this order). In particular, a quark-mass renormalization from further $\bar e_i$ as well as the mesonic LEC $\bar l_3$ have been absorbed into a redefinition of the $c_i$.
In the standard counting of ChPT, momenta and quark masses are counted in a common scheme 
$\Order(p)=\{p,\mpi\}/\Lambda_\text{b}$, motivated by the Gell-Mann--Oakes--Renner relation $\mpi^2=B(\muu+\md)+\Order(m_q^2)$ that relates quark and meson masses via the condensate $B=-\langle \bar q q\rangle/\Fpi^2$. The expansion in momenta $p$ proceeds relative to a breakdown scale $\Lambda_\text{b}$, estimates for which include
\beq
\Lambda_\text{b}\sim \Lambda_\chi\sim4\pi\Fpi\sim\mN\sim M_\rho\sim 1\GeV,
\eeq
where $\Lambda_\chi$ is the scale of chiral symmetry breaking and $M_\rho$ the mass of the $\rho$-meson. In ChEFT applications in multi-nucleon systems the breakdown scale tends to be 
lower, $\Lambda_\text{b}\sim 0.6\GeV$, so that 
relativistic corrections are often counted as 
$\{p,\mpi\}/\mN=\Order(p^2)$~\cite{Weinberg:1991um}. In the following, we will present results in both counting schemes, referring to the latter as $NN$ counting.

\begin{table}[t]
\footnotesize
\renewcommand{\arraystretch}{1.3}
\centering
\begin{tabular}{crrrrrrrr}
\toprule
& $c_1$ & $c_2$ & $c_3$ & $c_4$ & $\bar d_1+\bar d_2$ & $\bar d_3$ & $\bar d_5$ & $\bar d_{14}-\bar d_{15}$ \\
$c_1$ & $1$ & $-0.09$ & $0.50$ & $0.04$ & $-0.27$ & $0.13$ & $0.37$ & $0.31$\\
$c_2$ & & $1$ & $-0.85$ & $0.06$ & $0.64$ & $-0.40$ & $-0.24$ & $-0.48$\\
$c_3$ &&& $1$ & $-0.04$ & $-0.71$ & $0.55$ & $0.35$ & $0.67$\\
$c_4$ &&&& $1$ & $0.09$ & $-0.15$ & $-0.02$ & $-0.10$\\\midrule
$\bar d_1+\bar d_2$ &&&&& $1$ & $-0.61$ & $-0.58$ & $-0.86$\\
$\bar d_3$ &&&&&& $1$ & $-0.06$ & $0.86$\\
$\bar d_5$ &&&&&&& $1$ & $0.31$\\
$\bar d_{14}-\bar d_{15}$ &&&&&&&& $1$\\
\bottomrule
\end{tabular}
\caption{Correlation coefficients for the $\Order(p^3)$ extraction. The coefficients for the $\Order(p^2)$ case are identical to the $4\times 4$ $c_i$ submatrix.}
\label{tab:corr_p3}
\renewcommand{\arraystretch}{1.0}
\end{table}

\begin{table}
\footnotesize
\renewcommand{\arraystretch}{1.3}
\centering
\begin{tabular}{crrrrrrrrrrrrr}
\toprule
& $c_1$ & $c_2$ & $c_3$ & $c_4$ & $\bar d_1+\bar d_2$ & $\bar d_3$ & $\bar d_5$ & $\bar d_{14}-\bar d_{15}$ & $\bar e_{14}$ & $\bar e_{15}$ & $\bar e_{16}$ & $\bar e_{17}$ & $\bar e_{18}$ \\
$c_1$ & $1$ & $0.18$ & $0.58$ & $0.06$ & $-0.42$ & $0.71$ & $0.04$ & $0.47$ & $-0.59$ & $0.33$ & $-0.21$ & $-0.11$ & $-0.21$\\
$c_2$ & & $1$ & $-0.64$ & $-0.01$ & $0.67$ & $-0.36$ & $-0.27$ & $-0.55$ & $0.56$ & $-0.59$ & $0.59$ & $0.21$ & $0.47$\\
$c_3$ &&& $1$ & $0.04$ & $-0.86$ & $0.91$ & $0.16$ & $0.87$ & $-0.97$ & $0.68$ & $-0.60$ & $-0.24$ & $-0.46$\\
$c_4$ &&&& $1$ & $0.18$ & $-0.22$ & $0.03$ & $-0.31$ & $-0.02$ & $0.07$ & $-0.08$ & $-0.61$ & $-0.63$\\\midrule
$\bar d_1+\bar d_2$ &&&&& $1$ & $-0.83$ & $-0.40$ & $-0.94$ & $0.88$ & $-0.77$ & $0.74$ & $0.23$ & $0.34$\\
$\bar d_3$ &&&&&& $1$ & $0.05$ & $0.93$ & $-0.94$ & $0.53$ & $-0.47$ & $-0.07$ & $-0.17$\\
$\bar d_5$ &&&&&&& $1$ & $0.18$ & $-0.14$ & $0.40$ & $-0.29$ & $-0.18$ & $-0.29$\\
$\bar d_{14}-\bar d_{15}$ &&&&&&&& $1$ & $-0.91$ & $0.64$ & $-0.61$ & $-0.03$ & $-0.21$\\\midrule
$\bar e_{14}$ &&&&&&&&& $1$ & $-0.70$ & $0.65$ & $0.23$ & $0.43$\\
$\bar e_{15}$ &&&&&&&&&& $1$ & $-0.97$ & $-0.28$ & $-0.65$\\
$\bar e_{16}$ &&&&&&&&&&& $1$ & $0.29$ & $0.60$\\
$\bar e_{17}$ &&&&&&&&&&&& $1$ & $0.19$\\
$\bar e_{18}$ &&&&&&&&&&&&& $1$\\
\bottomrule
\end{tabular}
\caption{Correlation coefficients for the $\Order(p^4)$ extraction in standard counting.}
\label{tab:corr_p4}
\renewcommand{\arraystretch}{1.0}
\end{table}

\begin{table}[t]
\footnotesize
\renewcommand{\arraystretch}{1.3}
\centering
\begin{tabular}{crrrrrrrrrrrrr}
\toprule
& $c_1$ & $c_2$ & $c_3$ & $c_4$ & $\bar d_1+\bar d_2$ & $\bar d_3$ & $\bar d_5$ & $\bar d_{14}-\bar d_{15}$ & $\bar e_{14}$ & $\bar e_{15}$ & $\bar e_{16}$ & $\bar e_{17}$ & $\bar e_{18}$ \\
$c_1$ & $1$ & $-0.20$ & $0.58$ & $0.06$ & $-0.42$ & $0.68$ & $0.04$ & $0.47$ & $-0.60$ & $0.33$ & $-0.21$ & $-0.11$ & $-0.20$\\
$c_2$ & & $1$ & $-0.86$ & $-0.03$ & $0.83$ & $-0.63$ & $-0.28$ & $-0.73$ & $0.77$ & $-0.72$ & $0.67$ & $0.25$ & $0.55$\\
$c_3$ &&& $1$ & $0.04$ & $-0.86$ & $0.90$ & $0.16$ & $0.87$ & $-0.97$ & $0.68$ & $-0.60$ & $-0.24$ & $-0.46$\\
$c_4$ &&&& $1$ & $0.18$ & $-0.25$ & $0.03$ & $-0.31$ & $-0.02$ & $0.07$ & $-0.08$ & $-0.61$ & $-0.63$\\\midrule
$\bar d_1+\bar d_2$ &&&&& $1$ & $-0.83$ & $-0.40$ & $-0.94$ & $0.88$ & $-0.77$ & $0.74$ & $0.23$ & $0.34$\\
$\bar d_3$ &&&&&& $1$ & $0.03$ & $0.93$ & $-0.94$ & $0.52$ & $-0.45$ & $-0.05$ & $-0.14$\\
$\bar d_5$ &&&&&&& $1$ & $0.18$ & $-0.13$ & $0.40$ & $-0.29$ & $-0.18$ & $-0.29$\\
$\bar d_{14}-\bar d_{15}$ &&&&&&&& $1$ & $-0.91$ & $0.64$ & $-0.61$ & $-0.03$ & $-0.20$\\\midrule
$\bar e_{14}$ &&&&&&&&& $1$ & $-0.69$ & $0.65$ & $0.23$ & $0.42$\\
$\bar e_{15}$ &&&&&&&&&& $1$ & $-0.97$ & $-0.28$ & $-0.65$\\
$\bar e_{16}$ &&&&&&&&&&& $1$ & $0.29$ & $0.60$\\
$\bar e_{17}$ &&&&&&&&&&&& $1$ & $0.19$\\
$\bar e_{18}$ &&&&&&&&&&&&& $1$\\
\bottomrule
\end{tabular}
\caption{Correlation coefficients for the $\Order(p^4)$ extraction in $NN$ counting.}
\label{tab:corr_p4_NN}
\renewcommand{\arraystretch}{1.0}
\end{table}

Inverting the expressions for the subthreshold parameters from~\ref{app:ChPT_subthreshold}, we obtain the LECs summarized in Table~\ref{tab:LECs}, with correlation coefficients according to Tables~\ref{tab:corr_p3}--\ref{tab:corr_p4_NN}.
At $O(p^2)$ only the $c_i$ contribute, and only four subthreshold parameters are sensitive to these LECs. In particular, there is a LET for $d_{00}^-$
\beq
d_{00}^-\big|_\text{NLO}=\frac{1}{2\Fpi^2}=1.15\mpi^{-2},
\eeq
in fair agreement with $d_{00}^-=(1.41\pm 0.01)\mpi^{-2}$ from Table~\ref{tab:RS_subthr}.

At N$^2$LO four $\bar d_i$ appear, and eight subthreshold parameters receive contributions from LECs. In addition, there are five LETs
\begin{align}
d_{20}^+\big|_\text{N$^2$LO}&=0.22\mpi^{-5}, &
d_{11}^+\big|_\text{N$^2$LO}&=0.07\mpi^{-5}, &
d_{02}^+\big|_\text{N$^2$LO}&=0.034\mpi^{-5},\notag\\
b_{10}^-\big|_\text{N$^2$LO}&=0.92\mpi^{-4}, &
b_{01}^-\big|_\text{N$^2$LO}&=0.19\mpi^{-4},
\end{align} 
to be compared with the corresponding numbers in Table~\ref{tab:RS_subthr}. For all but $d_{11}^+$ these predictions are quite close to the full result.
Comparing the different extractions up to N$^3$LO, the convergence pattern for the $c_i$ looks reasonably stable. In contrast, while the N$^2$LO $\bar d_i$ are of natural size, their values increase by nearly an order of magnitude when going to N$^3$LO (except for $\bar d_5$). The origin of this behavior can be identified from the analytic expressions in~\eqref{subthr_ChPT}: $d_{00}^-$, $d_{10}^-$, $d_{01}^-$, and $b_{00}^+$ receive loop corrections involving terms that scale with $\ga^2(c_3-c_4)\sim-16\GeV^{-1}$, which are balanced by the large LECs in order to keep the subthreshold parameters at their physical values. 
The enhancement of the $c_i$, in turn, can be understood from resonance saturation, since, absent low-lying resonant states, they would be expected to scale as
$c_i \sim \ga/\Lambda_\text{b}=\Order(1\GeV^{-1})$~\cite{Bernard:2007zu}. While $t$-channel resonances are required as well to reproduce the physical values of the $c_i$, the most prominent enhancement for $c_{2-4}$ is generated by the $\Delta(1232)$~\cite{Bernard:1995dp,Bernard:1996gq,Becher:1999he}.
Given such large loop corrections the errors for the LECs at a given chiral order are negligible compared to the uncertainties to be attached to the chiral expansion itself.

The loop enhancement of the $\bar d_i$ can be made explicit by considering combinations of subthreshold parameters where the corresponding terms cancel. Those are
\begin{align}
 d_{00}^-+2\mpi^2 d_{01}^-&=\frac{8\mpi^2}{\Fpi^2}\bar d_5+\cdots, &
 2d_{10}^--5d_{01}^-&=\frac{2}{\Fpi^2}\Big[5\big(\bar d_1+\bar d_2\big)+4\bar d_3\Big]+\cdots,\notag\\
 b_{00}^+-4\mN d_{01}^-&=\frac{4\mN}{\Fpi^2}\Big[2\big(\bar d_1+\bar d_2\big)+\bar d_{14}-\bar d_{15}\Big]+\cdots,
\end{align}
which already explains why $\bar d_5$ is found to be of natural size. Similarly, the other two combinations become
\begin{align}
\frac{1}{9}\Big[5\big(\bar d_1+\bar d_2\big)+4\bar d_3\Big]&=(-0.52\pm 0.03)\GeV^{-2} \quad \big[(-0.53\pm 0.03)\GeV^{-2}\big],\notag\\
\frac{1}{3}\Big[2\big(\bar d_1+\bar d_2\big)+\bar d_{14}-\bar d_{15}\Big]&=(0.61\pm 0.02)\GeV^{-2} \quad \big[(0.06\pm 0.02)\GeV^{-2}\big],
\end{align}
where the first [second] number refers to standard [$NN$] counting, and the N$^2$LO values are $(0.37\pm 0.03)\GeV^{-2}$ and $(0.06\pm 0.02)\GeV^{-2}$, respectively. The huge cancellations amongst the individual terms precisely reflect the absence of the $\ga^2(c_3-c_4)$-enhanced contributions.

\subsection{Threshold parameters}
\label{sec:threshold_ChPT}

A powerful check on the convergence of the chiral expansion arises by comparing the series at two different kinematic points. Here, we take the LECs extracted from the subthreshold expansion and predict the coefficients in the expansion around threshold. The advantage of comparing scattering lengths and effective ranges as compared to phase shifts or cross sections is that the expressions can still be given analytically in rather compact form, see~\ref{app:ChPT_threshold}, so that the analysis becomes more transparent.
In the prediction of the threshold parameters we follow the strategy laid out in~\cite{Becher:1999he}: at a given order in the chiral expansion we use the LECs as determined from the subthreshold parameters at that chiral order. As already noted in~\cite{Gasser:1982ap}, the infrared singularities in the chiral expansion of the nucleon mass and the $\sigma$-term can in practice be absorbed into the LECs in such a way that their effect nearly cancels if relations between observables are considered. In this way, one effectively eliminates LECs in favor of physical quantities, in our case the $\pi N$ LECs in favor of the subthreshold parameters. Resonance enhancement from the $\Delta(1232)$ notwithstanding, this scheme amounts to a reordering of the perturbative expansion with the objective that the effects from higher-order infrared singularities be minimized. The corresponding results are shown in Table~\ref{tab:expansion_threshold}. 

The most striking feature of the predictions for the threshold parameters is that in some quantities the fourth-order results deteriorate substantially compared to the third-order ones, most notably in $a_{0+}^-$ and $b_{0+}^-$. The origin of this effect can be attributed to the $\ga^2(c_3-c_4)$-enhanced terms discussed in the previous section: while the LECs were adjusted in such a way as to produce the correct subthreshold parameters, the cancellation for the threshold parameters is incomplete and leads to a surprisingly large shift in these parameters. 
Another interesting feature is that the convergence in the $NN$ counting appears consistently improved compared to the standard counting, which suggests that also the $1/\mN$ corrections tend to slow down the convergence.

In view of these findings, the only way to improve the convergence globally in the full low-energy domain consists of including the $\Delta(1232)$ explicitly in the calculation to reduce the $c_i$ to more natural values and thereby mitigate the anomalously large loop effects. However, at subleading orders calculations with explicit $\Delta$ degrees of freedom quickly become extremely complex, so that often $\Delta$-less calculations can be carried out up to higher orders, and it is not evident which approach will ultimately prove more efficient. The main conclusion from the comparison of the convergence at subthreshold and threshold kinematics is that LECs for $\Delta$-less $NN$ applications need to be extracted from the subthreshold region in order to minimize the impact of $c_i$ loop effects.

\begin{table}[t]
\renewcommand{\arraystretch}{1.3}
\centering
\begin{tabular}{crrrrrrr}\toprule
& NLO & N$^2$LO & N$^3$LO & NLO$^{NN}$ & N$^2$LO$^{NN}$ & N$^3$LO$^{NN}$ & RS\\\midrule
$a_{0+}^+\ [10^{-3}\mpi^{-1}]$ & $-23.8$ & $0.2$ & $-7.9$ & $-14.2$ & $0.2$ & $-1.4$ & $-0.9\pm 1.4$\\
$a_{0+}^-\ [10^{-3}\mpi^{-1}]$ & $79.4$ & $92.9$ & $59.4$ & $79.4$ &  $92.2$ & $69.2$ & $85.4\pm 0.9$\\
$a_{1+}^+\ [10^{-3}\mpi^{-3}]$ & $102.6$ & $121.2$ & $131.8$ & $96.2$ & $113.4$ & $133.7$ & $131.2\pm1.7$\\
$a_{1+}^-\ [10^{-3}\mpi^{-3}]$ & $-65.2$ & $-75.3$ & $-89.6$ & $-60.8$ & $-74.6$ & $-80.5$ & $-80.3\pm1.1$\\
$a_{1-}^+\ [10^{-3}\mpi^{-3}]$ & $-45.0$ & $-47.0$ & $-72.7$ & $-32.3$ & $-54.1$ & $-55.0$ & $-50.9\pm1.9$\\
$a_{1-}^-\ [10^{-3}\mpi^{-3}]$ & $-11.2$ & $-2.8$ & $-23.2$ & $-6.8$ & $-14.2$ & $-9.8$ & $-9.9\pm1.2$\\
$b_{0+}^+\ [10^{-3}\mpi^{-3}]$ & $-70.4$ & $-23.3$ & $-44.9$ & $-80.0$ & $-46.5$ & $-42.9$ & $-45.0\pm1.0$\\
$b_{0+}^-\ [10^{-3}\mpi^{-3}]$ & $20.6$ & $23.3$ & $-63.0$ & $39.7$ & $36.4$ & $-29.7$ & $4.9\pm0.8$\\
\bottomrule
\end{tabular}
\caption{Predictions for the threshold parameters at NLO, N$^2$LO, and N$^3$LO, in standard and $NN$ counting.  At each order we use the LECs as given in Table~\ref{tab:LECs}.
The first three columns refer to standard counting, and the next three to $NN$ counting. 
We do not display the corresponding errors, which are much smaller than the uncertainties in the chiral expansion.}
\label{tab:expansion_threshold}
\end{table}

\subsection{Nucleon mass}
\label{sec:nucleon_mass}

At fourth order in the chiral expansion the nucleon mass can be expressed as~\cite{Steininger:1998ya,Kambor:1998pi,Becher:2001hv} 
\begin{align}
\label{mass}
\mN&=m-4c_1 \mpii^2\pm2Bc_5(\md-\muu)-\frac{e^2\Fpi^2}{2}(f_1\pm f_2+f_3)-\frac{\ga^2\big(2\mpi^3+\mpii^3\big)}{32\pi\Fpi^2}\notag\\
&-\frac{3}{32\pi^2\Fpi^2m}\Big(\ga^2+m(-8c_1+c_2+4c_3)\Big)\mpi^4\log\frac{\mpi}{m}+\bigg\{e_1-\frac{3}{128\pi^2\Fpi^2m}\big(2\ga^2-c_2m\big)\bigg\}\mpi^4+\Order\big(\mpi^5\big),
\end{align}
where we have included the third-order isospin-breaking corrections~\cite{Meissner:1997ii,Muller:1999ww,Gasser:2002am}.\footnote{The fourth-order corrections have been considered in~\cite{Muller:1999ww,Frink:2004ic,Tiburzi:2005na}, in particular the proton--neutron mass difference exhibits a chiral logarithm with large coefficient $(6\ga^2+1)/2\sim 5$. As discussed below, the main motivation for including the isospin-breaking terms here concerns the interpretation of the nucleon mass in the chiral limit. In this context, the fourth-order terms provide little additional insight due to the occurrence of new unknown LECs, so that for simplicity they will be ignored in the following.}
Here, $m$ denotes the nucleon mass in the chiral limit, the upper/lower sign refers to proton/neutron, and $\mpi$, $\mpii$, $\Fpi$, and $\ga$ are the physical quantities.
The renormalization of $\Fpi$ and $\ga$ is higher order in the chiral expansion, while the renormalization of $\mpii$ has been absorbed into a redefinition of $e_1=e_1(m)$, which represents a combination of $e_i$ from~\cite{Fettes:2000gb} (evaluated at renormalization scale $\mu=m$). Moreover, since we want to use the LECs $c_i$ as extracted from the RS analysis, they need to be redefined according to the quark-mass renormalization~\eqref{ci_ren}, an effect that can be absorbed into $e_1$ as well.

\begin{figure}[t]
\centering
\includegraphics[width=0.7\linewidth,clip]{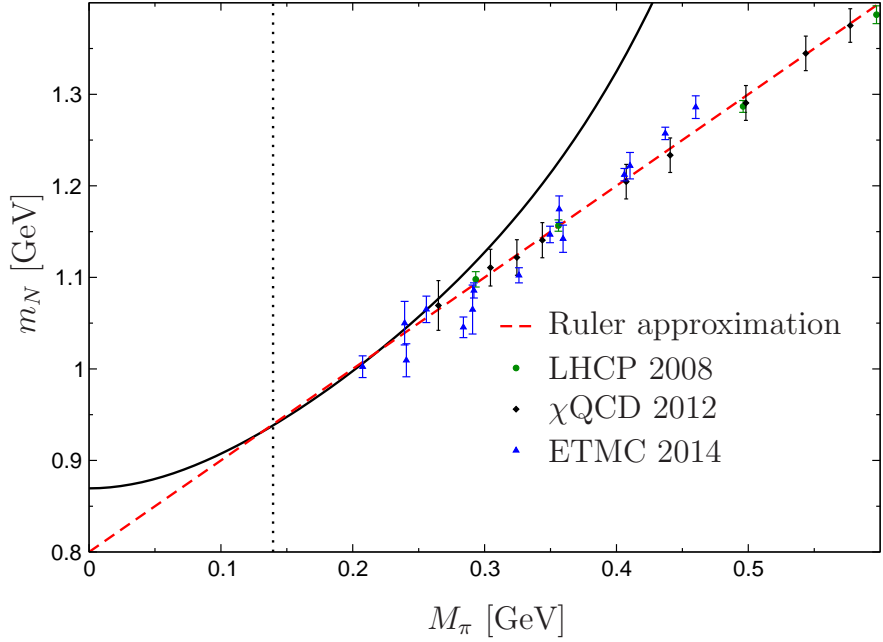}
\caption{Dependence of the nucleon mass as a function of $\mpi$ in comparison to lattice: LHCP~\cite{WalkerLoud:2008bp}, $\chi$QCD~\cite{Gong:2013vja}, and ETMC~\cite{Alexandrou:2014sha}. Calculations close to or even at the physical point do reproduce the physical mass of the nucleon: $953(41)\MeV$~\cite{Aoki:2008sm}, $936(25)(22)\MeV$~\cite{Durr:2008zz}, $933(8)(18)\MeV$~\cite{Abdel-Rehim:2015owa,Abdel-Rehim:2015pwa}. The black solid line refers to the fourth-order ChPT prediction with LECs determined as explained in the main text.}
\label{fig:ruler}
\end{figure}

The chiral expansion of the $\sigma$-term as defined in~\eqref{sigma_term_def} follows from~\eqref{mass} by means of the Feynman--Hellmann theorem~\cite{Hellmann:1937,Feynman:1939zza,Gasser:1979hf}
\begin{align}
\label{sigma_chiral}
\sigma_{\pi N}&=\hat m\frac{\partial \mN}{\partial\hat m}=\hat m \frac{\partial \mpii^2}{\partial\hat m}\frac{\partial \mN}{\partial\mpii^2}
=\frac{\mpii}{2}\bigg\{1-\frac{\mpii^2}{32\pi^2\Fpi^2}\big(\bar l_3-1\big)\bigg\}\frac{\partial \mN}{\partial\mpii}\notag\\
&=-4c_1\mpii^2 
-\frac{3\ga^2\mpii^2\big(2\mpi+\mpii\big)}{64\pi\Fpi^2}
-\frac{3}{64\pi^2\Fpi^2m}\Big(\ga^2+m(-8c_1+c_2+4c_3)\Big)\mpi^4\bigg\{4\log\frac{\mpi}{m}+1\bigg\}\notag\\
&+2\bigg\{e_1-\frac{3}{128\pi^2\Fpi^2m}\big(2\ga^2-c_2m\big)+\frac{c_1\big(\bar l_3-1\big)}{16\pi^2\Fpi^2}\bigg\}\mpi^4+\Order\big(\mpi^5\big),
\end{align}
where we have used
\beq
\mpii^2=2B\hat m\bigg\{1-\frac{2B\hat m}{32\pi^2\Fpi^2}\bar l_3\bigg\}.
\eeq
In the following, we will take $\bar l_3=3.41\pm 0.41$ from~\cite{Aoki:2013ldr}.
Note that $\bar l_3$ contains an implicit pion mass dependence in these conventions, see~\ref{app:ChPT}.
Since $e_1$ cannot be determined from the subthreshold parameters of $\pi N$ scattering, we fix it by demanding that~\eqref{sigma_chiral} reproduce~\eqref{sigma_term_result}.
With $e_1$ adjusted in this way, we can then predict the nucleon mass in the chiral limit. To facilitate the comparison to lattice we first rewrite~\eqref{mass} and~\eqref{sigma_chiral} in terms of isospin-limit quantities, which amounts to redefining $m$ according to
\beq
\tilde m=m \pm2Bc_5(\md-\muu)-\frac{e^2\Fpi^2}{2}(f_1\pm f_2+f_3)+\underbrace{2c_1\Delta_\pi-\frac{9\ga^2\mpi\Delta_\pi}{256\pi\Fpi^2}}_{-3.2\MeV},
\eeq
so that the identification of $m$ itself again involves a sizable shift due to the pion mass difference as well as the LECs $f_{1-3}$ and $c_5$. We obtain
\beq
\tilde m=869.5\MeV.
\eeq

The full pion-mass dependence is shown in Fig.~\ref{fig:ruler} compared to lattice results. The striking feature coined the ``ruler approximation''~\cite{WalkerLoud:2008pj,Walker-Loud:2013yua,Walker-Loud:2014iea} is that the straight line $800\MeV+\mpi$ reproduces lattice results over a wide range of pion masses, before around the physical region the curvature demanded by ChPT has to set in. This behavior has been confirmed in many more lattice calculations, see e.g.~\cite{Aoki:2008sm,Durr:2008zz,Alexandrou:2014sha}.
Figure~\ref{fig:ruler} demonstrates that the $\Order(p^4)$ prediction starts to deviate already at pion masses as low as $300\MeV$. As noted earlier~\cite{Beane:2004ks,Bernard:2007zu}, the range of convergence of the chiral expansion for the nucleon mass appears to be extremely limited. The fact that to a remarkably good approximation lattice results fall on a straight line implies that including higher chiral orders~\cite{McGovern:1998tm,McGovern:2006fm,Schindler:2006ha,Schindler:2007dr} in a fit to lattice data is not a solution:\footnote{In principle, we could have included the fifth-order term since it does not involve new unknown LECs. However, this would not be fully consistent because then $c_1$ and $e_1$ would need to be extracted from a fifth-order calculation of $\pi N$ scattering and the $\sigma$-term as well.} there have to be huge cancellations amongst the individual terms to produce the observed linear behavior.

We stress that this phenomenon solely concerns the range of convergence in $\mpi$, not the rate of convergence at the physical point. Based on the isospin-limit versions of~\eqref{mass} and~\eqref{sigma_chiral} (i.e.\ isospin-breaking effects absorbed into $e_1$ and $\tilde m$), we find 
\beq
\mN=\underbrace{869.5\MeV}_{\Order(\mpi^0)} \underbrace{+ 86.5\MeV}_{\Order(\mpi^2)} \underbrace{-15.4\MeV}_{\Order(\mpi^3)} \underbrace{-2.3\MeV}_{\Order(\mpi^4)} =938.3\MeV
\eeq
and
\beq
\sigma_{\pi N}=\underbrace{86.5\MeV}_{\Order(\mpi^2)} \underbrace{-23.2\MeV}_{\Order(\mpi^3)} \underbrace{-4.2\MeV}_{\Order(\mpi^4)} =59.1\MeV,
\eeq
both of which display a very reasonable convergence pattern.

\section{Summary and conclusions}
\label{sec:summary}

In the present paper we have provided a comprehensive review of Roy--Steiner equations for $\pi N$ scattering, regarding a solution of the $s$- and $t$-channel subsystems and, for the first time, a solution of the full set of equations, a detailed discussion of uncertainties, as well as consequences for the $\pi N$ $\sigma$-term and the matching to chiral perturbation theory. The main results can be summarized as follows:
\begin{enumerate}
 \item Due to the fact that crossing symmetry relates two different physical processes, $\pi N$ Roy--Steiner equations involve equations both for $\pi N\to\pi N$ as well as $\pi\pi\to\bar N N$. A full solution can be obtained without a complicated iteration process by noting that the interdependence of the two channels proceeds predominantly via the subtraction constants, here identified with subthreshold parameters, so that first the $t$-channel equations can be solved, whereupon the $s$-channel equations take a mathematical form equivalent to $\pi\pi$ Roy equations.
 \item Both subsystems are rigorously valid up to the maximal energies of $\sqrt{s}=1.38\GeV$ and $\sqrt{t}=2.00\GeV$, respectively, so that above these energies input from experiment is required. Moreover, the equations couple all partial waves, mandating a truncation in the partial-wave expansion.
 For the $s$-channel, we keep the $S$- and $P$-waves explicitly, while higher partial waves as well as the lower ones above the matching point chosen at its maximally allowed value of $\sqrt{\sm}=1.38\GeV$ are taken as input. Indeed, the driving terms that collect this $s$-channel input are found to be small at low energies and in no case are their uncertainties the dominant ones in the final solution.
 \item In contrast, for the $t$-channel data input is only available above the two-nucleon threshold, so that the imaginary parts in the pseudophysical region $4\mpi^2\leq t\leq 4\mN^2$ have to be reconstructed by other means. Unitarity in the form of Muskhelishvili--Omn\`es techniques provides that link at least up to the onset of inelastic contributions, and even beyond if the form of the inelasticities is sufficiently well constrained. We use a two-channel solution for the $S$-wave, and single-channel approximations for $1\leq J\leq 3$, where $F$-waves only enter the uncertainty estimates. A surprising feature of the solution is that the $D$-wave resonance $f_2(1270)$ plays such a prominent role that no satisfactory Roy--Steiner solution could be found if its contribution was omitted. Estimates for the uncertainties of the $t$-channel input and the role of inelastic channels were provided and propagated to the final solution. In the end, the uncertainties at intermediate energies overshadow any remaining contribution from the dispersive integrals above the two-nucleon threshold, so that we took $\sqrt{\tm}=2\mN$ and set the integrals above that energy to zero.
 \item The number of subtractions in the hyperbolic dispersion relations underlying the Roy--Steiner system was chosen in such a way as to match the number of degrees of freedom known from the mathematical properties of Roy equations. This implies that many more subtractions are performed than would be required for all dispersive integrals to converge. This strategy ensures both that the integrals converge quickly and that the equations can be solved in a fit by minimizing the difference between left- and right-hand side with respect to the subtraction constants and the parameters describing the low-energy $\pi N$ phase shifts.
 \item We found that the solution of the $s$-channel equations is greatly stabilized if values for the $S$-wave scattering lengths are imposed as additional constraints. We reviewed the extraction from pionic atoms in some detail, including the discussion of isospin-breaking corrections and the role of virtual photons. The final values used in this analysis are $a^{1/2}_{0+}=(169.8\pm 2.0)\times 10^{-3}\mpi^{-1}$ and $a^{3/2}_{0+}=(-86.3\pm 1.8)\times 10^{-3}\mpi^{-1}$.
 \item The output of the full solution is a set of consistent subthreshold parameters and low-energy $\pi N$ phase shifts, including robust uncertainty estimates. The error analysis includes the uncertainties in all input quantities, in particular the matching condition at $\sm$, the pionic-atom scattering lengths, and nearly-flat directions in the space of subtractions constants that can be resolved from sum rules for some of the higher subthreshold parameters. As expected, we found that for the phase shifts the uncertainty bands are dominated by the scattering-length errors close to threshold and by the matching condition close to $\sm$. Explicit parameterizations are provided in~\ref{app:numerical_sol}. 
 \item With subthreshold parameters, low-energy phase shifts, and $t$-channel partial waves determined, the same hyperbolic dispersion relations used for the derivation of the Roy--Steiner equations in the first place, now evaluated at the threshold point, yield sum rules for the parameters in the threshold expansion. We evaluated these sum rules for the $P$-wave scattering volumes and the $S$-wave effective ranges, with uncertainties propagated from the Roy--Steiner solution.
 \item The Cheng--Dashen low-energy theorem establishes a link between $\pi N$ scattering evaluated at subthreshold kinematics and the $\pi N$ $\sigma$-term. We reviewed the determination of the various corrections that enter the theorem, including isospin-breaking effects, which prove to be sizable in the isoscalar amplitude in question. Our final result $\sigma_{\pi N}=(59.1\pm 3.5)\MeV$ is higher than early determinations from $\pi N$ scattering, although already $4.2\MeV$ originate from a reappraisal of the corrections to the low-energy theorem, and thereof $3\MeV$ from isospin breaking.
 \item We match our results to chiral perturbation theory by identifying the subthreshold parameters with their respective chiral expansion. The matching thus reduces to equating a polynomial in the Mandelstam variables, and the corresponding low-energy constants are obtained by inverting a simple linear system of equations. We find that at a given order in the chiral expansion the uncertainties propagated from the Roy--Steiner solution are very small, but that chiral convergence for some subthreshold parameters is challenged by large loop corrections that involve the combination $\ga^2(c_3-c_4)$ of low-energy constants enhanced by saturation from the $\Delta(1232)$. One consequence of this behavior is that the low-energy constants fixed at subthreshold kinematics lead to an unexpectedly poor convergence pattern at next-to-next-to-next-to-leading order for some of the threshold parameters. We concluded that without explicit $\Delta$ degrees of freedom low-energy constants defined at one kinematic point do not necessarily guarantee convergence in the whole low-energy domain and argued that in $\Delta$-less applications in $NN$ scattering determinations of $\pi N$ low-energy constants at the subthreshold point are preferable for that reason.
 \item As a final application we investigated the chiral convergence of the quark-mass dependence of the nucleon mass. Combining information from the $\sigma$-term and the subthreshold matching, the expansion coefficients are determined up to next-to-next-to-next-to-leading order. We found that at the physical point the convergence pattern looks very convincing, but that the comparison to lattice data indicates a breakdown of the chiral expansion for pion masses as low as $300\MeV$.
\end{enumerate}

\section*{Note added in proof}

While this paper was under review, several lattice calculations of the $\sigma$-term at physical quark masses have been reported~\cite{Durr:2015dna,Yang:2015uis,Abdel-Rehim:2016won,Bali:2016lvx}. For a detailed comparison to the phenomenological determination discussed in Sect.~\ref{sec:sigma_term} see~\cite{Hoferichter:2016ocj}. 

\section*{Acknowledgements}

We thank C.~Ditsche for collaboration at early stages of this project,
and V.~Bernard, G.~Colangelo, H.~Leutwyler, B.~Moussallam, D.~Siemens, and A.~Walker-Loud for helpful discussions and communication, in particular for providing
the $\pi\pi$ phase shifts corresponding to~\cite{Caprini:2011ky} (GC), a version of the $\pi\pi\to\bar K K$ amplitude from~\cite{Buettiker:2003pp} consistent with~\cite{Caprini:2011ky} (BM),
and the lattice data shown in Fig.~\ref{fig:ruler} (AWL), as well as for pointing out an inconsistency in the chiral expansion of the threshold parameters (DS). 
Financial support by
 the Helmholtz Virtual Institute NAVI (VH-VI-417),
the DFG (SFB/TR 16, ``Subnuclear Structure of Matter''),
and 
the DOE (Grant No.\ DE-FG02-00ER41132) 
is gratefully acknowledged. 
The work of UGM was supported in part by the Chinese 
Academy of Sciences (CAS) President's International Fellowship 
Initiative (PIFI) (Grant No.\ 2015VMA076).

\appendix

\setcounter{table}{0}

\section{Integral kernels}
\label{app:kernels}

In this appendix we collect all the pole-term projections and kernel functions that appear in~\eqref{sRSpwhdr} and~\eqref{tRSpwhdr} as they follow from unsubtracted HDRs. The modifications necessary when introducing subtractions are provided in~\ref{app:kernel_subtractions}.

\subsection{$s$-channel projection}
\label{app:kernels_schannel}

\subsubsection{Nucleon pole}

The $s$-channel projection of the nucleon pole terms may be written as
\beq
\label{barNIpm}
N^I_{l+}(W)=\frac{g^2}{16\pi W}\bigg\{(E+\mN)(W-\mN)\bigg[\epsilon^I\frac{Q_l(y)}{\qq^2}+\frac{2\delta_{l0}}{\mN^2-s}\bigg]
+(E-\mN)(W+\mN)\epsilon^I\frac{Q_{l+1}(y)}{\qq^2}\Bigg\},
\eeq
$\epsilon^I=\pm1$ for $I=\pm$, and
\beq
\label{ydef}
y=1-\frac{s+\mN^2-\Sigma}{2\qq^2}.
\eeq
The Legendre functions of the second kind $Q_l(z)$ can be expressed for general complex argument as (see e.g.~\cite{Bateman:1953})
\beq
\label{qlneumann}
Q_l(z)=\frac{1}{2}\int\limits_{-1}^1\diff x\;\frac{P_l(x)}{z-x},
\eeq
which for $l=0$ reduces to
\beq
Q_0(z)=\frac{1}{2}\int\limits_{-1}^1\frac{\diff x}{z-x},\qquad
Q_0(z\pm i\eps)=\frac{1}{2}\log\bigg|\frac{1+z}{1-z}\bigg|\mp i\frac{\pi}{2}\theta(1-z^2).
\eeq
In the pseudophysical region of the $t$-channel reaction we also need the analytic continuation for purely imaginary argument. For $z=iy$, $y>1$, we have
\beq
Q_0(iy)=\frac{1}{2}\log\frac{iy+1}{iy-1}=\frac{1}{2}\log\frac{1+iy}{1-iy}-i\frac{\pi}{2}=i\left(\arctan y-\frac{\pi}{2}\right)=-Q_0(-iy).
\eeq

\subsubsection{$s$-channel}

The $s$-channel partial-wave expansion reads~\cite{Chew:1957zz}
\begin{align}
A^I(s,t)&=\sum\limits_{l=0}^\infty\Big\{ S^1_{l+1,l}(W,z_s)f^I_{l+}(W)-S^1_{l,l+1}(W,z_s)f^I_{(l+1)-}(W)\Big\},\notag\\
B^I(s,t)&=\sum\limits_{l=0}^\infty\Big\{ S^2_{l+1,l}(W,z_s)f^I_{l+}(W)-S^2_{l,l+1}(W,z_s)f^I_{(l+1)-}(W)\Big\},
\end{align}
with
\begin{align}
\label{sexpform}
S^1_{kn}(W,z_s)&=4\pi\bigg\{\frac{W+\mN}{E+\mN}P_k'(z_s)+\frac{W-\mN}{E-\mN}P_n'(z_s)\bigg\},\notag\\
S^2_{kn}(W,z_s)&=4\pi\bigg\{\frac{1}{E+\mN}\;P_k'(z_s)-\frac{1}{E-\mN}\;P_n'(z_s)\bigg\}.
\end{align}
Taken together with the partial-wave projection~\eqref{sprojform}, this ultimately leads to the kernel functions~\cite{Hite:1973pm}
\beq
\label{sskernelfunction}
K^I_{ll'}(W,W')=\frac{\varphi\big[U_{ll'}\big|\delta(W,W')\big]}{s'-s}-\epsilon^I\frac{\varphi\big[V_{ll'}\big|\varrho(W,W')\big]}{2\qq^2}-\frac{\varphi\big[U_{ll'}\big|\varkappa^I(W,W')\big]}{s'-a},
\eeq
where we have defined the following abbreviations: the structure
\beq
\varphi\big[a_{kn}\big|b(W,W')\big]=\frac{W'}{W}\Big\{b(W,-W')a_{kn}+b(W,W')a_{k,n+1}
+b(-W,-W')a_{k+1,n}+b(-W,W')a_{k+1,n+1}\Big\},
\eeq
for an arbitrary function $b(W,W')$, reflects MacDowell symmetry of $K^I_{ll'}(W,W')$ both with respect to $W$ and $W'$ and acts on
\begin{align}
\delta(W,W')&=\frac{E+\mN}{E'+\mN}\big[W'+W\big], & \varrho(W,W')&=\frac{E+\mN}{E'+\mN}\big[W'-W+2\mN\big],\notag\\
\varkappa^I(W,W')&=\frac{1}{2}\Big[\delta(W,W')+\epsilon^I\varrho(W,W')\Big],
\end{align}
as well as the angular kernels
\beq
\label{angularkernels}
U_{ll'}=\frac{1}{2}\int\limits_{-1}^1\diff z_s\;P_l(z_s)P_{l'}'(z_s'), \qquad
V_{ll'}=\frac{1}{2}\int\limits_{-1}^1\diff z_s\;\frac{P_l(z_s)P_{l'}'(z_s')}{x_s-z_s},
\eeq
with
\begin{align}
\label{alphabetadef}
z_s'&=\alpha z_s+\beta, & \alpha&=\frac{\qq^2}{\qq'^2}\frac{s-a}{s'-a}, & \beta&=1-\alpha-\frac{s'-s}{s'-a}\frac{s+s'-\Sigma}{2\qq'^2},\notag\\
x_s&=1-\frac{s+s'-\Sigma}{2\qq^2}, & x_s'&=\alpha x_s+\beta=1-\frac{s'+s-\Sigma}{2\qq'^2}.
\end{align}
For the lowest kernel functions one finds
\begin{align}
U_{l1}&=\delta_{l0}, & U_{l2}&=\alpha\delta_{l1}+3\beta\delta_{l0}, &
U_{l3}&=\alpha^2\delta_{l2}+5\alpha\beta\delta_{l1}+\frac{1}{2}\big\{5[\alpha^2+3\beta^2]-3\big\}\delta_{l0},\notag\\
V_{l1}&=Q_l(x_s), & V_{l2}&=3x_s'Q_l(x_s)-3\alpha\delta_{l0}, &
V_{l3}&=P_3'(x_s')Q_l(x_s)-\frac{5}{2}\alpha^2\delta_{l1}-\frac{15}{2}\alpha\big\{\alpha x_s+2\beta\big\}\delta_{l0},
\end{align}
and the calculation of higher kernel functions by means of~\eqref{angularkernels} is straightforward, but tedious. For the numerical implementation of $K^I_{ll'}(W,W')$ it is advantageous to analytically separate the Cauchy piece of the kernel functions according to
\begin{align}
\frac{\varphi\big[U_{ll'}\big|\delta(W,W')\big]}{s'-s}&=\frac{\gamma_{ll'}(W,W')}{W'-W}+\frac{1}{W'+W}\frac{W'}{W}\left\{\frac{E+\mN}{E'-\mN}U_{ll'}-\frac{E-\mN}{E'+\mN}U_{l+1,l'+1}\right\},\notag\\
\gamma_{ll'}(W,W')&=\frac{W'}{W}\left\{\frac{E+\mN}{E'+\mN}U_{l,l'+1}-\frac{E-\mN}{E'-\mN}U_{l+1,l'}\right\},\qquad \gamma_{ll'}(W,W)=\delta_{ll'}.
\end{align} 
Asymptotically, these kernel functions behave as
\begin{align}
K^I_{ll'}(W,W')&=\Order\big(\qq^{2l}\big),  &K^I_{ll'}(-W,W')&=\Order\big(\qq^{2l+2}\big) &&\text{for }\;|\qq|\rightarrow 0,\notag\\
K^I_{ll'}(W,W')&=\Order\big(\qq'^{-2l'}\big),  &K^I_{ll'}(W,-W')&=\Order\big(\qq^{-2l'-2}\big) &&\text{for }\;|\qq'|\rightarrow 0,\notag\\
K^I_{ll'}(W,W')&=\Order\big(|\qq'|^{-2l-1}\big)&& &&\text{for }\;|\qq'|\rightarrow\infty,
\end{align}
in accordance with MacDowell symmetry and the asymptotic properties of $f^I_{l\pm}(W)$.

\subsubsection{$t$-channel}

The $t$-channel partial-wave expansion takes the form~\cite{Frazer:1960zza}
\begin{align}
\label{texpform}
A^I(s,t)&=-\frac{4\pi}{p_t^2}\sum_J(2J+1)(p_tq_t)^J\bigg\{P_J(z_t)f^J_+(t)-\frac{\mN}{\sqrt{J(J+1)}}z_tP_J'(z_t)f^J_-(t)\bigg\},\notag\\
B^I(s,t)&=4\pi\sum_J\frac{2J+1}{\sqrt{J(J+1)}}(p_tq_t)^{J-1}P_J'(z_t)f^J_-(t).
\end{align}
Using the abbreviations
\beq
\label{psidef}
\psi\big[a_{kn}\big|d(W)\big]=d(W)a_{kn}+d(-W)a_{k+1,n}, \qquad \eta_J=\frac{2J+1}{4W\qq^2}\frac{(p_t'q_t')^J}{p_t'^2},
\eeq
for an arbitrary function $d(W)$, the corresponding kernels are~\cite{Hite:1973pm}
\begin{align}
\label{stkernelfunctions}
G_{lJ}(W,t')&=-\eta_J\psi\big[A_{lJ}\big|E+\mN\big],\notag\\
H_{lJ}(W,t')&=\frac{\eta_J}{\sqrt{J(J+1)}}\left\{\frac{p_t'}{q_t'}\psi\big[B_{lJ}\big|(W-\mN)(E+\mN)\big]+\mN\psi\big[C_{lJ}\big|E+\mN\big]\right\},
\end{align}
where $C_{lJ}=JA_{lJ}+B_{l,J-1}$, and the angular kernels for even $J$ are defined as
\beq
\label{angular_st_even}
A_{lJ}=\frac{1}{2}\int\limits_{-1}^1\diff z_s\;\frac{P_l(z_s)P_J(z_t')}{x_t-z_s},\qquad
B_{lJ}=\frac{\mu_1}{2}\int\limits_{-1}^1\diff z_s\;P_l(z_s)\frac{P_J'(z_t')}{z_t'}+\frac{\mu_2}{2}\int\limits_{-1}^1\diff z_s\;\frac{P_l(z_s)P_J'(z_t')/z_t'}{x_t-z_s},
\eeq
while for odd $J$ one finds 
\beq
\label{angular_st_odd}
A_{lJ}=\frac{\mu_1}{2}\int\limits_{-1}^1\diff z_s\;P_l(z_s)\frac{P_J(z_t')}{z_t'}+\frac{\mu_2}{2}\int\limits_{-1}^1\diff z_s\;\frac{P_l(z_s)P_J(z_t')/z_t'}{x_t-z_s},\qquad
B_{lJ}=\frac{1}{2}\int\limits_{-1}^1\diff z_s\;\frac{P_l(z_s)P_J'(z_t')}{x_t-z_s},
\eeq
with
\begin{align}
z_t'&=\sqrt{\gamma z_s+\delta}, & \gamma&=\frac{\qq^2(s-a)}{2p_t'^2q_t'^2}, &
\delta&=\frac{(t'-\Sigma+2a)^2-4(s-a)(2\qq^2+\Sigma-s-a)}{16p_t'^2q_t'^2},\notag\\
\mu_1&=-\frac{\qq^2}{2p_t'q_t'}, & \mu_2&=\frac{2s+t'-\Sigma}{4p_t'q_t'}, & x_t&=1+\frac{t'}{2\qq^2}.
\end{align}
It is important to note that the integrands in~\eqref{angular_st_even} and \eqref{angular_st_odd} only involve even powers of $z_t'$, so that no square roots of $z_s$ occur in the integrals. Explicitly, one finds for the lowest kernels
\begin{align}
\label{GH_explicit}
G_{l0}(W,t')&=-\frac{1}{4W\qq^2p_t'^2}\Big\{(E+\mN)Q_l(x_t)-(E-\mN)Q_{l+1}(x_t)\Big\},\notag\\
G_{l1}(W,t')&=\frac{3}{4}\bigg\{(2s+t'-\Sigma)G_{l0}(W,t')+\frac{E+\mN}{2Wp_t'^2}\delta_{l0}\bigg\},\notag\\
H_{l1}(W,t')&=\frac{1}{\sqrt{2}}\bigg\{\frac{3}{4}Z_l(W,t')-\mN G_{l1}(W,t')\bigg\},\notag\\
G_{l2}(W,t')&=\frac{5}{16}\bigg\{\Big[6s(s+t'-\Sigma)+(t'-\Sigma)^2+2(\mN^2-\mpi^2)^2\Big]G_{l0}(W,t')
+3\frac{(E+\mN)(s-a)}{Wp_t'^2}\delta_{l0}\bigg\},\notag\\
H_{l2}(W,t')&=\frac{15}{16\sqrt{6}}\bigg\{(2s+t'-\Sigma)Z_l(W,t')-\mN\Big[4s(s+t'-\Sigma)+(t'-\Sigma)^2\Big]G_{l0}(W,t')\notag\\
&\qquad\qquad\quad-2\frac{E+\mN}{W}\bigg[\frac{\mN(s-a)}{p_t'^2}+W-\mN\bigg]\delta_{l0}\bigg\},
\end{align}
where
\beq
Z_l(W,t')=\frac{1}{W\qq^2}\Big\{(E+\mN)(W-\mN)Q_l(x_t)+(E-\mN)(W+\mN)Q_{l+1}(x_t)\Big\}.
\eeq
In particular, they behave asymptotically according to the general relations
\begin{align}
\label{GHasymptotics}
G_{lJ}(W,t')&=\Order\big(\qq^{2l}\big),   &H_{lJ}(W,t')&=\Order\big(\qq^{2l}\big)  &&\text{for }\;|\qq|\rightarrow 0 ,\notag\\
G_{lJ}(-W,t')&=\Order\big(\qq^{2l+2}\big),  &H_{lJ}(-W,t')&=\Order\big(\qq^{2l+2}\big)  &&\text{for }\;|\qq|\rightarrow 0,\notag\\
G_{lJ}(W,t')&=\Order(1),   &H_{lJ}(W,t')&=\Order(1)  &&\text{for }\;q_t'\rightarrow 0,\notag\\
G_{lJ}(W,t')&=\Order\big(p_t'^{-2}\big),  &H_{lJ}(W,t')&=\Order\big(p_t'^{-2}\big) &&\text{for }\;p_t'\rightarrow 0,\notag\\
G_{lJ}(W,t')&=\Order\big(t'^{J-l-2}\big),  &H_{lJ}(W,t')&=\Order\big(t'^{J-l-2}\big)  &&\text{for }\;t'\rightarrow\infty.
\end{align}

\subsection{$t$-channel projection}
\label{app:kernels_tchannel}

\subsubsection{Nucleon pole}

The $t$-channel projection of the nucleon pole terms reads
\beq
\label{hatNJpm}
\tilde N^J_+(t)=\frac{g^2}{4\pi}\mN\bigg\{\frac{\tilde yQ_J(\tilde y)}{(p_tq_t)^{J}}-\delta_{J0}\bigg\},\qquad
\tilde N^J_-(t)=\frac{g^2}{4\pi}\frac{\sqrt{J(J+1)}}{2J+1}\frac{Q_{J-1}(\tilde y)-Q_{J+1}(\tilde y)}{(p_tq_t)^{J}},
\eeq
where
\beq
\tilde y=\frac{t-2\mpi^2}{4p_tq_t}.
\eeq
Although $\tilde y$ diverges at $\tpi$ and $\tN$, $\tilde N^J_\pm(t)$ itself remains finite. Expressions that allow for a stable numerical evaluation can be obtained by invoking the asymptotic form of the Legendre functions $Q_l(z)$ for $|z|\to\infty$~\cite{Bateman:1953}
\beq
Q_l(z)\sim\frac{2^l(l!)^2}{(2l+1)!}z^{-(l+1)},
\eeq
which leads to
\begin{align}
\label{tchannel_pole_asym}
\tilde N^J_+(t)&=
\frac{g^2}{4\pi}\frac{J!}{(2J+1)!!}\mN\bigg\{\left(\frac{4}{t-2\mpi^2}\right)^J-\delta_{J0}\bigg\}+\Order\big(p_t^2q_t^2\big),\notag\\
\tilde N^J_-(t)&=
\frac{g^2}{4\pi}\frac{J!}{(2J+1)!!}\sqrt{\frac{J+1}{J}}\left(\frac{4}{t-2\mpi^2}\right)^J+\Order\big(p_t^2q_t^2\big).
\end{align}
In contrast to the ostensible divergences for $p_tq_t\to 0$, the pole terms do involve a branch-point singularity at $\tpi-(\mpi^2/\mN)^2\approx 3.98\mpi^2$, where the branch cut of $Q_{J}(\tilde y)$ starts.

\subsubsection{$s$-channel}

Introducing
\beq
\tilde\psi\big[a_{kn}\big|d(W')\big]=d(W')a_{k,n+1}+d(-W')a_{kn}, \qquad \tilde\eta_J=\frac{2W'}{(p_tq_t)^{J-1}},
\eeq
in analogy to~\eqref{psidef}, the $s$-channel kernel functions become
\begin{align}
\label{tskernelfunctions}
\tilde G_{Jl}(t,W')&=\tilde\eta_J\left\{-\frac{p_t}{q_t}\tilde\psi\bigg[\tilde A_{Jl}\bigg|\frac{W'+\mN}{E'+\mN}\bigg]
+\mN\tilde\psi\bigg[\tilde B_{Jl}\bigg|\frac{1}{E'+\mN}\bigg]\right\},\notag\\
\tilde H_{Jl}(t,W')&=\tilde\eta_J\frac{\sqrt{J(J+1)}}{2J+1}\tilde\psi\bigg[\tilde C_{Jl}\bigg|\frac{1}{E'+\mN}\bigg],
\end{align}
with angular kernel functions
\beq
\tilde A_{Jl}=\frac{1}{p_tq_t}P_{l}'(\tilde z_s)Q_J(\tilde x_t)-\bar A_{Jl},\qquad
\tilde C_{Jl}=\tilde A_{J-1,l}-\tilde A_{J+1,l},\qquad
\tilde B_{Jl}=\frac{1}{p_tq_t}P_{l}'(\tilde z_s)\tilde x_tQ_J(\tilde x_t)-\bar B_{Jl},
\eeq
their polynomial parts
\begin{align}
\label{ABCbardef}
\bar A_{Jl}&=\frac{1}{2}\int\limits^1_{-1}\diff z_t\,P_J(z_t)\left\{\frac{1}{p_tq_t}\frac{P_l'(\tilde z_s)-P_l'(z_s')}{\tilde x_t-z_t}+\frac{1\pm1}{2(s'-a)}P_l'(z_s')\right\},\notag\\
\bar B_{Jl}&=\frac{1}{2}\int\limits^1_{-1}\diff z_t\,P_J(z_t)\left\{\frac{1}{p_tq_t}\frac{\tilde x_tP_l'(\tilde z_s)-z_tP_l'(z_s')}{\tilde x_t-z_t}+\frac{1\mp1}{2(s'-a)}z_tP_l'(z_s')\right\},
\end{align}
where the upper/lower sign refers to even/odd $J$, and
\begin{align}
\label{zsprimetozt}
z_s'&=\frac{z_t^2-\tilde\delta}{\tilde\gamma}, &
 \tilde\gamma&=\frac{\qq'^2(s'-a)}{2p_t^2q_t^2},&
\tilde\delta&=\frac{(t-\Sigma+2a)^2-4(s'-a)(2\qq'^2+\Sigma-s'-a)}{16p_t^2q_t^2},\notag\\
\tilde x_t&=\frac{t+2s'-\Sigma}{4p_tq_t}, & \tilde z_s&=\frac{\tilde x_t^2-\tdelta}{\tgamma}=1+\frac{t}{2\qq'^2}.
\end{align}
These kernel functions fulfill the asymptotic relations
\begin{align}
\label{tildeGHasymptotics}
\tilde G_{Jl}(t,W')&=\Order(1), &\tilde H_{Jl}(t,W')&=\Order(1) &&\text{for }\;p_tq_t\rightarrow 0,\notag\\
\tilde G_{Jl}(t,W')&=\Order\big(\qq'^{-2l}\big),&\tilde H_{Jl}(t,W')&=\Order\big(\qq'^{-2l}\big) &&\text{for }\;|\qq'|\rightarrow 0,\notag\\
\tilde G_{Jl}(t,-W')&=\Order\big(\qq'^{-2l-2}\big), &\tilde H_{Jl}(t,-W')&=\Order\big(\qq'^{-2l-2}\big) &&\text{for }\;|\qq'|\rightarrow 0,\notag\\
\tilde G_{Jl}(t,W')&=\Order\big(\qq'^{-2J}\big), &\tilde H_{Jl}(t,W')&=\Order\big(\qq'^{-2J}\big)&&\text{for }\;|\qq'|\rightarrow\infty.
\end{align}
In particular, the finite pieces for $p_tq_t\to 0$ can be derived along the lines that led to~\eqref{tchannel_pole_asym}.  

\subsubsection{$t$-channel}

The $t$-channel kernel functions follow from
\beq
\begin{pmatrix}\tilde K^1_{JJ'}(t,t')&\tilde K^2_{JJ'}(t,t')\\0&\tilde K^3_{JJ'}(t,t')\end{pmatrix}
 =\frac{\zeta_{JJ'}}{t'-t}\begin{pmatrix}u_{JJ'}&v_{JJ'}\\0&w_{JJ'}\end{pmatrix},\qquad
\zeta_{JJ'}=(2J'+1)\frac{(p_t'q_t')^{J'-1}}{(p_tq_t)^{J-1}},
\eeq
with angular kernels for even $J$ and $J'$
\begin{align}
\label{uvw_even}
u_{JJ'}&=\frac{p_tq_t'}{q_tp_t'}\int\limits_0^1\diff z_t\;P_J(z_t)P_{J'}(z_t'),\qquad
v_{JJ'}=\frac{\mN}{\sqrt{J'(J'+1)}}\frac{p_t}{q_tp_t'q_t'}\int\limits^1_0\diff z_t\;P_J(z_t)\Big\{q_t^2z_t^2-q_t'^2z_t'^2\Big\}\frac{P_{J'}'(z_t')}{z_t'},\notag\\
w_{JJ'}&=\frac{1}{2J+1}\sqrt{\frac{J(J+1)}{J'(J'+1)}}\frac{p_tq_t}{p_t'q_t'}\int\limits^1_0\diff z_t\;\Big\{P_{J-1}(z_t)-P_{J+1}(z_t)\Big\}z_t\frac{P_{J'}'(z_t')}{z_t'},
\end{align}
and for odd $J$ and $J'$
\begin{align}
\label{uvw_odd}
u_{JJ'}&=\frac{p_t^2}{p_t'^2}\int\limits_0^1\diff z_t\;P_J(z_t)z_t\frac{P_{J'}(z_t')}{z_t'}, \qquad
v_{JJ'}=\frac{\mN}{\sqrt{J'(J'+1)}}\bigg\{1-\frac{p_t^2}{p_t'^2}\bigg\}\int\limits^1_0\diff z_t\;P_J(z_t)z_tP_{J'}'(z_t'),\notag\\
w_{JJ'}&=\frac{1}{2J+1}\sqrt{\frac{J(J+1)}{J'(J'+1)}}\int\limits^1_0\diff z_t\;\Big\{P_{J-1}(z_t)-P_{J+1}(z_t)\Big\}P_{J'}'(z_t'),
\end{align}
where the angles are related by
\beq
\label{tildealphabetadef}
z_t'=\sqrt{\tilde\alpha z_t^2+\tilde\beta}, \qquad \tilde\alpha=\frac{p_t^2q_t^2}{p_t'^2q_t'^2},\qquad
\tilde\beta=\frac{t'-t}{16p_t'^2q_t'^2}(t+t'-2\Sigma+4a).
\eeq
Again, only even powers of $z_t'^2$ appear in the angular integrals~\eqref{uvw_even} and \eqref{uvw_odd}.
Starting from these equations, one can show that for $J'<J$ all kernel functions vanish
\beq
 \label{vanishingtildeK}
\tilde K^1_{JJ'}(t,t')=\tilde K^2_{JJ'}(t,t')=\tilde K^3_{JJ'}(t,t')=0\qquad \text{for } J'<J,
\eeq
while the diagonal kernels fulfill the general relations
\beq
\tilde K_{JJ}^1(t,t')=\frac{p_t^2}{p_t'^2}\frac{1}{t'-t},\qquad
\tilde K_{JJ}^2(t,t')=\sqrt{\frac{J}{J+1}}\frac{\mN}{4p_t'^2},\qquad \tilde K_{JJ}^3(t,t')=\frac{1}{t'-t}.
\eeq
Finally, the non-vanishing, non-diagonal kernel functions with $J\leq3$ and $J'\leq3$ read
\begin{align}
\label{tildeK02}
\tilde K_{02}^1(t,t')&=\frac{5}{16}\frac{p_t^2}{p_t'^2}\Big\{t+t'-2\Sigma+6a\Big\},\qquad\;\,
\tilde K_{02}^2(t,t')=\frac{5\mN}{16\sqrt{6}}\frac{p_t^2}{p_t'^2}\Big\{4q_t^2-3(t+t'-2\Sigma+4a)\Big\}\notag,\\
\tilde K_{13}^1(t,t')&=\frac{7}{48}\frac{p_t^2}{p_t'^2}\Big\{t+t'-2\Sigma+10a\Big\},\qquad 
\tilde K_{13}^3(t,t')=\frac{7}{8\sqrt{6}}\Big\{t+t'-2\Sigma+5a\Big\},\notag\\
\tilde K_{13}^2(t,t')&=\frac{7\mN}{64\sqrt{3}}\frac{1}{p_t'^2}\Big\{8p_t^2q_t^2+(t'-t)(t+t'-2\Sigma+5a)\Big\}.
\end{align} 
The general asymptotic properties of the kernels are given in Table~\ref{table:asym_K}. Note, however, that exceptionally
\beq
\label{tilde2K02asymptotics}
\tilde K^2_{02}(t,t')=\Order\big(p_t^2\big) \qquad \text{for }\;p_t \rightarrow 0, \qquad\quad \tilde K^2_{02}(t,t')=\Order(1) \qquad \text{for }\;t'\rightarrow\infty.
\eeq

\begin{table}
\centering
\renewcommand{\arraystretch}{1.3}
\begin{tabular}{ccccccc}
\toprule
& $p_t\to 0$ & $q_t\to 0$ & $t\to\infty$ & $p_t'\to 0$ & $q_t'\to 0$ & $t'\to\infty$ \\\midrule
$\tilde K^1_{JJ'}(t,t')$ & $\Order\big(p_t^2\big)$ & $\Order(1)$ & $\Order\big(t^{J'-J}\big)$ & $\Order\big(p_t'^{-2}\big)$ & $\Order(1)$ & $\Order\big(t'^{J'-J-2}\big)$  \\
$\tilde K^2_{JJ'}(t,t')$ & $\Order(1)$ & $\Order(1)$ & $\Order\big(t^{J'-J}\big)$ & $\Order\big(p_t'^{-2}\big)$ & $\Order(1)$ & $\Order\big(t'^{J'-J-1}\big)$  \\
$\tilde K^3_{JJ'}(t,t')$ & $\Order(1)$ & $\Order(1)$ & $\Order\big(t^{J'-J-1}\big)$ & $\Order(1)$ & $\Order(1)$ & $\Order\big(t'^{J'-J-1}\big)$  \\
\bottomrule
\end{tabular}
\caption{Asymptotic properties of the $t$-channel kernel functions.}
\label{table:asym_K}
\renewcommand{\arraystretch}{1.0}
\end{table}

\section{Subtractions}
\label{app:subtractions}

\subsection{Sum rules for subthreshold parameters}
\label{app:subthr_sum_rules}

The sum rules based on HDRs for all the subthreshold parameters relevant for the matching to ChPT are
\begin{align}
\label{sum_rules_subthr}
d_{00}^+&=-\frac{g^2}{\mN}+\frac{1}{\pi}\int\limits_{s_+}^\infty\diff s'\,h_0(s')\ste{\Im A^+}
+\frac{1}{\pi}\int\limits_{\tpi}^\infty\frac{\diff t'}{t'}\ste{\Im A^+},\notag\\
b_{00}^-&=\frac{g^2}{2\mN^2}+\frac{1}{\pi}\int\limits_{s_+}^\infty\diff s'\,h_0(s')\ste{\Im B^-}
+\frac{1}{\pi}\int\limits_{\tpi}^\infty\frac{\diff t'}{t'}\ste{\Im B^-},\notag\\
d_{01}^+&=\frac{1}{\pi}\int\limits_{s_+}^\infty\diff s'\bigg\{h_0(s')\ste{\partial_t\Im A^+}-h_2^0(s')\ste{\Im A^+}\bigg\}
+\frac{1}{\pi}\int\limits_{\tpi}^\infty\frac{\diff t'}{t'}\bigg\{\ste{\partial_t\Im A^+}+\frac{1}{t'}\ste{\Im A^+}\bigg\},\notag\\
a_{00}^-&=\frac{4\mN}{\pi}\int\limits_{s_+}^\infty\diff s'\,h_2^0(s')\ste{\Im A^-}
+\frac{1}{\pi}\int\limits_{\tpi}^\infty\frac{\diff t'}{t'}\ste{\Im A^-/\nu'},\notag\\
b_{00}^+&=\frac{4\mN}{\pi}\int\limits_{s_+}^\infty\diff s'\,h_2^0(s')\ste{\Im B^+}
+\frac{1}{\pi}\int\limits_{\tpi}^\infty\frac{\diff t'}{t'}\ste{\Im B^+/\nu'},\notag\\
b_{01}^-&=\frac{1}{\pi}\int\limits_{s_+}^\infty\diff s'\bigg\{h_0(s')\ste{\partial_t\Im B^-}-h_2^0(s')\ste{\Im B^-}\bigg\}
+\frac{1}{\pi}\int\limits_{\tpi}^\infty\frac{\diff t'}{t'}\bigg\{\ste{\partial_t\Im B^-}+\frac{1}{t'}\ste{\Im B^-}\bigg\},\notag\\
a_{10}^+&=\frac{1}{\pi}\int\limits_{s_+}^\infty\diff s'\bigg\{h_0(s')\ste{\partial_{\nu^2}\Im A^+}+8\mN^2h_3^0(s')\ste{\Im A^+}\bigg\}
+\frac{1}{\pi}\int\limits_{\tpi}^\infty\frac{\diff t'}{t'}\ste{\partial_{\nu^2}\Im A^+},\notag\\
a_{01}^-&=\frac{4\mN}{\pi}\int\limits_{s_+}^\infty\diff s'\,\bigg\{h_2^0(s')\ste{\partial_t\Im A^-}-h_3^0(s')\ste{\Im A^-}\bigg\}
+\frac{1}{\pi}\int\limits_{\tpi}^\infty\frac{\diff t'}{t'}\bigg\{\ste{\partial_t\Im A^-/\nu'}+\frac{1}{t'}\ste{\Im A^-/\nu'}\bigg\},\notag\\
a_{10}^-&=\frac{4\mN}{\pi}\int\limits_{s_+}^\infty\diff s'\,\bigg\{h^0_2(s')\ste{\partial_{\nu^2}\Im A^-}+4\mN^2h^0_4(s')\ste{\Im A^-}\bigg\}
+\frac{1}{\pi}\int\limits_{\tpi}^\infty\frac{\diff t'}{t'}\ste{\partial_{\nu^2}\Im A^-/\nu'},\notag\\
b_{10}^-&=\frac{1}{\pi}\int\limits_{s_+}^\infty\diff s'\bigg\{h_0(s')\ste{\partial_{\nu^2}\Im B^-}+8\mN^2h_3^0(s')\ste{\Im B^-}\bigg\}
+\frac{1}{\pi}\int\limits_{\tpi}^\infty\frac{\diff t'}{t'}\ste{\partial_{\nu^2}\Im B^-},\notag\\
b_{01}^+&=\frac{4\mN}{\pi}\int\limits_{s_+}^\infty\diff s'\,\bigg\{h_2^0(s')\ste{\partial_t\Im B^+}-h_3^0(s')\ste{\Im B^+}\bigg\}
+\frac{1}{\pi}\int\limits_{\tpi}^\infty\frac{\diff t'}{t'}\bigg\{\ste{\partial_t\Im B^+/\nu'}+\frac{1}{t'}\ste{\Im B^+/\nu'}\bigg\},\notag\\
b_{10}^+&=\frac{4\mN}{\pi}\int\limits_{s_+}^\infty\diff s'\,\bigg\{h^0_2(s')\ste{\partial_{\nu^2}\Im B^+}+4\mN^2h^0_4(s')\ste{\Im B^+}\bigg\}
+\frac{1}{\pi}\int\limits_{\tpi}^\infty\frac{\diff t'}{t'}\ste{\partial_{\nu^2}\Im B^+/\nu'},\notag\\
2a_{20}^+&=\frac{1}{\pi}\int\limits_{s_+}^\infty\diff s'\bigg\{h_0(s')\ste{\partial^2_{\nu^2}\Im A^+}+16\mN^2h_3^0(s')\ste{\partial_{\nu^2}\Im A^+}+64\mN^4h_5^0(s')\ste{\Im A^+}\bigg\}\notag\\
&+\frac{1}{\pi}\int\limits_{\tpi}^\infty\frac{\diff t'}{t'}\ste{\partial^2_{\nu^2}\Im A^+},\notag\\
a_{11}^+&=\frac{1}{\pi}\int\limits_{s_+}^\infty\diff s'\bigg\{h_0(s')\ste{\partial_{\nu^2}\partial_t\Im A^+}+8\mN^2h_3^0(s')\ste{\partial_t\Im A^+}-h_2^0(s')\ste{\partial_{\nu^2}\Im A^+}-12\mN^2h_4^0(s')\ste{\Im A^+}\bigg\}\notag\\
&+\frac{1}{\pi}\int\limits_{\tpi}^\infty\frac{\diff t'}{t'}\bigg\{\ste{\partial_{\nu^2}\partial_t\Im A^+}+\frac{1}{t'}\ste{\partial_{\nu^2}\Im A^+}\bigg\},\notag\\
2a_{02}^+&=\frac{1}{\pi}\int\limits_{s_+}^\infty\diff s'\bigg\{h_0(s')\ste{\partial^2_t\Im A^+}-2h_2^0(s')\ste{\partial_t\Im A^+}+h_3^0(s')\ste{\Im A^+}\bigg\}\notag\\
&+\frac{1}{\pi}\int\limits_{\tpi}^\infty\frac{\diff t'}{t'}\bigg\{\ste{\partial^2_t\Im A^+}+\frac{2}{t'}\ste{\partial_t\Im A^+}+\frac{2}{t'^2}\ste{\Im A^+}\bigg\},
\end{align}
where we have introduced the notation
\beq
\label{def_h0}
h_0(s')=\frac{2}{s'-s_0}-\frac{1}{s'-a},\qquad h_n^0(s')=\frac{1}{(s'-s_0)^n}.
\eeq
In~\eqref{sum_rules_subthr} the dependence of the amplitudes on the internal kinematics $(s',t')$ is suppressed  and the subscript $(0,0)$ indicates evaluation at $(\nu=0,t=0)$. To this end, one may use the relations
\begin{align}
\label{zsdeltzs}
\ste{z_s'}&=1-\frac{(s'-s_0)^2}{2\qq'^2(s'-a)}, & 
\ste{\partial_tz_s'}&=\frac{s_0-a}{2\qq'^2(s'-a)},&
\ste{\partial_{\nu^2}z_s'}&=\frac{2\mN^2}{\qq'^2(s'-a)},\notag\\
\ste{\partial^2_{\nu^2}z_s'}&=0, &
\ste{\partial_{\nu^2}\partial_tz_s'}&=0, &
\ste{\partial^2_tz_s'}&=-\frac{1}{4\qq'^2(s'-a)},\notag\\
\ste{z_t'^2}&=1+\frac{at'-4\mN^2\mpi^2}{4p_t'^2q_t'^2}, & 
\ste{\partial_tz_t'^2}&=\frac{s_0-a}{4p_t'^2q_t'^2}, &
\ste{\partial_{\nu^2}z_t'^2}&=\frac{\mN^2}{p_t'^2q_t'^2},\notag\\
\ste{\partial^2_{\nu^2}z_t'^2}&=0, &
\ste{\partial_{\nu^2}\partial_tz_t'^2}&=0, &
\ste{\partial^2_tz_t'^2}&=-\frac{1}{8p_t'^2q_t'^2}.
\end{align} 

\subsection{Subtracted kernel functions}
\label{app:kernel_subtractions}

In this appendix we list the additional contributions from subtraction terms as well as the modifications to the kernel functions required when introducing subtractions. We do not provide an exhaustive list of all possible variants (depending on which subthreshold parameters are included, see Sect.~\ref{sec:piN_subtractions}), but present the version actually used in the calculation. 

\subsubsection{$s$-channel projection}

For the $s$-channel we include all subthreshold parameters listed in Sect.~\ref{sec:piN_subtractions}.
For convenience, the contributions to the partial waves that originate from these subtraction constants
are collected by modifying the pole terms according to
\beq
N_{l+}^I(W)\to N_{l+}^I(W)+\Delta N_{l+}^I(W),\qquad
\Delta N_{l+}^I(W)=\widehat{\Delta N}_{l}^I(W)-\widehat{\Delta N}_{l+1}^I(-W),
\eeq
and
\begin{align}
\widehat{\Delta N}_{l}^+(W)&=\frac{E+\mN}{4\pi W}\Bigg\{\frac{\delta_{l0}}{2}\bigg(d_{00}^++\frac{g^2}{\mN}\bigg)
-d_{01}^+\qq^2\chi_l^t
+\frac{b_{00}^+\chi_l^{\nu}}{4 \mN}(W-\mN)
+\frac{a_{10}^+\chi_l^{\nu^2}}{8\mN^2}\Bigg\},\notag\\
\widehat{\Delta N}_{l}^-(W)&=\frac{E+\mN}{4\pi W}\Bigg\{
(W-\mN)\Bigg[\frac{\delta_{l0}}{2}\bigg(b_{00}^--\frac{g^2}{2\mN^2}\bigg)
-b_{01}^-\qq^2\chi_l^t
+\frac{b_{10}^-\chi_l^{\nu^2}}{8 \mN^2}\Bigg]
+\frac{a_{00}^-\chi_l^{\nu}}{4 \mN}
-\frac{a_{01}^-\qq^2\chi_l^{\nu t}}{2 \mN}
+\frac{a_{10}^-\chi_l^{\nu^3}}{16 \mN^3}
\Bigg\},
\end{align}
with 
\begin{align}
\chi_l^{\nu}&=(s-s_0-\qq^2)\delta_{l0}+\frac{\qq^2}{3}\delta_{l1},\qquad\hspace{-2pt}
\chi_l^{\nu t}=\bigg(s-s_0-\frac{4}{3}\qq^2\bigg)\delta_{l0}+(2\qq^2-s+s_0)\frac{\delta_{l1}}{3}-\frac{2}{15}\qq^2\delta_{l2},\notag\\
\chi_l^{\nu^2}&=\bigg((s-s_0-\qq^2)^2+\frac{\qq^4}{3}\bigg)\delta_{l0}+\frac{2}{3}\qq^2(s-s_0-\qq^2)\delta_{l1}+\frac{2}{15}\qq^4\delta_{l2},\qquad
\chi_l^t=\delta_{l0}-\frac{\delta_{l1}}{3},\notag\\
\chi_l^{\nu^3}&=(s-s_0-\qq^2)\Big((s-s_0-\qq^2)^2+\qq^4\Big)\delta_{l0}+\qq^2\bigg((s-s_0-\qq^2)^2+\frac{\qq^4}{5}\bigg)\delta_{l1}
+\frac{2}{5}\qq^4(s-s_0-\qq^2)\delta_{l2}+\frac{2}{35}\qq^6\delta_{l3}.
\end{align}
Next, the $s$-channel kernels become
\beq
K_{ll'}^I(W,W')\to K_{ll'}^I(W,W')+\Delta K_{ll'}^I(W,W'),\qquad
\Delta K_{ll'}^I(W,W')=\widehat{\Delta K}_{ll'}^I(W,W')-\widehat{\Delta K}_{l+1,l'}^I(-W,W'),
\eeq
with
\begin{align}
\widehat{\Delta K}_{ll'}^+(W,W')&=\frac{E+\mN}{4\pi W}\Bigg[-W'h_0(s')\ste{S^1_{l'+1,l'}(W',z_s')}\delta_{l0}\notag\\
&+2\qq^2W'\chi_l^t\bigg\{h_0(s')\ste{\partial_tS^1_{l'+1,l'}(W',z_s')}-h^0_2(s')\ste{S^1_{l'+1,l'}(W',z_s')}\bigg\}\notag\\
&-2W'\chi_l^{\nu}(W-\mN)h^0_2(s')\ste{S^2_{l'+1,l'}(W',z_s')}\notag\\
&-\frac{W'}{4 \mN^2}\chi_l^{\nu^2}
\bigg\{h_0(s')\ste{\partial_{\nu^2}S^1_{l'+1,l'}(W',z_s')}+8\mN^2h^0_3(s')\ste{S^1_{l'+1,l'}(W',z_s')}\bigg\}\Bigg],\notag\\
\widehat{\Delta K}_{ll'}^-(W,W')&=\frac{E+\mN}{4\pi W}\Bigg[-W'(W-\mN)h_0(s')\ste{S^2_{l'+1,l'}(W',z_s')}\delta_{l0}\notag\\
&+2\qq^2W'\chi_l^t(W-\mN)
\bigg\{h_0(s')\ste{\partial_tS^2_{l'+1,l'}(W',z_s')}-h^0_2(s')\ste{S^2_{l'+1,l'}(W',z_s')}\bigg\}\notag\\
&-2W'\chi_l^{\nu}h^0_2(s')\ste{S^1_{l'+1,l'}(W',z_s')}
-4\qq^2 W'\chi_l^{\nu t}
\bigg\{h^0_3(s')\ste{S^1_{l'+1,l'}(W',z_s')}-h^0_2(s')\ste{\partial_tS^1_{l'+1,l'}(W',z_s')}\bigg\}\notag\\
&-\frac{W'}{2 \mN^2}\chi_l^{\nu^3}
\bigg\{h^0_2(s')\ste{\partial_{\nu^2}S^1_{l'+1,l'}(W',z_s')}+4\mN^2h^0_4(s')\ste{S^1_{l'+1,l'}(W',z_s')}\bigg\}\notag\\
&-\frac{W'}{4 \mN^2}\chi_l^{\nu^2}(W-\mN)
\bigg\{h_0(s')\ste{\partial_{\nu^2}S^2_{l'+1,l'}(W',z_s')}+8\mN^2h^0_3(s')\ste{S^2_{l'+1,l'}(W',z_s')}\bigg\}\Bigg].
\end{align}
Similarly, we define
\beq
G_{lJ}(W,t')\to G_{lJ}(W,t')+\Delta G_{lJ}(W,t'),\qquad
\Delta G_{lJ}(W,t')=\widehat{\Delta G}_{lJ}(W,t')-\widehat{\Delta G}_{l+1,J}(-W,t'),
\eeq
and accordingly for $H_{lJ}(W,t')$, which yields for even $J$
\begin{align}
\widehat{\Delta G}_{lJ}(W,t')&=\frac{E+\mN}{2W}(2J+1)\frac{(p_t'q_t')^J}{t'p_t'^2}\Bigg\{\ste{P_J(z_t')}\delta_{l0}
-2\qq^2
\chi_l^t\ste{\frac{1}{t'}P_J(z_t')+\partial_tP_J(z_t')}
+\frac{\chi_l^{\nu^2}}{4\mN^2}\ste{\partial_{\nu^2}P_J(z_t')}\Bigg\},\notag\\
\widehat{\Delta H}_{lJ}(W,t')&=-\frac{E+\mN}{2W}\frac{2J+1}{\sqrt{J(J+1)}}\frac{(p_t'q_t')^J}{t'p_t'^2}\Bigg\{\mN\ste{z_t'P'_J(z_t')}\delta_{l0}+\frac{W-\mN}{2q_t'^2}\chi_l^{\nu}\ste{\frac{P'_J(z_t')}{z_t'}}\notag\\
&-2\mN \qq^2\chi_l^t\ste{\frac{1}{t'}z_t'P'_J(z_t')+\partial_t(z_t'P'_J(z_t'))}
+\frac{\chi_l^{\nu^2}}{4\mN}\ste{\partial_{\nu^2}z_t'P'_J(z_t')}\Bigg\},
\end{align}
and for odd $J$
\begin{align}
 \widehat{\Delta G}_{lJ}(W,t')&=\frac{E+\mN}{2W}(2J+1)\frac{(p_t'q_t')^{J-1}}{t'p_t'^2}\Bigg\{\frac{\chi_l^\nu}{2}\ste{\frac{P_J(z_t')}{z_t'}}
 -\qq^2\chi_l^{\nu t}\ste{\frac{1}{t'}\frac{P_J(z_t')}{z_t'}+\partial_t\frac{P_J(z_t')}{z_t'}}
 +\frac{\chi_l^{\nu^3}}{8\mN^2}\ste{\partial_{\nu^2}\frac{P_J(z_t')}{z_t'}}\Bigg\},\notag\\
\widehat{\Delta H}_{lJ}(W,t')&=-\frac{E+\mN}{2W}\frac{2J+1}{\sqrt{J(J+1)}}\frac{(p_t'q_t')^{J-1}}{t'p_t'^2}\Bigg\{
(W-\mN)p_t'^2\ste{P'_J(z_t')}\delta_{l0}+\frac{\mN}{2}\chi_l^{\nu}\ste{P'_J(z_t')}\notag\\
&-\mN \qq^2\bigg[\chi_l^{\nu t}+\frac{W-\mN}{\mN}2p_t'^2\chi_l^t\bigg]\ste{\frac{1}{t'}P_J'(z_t')+\partial_tP_J'(z_t')}
+\frac{1}{8\mN}\ste{\partial_{\nu^2}P'_J(z_t')}
\bigg[\chi_l^{\nu^3}+\frac{W-\mN}{\mN}2p_t'^2\chi_l^{\nu^2}\bigg]\Bigg\}.
\end{align}

\subsubsection{$t$-channel projection}

For the $t$-channel projection we do not consider all subtractions implemented for the $s$-channel case, since some subthreshold parameters do not further improve the convergence of the ensuing dispersion relations for the $t$-channel partial waves. In this appendix, we summarize the version used for the $P$- and $D$-waves (with all parameters but $a_{10}^-$ and $b_{10}^-$). For the $S$-wave, one can use a variant with even less subtractions, see~\cite{Hoferichter:2012wf}.

The subthreshold-parameter contributions to the $t$-channel amplitudes are included by modifying the nucleon pole terms according to
\beq
\tilde N^J_\pm(t)\to\tilde N^J_\pm(t)+\Delta\tilde N^J_\pm(t),
\eeq
with
\begin{align}
\label{tchannel_pole_corr}
\Delta\tilde N^J_+(t)&=-\frac{p_t^2}{4\pi}\bigg(d_{00}^++\frac{g^2}{\mN}+d_{01}^+t-b_{00}^+\frac{q_t^2}{3}\bigg)\delta_{J0}
+\frac{\mN}{12\pi}\bigg(b_{00}^--\frac{g^2}{2\mN^2}+b_{01}^-t-\big(a_{00}^-+a_{01}^-t\big)\frac{p_t^2}{\mN^2}\bigg)\delta_{J1}
+\frac{b_{00}^+}{30\pi}\delta_{J2}\notag\\
&-\frac{p_t^2}{12\pi \mN^2}a_{10}^+\Big(p_t^2q_t^2\delta_{J0}+\frac{2}{5}\delta_{J2}\Big),\notag\\
\Delta\tilde N^J_-(t)&=\frac{\sqrt{2}}{12\pi}\bigg(b_{00}^--\frac{g^2}{2\mN^2}+b_{01}^-t\bigg)\delta_{J1}
+\frac{\sqrt{6}}{60\pi\mN}b_{00}^+\delta_{J2}.
\end{align}
The subtracted versions of $\tilde G_{Jl}(t,W')$ and $\tilde H_{Jl}(t,W')$ can be expressed by redefining the polynomial parts of the pertinent angular kernels according to
\beq
\bar A_{Jl}\to\bar A_{Jl}+\Delta\bar A_{Jl},
\eeq
and analogously for $\bar B_{Jl}$, $\bar C_{Jl}$. We find
\begin{align}
\Delta\bar A_{Jl}&=\bigg\{\Big(h_0(s')-t\,h^0_2(s')\Big)\ste{P_l'(z_s')}+t\,h_0(s')\ste{\partial_tP_l'(z_s')}\bigg\}\delta_{J0}\notag\\
&+\frac{4}{3}p_tq_t\bigg\{\Big(h^0_2(s')-t\, h^0_3(s')\Big)\ste{P_l'(z_s')}
+t\, h^0_2(s')\ste{\partial_tP_l'(z_s')}\bigg\}\delta_{J1}\notag\\
&+\frac{p_t^2q_t^2}{3\mN^2}
\bigg\{h_0(s')\ste{\partial_{\nu^2}P_l'(z_s')}+8\mN^2h^0_3(s')\ste{P_l'(z_s')}\bigg\}\Big(\delta_{J0}+\frac{2}{5}\delta_{J2}\Big),\notag\\
\Delta\bar B_{Jl}&=\frac{1}{3}\bigg\{\Big(h_0(s')-t\,h^0_2(s')\Big)\ste{P_l'(z_s')}
+t\,h_0(s')\ste{\partial_tP_l'(z_s')}\bigg\}\delta_{J1}
+\frac{4}{3}p_tq_th^0_2(s')\ste{P_l'(z_s')}\Big(\delta_{J0}+\frac{2}{5}\delta_{J2}\Big),\notag\\
\Delta\bar C_{Jl}&=\bigg\{\Big(h_0(s')-t\,h^0_2(s')\Big)\ste{P_l'(z_s')}+t\,h_0(s')\ste{\partial_tP_l'(z_s')}\bigg\}\delta_{J1}+\frac{4}{3}p_tq_th^0_2(s')\ste{P_l'(z_s')}\delta_{J2}.
\end{align}
For the $t$-channel we list the final kernel functions explicitly for $J,J'\leq 3$
\begin{align}
 \tilde K^1_{00}(t,t')&= \tilde K^1_{11}(t,t')=\frac{p_t^2}{p_t'^2}\frac{t^2}{t'^2(t'-t)},\qquad
\tilde K^1_{22}(t,t')=\frac{p_t^2}{p_t'^2}\frac{t}{t'(t'-t)},\qquad
\tilde K^1_{02}(t,t')=\frac{5}{32}\frac{p_t^2}{p_t'^2}\frac{t(t_Nt_\pi-t t')}{t'^2},\notag\\
\tilde K^1_{13}(t,t')&=\frac{7}{32}\frac{p_t^2}{p_t'^2}\frac{1}{t'^2}\Big\{\tpi\tN(t+t')-tt'(\tpi+\tN)\Big\},\qquad
\tilde K^2_{11}(t,t')=\frac{t^2}{t'^2}\frac{\mN}{4\sqrt{2}p_t'^2},\qquad
\tilde K^2_{22}(t,t')=\frac{t}{t'}\sqrt{\frac{2}{3}}\frac{\mN}{4p_t'^2},\notag\\
\tilde K^2_{02}(t,t')&=\frac{5\mN}{4\sqrt{6}}\frac{p_t^2q_t^2}{p_t'^2}\frac{t}{t'},\qquad
\tilde K^2_{13}(t,t')=\frac{7\mN}{16\sqrt{3}}\bigg\{3\frac{p_t^2q_t^2}{p_t'^2}-\frac{t^2}{t'^2}q_t'^2\bigg\},\notag\\
\tilde K^3_{11}(t,t')&=\frac{t^2}{t'^2}\frac{1}{t'-t},\qquad
\tilde K^3_{22}(t,t')=\frac{t}{t'}\frac{1}{t'-t},\qquad
\tilde K^3_{13}(t,t')=\frac{7}{8\sqrt{6}}\bigg\{\bigg(1+\frac{t}{t'}\bigg)\frac{\tN\tpi}{4t'}-\frac{t}{t'}s_0\bigg\}.
\end{align}

\section{Coupled-channel unitarity for $\boldsymbol{\pi\pi\to\bar N N}$}
\label{app:tchannel_unitarity}

In this appendix we derive the unitarity relation for a coupled system of $\pi\pi$, $\bar K K$, and $\bar N N$ states.
First, we collect all the partial-wave amplitudes involved in the full system. The precise definition of the $\pi\pi$ partial waves $t^{I_t}_J(t)$ is provided by
\beq
\label{pipiexpansion}
T^{I_t}(s,t)=32\pi\sum\limits_{J=0}^\infty(2J+1)t^{I_t}_J(t)P_J(\cos\theta^{\pi\pi}),
\eeq
with CMS scattering angle $\theta^{\pi\pi}$ and normalization
\beq
\frac{\diff\sigma^{I_t}_{\pi\pi\to\pi\pi}}{\diff\Omega}=\left|\frac{T^{I_t}(s,t)}{8\pi\sqrt{t}}\right|^2.
\eeq
Elastic unitarity then implies
\beq
\label{pipielunitrel}
\Im t^{I_t}_J(t)=\sigma^\pi_t\big|t^{I_t}_J(t)\big|^2\,\theta\big(t-\tpi\big), \qquad 
t^{I_t}_J(t)=\frac{e^{i\delta^{I_t}_J(t)}\sin\delta^{I_t}_J(t)}{\sigma^\pi_t}.
\eeq
The $t$-channel partial waves $f^J_{\pm}(t)$ are related to the helicity amplitudes $F_{\bar\lambda\lambda}(s,t)$ by~\cite{Frazer:1960zza}
\begin{align}
\label{barNNtopipiJWpwe}
F_{++}(s,t)&=F_{--}(s,t)=\frac{4\pi\sqrt{t}}{q_t}\sum\limits_{J=0}^\infty(2J+1)F^J_+(t)P_J(\cos\theta_t),\notag\\
F_{+-}(s,t)&=-F_{-+}(s,t)=\frac{4\pi\sqrt{t}}{q_t}\sum\limits_{J=1}^\infty\frac{2J+1}{\sqrt{J(J+1)}}F^J_-(t)\sin\theta_tP_J'(\cos\theta_t), \notag\\
F_+^J(t)&=\frac{q_t}{p_t}(p_tq_t)^J\frac{2}{\sqrt{t}}f_+^J(t),\qquad F_-^J(t)=\frac{q_t}{p_t}(p_tq_t)^Jf_-^J(t),
\end{align}
with normalization of the spin-averaged cross section
\beq
\label{barNNtopipidiffcrosssect}
\frac{\diff\bar\sigma_{\pi\pi\to\bar NN}}{\diff\Omega}=
\frac{p_t}{q_t}\sum\limits_{\bar\lambda,\lambda}\left|\frac{F_{\bar\lambda\lambda}(s,t)}{8\pi\sqrt{t}}\right|^2=
\frac{2p_t}{q_t}\left\{\left|\frac{F_{++}(s,t)}{8\pi\sqrt{t}}\right|^2+\left|\frac{F_{+-}(s,t)}{8\pi\sqrt{t}}\right|^2\right\}.
\eeq
The partial waves for $\pi\pi\to\bar K K$ are defined by
\beq
\label{piKexpansion}
G^{I_t}(s,t)=16\pi\sqrt{2}\sum\limits_{J=0}^\infty(2J+1)(k_tq_t)^Jg^{I_t}_J(t)P_J(\cos\theta_t^{\pi K}), \qquad k_t=\sqrt{\frac{t}{4}-\mK^2}=\frac{\sqrt{t}}{2}\sigma^K_t,
\eeq
and
\beq
\frac{\diff\sigma^{I_t}_{\pi\pi\to\bar KK}}{\diff\Omega}=\frac{k_t}{q_t}\left|\frac{G^{I_t}(s,t)}{8\pi\sqrt{t}}\right|^2.
\eeq
Finally, we need the $\bar K K$ partial waves $r^{I_t}_J(t)$
\beq
R^{I_t}(s,t)=16\pi\sum\limits_{J=0}^\infty(2J+1)r^{I_t}_J(t)P_J(\cos\theta_t^{\bar K K}),\qquad 
\frac{\diff\sigma^{I_t}_{\bar K K\to\bar KK}}{\diff\Omega}=\left|\frac{R^{I_t}(s,t)}{8\pi\sqrt{t}}\right|^2,
\eeq
 and the $\bar K K\to\bar N N$ amplitudes $h^J_{\pm}(t)$, the $KN$ analogs of $f^J_{\pm}(t)$ (similarly, $F^J_\pm(t)\to H^J_\pm(t)$ in the $K N$ system). Our conventions for these partial waves are similar to the $\pi N$ ones (see~\cite{Hoferichter:2012wf} for details), the important difference being that due to the different isospin of the kaon the isospin crossing coefficients~\eqref{crossingcoefficients} simply become $c_J\to c_J^{KN}=1/2$.

With the $T$-matrix elements $T_{11}=T_{\pi\pi\to\pi\pi}$, $T_{12}=T_{\bar KK\to\pi\pi}$, $T_{13}=T_{\bar NN\to\pi\pi}$ etc., unitarity in the multi-channel case requires
\beq
S_{fj}^{*}S_{ji}^{}=\delta_{fi}, \qquad S_{fi}=\delta_{fi}+iT_{fi}=\delta_{if}+iT_{if}=S_{if}.
\eeq
In particular, one finds
\beq
\label{stmatrixelemunitrel1}
|S_{11}|^2+|S_{12}|^2+|S_{13}|^2=1\quad\Rightarrow\quad 2\,\Im T_{11}=|T_{11}|^2+|T_{12}|^2+|T_{13}|^2,
\eeq
and
\beq
\label{stmatrixelemunitrel2}
S_{11}^{*}S_{13}^{}+S_{12}^{*}S_{23}^{}+S_{13}^{*}S_{33}^{}=0\quad\Rightarrow\quad
 2\,\Im T_{13}=T_{11}^{*}T_{13}^{}+T_{12}^{*}T_{23}^{}+T_{13}^{*}T_{33}^{}.
\eeq
Taking into account the different helicity projections, \eqref{stmatrixelemunitrel1} in the $t$-channel isospin basis becomes
\beq
\label{smatrixpwaunitrel}
\big|\big[S^{I_t}_J(t)\big]_{\pi\pi\to\pi\pi}\big|^2+\big|\big[S^{I_t}_J(t)\big]_{\pi\pi\to\bar KK}\big|^2
+2\left\{\big|\big[S^J_+(t)\big]^{I_t}_{\pi\pi\to\bar NN}\big|^2+\big|\big[S^J_-(t)\big]^{I_t}_{\pi\pi\to\bar NN}\big|^2\right\}=1,
\eeq
while the analog of~\eqref{selunitrel} for $\pi\pi$ scattering with $\pi\pi$, $\bar K K$, and $\bar N N$ intermediate states leads to
\begin{align}
\label{pipiextelunitrel}
\Im t^{I_t}_J(t)&=\sigma^\pi_t\big|t_J^{I_t}(t)\big|^2\,\theta\big(t-\tpi\big)+(k_tq_t)^{2J}\sigma^K_t\big|g_J^{I_t}(t)\big|^2\,\theta\big(t-\tK\big)
+\frac{t}{16q_t^2}\frac{\sigma^N_t}{c_J^2}\bigg\{\big|F_+^J(t)\big|^2+\big|F_-^J(t)\big|^2\bigg\}\theta\big(t-\tN\big).
\end{align}
In this way, the comparison of~\eqref{smatrixpwaunitrel} and~\eqref{pipiextelunitrel}, successively for $t<\tK$, $t<\tN$, and $t>\tN$ determines the relation between $S$-matrix elements and partial waves
\begin{align}
\big[S^{I_t}_J(t)\big]_{\pi\pi\to\pi\pi}&=1+i\frac{4q_t}{\sqrt{t}}t^{I_t}_J(t)\,\theta\big(t-\tpi\big), &
\big[S^{I_t}_J(t)\big]_{\pi\pi\to\bar KK}&=i\frac{4(k_tq_t)^{J+\frac{1}{2}}}{\sqrt{t}}g^{I_t}_J(t)\,\theta\big(t-\tK\big),\notag\\
\big[S^J_\pm(t)\big]^{I_t}_{\pi\pi\to\bar NN}&=\frac{i}{c_J\sqrt{2}}\sqrt{\frac{p_t}{q_t}}F^J_\pm(t)\,\theta\big(t-\tN\big).
\end{align}
In fact, these relations already imply interesting constraints on the partial waves that follow from the unitarity bound $|S_{ij}|\leq 1$, e.g.\ the asymptotic behavior of $f^J_{\pm}(t)$ for $t\to\infty$ will be restricted by 
\beq
f^J_+(t)=\Order\big(t^{-J+\frac{1}{2}}\big), \qquad f^J_-(t)=\Order\big(t^{-J}\big).
\eeq
Similarly, unitarity for $\bar K K$ scattering
\begin{align}
\Im r^{I_t}_J(t)&=(k_tq_t)^{2J}\sigma^\pi_t\big|g_J^{I_t}(t)\big|^2\,\theta\big(t-\tpi\big)+\sigma^K_t\big|r_J^{I_t}(t)\big|^2\,\theta\big(t-\tK\big)
+\frac{t}{8k_t^2}\frac{\sigma^N_t}{(c^{KN}_J)^2}\bigg\{\big|H_+^J(t)\big|^2+\big|H_-^J(t)\big|^2\bigg\}\theta\big(t-\tN\big)
\end{align} 
settles the normalization of the remaining partial waves
\beq
\big[S^{I_t}_J(t)\big]_{\bar KK\to\bar KK}=1+i\frac{4k_t}{\sqrt{t}}r^{I_t}_J(t)\,\theta\big(t-\tK\big),\qquad
\big[S^J_\pm(t)\big]^{I_t}_{\bar K K\to\bar NN}=\frac{i}{c_J^{KN}}\sqrt{\frac{p_t}{k_t}}H^J_\pm(t)\,\theta\big(t-\tN\big).
\eeq
Once the relations between $S$-matrix elements and partial-wave amplitudes are established, the unitarity relations for $f^J_{\pm}(t)$ and $h^J_{\pm}(t)$ including $\pi\pi$ and $\bar K K$ intermediate states follow from~\eqref{stmatrixelemunitrel2} and its analog for $\Im T_{23}$
\begin{align}
\label{exttunitrel}
\Im f^J_{\pm}(t)&=\sigma^\pi_t\big(t^{I_t}_J(t)\big)^*f^J_{\pm}(t)\,\theta\big(t-\tpi\big)
+\frac{\sqrt{2}\,c_J}{c_J^{KN}}k_t^{2J}\sigma^K_t\big(g^{I_t}_J(t)\big)^*h^J_\pm(t)\,\theta\big(t-\tK\big),\notag\\
\Im h^J_{\pm}(t)&=\sigma^K_t\big(r^{I_t}_J(t)\big)^*h^J_{\pm}(t)\,\theta\big(t-\tK\big)
+\frac{c_J^{KN}}{\sqrt{2}\,c_J}\,\sigma^\pi_tq_t^{2J}\big(g^{I_t}_J(t)\big)^*f_\pm^J(t)\,\theta\big(t-\tpi\big).
\end{align}
The factors $c_J$ and $c_J^{KN}$ originate from the transition between the $I_t=0,1$ and the $I=\pm$ bases in $\pi N$ and $KN$ scattering, since the derivation of the unitarity relations proceeds in the isospin basis, whereas the $t$-channel $\pi N$ and $KN$ partial waves are defined from the $I=\pm$ amplitudes. Moreover, the additional factor $\sqrt{2}$ is a remnant from a symmetry factor in the $\pi\pi$ system that occurs since this factor is not included in the conventional definition of the $\pi\pi\to\bar N N$ partial waves~\eqref{barNNtopipiJWpwe}. In contrast, these symmetry factors as required by the general result~\eqref{diffmodJWpwe} are included in the definition of the $\pi\pi$ and $\pi\pi\to\bar K K$ partial waves in~\eqref{pipiexpansion} and~\eqref{piKexpansion}, otherwise they would occur in~\eqref{exttunitrel} as well.

The structure of~\eqref{exttunitrel} can be made more apparent by noting that by virtue of unitarity in the $\pi\pi$/$\bar K K$ system the pertinent $T$-matrix becomes 
\beq
\label{Tmatrix}
T_J(t)=\begin{pmatrix}
      \frac{\eta^{I_t}_J(t) e^{2i\delta^{I_t}_J(t)}-1}{2i\sigma_t^\pi q_t^{2J}} & \big|g^{I_t}_J(t)\big|e^{i\psi^{I_t}_J(t)}\\ 
  \big|g^{I_t}_J(t)\big|e^{i\psi^{I_t}_J(t)} & \frac{\eta^{I_t}_J(t) e^{2i\big(\psi^{I_t}_J(t)-\delta^{I_t}_J(t)\big)}-1}{2i\sigma_t^K k_t^{2J}}
\end{pmatrix},
\eeq
where $\psi^{I_t}_J(t)$ denotes the phase of $g^{I_t}_J(t)$, the inelasticity parameter is given by
\beq
\eta_J^{I_t}(t)=\sqrt{1-\big|\big[S^{I_t}_J(t)\big]_{\pi\pi\to\bar KK}\big|^2}=\sqrt{1-4\sigma_t^\pi\sigma_t^K(k_tq_t)^{2J}\big|g^{I_t}_J(t)\big|^2\,\theta\big(t-\tK\big)},
\eeq
and the $\bar K K$ partial waves may be identified as
\beq
r^{I_t}_J(t)=\frac{\eta^{I_t}_J(t) e^{2i\big(\psi^{I_t}_J(t)-\delta^{I_t}_J(t)\big)}-1}{2i\sigma_t^K}.
\eeq
Together with the phase-space factor
\beq
\Sigma_J(t)=\text{diag}\Big(\sigma_t^\pi q_t^{2J}\theta\big(t-\tpi\big), \sigma_t^Kk_t^{2J}\theta\big(t-\tK\big)\Big),
\eeq
the $t$-channel unitarity relation~\eqref{exttunitrel} then takes the form
\beq
\label{two_channel_unitarity}
\Im \ff^J_\pm(t)=T_J^*(t)\Sigma_J(t)\ff^J_\pm(t),\qquad 
\ff^J_\pm(t)=\begin{pmatrix}
               f^J_\pm(t)\\
\frac{\sqrt{2}\,c_J}{c_J^{KN}}h^J_\pm(t)
              \end{pmatrix}.
\eeq
This equation serves as the starting point for a full two-channel treatment of the coupled system of $f^0_+(t)$ and $h^0_+(t)$ as performed in~\cite{Hoferichter:2012wf}. 

\setcounter{table}{0}

\section{Explicit numerical solutions for the $\boldsymbol{\pi N}$ phase shifts}
\label{app:numerical_sol}

\begin{table}
\centering
\renewcommand{\arraystretch}{1.3}
\begin{tabular}{crr}
\toprule
 & $I_s=1/2$ & $I_s=3/2$  \\\midrule
$A_{0+}$ & $1.217$               & $-0.6183$ \\
$B_{0+}$ & $-1.879\times 10$   & $-1.831 \times 10$\\
$C_{0+}$ & $1.958\times 10^2$  & $3.090\times 10^2$ \\
$D_{0+}$ & $-1.235\times 10^3$ & $-2.846\times 10^3$  \\
$E_{0+}$ & $3.350\times 10^3$  & $9.529\times 10^3$ \\
$s_{0+}$ & $2.494$   &   $-1.809\times 10^3$\\\midrule
$\Delta A_{0+}$ & $1.433\times 10^{-2}$  &   $1.289\times 10^{-2}$\\ 
$\Delta B_{0+}$ & $8.592\times 10^{-2}$  &   $0.1744$\\
$\rho_{0+}$     & $1.000$      &  $-0.2584$  \\\bottomrule
\end{tabular}
\caption{$S$-wave parameters for the solution of the RS equations. All parameters are given in appropriate powers of $\GeV$.}
\label{app:Swave-parameters}
\renewcommand{\arraystretch}{1.0}
\end{table}

In this appendix, we summarize the parameterizations we use for each partial wave and provide the explicit values of the parameters for our solution of the RS equations. 
We use the Schenk-like parameterization 
\beq
\label{app:schenk-Spar}
\tan{\delta_{0+}^{I_s}}=|\qq|\left(A_{0+}^{I_s}+B_{0+}^{I_s}\qq^2+C_{0+}^{I_s}\qq^4+D_{0+}^{I_s}\qq^6+E_{0+}^{I_s}\qq^8\right)\frac{s_+-s_{0+}^{I_s}}{s-s_{0+}^{I_s}}
\eeq
for the $S$-wave phase shifts, where the first parameters $A_{0+}^{I_s}$ are fixed by the scattering-length values extracted from pionic atoms.
In order to keep the same number of free parameters, we use for the $P_{13}$-, $P_{11}$-, and $P_{33}$-wave phase shifts a Schenk-like parameterization with one parameter less in the momentum expansion,
\beq
\label{app:schenk-Ppar}
\tan{\delta_{1{\pm}}^{I_s}}=|\qq|^{3}\left(A_{1\pm}^{I_s}+B_{1\pm}^{I_s}\qq^2+C_{1\pm}^{I_s}\qq^4+D_{1\pm}^{I_s}\qq^6\right)\frac{s_+-s_{1\pm}^{I_s}}{s-s_{1\pm}^{I_s}}.
\eeq

In the case of the $P_{33}$-wave, we describe its phase shift by the conformal parameterization
\beq
\label{app:conf-par}
\cot{\delta_{1+}^{3/2}}=\frac{1}{|\qq|^{3}}\frac{s-s_{1+}^{3/2}}{s_+-s_{1+}^{3/2}}\left(\frac{1}{ \tilde A_{1+}^{3/2}}+ \tilde B_{1+}^{3/2}\big[w(s)-w_+\big]+ \tilde C_{1+}^{3/2}\big[w(s)-w_+\big]^2\right),\quad
w(s)=\frac{\sqrt{s}-\sqrt{\bar s_{1+}^{3/2}-s}}{\sqrt{s}+\sqrt{\bar s_{1+}^{3/2}-s}},\quad
w_+=w(s_+).
\eeq

The last two parameters in the respective expansion in~\eqref{app:schenk-Spar}, \eqref{app:schenk-Ppar}, and \eqref{app:conf-par} are fixed by the matching conditions~\eqref{eq:schenk-par-Sw}. 
In order to facilitate the use of these parameterizations, we give explicitly their numerical values instead of the SAID input quantities. 
The value of each coefficient for our central solution is given in Tables~\ref{app:Swave-parameters} and \ref{app:Pwave-parameters}.

The phase-shift error bands depicted in Fig.~\ref{bands} can be reproduced using the formula
\beq\label{app:phase-errors}
\Delta\delta_{l{\pm}}^{I_s}=\sqrt{\left(\frac{\partial \delta_{l{\pm}}^{I_s}}{\partial A_{l\pm}^{I_s}}\right)^2\left(\Delta A_{l\pm}^{I_s}\right)^2+\left(\frac{\partial \delta_{l{\pm}}^{I_s}}{\partial B_{l\pm}^{I_s}}\right)^2\left(\Delta B_{l\pm}^{I_s}\right)^2+2\left(\frac{\partial \delta_{l{\pm}}^{I_s}}{\partial A_{l\pm}^{I_s}}\right)\left(\frac{\partial \delta_{l{\pm}}^{I_s}}{\partial B_{l\pm}^{I_s}}\right)\rho^{I_s}_{l\pm}\Delta A_{l\pm}^{I_s}\Delta B_{l\pm}^{I_s}},
\eeq
where $\Delta A_{l\pm}^{I_s}$ and $\Delta B_{l\pm}^{I_s}$ are the coefficient errors and $\rho^{I_s}_{l\pm}$ their correlation coefficient (and analogously for $\tilde A_{l\pm}^{I_s}$ and $\tilde B_{l\pm}^{I_s}$). 
Their values are also included in Tables~\ref{app:Swave-parameters} and \ref{app:Pwave-parameters}.
Note that in the case of the $S$-waves, the coefficients $\Delta A_{0+}^{I_s}$ are fixed by the uncertainties of the pionic-atom scattering lengths. 

\begin{table}
\centering
\renewcommand{\arraystretch}{1.3}
\begin{tabular}{cr}
\toprule
 & $I_s=1/2$   \\\midrule
$A_{1+}$          &    $-1.085\times 10$ \\
$B_{1+}$          &    $-1.145\times 10$ \\
$C_{1+}$          &    $3.651\times 10^2$ \\
$D_{1+}$          &    $-1.052\times 10^3$ \\
$s_{1+}$          &    $0.9639$ \\\midrule
$\Delta A_{1+}$   &    $0.5649$ \\ 
$\Delta B_{1+}$   &     $1.896$ \\
$\rho_{1+}$       &  $-1.000$  \\\bottomrule
\end{tabular}
\quad\begin{tabular}{cr}
\toprule
 & $I_s=3/2$   \\\midrule
$\tilde A_{1+}$          &     $7.781\times 10$    \\
$\tilde B_{1+}$          &   $-3.986\times10^{-2}$  \\
$\tilde C_{1+}$          &   $-0.3098$               \\
$s_{1+}$          &    $0.4509$               \\
$\sqrt{\bar s_{1+}}$     &     $1.540$               \\\midrule
$\Delta \tilde A_{1+}$   &     $1.257$               \\ 
$\Delta \tilde B_{1+}$   &    $1.113\times10^{-4}$  \\
$\rho_{1+}$       &    $0.2094$                \\\bottomrule
\end{tabular}
\quad\begin{tabular}{crr}
\toprule
 & $I_s=1/2$ & $I_s=3/2$                       \\\midrule
$A_{1-}$        &     $-2.569\times 10$    &        $-1.477\times 10$            \\
$B_{1-}$        &     $8.062\times 10^2$   &        $1.467\times 10^2$            \\
$C_{1-}$        &     $-4.214\times 10^3$  &        $-1.633\times 10^3$            \\
$D_{1-}$        &     $3.986 \times 10^4$  &        $6.508\times 10^3$            \\
$s_{1-}$        &     $0.9340$                &         $0.4081$            \\\midrule
$\Delta A_{1-}$ &     $1.800$                &         $0.7257$           \\ 
$\Delta B_{1-}$ &     $2.650\times 10$     &         $5.507$           \\
$\rho_{1-}$     &     $-0.2510$               &        $-0.9882$              \\\bottomrule
\end{tabular}
\caption{$P$-wave parameters for the solution of the RS equations. All parameters are given in appropriate powers of $\GeV$.}
\label{app:Pwave-parameters}
\renewcommand{\arraystretch}{1.0}
\end{table}

\section{Sum rules for threshold parameters}
\label{app:thr_sum_rules}

The sum rules for the covariant amplitudes based on HDRs relevant for the threshold parameters discussed in Sects.~\ref{sec:threshold} and \ref{sec:threshold_ChPT} are
\begin{align}
\zeq{A^+(s,0)}&=d_{00}^++\frac{g^2}{\mN}+a_{10}^+\mpi^2\notag\\
&+\frac{1}{\pi}\int\limits_{\tpi}^\infty\frac{\diff t'}{t'}
\bigg\{\te{\Im A^+}-\ste{\Im A^+}-\mpi^2\ste{\partial_{\nu^2}\Im A^+}\bigg\}\notag\\
&+\frac{1}{\pi}\int\limits_{s_+}^\infty\diff s'\Bigg\{h_+(s')\te{\Im A^+}-\mpi^2h_0(s')\ste{\partial_{\nu^2}\Im A^+} 
-\big(h_0(s')+8\mN^2\mpi^2 h^0_3(s')\big)\ste{\Im A^+}\Bigg\},\notag\\
\zeq{A^-(s,0)}&=a_{00}^-\mpi+a_{10}^-\mpi^3 
+\frac{\mpi}{\pi}\int\limits_{\tpi}^\infty\frac{\diff t'}{t'}
\Bigg\{\te{\Im A^-/\nu'}
-\ste{\Im A^-/\nu'}-\mpi^2\ste{\partial_{\nu^2}\Im A^-/\nu'}\Bigg\}\notag\\
&+\frac{1}{\pi}\int\limits_{s_+}^\infty\diff s'\Bigg\{h_-(s')\te{\Im A^-}
-4\mN\mpi^3h^0_2(s')\ste{\partial_{\nu^2}\Im A^-}\notag\\
&\qquad-4\mN\mpi \big(h^0_2(s')+4\mN^2\mpi^2h^0_4(s')\big)\ste{\Im A^-}\Bigg\},\notag\\
\zeq{B^+(s,0)}&=N_-+ b_{00}^+\mpi
+\frac{\mpi}{\pi}\int\limits_{\tpi}^\infty\frac{\diff t'}{t'}
\bigg\{\te{\Im B^+/\nu'}-\ste{\Im B^+/\nu'}\bigg\}\notag\\
&+\frac{1}{\pi}\int\limits_{s_+}^\infty\diff s'\bigg\{h_-(s')\te{\Im B^+}
-4\mN\mpi h^0_2(s')\ste{\Im B^+}\bigg\},\notag\\
\zeq{ B^-(s,0)}&=N_++b_{00}^--\frac{g^2}{2\mN^2}+b_{10}^-\mpi^2+\frac{1}{\pi}\int\limits_{\tpi}^\infty\frac{\diff t'}{t'}
\bigg\{\te{\Im B^-}-\ste{\Im B^-}-\mpi^2\ste{\partial_{\nu^2}\Im B^-}\bigg\}\notag\\
&+\frac{1}{\pi}\int\limits_{s_+}^\infty\diff s'\Bigg\{h_+(s')\te{\Im B^-}-\mpi^2h_0(s')\ste{\partial_{\nu^2}\Im B^-}
-\big(h_0(s')+8\mN^2\mpi^2h^0_3(s')\big)\ste{\Im B^-}\Bigg\},\notag\\
\zetq{\partial_tA^+(s,t)}&=d_{01}^++a_{10}^+\zeta_{\nu^2,t}+\frac{1}{\pi}\int\limits_{\tpi}^\infty\frac{\diff t'}{t'}
\Bigg\{\frac{1}{t'}\te{\Im A^+}+\te{\partial_t\Im A^+}\notag\\
&\qquad\qquad-\frac{1}{t'}\ste{\Im A^+}-\ste{\partial_t\Im A^+}-\zeta_{\nu^2,t}\ste{\partial_{\nu^2}\Im A^+}\Bigg\}\notag\\
&+\frac{1}{\pi}\int\limits_{s_+}^\infty\diff s'\Bigg\{-h^-_2(s')\te{\Im A^+}+
h_+(s')\te{\partial_t\Im A^+}\notag\\
&+h^0_2(s')\ste{\Im A^+}-
h_0(s')\ste{\partial_t\Im A^+}-\zeta_{\nu^2,t}\Big(8\mN^2h^0_3(s')\ste{\Im A^+}+h_0(s')\ste{\partial_{\nu^2}\Im A^+}\Big)\Bigg\},\notag\\
 \zetq{\partial_tA^-(s,t)}&=\frac{a_{00}^-}{4\mN}+a_{01}^-\mpi+a_{10}^-\zeta_{\nu^3,t}
+\frac{1}{\pi}\int\limits_{\tpi}^\infty\frac{\diff t'}{t'}\Bigg\{\bigg(\frac{1}{4\mN}+\frac{\mpi}{t'}\bigg)
\te{\Im A^-/\nu'}+\mpi\te{\partial_t\Im A^-/\nu'}\notag\\
&\qquad-\mpi\ste{\partial_t\Im A^-/\nu'}
-\bigg(\frac{1}{4\mN}+\frac{\mpi}{t'}\bigg)\ste{\Im A^-/\nu'}
-\zeta_{\nu^3,t}\ste{\partial_{\nu^2}\Im A^-/\nu'}\Bigg\}\notag\\
&+\frac{1}{\pi}\int\limits_{s_+}^\infty\diff s'\Bigg\{h^-_2(s')\te{\Im A^-}+h_-(s')\te{\partial_t\Im A^-}-h^0_2(s')\ste{\Im A^-}\notag\\
&\qquad+4\mN\mpi \Big(h^0_3(s')\ste{\Im A^-}-h^0_2(s')\ste{\partial_t\Im A^-}\Big)\notag\\
&\qquad-4\mN\zeta_{\nu^3,t} \Big(4\mN^2h^0_4(s')\ste{\Im A^-}+h^0_2(s')\ste{\partial_{\nu^2}\Im A^-}\Big)\Bigg\},\notag\\
\zetq{\partial_tB^+(s,t)}&=N^-_2+\frac{b_{00}^+}{4\mN}
+\frac{1}{\pi}\int\limits_{\tpi}^\infty\frac{\diff t'}{t'}\Bigg\{\bigg(\frac{1}{4\mN}+\frac{\mpi}{t'}\bigg)
\te{\Im B^+/\nu'}\notag\\
&\qquad\qquad+\mpi\te{\partial_t\Im B^+/\nu'}-\frac{1}{4\mN}\ste{\Im B^+/\nu'}\Bigg\}\notag\\
&+\frac{1}{\pi}\int\limits_{s_+}^\infty\diff s'\Bigg\{h^-_2(s')\te{\Im B^+}+h_-(s')\te{\partial_t\Im B^+}
-h^0_2(s')\ste{\Im B^+}\Bigg\},\notag\\
\zetq{\partial_tB^-(s,t)}&=-N^-_2+b_{01}^-+b_{10}^-\zeta_{\nu^2,t}+\frac{1}{\pi}\int\limits_{\tpi}^\infty\frac{\diff t'}{t'}
\Bigg\{\frac{1}{t'}\te{\Im B^-}+\te{\partial_t\Im B^-}\notag\\
&\qquad\qquad-\frac{1}{t'}\ste{\Im B^-}-\ste{\partial_t\Im B^-}-\zeta_{\nu^2,t}\ste{\partial_{\nu^2}\Im B^-}\Bigg\}\notag\\
&+\frac{1}{\pi}\int\limits_{s_+}^\infty\diff s'\Bigg\{-h^-_2(s')\te{\Im B^-}+
h_+(s')\te{\partial_t\Im B^-}\notag\\
&+h^0_2(s')\ste{\Im B^-}-h_0(s')\ste{\partial_t\Im B^-}
-\zeta_{\nu^2,t} \Big(8\mN^2h^0_3(s')\ste{\Im B^-}+h_0(s')\ste{\partial_{\nu^2}\Im B^-}\Big)\Bigg\},\notag\\
 \zeq{\partial_{\qq^2}A^+(s,0)}&=a_{10}^+\zeta_{\nu^2,\qq^2}+\frac{1}{\pi}\int\limits_{\tpi}^\infty\frac{\diff t'}{t'}
\bigg\{\te{\partial_{\qq^2}\Im A^+}-\zeta_{\nu^2,\qq^2}\ste{\partial_{\nu^2}\Im A^+}\bigg\}\notag\\
&+\frac{1}{\pi}\int\limits_{s_+}^\infty\diff s'\Bigg\{\zeta_s(h^+_2(s')-h^-_2(s'))\te{\Im A^+}+h_+(s')\te{\partial_{\qq^2}\Im A^+}\notag\\
&-\zeta_{\nu^2,\qq^2}\Big(8\mN^2h^0_3(s')\ste{\Im A^+}+h_0(s')\ste{\partial_{\nu^2}\Im A^+}\Big)
-\frac{\zeta_{A^+}}{(s'-s_+)^{3/2}}\Bigg\},\notag\\
 \zeq{\partial_{\qq^2}A^-(s,0)}&=a_{00}^-\frac{\zeta_s}{2\mN}+a_{10}^-\zeta_{\nu^3,\qq^2}
+\frac{1}{\pi}\int\limits_{\tpi}^\infty\frac{\diff t'}{t'}\Bigg\{\mpi\te{\partial_{\qq^2}\Im A^-/\nu'}\notag\\
&\qquad\qquad+\frac{\zeta_s}{2\mN}\Big(\te{\Im A^-/\nu'}-\ste{\Im A^-/\nu'}\Big)-\zeta_{\nu^3,\qq^2}\ste{\partial_{\nu^2}\Im A^-/\nu'}\Bigg\}\notag\\
&+\frac{1}{\pi}\int\limits_{s_+}^\infty\diff s'\Bigg\{\zeta_s(h^+_2(s')+h^-_2(s'))\te{\Im A^-}+h_-(s')\te{\partial_{\qq^2}\Im A^-}\notag\\
&-4\mN\zeta_{\nu^3,\qq^2}\Big(4\mN^2h^0_4(s')\ste{\Im A^-}+h^0_2(s')\ste{\partial_{\nu^2}\Im A^-}\Big)
-2\zeta_sh^0_2(s')\ste{\Im A^-}-\frac{\zeta_{A^-}}{(s'-s_+)^{3/2}}\Bigg\},\notag\\
 \zeq{\partial_{\qq^2}B^+(s,0)}&=\zeta_s(N^+_2+N^-_2)+b_{00}^+\frac{\zeta_s}{2\mN}
+\frac{1}{\pi}\int\limits_{\tpi}^\infty\frac{\diff t'}{t'}\Bigg\{\mpi\te{\partial_{\qq^2}\Im B^+/\nu'}\notag\\
&\qquad\qquad+\frac{\zeta_s}{2\mN}\Big(\te{\Im B^+/\nu'}-\ste{\Im B^+/\nu'}\Big)\Bigg\}\notag\\
&+\frac{1}{\pi}\int\limits_{s_+}^\infty\diff s'\Bigg\{\zeta_s(h^+_2(s')+h^-_2(s'))\te{\Im B^+}+h_-(s')\te{\partial_{\qq^2}\Im B^+}\notag\\
&\qquad-2\zeta_sh^0_2(s')\ste{\Im B^+}
-\frac{\zeta_{B^+}}{(s'-s_+)^{3/2}}\Bigg\},\notag\\
 \zeq{\partial_{\qq^2}B^-(s,0)}&=\zeta_s(N^+_2-N^-_2)+b_{10}^-\zeta_{\nu^2,\qq^2}
 +\frac{1}{\pi}\int\limits_{\tpi}^\infty\frac{\diff t'}{t'}
\bigg\{\te{\partial_{\qq^2}\Im B^-}-\zeta_{\nu^2,\qq^2}\ste{\partial_{\nu^2}\Im B^-}\bigg\}\notag\\
&+\frac{1}{\pi}\int\limits_{s_+}^\infty\diff s'\Bigg\{\zeta_s(h^+_2(s')-h^-_2(s'))\te{\Im B^-}+h_+(s')\te{\partial_{\qq^2}\Im B^-}\notag\\
&-\zeta_{\nu^2,\qq^2} \Big(8\mN^2h^0_3(s')\ste{\Im B^-}+h_0(s')\ste{\partial_{\nu^2}\Im B^-}\Big)-\frac{\zeta_{B^-}}{(s'-s_+)^{3/2}}\Bigg\},
\end{align}
where the subscript $(\mpi,0)$ denotes evaluation at threshold $(\nu=\mpi,\,t=0)$ and
\begin{align}
h_+(s')&=\frac{1}{s'-s_+}+\frac{1}{s'-s_-}-\frac{1}{s'-a},\qquad h_-(s')=\frac{1}{s'-s_+}-\frac{1}{s'-s_-},\notag\\
N_\pm&=g^2\bigg(\frac{1}{\mN^2-s_+}\pm \frac{1}{\mN^2-s_-}\bigg),\qquad N_n^\pm=\frac{g^2}{(\mN^2-s_\pm)^n},\qquad h_n^\pm(s')=\frac{1}{(s'-s_\pm)^n}.
\end{align}
The amplitudes may be calculated by summing the pertinent partial waves using
\begin{align}
\te{z_s'}&=-\frac{s'+a}{s'-a}, &  
\te{\partial_tz_s'}&=\frac{s_+-a}{2\qq'^2(s'-a)},
&\te{\partial_{\qq^2}z_s'}&=\frac{2s_+}{\qq'^2(s'-a)},\notag\\
\te{z_t'^2}&=1+\frac{at'}{4p_t'^2q_t'^2}, & 
\te{\partial_tz_t'^2}&=\frac{s_+-a}{4p_t'^2q_t'^2},
&\te{\partial_{\qq^2}z_t'^2}&=\frac{s_+}{p_t'^2q_t'^2}.
\end{align} 
Finally, we have defined the derivatives
\begin{align}
\zeta_s&=\te{\partial_{\qq^2}s}=\frac{4s_+}{s_+-s_-}=\frac{s_+}{\mN\mpi}, &&\notag\\
\zeta_{\nu^2,t}&=\te{\partial_t\nu^2}=\frac{\mpi}{2\mN}, &\zeta_{\nu^2,\qq^2}&=\te{\partial_{\qq^2}\nu^2}=\frac{s_+}{\mN^2},\notag\\
\zeta_{\nu^3,t}&=\te{\partial_t\nu^3}=\frac{3\mpi^2}{4\mN}, & \zeta_{\nu^3,\qq^2}&=\te{\partial_{\qq^2}\nu^3}=\frac{3\mpi s_+}{2\mN^2},
\end{align}
and removed the threshold divergence by subtracting the terms involving
\begin{align}
 \zeta_{A^+}&=(2\mN+\mpi)\zeta_{B^+}=\frac{2\pi W_+(2\mN+\mpi)}{3\mN\sqrt{\mN\mpi}}\Big[\big(a_{0+}^{1/2}\big)^2+2\big(a_{0+}^{3/2}\big)^2\Big],\notag\\
\zeta_{A^-}&=(2\mN+\mpi)\zeta_{B^-}=\frac{2\pi W_+(2\mN+\mpi)}{3\mN\sqrt{\mN\mpi}}\Big[\big(a_{0+}^{1/2}\big)^2-\big(a_{0+}^{3/2}\big)^2\Big].
\end{align}

\section{Chiral expansion}
\label{app:ChPT}

In this appendix we collect the expressions for the chiral expansion of $\pi N$ threshold and subthreshold parameters up to N$^3$LO,  $\Order(p^4)$, in ChPT.
The NLO, N$^2$LO, N$^3$LO LECs are denoted by $c_i$, $d_i$, and $e_i$, respectively, corresponding to the chiral Lagrangian of~\cite{Fettes:2000gb}. 
The notation for the renormalized $d_i$ and $e_i$ is
\begin{align}
 \bar d_i&=d_i-\frac{\delta_i}{\Fpi^2}\lambda_\pi=d_i^r(\mu)-\frac{\delta_i}{16\pi^2\Fpi^2}\log\frac{\mpi}{\mu},\notag\\
 \bar e_i&=e_i-\frac{\eps_i}{\Fpi^2}\lambda_\pi=e_i^r(\mu)-\frac{\eps_i}{16\pi^2\Fpi^2}\log\frac{\mpi}{\mu},
\end{align}
where $\delta_i$ and $\eps_i$ denote the $\beta$-functions, $\mu$ the renormalization scale, and
\beq
\lambda_\pi=\lambda+\frac{1}{16\pi^2}\log\frac{\mpi}{\mu},\qquad \lambda=\frac{\mu^{d-4}}{16\pi^2}\bigg\{\frac{1}{d-4}-\frac{1}{2}\bigg(\log (4\pi)-\gamma_E+1\bigg)\Bigg\}.
\eeq
Due to the renormalization of the pion mass also mesonic LECs $l_i$ enter. We use the convention from~\cite{Gasser:1983yg}
\beq
l_i^r(\mu)=\frac{\gamma_i}{32\pi^2}\bigg(\bar l_i+\log \frac{\mpi^2}{\mu^2}\bigg),
\eeq
with $\beta$-functions $\gamma_i$.

\subsection{Subthreshold parameters}
\label{app:ChPT_subthreshold}

The expressions for the subthreshold parameters in the above conventions can be reconstructed from~\cite{Becher:2001hv,Fettes:2000xg,Krebs:2012yv,Siemens:2016hdi}, in the standard counting of heavy-baryon ChPT they read
\begin{align}
\label{subthr_ChPT}
 d_{00}^+&=-\frac{2 \mpi^2 (2 \tilde c_1-\tilde c_3)}{\Fpi^2}
 +\frac{\ga^2 \big(3+8 \ga^2\big) \mpi^3}{64 \pi  \Fpi^4}
 +\mpi^4 \Bigg\{\frac{16 \bar e_{14}}{\Fpi^2}+\frac{3\ga^2 \big(1+6 \ga^2\big)}{64 \pi ^2 \Fpi^4 \mN}-\frac{2 c_1-c_3}{16 \pi ^2 \Fpi^4}\Bigg\},\notag\\
 d_{10}^+&=\frac{2 \tilde c_2}{\Fpi^2}-\frac{\big(4+5 \ga^4\big) \mpi}{32 \pi  \Fpi^4}
 +\mpi^2 \Bigg\{\frac{16 \bar e_{15}}{\Fpi^2}-\frac{16 c_1 c_2}{\Fpi^2 \mN}-\frac{1+\ga^2}{4 \pi ^2 \Fpi^4 \mN}-\frac{197 \ga^4}{240 \pi ^2 \Fpi^4 \mN}\Bigg\},\notag\\
 d_{01}^+&=-\frac{\tilde c_3}{\Fpi^2}-\frac{\ga^2\big(77+48 \ga^2\big)  \mpi}{768 \pi  \Fpi^4}
 -\mpi^2 \Bigg\{\frac{16 \bar e_{14}}{\Fpi^2}-\frac{52 c_1-c_2-32 c_3}{192 \pi ^2 \Fpi^4}+\frac{\ga^2\big(47+66 \ga^2\big) }{384 \pi ^2 \Fpi^4 \mN}\Bigg\},\notag\\
 d_{20}^+&=\frac{12+5 \ga^4}{192 \pi  \Fpi^4 \mpi}+\frac{16 \bar e_{16}}{\Fpi^2}+\frac{17+10 \ga^2}{24 \pi ^2 \Fpi^4 \mN}+\frac{173 \ga^4}{280 \pi ^2 \Fpi^4 \mN},\notag\\
 d_{11}^+&=\frac{\ga^4}{64 \pi  \Fpi^4 \mpi}-\frac{8 \bar e_{15}}{\Fpi^2}+\frac{9+2 \ga^2}{96 \pi ^2 \Fpi^4 \mN}+\frac{67 \ga^4}{240 \pi ^2 \Fpi^4 \mN},\notag\\
 d_{02}^+&=\frac{193 \ga^2}{15360 \pi  \Fpi^4 \mpi}+\frac{4 \bar e_{14}}{\Fpi^2}-\frac{c_2}{8 \Fpi^2 \mN^2}+\frac{29 \ga^2}{480 \pi ^2 \Fpi^4 \mN}+\frac{\ga^4}{64 \pi ^2 \Fpi^4 \mN}-\frac{19 c_1}{480 \pi ^2 \Fpi^4}+\frac{7 c_2}{640 \pi ^2 \Fpi^4}+\frac{7 c_3}{80 \pi ^2 \Fpi^4},\notag\\
 d_{00}^-&=\frac{1}{2 \Fpi^2}
 +\frac{4 \mpi^2 (\bar d_1+\bar d_2+2 \bar d_5)}{\Fpi^2}+\frac{\ga^4 \mpi^2}{48 \pi ^2 \Fpi^4}
 -\mpi^3 \Bigg\{\frac{8+12 \ga^2+11 \ga^4}{128 \pi  \Fpi^4 \mN}-\frac{4 c_1+\ga^2 (c_3-c_4)}{4 \pi  \Fpi^4}\Bigg\},\notag\\
 d_{10}^-&=\frac{4 \bar d_3}{\Fpi^2}-\frac{15+7 \ga^4}{240 \pi ^2 \Fpi^4}
 +\frac{\big(168+138 \ga^2+85 \ga^4\big) \mpi}{768 \pi  \Fpi^4 \mN}
 -\frac{\mpi \big[8 (c_1+ c_2+ c_3)+5 (c_3-c_4) \ga^2\big]}{16 \pi  \Fpi^4},\notag\\
 d_{01}^-&=-\frac{2 \big(\bar d_1+\bar d_2\big)}{\Fpi^2}-\frac{1+7 \ga^2+2 \ga^4}{192 \pi ^2 \Fpi^4}
 +\frac{\big(12+53 \ga^2+24 \ga^4\big) \mpi}{384 \pi  \Fpi^4 \mN}
 -\frac{\ga^2 \mpi (c_3-c_4)}{8 \pi  \Fpi^4},\notag\\
 b_{00}^+&=\frac{4\mN \big(\bar d_{14}-\bar d_{15}\big) }{\Fpi^2}
 -\frac{\ga^4 \mN}{8 \pi ^2 \Fpi^4}+\frac{\big(8+7 \ga^2\big) \ga^2 \mpi}{64 \pi  \Fpi^4}-\frac{\ga^2 \mN \mpi (c_3-c_4)}{2 \pi  \Fpi^4},\notag\\
 b_{00}^-&=\frac{1}{2 \Fpi^2}+\frac{2 \tilde c_4 \mN}{\Fpi^2}
 -\frac{\ga^2\big(1+\ga^2\big)  \mN \mpi}{8 \pi  \Fpi^4}
 +\frac{16 \bar e_{17} \mN \mpi^2}{\Fpi^2}
 -\frac{\ga^2 \mpi^2 \big(3+2\ga^2+9 c_4 \mN\big)}{12 \pi ^2 \Fpi^4},\notag\\
 b_{10}^-&=\frac{\ga^4 \mN}{32 \pi  \Fpi^4 \mpi}+\frac{16 \bar e_{18} \mN}{\Fpi^2}+\frac{\ga^2 \big(25+36\ga^2+80 c_4 \mN\big)}{120 \pi ^2 \Fpi^4},\notag\\
 b_{01}^-&=\frac{\ga^2 \mN}{96 \pi  \Fpi^4 \mpi}-\frac{8 \bar e_{17} \mN}{\Fpi^2}-\frac{1-9\ga^2-4\ga^4+4 c_4 \mN}{192 \pi ^2 \Fpi^4},
\end{align}
where
\begin{align}
\label{ci_ren}
 \tilde c_1&= c_1-2\mpi^2\bigg(\bar e_{22}-4\bar e_{38}-\frac{\bar l_3 c_1}{64\pi^2\Fpi^2}\bigg), &
  \tilde c_3&=c_3+4\mpi^2\big(2\bar e_{19}-\bar e_{22}-\bar e_{36}\big), \notag\\
 \tilde c_2&=c_2+8\mpi^2\big(\bar e_{20}+\bar e_{35}\big), &
 \tilde c_4&=c_4+4\mpi^2\big(2\bar e_{21}-\bar e_{37}\big),
\end{align}
subsume the quark-mass renormalization of the $c_i$. Since such effects are observable in neither $\pi N$ nor $NN$ physics, we will follow~\cite{Fettes:2000xg,Krebs:2012yv} and use conventions where the correction terms in~\eqref{ci_ren} are put to zero. Accordingly, the distinction between $c_i$ and $\tilde c_i$ will be dropped in the following.

In the $NN$ counting of the nucleon mass~\cite{Weinberg:1991um}, the subthreshold parameters become
\begin{align}
\label{subthr_ChPT_NN}
 d_{00}^+&=-\frac{2 \mpi^2 (2  c_1- c_3)}{\Fpi^2}
 +\frac{\ga^2 \big(3+8 \ga^2\big) \mpi^3}{64 \pi  \Fpi^4}
 +\mpi^4 \Bigg\{\frac{16 \bar e_{14}}{\Fpi^2}-\frac{2 c_1-c_3}{16 \pi ^2 \Fpi^4}\Bigg\},\notag\\
 d_{10}^+&=\frac{2  c_2}{\Fpi^2}-\frac{\big(4+5 \ga^4\big) \mpi}{32 \pi  \Fpi^4}
 +\frac{16\mpi^2 \bar e_{15}}{\Fpi^2},\notag\\
 d_{01}^+&=-\frac{ c_3}{\Fpi^2}-\frac{\ga^2\big(77+48 \ga^2\big)  \mpi}{768 \pi  \Fpi^4}
 -\mpi^2 \Bigg\{\frac{16 \bar e_{14}}{\Fpi^2}-\frac{52 c_1-c_2-32 c_3}{192 \pi ^2 \Fpi^4}\Bigg\},\notag\\
 d_{20}^+&=\frac{12+5 \ga^4}{192 \pi  \Fpi^4 \mpi}+\frac{16 \bar e_{16}}{\Fpi^2},\notag\\
 d_{11}^+&=\frac{\ga^4}{64 \pi  \Fpi^4 \mpi}-\frac{8 \bar e_{15}}{\Fpi^2},\notag\\
 d_{02}^+&=\frac{193 \ga^2}{15360 \pi  \Fpi^4 \mpi}+\frac{4 \bar e_{14}}{\Fpi^2}-\frac{19 c_1}{480 \pi ^2 \Fpi^4}+\frac{7 c_2}{640 \pi ^2 \Fpi^4}+\frac{7 c_3}{80 \pi ^2 \Fpi^4},\notag\\
 d_{00}^-&=\frac{1}{2 \Fpi^2}
 +\frac{4 \mpi^2 (\bar d_1+\bar d_2+2 \bar d_5)}{\Fpi^2}+\frac{\ga^4 \mpi^2}{48 \pi ^2 \Fpi^4}
 +\frac{4 c_1+\ga^2 (c_3-c_4)}{4 \pi  \Fpi^4}\mpi^3,\notag\\
 d_{10}^-&=\frac{4 \bar d_3}{\Fpi^2}-\frac{15+7 \ga^4}{240 \pi ^2 \Fpi^4}
 -\frac{\mpi \big[8 (c_1+ c_2+ c_3)+5 (c_3-c_4) \ga^2\big]}{16 \pi  \Fpi^4},\notag\\
 d_{01}^-&=-\frac{2 \big(\bar d_1+\bar d_2\big)}{\Fpi^2}-\frac{1+7 \ga^2+2 \ga^4}{192 \pi ^2 \Fpi^4}
 -\frac{\ga^2 \mpi (c_3-c_4)}{8 \pi  \Fpi^4},\notag\\
 b_{00}^+&=\frac{4\mN \big(\bar d_{14}-\bar d_{15}\big) }{\Fpi^2}
 -\frac{\ga^4 \mN}{8 \pi ^2 \Fpi^4}-\frac{\ga^2 \mN \mpi (c_3-c_4)}{2 \pi  \Fpi^4},\notag\\
 b_{00}^-&=\frac{1}{2 \Fpi^2}+\frac{2  c_4 \mN}{\Fpi^2}
 -\frac{\ga^2\big(1+\ga^2\big)  \mN \mpi}{8 \pi  \Fpi^4}
 +\frac{16 \bar e_{17} \mN \mpi^2}{\Fpi^2}
 -\frac{ 3\ga^2 \mpi^2 c_4 \mN}{4 \pi ^2 \Fpi^4},\notag\\
 b_{10}^-&=\frac{\ga^4 \mN}{32 \pi  \Fpi^4 \mpi}+\frac{16 \bar e_{18} \mN}{\Fpi^2}+\frac{2\ga^2 c_4 \mN}{3 \pi ^2 \Fpi^4},\notag\\
 b_{01}^-&=\frac{\ga^2 \mN}{96 \pi  \Fpi^4 \mpi}-\frac{8 \bar e_{17} \mN}{\Fpi^2}-\frac{c_4 \mN}{48 \pi ^2 \Fpi^4}.
\end{align}

\subsection{Threshold parameters}
\label{app:ChPT_threshold}

The expressions for the chiral expansion of the threshold parameters in standard counting read~\cite{Fettes:2000xg,Becher:2001hv}
\begin{align}
a_{0+}^+&=-\frac{\mpi^2 \big[\ga^2+8 \mN (2 c_1-c_2-c_3)\big]}{16 \pi  \Fpi^2 (\mN+\mpi)}+\frac{3 \ga^2 \mN \mpi^3}{256 \pi ^2 \Fpi^4 (\mN+\mpi)}-\frac{\ga^2 \mpi^4}{64 \pi  \Fpi^2 \mN^2 (\mN+\mpi)}\\
 &+\frac{\mpi^4 \big[-16 c_1 c_2+\bar d_{18} \ga+16 \mN (\bar e_{14}+\bar e_{15}+\bar e_{16})\big]}{4 \pi  \Fpi^2 (\mN+\mpi)}
 -\frac{\mpi^4 \big[8-3\ga^2+2 \ga^4+4 \mN (2 c_1-c_3)\big]}{256 \pi ^3 \Fpi^4 (\mN+\mpi)},\notag\\
 a_{0+}^-&=\frac{\mN \mpi}{8 \pi  \Fpi^2 (\mN+\mpi)}+\frac{\mN \mpi^3}{64 \pi ^3 \Fpi^4 (\mN+\mpi)}+\frac{\ga^2 \mpi^3}{32 \pi  \Fpi^2 \mN (\mN+\mpi)}+\frac{\mN \mpi^3 (\bar d_1+\bar d_2+\bar d_3+2 \bar d_5)}{\pi  \Fpi^2 (\mN+\mpi)},\notag\\
 a_{1+}^+&=\frac{\ga^2 \mN}{24 \pi  \Fpi^2 \mpi (\mN+\mpi)}+\frac{\ga^2-4 c_3 \mN}{24 \pi  \Fpi^2 (\mN+\mpi)}
 +\frac{\ga^2 \mpi}{32 \pi  \Fpi^2 \mN (\mN+\mpi)}
 +\frac{\mpi \big[c_2-\mN (\bar d_{14}-\bar d_{15}+\bar d_{18} \ga)\big]}{6 \pi  \Fpi^2 (\mN+\mpi)}\notag\\
 &-\frac{ \ga^2 \mN \mpi\big[231 \pi+8 (12 \pi -7) \ga^2 \big]}{13824 \pi ^3 \Fpi^4 (\mN+\mpi)}
 -\frac{\mpi^2 \big[\bar d_{18} \ga+8 \mN (2 \bar e_{14}+\bar e_{15})\big]}{6 \pi  \Fpi^2 (\mN+\mpi)}+\frac{\ga^2 \mpi^2}{48 \pi  \Fpi^2 \mN^2 (\mN+\mpi)}\notag\\
 &+\frac{\mpi^2 \big\{36-(133+24 \pi ) \ga^2+2 (5-33 \pi ) \ga^4+6 \mN \big[52 c_1-c_2-32 c_3+16 \pi  \ga^2 (c_3-c_4)\big]\big\}}{6912 \pi ^3 \Fpi^4 (\mN+\mpi)},\notag\\
 a_{1+}^-&=-\frac{\ga^2 \mN}{24 \pi  \Fpi^2 \mpi (\mN+\mpi)}-\frac{\ga^2+2 c_4 \mN}{24 \pi  \Fpi^2 (\mN+\mpi)}
 -\frac{\mN \mpi \big[2 (\bar d_1+\bar d_2)-\bar d_{18} \ga\big]}{6 \pi  \Fpi^2 (\mN+\mpi)}
 -\frac{\ga^2 \mpi}{32 \pi  \Fpi^2 \mN (\mN+\mpi)}\notag\\
 &-\frac{\mN \mpi\big[3+3 (7-6 \pi ) \ga^2+2 (1-6 \pi ) \ga^4\big]}{3456 \pi ^3 \Fpi^4 (\mN+\mpi)}
  +\frac{\mpi^2 \big[\bar d_1+\bar d_2+3 \bar d_3+2 \bar d_5+\bar d_{18} \ga-4 \mN (\bar e_{17}+\bar e_{18})\big]}{6 \pi  \Fpi^2 (\mN+\mpi)}\notag\\
 &-\frac{\ga^2 \mpi^2}{48 \pi  \Fpi^2 \mN^2 (\mN+\mpi)}
 +\frac{\mpi^2 \big\{-18+3 (12+47 \pi ) \ga^2+2(21 \pi -2) \ga^4+8 \ga^2 \mN \big[11 c_4-12 \pi  (c_3-c_4)\big]\big\}}{6912 \pi ^3 \Fpi^4 (\mN+\mpi)},\notag\\
 a_{1-}^+&=-\frac{\ga^2 \mN}{12 \pi  \Fpi^2 \mpi (\mN+\mpi)}-\frac{\ga^2+2 c_3 \mN}{12 \pi  \Fpi^2 (\mN+\mpi)}
 +\frac{\mpi \big[c_2+2 \mN (\bar d_{14}-\bar d_{15}+\bar d_{18} \ga)\big]}{6 \pi  \Fpi^2 (\mN+\mpi)}\notag\\
 &-\frac{ \ga^2 \mN \mpi\big[231 \pi+16 (7+6 \pi ) \ga^2 \big]}{13824 \pi ^3 \Fpi^4 (\mN+\mpi)}+\frac{\mpi^2 (2 c_1-c_2-c_3)}{8 \pi  \Fpi^2 \mN (\mN+\mpi)}
 +\frac{\mpi^2 \big[3 (\bar d_{14}-\bar d_{15})+2 \bar d_{18} \ga-8 \mN (2 \bar e_{14}+\bar e_{15})\big]}{6 \pi  \Fpi^2 (\mN+\mpi)}\notag\\
 &+\frac{\ga^2 \mpi^2}{192 \pi  \Fpi^2 \mN^2 (\mN+\mpi)}
 +\frac{\mpi^2 \big\{36+(48 \pi -133) \ga^2-2 (37+6 \pi ) \ga^4+6 \mN \big[52 c_1-c_2-32 c_3-32 \pi  \ga^2 (c_3-c_4)\big]\big\}}{6912 \pi ^3 \Fpi^4 (\mN+\mpi)},\notag\\
 a_{1-}^-&=-\frac{\ga^2 \mN}{24 \pi  \Fpi^2 \mpi (\mN+\mpi)}+\frac{3-2 \ga^2+8 c_4 \mN}{48 \pi  \Fpi^2 (\mN+\mpi)}
 +\frac{\mpi \big[3-3 \ga^2+24 c_4 \mN-16 \mN^2 (2 (\bar d_1+\bar d_2)-\bar d_{18} \ga)\big]}{96 \pi  \Fpi^2 \mN (\mN+\mpi)}\notag\\
 &-\frac{ \mN \mpi\big[3+3 (7+12 \pi ) \ga^2+2 (1+12 \pi ) \ga^4\big]}{3456 \pi ^3 \Fpi^4 (\mN+\mpi)}
 +\frac{\mpi^2 \big[\bar d_1+\bar d_2+3 \bar d_3+2 \bar d_5+\bar d_{18} \ga+8 \mN (\bar e_{17}+\bar e_{18})\big]}{6 \pi  \Fpi^2 (\mN+\mpi)}\notag\\
 &-\frac{\ga^2 \mpi^2}{192 \pi  \Fpi^2 \mN^2 (\mN+\mpi)}
 -\frac{\mpi^2 \big\{18+3 (24-11 \pi ) \ga^2+2(2+15 \pi ) \ga^4+16 \ga^2 \mN \big[11 c_4+6 \pi  (c_3-c_4)\big]\big\}}{6912 \pi ^3 \Fpi^4 (\mN+\mpi)},\notag\\
 b_{0+}^+&=\frac{\ga^2+8 \mN (c_2+c_3)}{16 \pi  \Fpi^2 (\mN+\mpi)}+\frac{77 \ga^2 \mN \mpi}{1536 \pi ^2 \Fpi^4 (\mN+\mpi)}+\frac{\mpi \big[3 \ga^2+8 \mN (2 c_1+c_2-c_3)\big]}{32 \pi  \Fpi^2 \mN (\mN+\mpi)}\notag\\
 &+\frac{\mpi^2 \big[3 \ga^2+8 \mN (-2 c_1+5 c_2+c_3)\big]}{64 \pi  \Fpi^2 \mN^2 (\mN+\mpi)}
 -\frac{\mpi^2 \big[16 c_1 c_2+2 (\bar d_{14}-\bar d_{15})+\bar d_{18} \ga-32 \mN (\bar e_{14}+\bar e_{15}+\bar e_{16})\big]}{4 \pi  \Fpi^2 (\mN+\mpi)}\notag\\
 &+\frac{\mpi^2 \big[216-(22+27 \pi ) \ga^2+4 (11-24 \pi ) \ga^4-12 \mN (52 c_1-c_2-32 c_3)\big]}{4608 \pi ^3 \Fpi^4 (\mN+\mpi)},\notag\\
 b_{0+}^-&=\frac{\mN}{16 \pi  \Fpi^2 \mpi (\mN+\mpi)}-\frac{\ga^2}{8 \pi  \Fpi^2 (\mN+\mpi)}
 +\frac{\mpi (2-5 \ga^2-16 c_4 \mN)}{64 \pi  \Fpi^2 \mN (\mN+\mpi)}+\frac{\mN \mpi \big[3 (\bar d_1+\bar d_2+\bar d_3)+2 \bar d_5\big]}{2 \pi  \Fpi^2 (\mN+\mpi)}\notag\\
 &-\frac{\mN \mpi(2-7 \ga^2)}{384 \pi ^3 \Fpi^4 (\mN+\mpi)}
 +\frac{\mpi^2 (2 \bar d_3+\bar d_{18} \ga)}{2 \pi  \Fpi^2 (\mN+\mpi)}-\frac{\ga^2 \mpi^2}{16 \pi  \Fpi^2 \mN^2 (\mN+\mpi)}
 -\frac{\mpi^2\big[36+69 \pi  \ga^2+4(1-3 \pi ) \ga^4\big]}{2304 \pi ^3 \Fpi^4 (\mN+\mpi)},\notag
\end{align}
and in $NN$ counting
\begin{align}
a_{0+}^+&=-\frac{\mpi^2 \big[\ga^2+8 \mN (2 c_1-c_2-c_3)\big]}{16 \pi  \Fpi^2 (\mN+\mpi)}+\frac{3 \ga^2 \mN \mpi^3}{256 \pi ^2 \Fpi^4 (\mN+\mpi)}
+\frac{4\mN\mpi^4 (\bar e_{14}+\bar e_{15}+\bar e_{16})}{\pi  \Fpi^2 (\mN+\mpi)}
 -\frac{\mN\mpi^4 (2 c_1-c_3)}{64 \pi ^3 \Fpi^4 (\mN+\mpi)},\notag\\
 a_{0+}^-&=\frac{\mN \mpi}{8 \pi  \Fpi^2 (\mN+\mpi)}+\frac{\mN \mpi^3}{64 \pi ^3 \Fpi^4 (\mN+\mpi)}+\frac{\mN \mpi^3 (\bar d_1+\bar d_2+\bar d_3+2 \bar d_5)}{\pi  \Fpi^2 (\mN+\mpi)},\notag\\
 a_{1+}^+&=\frac{\ga^2 \mN}{24 \pi  \Fpi^2 \mpi (\mN+\mpi)}+\frac{\ga^2-4 c_3 \mN}{24 \pi  \Fpi^2 (\mN+\mpi)}
 +\frac{\mpi \big[c_2-\mN (\bar d_{14}-\bar d_{15}+\bar d_{18} \ga)\big]}{6 \pi  \Fpi^2 (\mN+\mpi)}\notag\\
 &
 -\frac{ \ga^2 \mN \mpi\big[231 \pi+8 (12 \pi -7) \ga^2 \big]}{13824 \pi ^3 \Fpi^4 (\mN+\mpi)}-\frac{4\mN\mpi^2 (2 \bar e_{14}+\bar e_{15})}{3 \pi  \Fpi^2 (\mN+\mpi)}
 +\frac{\mN\mpi^2\big[52 c_1-c_2-32 c_3+16 \pi  \ga^2 (c_3-c_4)\big]}{1152 \pi ^3 \Fpi^4 (\mN+\mpi)},\notag\\
 a_{1+}^-&=-\frac{\ga^2 \mN}{24 \pi  \Fpi^2 \mpi (\mN+\mpi)}-\frac{\ga^2+2 c_4 \mN}{24 \pi  \Fpi^2 (\mN+\mpi)}
 -\frac{\mN \mpi \big[2 (\bar d_1+\bar d_2)-\bar d_{18} \ga\big]}{6 \pi  \Fpi^2 (\mN+\mpi)}
 \notag\\
  &
 -\frac{\mN \mpi\big[3+3 (7-6 \pi ) \ga^2+2 (1-6 \pi ) \ga^4\big]}{3456 \pi ^3 \Fpi^4 (\mN+\mpi)}
  -\frac{2\mN\mpi^2 (\bar e_{17}+\bar e_{18})}{3 \pi  \Fpi^2 (\mN+\mpi)}
  +\frac{\mN\mpi^2\ga^2\big[11 c_4-12 \pi  (c_3-c_4)\big]}{864 \pi ^3 \Fpi^4 (\mN+\mpi)},\notag\\
 a_{1-}^+&=-\frac{\ga^2 \mN}{12 \pi  \Fpi^2 \mpi (\mN+\mpi)}-\frac{\ga^2+2 c_3 \mN}{12 \pi  \Fpi^2 (\mN+\mpi)}
 +\frac{\mpi \big[c_2+2 \mN (\bar d_{14}-\bar d_{15}+\bar d_{18} \ga)\big]}{6 \pi  \Fpi^2 (\mN+\mpi)}\notag\\
 &
 -\frac{ \ga^2 \mN \mpi\big[231 \pi+16 (7+6 \pi ) \ga^2 \big]}{13824 \pi ^3 \Fpi^4 (\mN+\mpi)}
 -\frac{4\mN\mpi^2 (2 \bar e_{14}+\bar e_{15})}{3 \pi  \Fpi^2 (\mN+\mpi)}
 +\frac{\mN\mpi^2\big[52 c_1-c_2-32 c_3-32 \pi  \ga^2 (c_3-c_4)\big]}{1152 \pi ^3 \Fpi^4 (\mN+\mpi)},\notag\\
 a_{1-}^-&=-\frac{\ga^2 \mN}{24 \pi  \Fpi^2 \mpi (\mN+\mpi)}+\frac{3-2 \ga^2+8 c_4 \mN}{48 \pi  \Fpi^2 (\mN+\mpi)}
 +\frac{\mpi \big[3 c_4-2 \mN (2 (\bar d_1+\bar d_2)-\bar d_{18} \ga)\big]}{12 \pi  \Fpi^2 (\mN+\mpi)}\notag\\
 &
 -\frac{ \mN \mpi\big[3+3 (7+12 \pi ) \ga^2+2 (1+12 \pi ) \ga^4\big]}{3456 \pi ^3 \Fpi^4 (\mN+\mpi)}
 +\frac{4\mN\mpi^2(\bar e_{17}+\bar e_{18})}{3 \pi  \Fpi^2 (\mN+\mpi)}
 -\frac{\mN\mpi^2\ga^2 \big[11 c_4+6 \pi  (c_3-c_4)\big]}{432 \pi ^3 \Fpi^4 (\mN+\mpi)},\notag\\
 b_{0+}^+&=\frac{\ga^2+8 \mN (c_2+c_3)}{16 \pi  \Fpi^2 (\mN+\mpi)}+\frac{77 \ga^2 \mN \mpi}{1536 \pi ^2 \Fpi^4 (\mN+\mpi)}+\frac{\mpi (2 c_1+c_2-c_3)}{4 \pi  \Fpi^2 (\mN+\mpi)}\notag\\
 &
 +\frac{8\mN\mpi^2(\bar e_{14}+\bar e_{15}+\bar e_{16})}{\pi  \Fpi^2 (\mN+\mpi)}
 -\frac{\mN\mpi^2 (52 c_1-c_2-32 c_3)}{384 \pi ^3 \Fpi^4 (\mN+\mpi)},\notag\\
 b_{0+}^-&=\frac{\mN}{16 \pi  \Fpi^2 \mpi (\mN+\mpi)}-\frac{\ga^2}{8 \pi  \Fpi^2 (\mN+\mpi)}\notag\\
 &-\frac{\mpi c_4}{4 \pi  \Fpi^2(\mN+\mpi)}+\frac{\mN \mpi \big[3 (\bar d_1+\bar d_2+\bar d_3)+2 \bar d_5\big]}{2 \pi  \Fpi^2 (\mN+\mpi)}-\frac{\mN \mpi(2-7 \ga^2)}{384 \pi ^3 \Fpi^4 (\mN+\mpi)}.
\end{align}

\end{document}